%% file: main.tex
\documentclass[11pt,a4paper]{report}

\usepackage{epsfig}

\usepackage{nicefrac}

\usepackage{float}

\usepackage{rotating}
\usepackage{ifthen} 
\newboolean{pdflatex}
\setboolean{pdflatex}{true} 

\newboolean{articletitles}
\setboolean{articletitles}{true} 

\usepackage{bm}
\usepackage{caption}
\newboolean{uprightparticles}
\setboolean{uprightparticles}{false} 
\newboolean{inbibliography}
\setboolean{inbibliography}{false} 

\usepackage{multirow}

\usepackage[utf8]{inputenc}

\def\papercopyright{\the\year\ CERN for the benefit of the LHCb collaboration} 
\def\paperlicence{CC-BY-4.0 licence}
\def\paperlicenceurl{https://creativecommons.org/licenses/by/4.0/}
\def\paperasciititle{Physics case for an LHCb Upgrade II} 
\def\paperauthors{LHCb collaboration} 
\def\paperkeywords{{High Energy Physics}, {LHCb}, {HL-LHC}} 

\input{preamble}

\graphicspath{{./CONTRIBUTIONS/1_Executive_Summary/figs/}
              {./CONTRIBUTIONS/2_Introduction/figs/}
              {./CONTRIBUTIONS/3_Time_dependent_CP_violation_measurements/figs/}
              {./CONTRIBUTIONS/4_Time_integrated_CP_violation_measurements/figs/}
              {./CONTRIBUTIONS/5_Measurements_of_unitarity_triangle_sides_and_semileptonic_decays/figs/}
              {./CONTRIBUTIONS/6_CP_violation_and_mixing_in_charm/figs/}
              {./CONTRIBUTIONS/7_Rare_and_radiative_decays/figs/}
              {./CONTRIBUTIONS/8_Forward_and_high_pT_physics/figs/}
              {./CONTRIBUTIONS/9_Exotic_hadrons_and_spectroscopy_with_heavy_flavours/figs/}
              {./CONTRIBUTIONS/10_Further_opportunities/figs/}
              {./CONTRIBUTIONS/11_Summary_and_conclusions/figs/}}

\begin{document}

\input{titlepage}

\newpage
\input{LHCb_authorlist}

\newpage
\addtocontents{toc}{\protect\enlargethispage{\baselineskip}}
\tableofcontents
\cleardoublepage
\onecolumn
\linewidth=\columnwidth
\def\LHCb{LHCb}
\def\GeVc{GeV$/c$}
\def\GeVcc{GeV$/c^2$}
\def\MeVcc{MeV$/c^2$}
\def\microns{$\mu$m}
\def\BJLK{$\rm B^0_d \rightarrow J/\psi K_S $}
\def\BJELK{$\rm B^0_d \rightarrow J/\psi(e^+ e^-) K_S $}
\def\BJMLK{$\rm B^0_d \rightarrow J/\psi(\mu^+ \mu^-) K_S $}
\def\BJpsiKs{$\rm B^0_d \rightarrow J/\psi K_S$}
\def\Bpipi{$\rm B^0_d\rightarrow\pi^+\pi^-$}
\def\BKpi{$\rm B^0_d\rightarrow K^+\pi^-$}
\def\Kpinunu{$\rm K\rightarrow \pi\nu\overline{\nu}$}
\def\Brhoppim{$\rm B^0_d\rightarrow \rho^+\pi^-$}
\def\Brhompip{$\rm B^0_d\rightarrow \rho^-\pi^+$}
\def\Brhopi{$\rm B^0_d\rightarrow \rho\pi \: $}
\def\Brho0pi0{$\rm B^0_d\rightarrow \rho^0\pi^0$}
\def\BDKs{$\rm B^0_d\rightarrow \overline{D}{}^0 K^{*0}$}
\def\BDstarpi{$\rm  B^0_d\rightarrow D{}^{*-} \pi^+$}
\def\BsKK{$\rm B^0_s\rightarrow K^+K^-$}
\def\BsKpi{$\rm B^0_s\rightarrow K^-\pi^+$}
\def\BsDspi{$\rm B^0_s \rightarrow D_s^- \pi^+$}
\def\BsDsK{$\rm B^0_s\rightarrow D_s^+K^-$}
\def\BsDs3pi{$\rm B^0_s\rightarrow D_s^+\pi^+\pi^-\pi^-$}
\def\BsJpsiphi{$\rm B^0_s \rightarrow J/\psi \phi$}
\def\BsKsgamma{$\rm B^0_s\rightarrow K^{*0} \gamma$}
\def\BdKsgamma{$\rm B^0_d\rightarrow K^{*0} \gamma$}
\def\bjpsiee{$\rm 
     B^0\rightarrow J/\psi \rightarrow e^+ e^-$}
\def\bjpsimumu{$\rm 
     B^0\rightarrow J/\psi \rightarrow \mu^+ \mu^-$}
\def\ptran{$p_{\rm T}$}
\def\etran{$E_{\rm T}$}
\def\C4F10{$\rm C_4F_{10}$}
\def\CF4{$\rm CF_4$}

\setcounter{page}{0}
\pagenumbering{arabic}
%
%
%
%
\chapter{Executive summary}

\input{CONTRIBUTIONS/1_Executive_Summary/1.tex}

\chapter{Introduction}
\input{CONTRIBUTIONS/2_Introduction/2.tex}

\chapter{Time-dependent \boldmath $\CP$-violation measurements}
\input{CONTRIBUTIONS/3_Time_dependent_CP_violation_measurements/3.tex}

\chapter{Time-integrated \boldmath $\CP$-violation measurements}
\input{CONTRIBUTIONS/4_Time_integrated_CP_violation_measurements/4.tex}

\chapter{Measurements of unitarity triangle sides and semileptonic decays}
\input{CONTRIBUTIONS/5_Measurements_of_unitarity_triangle_sides_and_semileptonic_decays/5.tex}

\chapter{Mixing and \boldmath $\CP$ violation in charm}
\input{CONTRIBUTIONS/6_CP_violation_and_mixing_in_charm/6.tex}

\chapter{Rare decays}
\input{CONTRIBUTIONS/7_Rare_and_radiative_decays/7.tex}

\chapter{Forward and high-\boldmath\pt physics}
\input{CONTRIBUTIONS/8_Forward_and_high_pT_physics/8.tex}

\chapter{Exotic hadrons and spectroscopy with heavy flavours}
\input{CONTRIBUTIONS/9_Exotic_hadrons_and_spectroscopy_with_heavy_flavours/9.tex}


\chapter{Summary and conclusions}
\input{CONTRIBUTIONS/11_Summary_and_conclusions/11.tex}

\newpage 
\section*{Acknowledgements}
\noindent We express our gratitude to our colleagues in the CERN
accelerator departments for the excellent performance of the LHC and
for their studies on the feasibility of achieving the luminosity for \upgradetwo at the LHCb
Interaction point. We thank all the members of the theoretical and
experimental physics communities that have attended the LHCb
\upgradetwo workshops or otherwise contributed to this activity.
We
thank the technical and administrative staff at the LHCb
institutes.
We acknowledge support from CERN and from the national agencies:
CAPES, CNPq, FAPERJ and FINEP (Brazil); 
MOST and NSFC (China); 
CNRS/IN2P3 (France); 
BMBF, DFG and MPG (Germany); 
INFN (Italy); 
NWO (Netherlands); 
MNiSW and NCN (Poland); 
MEN/IFA (Romania); 
MSHE (Russia); 
MinECo (Spain); 
SNSF and SER (Switzerland); 
NASU (Ukraine); 
STFC (United Kingdom); 
NSF (USA).
We acknowledge the computing resources that are provided by CERN, IN2P3
(France), KIT and DESY (Germany), INFN (Italy), SURF (Netherlands),
PIC (Spain), GridPP (United Kingdom), RRCKI and Yandex
LLC (Russia), CSCS (Switzerland), IFIN-HH (Romania), CBPF (Brazil),
PL-GRID (Poland) and OSC (USA).
We are indebted to the communities behind the multiple open-source
software packages on which we depend.
Individual groups or members have received support from
AvH Foundation (Germany);
EPLANET, Marie Sk\l{}odowska-Curie Actions and ERC (European Union);
ANR, Labex P2IO and OCEVU, and R\'{e}gion Auvergne-Rh\^{o}ne-Alpes (France);
Key Research Program of Frontier Sciences of CAS, CAS PIFI, and the Thousand Talents Program (China);
RFBR, RSF and Yandex LLC (Russia);
GVA, XuntaGal and GENCAT (Spain);
the Royal Society
and the Leverhulme Trust (United Kingdom);
Laboratory Directed Research and Development program of LANL (USA).

\appendix
\chapter{Further opportunities}
\input{CONTRIBUTIONS/10_Further_opportunities/10.tex}

\input{CONTRIBUTIONS/11_Summary_and_conclusions/wilson.tex}

\addcontentsline{toc}{chapter}{References}
\setboolean{inbibliography}{true}
\bibliographystyle{LHCb}
\mciteErrorOnUnknownfalse
\bibliography{LHCb-PAPER,LHCb-CONF,LHCb-DP,LHCb-TDR,standard,UpgradeII_Physics_LHCC,CONTRIBUTIONS/1_Executive_Summary/1,CONTRIBUTIONS/2_Introduction/2,CONTRIBUTIONS/3_Time_dependent_CP_violation_measurements/3,CONTRIBUTIONS/4_Time_integrated_CP_violation_measurements/4,CONTRIBUTIONS/5_Measurements_of_unitarity_triangle_sides_and_semileptonic_decays/5,CONTRIBUTIONS/6_CP_violation_and_mixing_in_charm/6,CONTRIBUTIONS/7_Rare_and_radiative_decays/7,CONTRIBUTIONS/8_Forward_and_high_pT_physics/8,CONTRIBUTIONS/9_Exotic_hadrons_and_spectroscopy_with_heavy_flavours/9,CONTRIBUTIONS/10_Further_opportunities/10,CONTRIBUTIONS/11_Summary_and_conclusions/11}



\end{document}

%% file: preamble.tex
\usepackage[top=1in, bottom=1.25in, left=1in, right=1in]{geometry}

\usepackage{microtype}
\usepackage{lineno}  
\usepackage{xspace} 

\usepackage{graphicx}  
\usepackage{overpic}
\usepackage{color}
\usepackage{colortbl}
\graphicspath{{./figs/}} 

\usepackage{amsmath} 
\usepackage{amssymb}
\usepackage{amsfonts}
\usepackage{upgreek} 

\newcommand*\patchAmsMathEnvironmentForLineno[1]{%
\expandafter\let\csname old#1\expandafter\endcsname\csname #1\endcsname
\expandafter\let\csname oldend#1\expandafter\endcsname\csname
end#1\endcsname
 \renewenvironment{#1}%
   {\linenomath\csname old#1\endcsname}%
   {\csname oldend#1\endcsname\endlinenomath}%
}
\newcommand*\patchBothAmsMathEnvironmentsForLineno[1]{%
  \patchAmsMathEnvironmentForLineno{#1}%
  \patchAmsMathEnvironmentForLineno{#1*}%
}
\AtBeginDocument{%
\patchBothAmsMathEnvironmentsForLineno{equation}%
\patchBothAmsMathEnvironmentsForLineno{align}%
\patchBothAmsMathEnvironmentsForLineno{flalign}%
\patchBothAmsMathEnvironmentsForLineno{alignat}%
\patchBothAmsMathEnvironmentsForLineno{gather}%
\patchBothAmsMathEnvironmentsForLineno{multline}%
\patchBothAmsMathEnvironmentsForLineno{eqnarray}%
}

\usepackage{hyperxmp}
\usepackage[pdftex,
            pdfauthor={\paperauthors},
            pdftitle={\paperasciititle},
            pdfkeywords={\paperkeywords},
            pdfcopyright={Copyright (C) \papercopyright},
            pdflicenseurl={\paperlicenceurl},
            plainpages=false,
            pdfpagelabels]{hyperref}
\usepackage[all]{hypcap} 

\input{lhcb-symbols-def} 
\input{additional-symbols-def} 

\usepackage{cite} 
\usepackage{mciteplus}

\usepackage{titlesec}
\titleformat{\section}[block]{\Large\bfseries\boldmath}{\thesection}{0.5em}{}
\titleformat{\subsection}[block]{\large\bfseries\boldmath}{\thesubsection}{0.5em}{}
\titleformat{\subsubsection}[block]{\normalsize\bfseries\boldmath}{\thesubsubsection}{0.5em}{}
\titleformat{\chapter}[block]{\Huge\bfseries\boldmath}{\thechapter}{0.5em}{}

%% file: lhcb-symbols-def.tex
\usepackage{xspace} 
\usepackage{upgreek}

\def\lhcb   {\mbox{LHCb}\xspace}

\def\babar  {\mbox{BaBar}\xspace}
\def\belle  {\mbox{Belle}\xspace}
\def\belletwo {\mbox{Belle~II}\xspace}

\def\lhc    {\mbox{LHC}\xspace}

\def\tevatron {Tevatron\xspace}
\def\bfactories {\mbox{\B Factories}\xspace}

\def\upgradeone {\mbox{Upgrade~I}\xspace}
\def\upgradetwo {\mbox{Upgrade~II}\xspace}

\def\ecal   {ECAL\xspace}

\def\MagUp {\mbox{\em Mag\kern -0.05em Up}\xspace}

\ifthenelse{\boolean{uprightparticles}}%
{
 \def\Pbeta       {\ensuremath{\upbeta}\xspace}
 \def\Pgamma      {\ensuremath{\upgamma}\xspace}

 \def\Peta        {\ensuremath{\upeta}\xspace}

 \def\Pmu         {\ensuremath{\upmu}\xspace}                 
 \def\Pnu         {\ensuremath{\upnu}\xspace}                 
                  
 \def\Ppi         {\ensuremath{\uppi}\xspace}                 
                  
 \def\Prho        {\ensuremath{\uprho}\xspace}                 
                  
 \def\Ptau        {\ensuremath{\uptau}\xspace}                 
                  
 \def\Pphi        {\ensuremath{\upphi}\xspace}                 
                  
 \def\Pchi        {\ensuremath{\upchi}\xspace}                 
 \def\Ppsi        {\ensuremath{\uppsi}\xspace}                 
 \def\Pomega      {\ensuremath{\upomega}\xspace}                 

 \def\PDelta      {\ensuremath{\Delta}\xspace}                 
 \def\PXi         {\ensuremath{\Xi}\xspace}                 
 \def\PLambda     {\ensuremath{\Lambda}\xspace}                 
 \def\PSigma      {\ensuremath{\Sigma}\xspace}                 
 \def\POmega      {\ensuremath{\Omega}\xspace}                 
 \def\PUpsilon    {\ensuremath{\Upsilon}\xspace}

 \def\PB      {\ensuremath{\mathrm{B}}\xspace}                 
                  
 \def\PD      {\ensuremath{\mathrm{D}}\xspace}

 \def\PJ      {\ensuremath{\mathrm{J}}\xspace}                 
 \def\PK      {\ensuremath{\mathrm{K}}\xspace}

 \def\PP      {\ensuremath{\mathrm{P}}\xspace}                 
 \def\PQ      {\ensuremath{\mathrm{Q}}\xspace}

 \def\PW      {\ensuremath{\mathrm{W}}\xspace}                 
 \def\PX      {\ensuremath{\mathrm{X}}\xspace}                 
                  
 \def\PZ      {\ensuremath{\mathrm{Z}}\xspace}                 
                  
 \def\Pb      {\ensuremath{\mathrm{b}}\xspace}                 
 \def\Pc      {\ensuremath{\mathrm{c}}\xspace}                 
 \def\Pd      {\ensuremath{\mathrm{d}}\xspace}                 
 \def\Pe      {\ensuremath{\mathrm{e}}\xspace}

 \def\Ph      {\ensuremath{\mathrm{h}}\xspace}                 
 \def\Pi      {\ensuremath{\mathrm{i}}\xspace}

 \def\Pn      {\ensuremath{\mathrm{n}}\xspace}                 
                  
 \def\Pp      {\ensuremath{\mathrm{p}}\xspace}                 
 \def\Pq      {\ensuremath{\mathrm{q}}\xspace}                 
                  
 \def\Ps      {\ensuremath{\mathrm{s}}\xspace}                 
 \def\Pt      {\ensuremath{\mathrm{t}}\xspace}                 
 \def\Pu      {\ensuremath{\mathrm{u}}\xspace}

}
{
 \def\Pbeta       {\ensuremath{\beta}\xspace}
 \def\Pgamma      {\ensuremath{\gamma}\xspace}

 \def\Peta        {\ensuremath{\eta}\xspace}

 \def\Pmu         {\ensuremath{\mu}\xspace}                 
 \def\Pnu         {\ensuremath{\nu}\xspace}                 
                  
 \def\Ppi         {\ensuremath{\pi}\xspace}                 
                  
 \def\Prho        {\ensuremath{\rho}\xspace}                 
                  
 \def\Ptau        {\ensuremath{\tau}\xspace}                 
                  
 \def\Pphi        {\ensuremath{\phi}\xspace}                 
                  
 \def\Pchi        {\ensuremath{\chi}\xspace}                 
 \def\Ppsi        {\ensuremath{\psi}\xspace}                 
 \def\Pomega      {\ensuremath{\omega}\xspace}                 
 \mathchardef\PDelta="7101
 \mathchardef\PXi="7104
 \mathchardef\PLambda="7103
 \mathchardef\PSigma="7106
 \mathchardef\POmega="710A
 \mathchardef\PUpsilon="7107
                  
 \def\PB      {\ensuremath{B}\xspace}                 
                  
 \def\PD      {\ensuremath{D}\xspace}

 \def\PJ      {\ensuremath{J}\xspace}                 
 \def\PK      {\ensuremath{K}\xspace}

 \def\PP      {\ensuremath{P}\xspace}                 
 \def\PQ      {\ensuremath{Q}\xspace}

 \def\PW      {\ensuremath{W}\xspace}                 
 \def\PX      {\ensuremath{X}\xspace}                 
                  
 \def\PZ      {\ensuremath{Z}\xspace}                 
                  
 \def\Pb      {\ensuremath{b}\xspace}                 
 \def\Pc      {\ensuremath{c}\xspace}                 
 \def\Pd      {\ensuremath{d}\xspace}                 
 \def\Pe      {\ensuremath{e}\xspace}

 \def\Ph      {\ensuremath{h}\xspace}                 
 \def\Pi      {\ensuremath{i}\xspace}

 \def\Pn      {\ensuremath{n}\xspace}                 
                  
 \def\Pp      {\ensuremath{p}\xspace}                 
 \def\Pq      {\ensuremath{q}\xspace}                 
                  
 \def\Ps      {\ensuremath{s}\xspace}                 
 \def\Pt      {\ensuremath{t}\xspace}                 
 \def\Pu      {\ensuremath{u}\xspace}

}

\makeatletter
\ifcase \@ptsize \relax
  \newcommand{\miniscule}{\@setfontsize\miniscule{4}{5}}
\or
  \newcommand{\miniscule}{\@setfontsize\miniscule{5}{6}}
\or
  \newcommand{\miniscule}{\@setfontsize\miniscule{5}{6}}
\fi
\makeatother

\DeclareRobustCommand{\optbar}[1]{\shortstack{{\miniscule (\rule[.5ex]{1.25em}{.18mm})}
  \\ [-.7ex] $#1$}}

\def\en         {{\ensuremath{\Pe^-}}\xspace}   
\def\ep         {{\ensuremath{\Pe^+}}\xspace}
\def\epm        {{\ensuremath{\Pe^\pm}}\xspace} 
\def\emp        {{\ensuremath{\Pe^\mp}}\xspace} 
\def\epem       {{\ensuremath{\Pe^+\Pe^-}}\xspace}

\def\mup        {{\ensuremath{\Pmu^+}}\xspace}
\def\mun        {{\ensuremath{\Pmu^-}}\xspace} 
\def\mupm       {{\ensuremath{\Pmu^\pm}}\xspace} 
\def\mump       {{\ensuremath{\Pmu^\mp}}\xspace} 
\def\mumu       {{\ensuremath{\Pmu^+\Pmu^-}}\xspace}

\def\taup       {{\ensuremath{\Ptau^+}}\xspace}
\def\taum       {{\ensuremath{\Ptau^-}}\xspace}
\def\taupm      {{\ensuremath{\Ptau^\pm}}\xspace}
\def\taump      {{\ensuremath{\Ptau^\mp}}\xspace}
\def\tautau     {{\ensuremath{\Ptau^+\Ptau^-}}\xspace}

\def\ellm       {{\ensuremath{\ell^-}}\xspace}
\def\ellp       {{\ensuremath{\ell^+}}\xspace}
\def\ellell     {\ensuremath{\ell^+ \ell^-}\xspace}

\def\neu        {{\ensuremath{\Pnu}}\xspace}
\def\neub       {{\ensuremath{\overline{\Pnu}}}\xspace}
\def\neum       {{\ensuremath{\neu_\mu}}\xspace}
\def\neumb      {{\ensuremath{\neub_\mu}}\xspace}

\def\neulb      {{\ensuremath{\neub_\ell}}\xspace}

\def\g      {{\ensuremath{\Pgamma}}\xspace}

\def\Z      {{\ensuremath{\PZ}}\xspace}

\def\quark     {{\ensuremath{\Pq}}\xspace}
\def\quarkbar  {{\ensuremath{\overline \quark}}\xspace}
\def\qqbar     {{\ensuremath{\quark\quarkbar}}\xspace}
\def\uquark    {{\ensuremath{\Pu}}\xspace}
\def\uquarkbar {{\ensuremath{\overline \uquark}}\xspace}

\def\dquark    {{\ensuremath{\Pd}}\xspace}
\def\dquarkbar {{\ensuremath{\overline \dquark}}\xspace}

\def\squark    {{\ensuremath{\Ps}}\xspace}
\def\squarkbar {{\ensuremath{\overline \squark}}\xspace}
\def\ssbar     {{\ensuremath{\squark\squarkbar}}\xspace}
\def\cquark    {{\ensuremath{\Pc}}\xspace}
\def\cquarkbar {{\ensuremath{\overline \cquark}}\xspace}
\def\ccbar     {{\ensuremath{\cquark\cquarkbar}}\xspace}
\def\bquark    {{\ensuremath{\Pb}}\xspace}
\def\bquarkbar {{\ensuremath{\overline \bquark}}\xspace}
\def\bbbar     {{\ensuremath{\bquark\bquarkbar}}\xspace}
\def\tquark    {{\ensuremath{\Pt}}\xspace}

\def\hadron {{\ensuremath{\Ph}}\xspace}
\def\pion   {{\ensuremath{\Ppi}}\xspace}
\def\piz    {{\ensuremath{\pion^0}}\xspace}
\def\pip    {{\ensuremath{\pion^+}}\xspace}
\def\pim    {{\ensuremath{\pion^-}}\xspace}
\def\pipm   {{\ensuremath{\pion^\pm}}\xspace}
\def\pimp   {{\ensuremath{\pion^\mp}}\xspace}

\def\rhomeson {{\ensuremath{\Prho}}\xspace}
\def\rhoz     {{\ensuremath{\rhomeson^0}}\xspace}
\def\rhop     {{\ensuremath{\rhomeson^+}}\xspace}
\def\rhom     {{\ensuremath{\rhomeson^-}}\xspace}
\def\rhopm    {{\ensuremath{\rhomeson^\pm}}\xspace}

\def\kaon    {{\ensuremath{\PK}}\xspace}
  \def\Kbar    {{\kern 0.2em\overline{\kern -0.2em \PK}{}}\xspace}

\def\KorKbar    {\kern 0.18em\optbar{\kern -0.18em K}{}\xspace}
\def\Kz      {{\ensuremath{\kaon^0}}\xspace}
\def\Kzb     {{\ensuremath{\Kbar{}^0}}\xspace}
\def\Kp      {{\ensuremath{\kaon^+}}\xspace}
\def\Km      {{\ensuremath{\kaon^-}}\xspace}
\def\Kpm     {{\ensuremath{\kaon^\pm}}\xspace}
\def\Kmp     {{\ensuremath{\kaon^\mp}}\xspace}
\def\KS      {{\ensuremath{\kaon^0_{\mathrm{ \scriptscriptstyle S}}}}\xspace}
\def\KL      {{\ensuremath{\kaon^0_{\mathrm{ \scriptscriptstyle L}}}}\xspace}
\def\Kstarz  {{\ensuremath{\kaon^{*0}}}\xspace}
\def\Kstarzb {{\ensuremath{\Kbar{}^{*0}}}\xspace}
\def\Kstar   {{\ensuremath{\kaon^*}}\xspace}
\def\Kstarb  {{\ensuremath{\Kbar{}^*}}\xspace}
\def\Kstarp  {{\ensuremath{\kaon^{*+}}}\xspace}

\def\Kstarpm {{\ensuremath{\kaon^{*\pm}}}\xspace}

\newcommand{\etaz}{\ensuremath{\Peta}\xspace}
\newcommand{\etapr}{\ensuremath{\Peta^{\prime}}\xspace}

\newcommand{\omegaz}{\ensuremath{\Pomega}\xspace}

  \def\Dbar    {{\kern 0.2em\overline{\kern -0.2em \PD}{}}\xspace}
\def\D       {{\ensuremath{\PD}}\xspace}
\def\Db      {{\ensuremath{\Dbar}}\xspace}
\def\DorDbar    {\kern 0.18em\optbar{\kern -0.18em D}{}\xspace}
\def\Dz      {{\ensuremath{\D^0}}\xspace}
\def\Dzb     {{\ensuremath{\Dbar{}^0}}\xspace}
\def\Dp      {{\ensuremath{\D^+}}\xspace}
\def\Dm      {{\ensuremath{\D^-}}\xspace}

\def\Dmp     {{\ensuremath{\D^\mp}}\xspace}
\def\Dstar   {{\ensuremath{\D^*}}\xspace}
\def\Dstarb  {{\ensuremath{\Dbar{}^*}}\xspace}
\def\Dstarz  {{\ensuremath{\D^{*0}}}\xspace}
\def\Dstarzb {{\ensuremath{\Dbar{}^{*0}}}\xspace}

\def\Dstarp  {{\ensuremath{\D^{*+}}}\xspace}
\def\Dstarm  {{\ensuremath{\D^{*-}}}\xspace}
\def\Dstarpm {{\ensuremath{\D^{*\pm}}}\xspace}

\def\Ds      {{\ensuremath{\D^+_\squark}}\xspace}
\def\Dsp     {{\ensuremath{\D^+_\squark}}\xspace}
\def\Dsm     {{\ensuremath{\D^-_\squark}}\xspace}
\def\Dspm    {{\ensuremath{\D^{\pm}_\squark}}\xspace}
\def\Dsmp    {{\ensuremath{\D^{\mp}_\squark}}\xspace}

\def\Dssp    {{\ensuremath{\D^{*+}_\squark}}\xspace}

\def\B       {{\ensuremath{\PB}}\xspace}
\def\Bbar    {{\ensuremath{\kern 0.18em\overline{\kern -0.18em \PB}{}}}\xspace}
\def\Bb      {{\ensuremath{\Bbar}}\xspace}
\def\BorBbar    {\kern 0.18em\optbar{\kern -0.18em B}{}\xspace}
\def\Bz      {{\ensuremath{\B^0}}\xspace}
\def\Bzb     {{\ensuremath{\Bbar{}^0}}\xspace}
\def\Bu      {{\ensuremath{\B^+}}\xspace}
\def\Bub     {{\ensuremath{\B^-}}\xspace}
\def\Bp      {{\ensuremath{\Bu}}\xspace}
\def\Bm      {{\ensuremath{\Bub}}\xspace}
\def\Bpm     {{\ensuremath{\B^\pm}}\xspace}

\def\Bd      {{\ensuremath{\B^0}}\xspace}
\def\Bs      {{\ensuremath{\B^0_\squark}}\xspace}
\def\Bsb     {{\ensuremath{\Bbar{}^0_\squark}}\xspace}
\def\BdorBs  {{\ensuremath{\B^0_{(\squark)}}}\xspace}
\def\Bdb     {{\ensuremath{\Bbar{}^0}}\xspace}
\def\Bc      {{\ensuremath{\B_\cquark^+}}\xspace}
\def\Bcp     {{\ensuremath{\B_\cquark^+}}\xspace}

\def\Bcpm    {{\ensuremath{\B_\cquark^\pm}}\xspace}
\def\Bds     {{\ensuremath{\B_{(\squark)}^0}}\xspace}
\def\Bdsb    {{\ensuremath{\Bbar{}_{(\squark)}^0}}\xspace}

\def\jpsi     {{\ensuremath{{\PJ\mskip -3mu/\mskip -2mu\Ppsi\mskip 2mu}}}\xspace}
\def\psitwos  {{\ensuremath{\Ppsi{(2S)}}}\xspace}

\def\etac     {{\ensuremath{\Peta_\cquark}}\xspace}
\def\chiczero {{\ensuremath{\Pchi_{\cquark 0}}}\xspace}
\def\chicone  {{\ensuremath{\Pchi_{\cquark 1}}}\xspace}

\def\Y#1S{\ensuremath{\PUpsilon{(#1S)}}\xspace}
\def\OneS  {{\Y1S}}

\def\FourS {{\Y4S}}

\def\chic  {{\ensuremath{\Pchi_{c}}}\xspace}

\def\proton      {{\ensuremath{\Pp}}\xspace}
\def\antiproton  {{\ensuremath{\overline \proton}}\xspace}
\def\neutron     {{\ensuremath{\Pn}}\xspace}

\def\Xires       {{\ensuremath{\PXi}}\xspace}
\def\Xiresbar    {{\ensuremath{\overline \Xires}}\xspace}
\def\Lz          {{\ensuremath{\PLambda}}\xspace}
\def\Lbar        {{\ensuremath{\kern 0.1em\overline{\kern -0.1em\PLambda}}}\xspace}
\def\LorLbar     {\kern 0.18em\optbar{\kern -0.18em \PLambda}{}\xspace}
\def\Lambdares   {{\ensuremath{\PLambda}}\xspace}

\def\Sigmares    {{\ensuremath{\PSigma}}\xspace}

\def\Omegares    {{\ensuremath{\POmega}}\xspace}
\def\Omegaresbar {{\ensuremath{\overline \POmega}}\xspace}

\def\Lb      {{\ensuremath{\Lz^0_\bquark}}\xspace}

\def\Lc      {{\ensuremath{\Lz^+_\cquark}}\xspace}
\def\Lcbar   {{\ensuremath{\Lbar{}^-_\cquark}}\xspace}
\def\Xib     {{\ensuremath{\Xires_\bquark}}\xspace}
\def\Xibz    {{\ensuremath{\Xires^0_\bquark}}\xspace}
\def\Xibm    {{\ensuremath{\Xires^-_\bquark}}\xspace}

\def\Xibbarp {{\ensuremath{\Xiresbar{}_\bquark^+}}\xspace}

\def\Xicz    {{\ensuremath{\Xires^0_\cquark}}\xspace}
\def\Xicp    {{\ensuremath{\Xires^+_\cquark}}\xspace}

\def\Omegac    {{\ensuremath{\Omegares^0_\cquark}}\xspace}

\def\Omegab    {{\ensuremath{\Omegares^-_\bquark}}\xspace}
\def\Omegabbar {{\ensuremath{\Omegaresbar{}_\bquark^+}}\xspace}
\def\Xicc      {{\ensuremath{\Xires_{\cquark\cquark}}}\xspace}

\def\Xiccp     {{\ensuremath{\Xires^+_{\cquark\cquark}}}\xspace}
\def\Xiccpp    {{\ensuremath{\Xires^{++}_{\cquark\cquark}}}\xspace}

\def\Omegacc     {{\ensuremath{\Omegares^+_{\cquark\cquark}}}\xspace}

\def\BF         {{\ensuremath{\mathcal{B}}}\xspace}

\def\BR         {\BF}
\newcommand{\decay}[2]{\mbox{\ensuremath{#1\!\to #2}}\xspace}         
\def\ra                 {\ensuremath{\rightarrow}\xspace}
\def\to                 {\ensuremath{\rightarrow}\xspace}

\def\grpsuthree {{\ensuremath{\mathrm{SU}(3)}}\xspace}

\def\ssqtwef {{\ensuremath{{\sin}^{2}\theta_{\mathrm{W}}^{\mathrm{eff}}}}\xspace}

\def\order   {{\ensuremath{\mathcal{O}}}\xspace}

\newcommand{\as}{{\ensuremath{\alpha_s}}\xspace}

\def\qsq       {{\ensuremath{q^2}}\xspace}

\def\CP                {{\ensuremath{C\!P}}\xspace}
\def\CPT               {{\ensuremath{C\!PT}}\xspace}

\def\Vtd  {{\ensuremath{V_{\tquark\dquark}}}\xspace}
\def\Vus  {{\ensuremath{V_{\uquark\squark}}}\xspace}

\def\Vts  {{\ensuremath{V_{\tquark\squark}}}\xspace}
\def\Vub  {{\ensuremath{V_{\uquark\bquark}}}\xspace}
\def\Vcb  {{\ensuremath{V_{\cquark\bquark}}}\xspace}

\newcommand{\DG}{{\ensuremath{\Delta\Gamma}}\xspace}
\newcommand{\DGs}{{\ensuremath{\Delta\Gamma_{\squark}}}\xspace}

\newcommand{\Gs}{{\ensuremath{\Gamma_{\squark}}}\xspace}

\newcommand{\phid}{{\ensuremath{\phi_{\dquark}}}\xspace}

\newcommand{\phis}{{\ensuremath{\phi_{\squark}}}\xspace}
\newcommand{\betas}{{\ensuremath{\beta_{\squark}}}\xspace}

\def\BTohh        {\decay{\B}{\Ph^+ \Ph'^-}}
\def\BdTopipi     {\decay{\Bd}{\pip\pim}}

\def\BsToKK       {\decay{\Bs}{\Kp\Km}}

\def\Cpipi        {\ensuremath{C_{\pip\pim}}\xspace}
\def\Spipi        {\ensuremath{S_{\pip\pim}}\xspace}
\def\CKK          {\ensuremath{C_{\Kp\Km}}\xspace}
\def\SKK          {\ensuremath{S_{\Kp\Km}}\xspace}
\def\ADGKK        {\ensuremath{A^{\DG}_{\Kp\Km}}\xspace}

\def\FL       {\ensuremath{F_{\mathrm{L}}}\xspace}
\def\AT#1     {\ensuremath{A_{\mathrm{T}}^{#1}}\xspace}           

\def\Bsmm     {\decay{\Bs}{\mup\mun}}
\def\Bdmm     {\decay{\Bd}{\mup\mun}}

\def\C#1      {\ensuremath{\mathcal{C}_{#1}}\xspace}                       
\def\Cp#1     {\ensuremath{\mathcal{C}_{#1}^{'}}\xspace}                    
\def\Ceff#1   {\ensuremath{\mathcal{C}_{#1}^{\mathrm{(eff)}}}\xspace}        
\def\Cpeff#1  {\ensuremath{\mathcal{C}_{#1}^{'\mathrm{(eff)}}}\xspace}       
\def\Ope#1    {\ensuremath{\mathcal{O}_{#1}}\xspace}                       
\def\Opep#1   {\ensuremath{\mathcal{O}_{#1}^{'}}\xspace}                    

\def\agamma     {\ensuremath{A_{\Gamma}}\xspace}

\newcommand{\ket}[1]{\ensuremath{|#1\rangle}}              

\newcommand{\tev}{\ifthenelse{\boolean{inbibliography}}{\ensuremath{~T\kern -0.05em eV}}{\ensuremath{\mathrm{\,Te\kern -0.1em V}}}\xspace}
\newcommand{\gev}{\ensuremath{\mathrm{\,Ge\kern -0.1em V}}\xspace}
\newcommand{\mev}{\ensuremath{\mathrm{\,Me\kern -0.1em V}}\xspace}
\newcommand{\kev}{\ensuremath{\mathrm{\,ke\kern -0.1em V}}\xspace}
\newcommand{\ev}{\ensuremath{\mathrm{\,e\kern -0.1em V}}\xspace}
\newcommand{\gevc}{\ensuremath{{\mathrm{\,Ge\kern -0.1em V\!/}c}}\xspace}
\newcommand{\mevc}{\ensuremath{{\mathrm{\,Me\kern -0.1em V\!/}c}}\xspace}
\newcommand{\gevcc}{\ensuremath{{\mathrm{\,Ge\kern -0.1em V\!/}c^2}}\xspace}
\newcommand{\gevgevcccc}{\ensuremath{{\mathrm{\,Ge\kern -0.1em V^2\!/}c^4}}\xspace}
\newcommand{\mevcc}{\ensuremath{{\mathrm{\,Me\kern -0.1em V\!/}c^2}}\xspace}

\def\ma   {\ensuremath{{\mathrm{ \,m}}^2}\xspace}
\def\cm   {\ensuremath{\mathrm{ \,cm}}\xspace}

\def\mm   {\ensuremath{\mathrm{ \,mm}}\xspace}

\def\mum  {\ensuremath{{\,\upmu\mathrm{m}}}\xspace}

\def\nb {\ensuremath{\mathrm{ \,nb}}\xspace}
\def\invnb {\ensuremath{\mbox{\,nb}^{-1}}\xspace}

\def\invpb {\ensuremath{\mbox{\,pb}^{-1}}\xspace}
\def\fb   {\ensuremath{\mbox{\,fb}}\xspace}
\def\invfb   {\ensuremath{\mbox{\,fb}^{-1}}\xspace}

\def\invab   {\ensuremath{\mbox{\,ab}^{-1}}\xspace}

\def\sec  {\ensuremath{\mathrm{{\,s}}}\xspace}

\def\ps   {\ensuremath{{\mathrm{ \,ps}}}\xspace}
\def\fs   {\ensuremath{\mathrm{ \,fs}}\xspace}

\def\mhz  {\ensuremath{{\mathrm{ \,MHz}}}\xspace}

\newcommand{\stat}{\ensuremath{\mathrm{\,(stat)}}\xspace}
\newcommand{\syst}{\ensuremath{\mathrm{\,(syst)}}\xspace}

\def\order{{\ensuremath{\mathcal{O}}}\xspace}
\newcommand{\chisq}{\ensuremath{\chi^2}\xspace}

\def\deriv {\ensuremath{\mathrm{d}}}

\def\gsim{{~\raise.15em\hbox{$>$}\kern-.85em
          \lower.35em\hbox{$\sim$}~}\xspace}
\def\lsim{{~\raise.15em\hbox{$<$}\kern-.85em
          \lower.35em\hbox{$\sim$}~}\xspace}

\newcommand{\Real}{\ensuremath{\mathcal{R}e}\xspace}
\newcommand{\Imag}{\ensuremath{\mathcal{I}m}\xspace}

\def\sqs   {\ensuremath{\protect\sqrt{s}}\xspace}
\def\sqsnn {\ensuremath{\protect\sqrt{s_{\scriptscriptstyle\rm NN}}}\xspace}
\def\pt         {\ensuremath{p_{\mathrm{T}}}\xspace}

\def\degrees{\ensuremath{^{\circ}}\xspace}

\def\mrad{\ensuremath{\mathrm{ \,mrad}}\xspace}
\def\rad{\ensuremath{\mathrm{ \,rad}}\xspace}

\newcommand{\lum} {\ensuremath{\mathcal{L}}\xspace}

\def\tell1  {TELL1\xspace}
\def\ukl1   {UKL1\xspace}

\newcommand{\eg}{\mbox{\itshape e.g.}\xspace}
\newcommand{\ie}{\mbox{\itshape i.e.}\xspace}

\newcommand{\etc}{\mbox{\itshape etc.}\xspace}

\newcommand{\vs}{\mbox{\itshape vs.}\xspace}

%% file: additional-symbols-def.tex
\def\Xibz    {\ensuremath{\Xires^0_\bquark}\xspace}
\def\Xibzbar {\ensuremath{\Xiresbar^0_\bquark}\xspace}
\def\XibPrimeMinus  {{\ensuremath{\Xires^{\prime -}_\bquark}}\xspace}
\def\XibStarMinus  {{\ensuremath{\Xires^{*-}_\bquark}}\xspace}

\def\Xiccpp  {\ensuremath{\PXi_{\cquark\cquark}^{++}}\xspace}

\newcommand{\BRof}[1]{\ensuremath{{\cal B}(#1)}\xspace}
\newcommand{\phisPP}{\ensuremath{{\phi_{s}^{\squark\squarkbar\squark}}}\xspace}

\newcommand{\Ksmm}{\ensuremath{\KS\to\mu^+\mu^-}\xspace}
\newcommand{\Kspizmm}{\ensuremath{\KS\to\pi^0\mu^+\mu^-}\xspace}

\newcommand{\figref}[1]{Fig.~\ref{#1}}
\newcommand{\tabref}[1]{Table~\ref{#1}}

\def\lumunit {\ensuremath{\rm \, cm^{-2}s^{-1}}\xspace}

%% file: titlepage.tex
\begin{titlepage}
\centerline{\large EUROPEAN ORGANIZATION FOR NUCLEAR RESEARCH (CERN)}
\vspace*{1.5cm}
\noindent
\begin{tabular*}{\linewidth}{lc@{\extracolsep{\fill}}r@{\extracolsep{0pt}}}
\ifthenelse{\boolean{pdflatex}}
{\vspace*{-2.7cm}\mbox{\!\!\!\includegraphics[width=.14\textwidth]{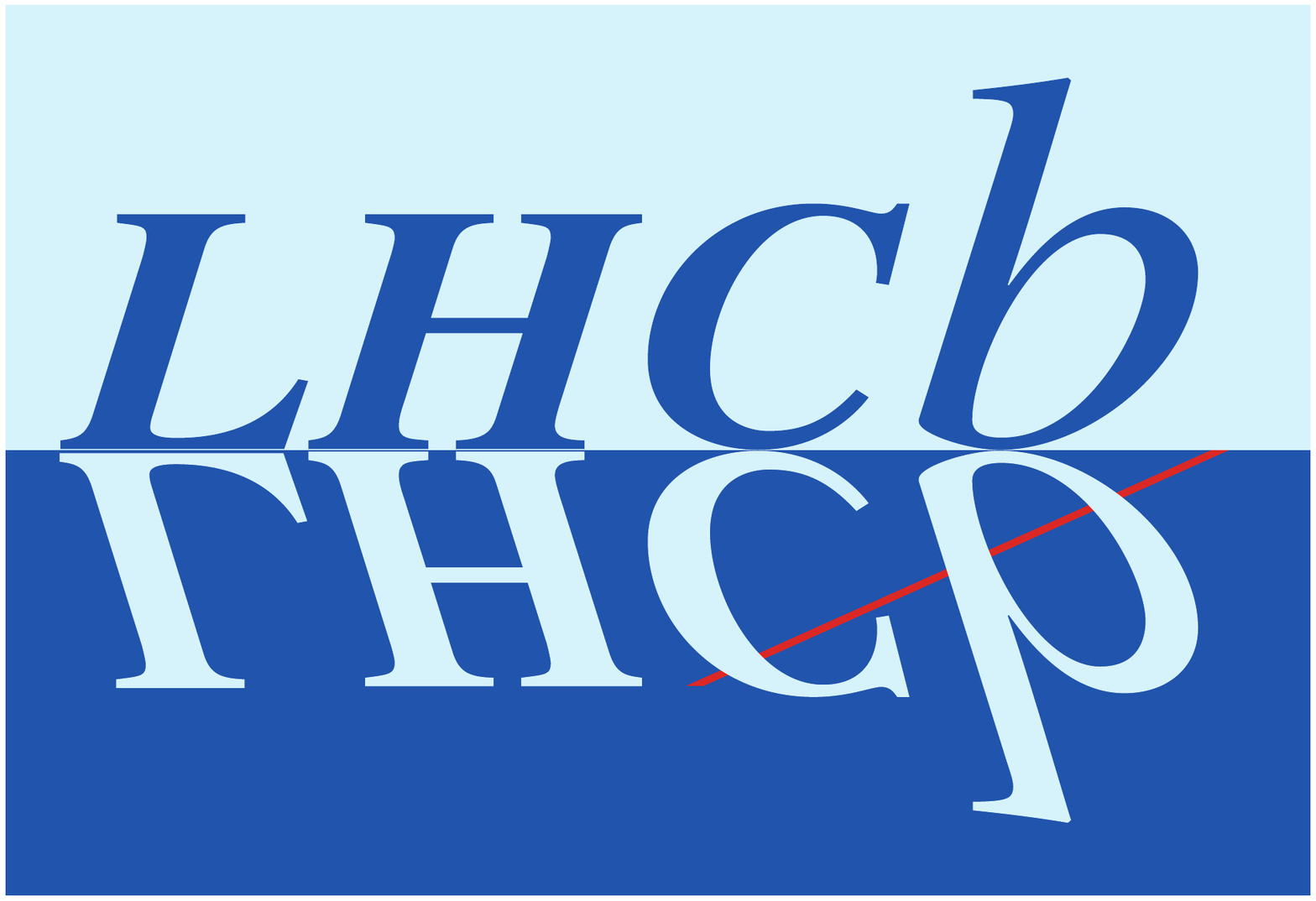}} & &}%
{\vspace*{-1.2cm}\mbox{\!\!\!\includegraphics[width=.12\textwidth]{lhcb-logo.eps}} & &}%
\\
 & & CERN-LHCC-2018-027  \\  
 & & LHCB-PUB-2018-009 \\
 & & 27 August 2018 \\ 
 & & \\
\end{tabular*}

\vspace*{2cm}

{\bf\boldmath\huge
\begin{center}
Physics case for an LHCb \upgradetwo\\
\vspace*{0.2cm} \LARGE Opportunities in flavour physics, \\ and beyond, in the HL-LHC era
\end{center}
}


\begin{center}
The LHCb collaboration
\end{center}


\vspace*{2cm}
\begin{center}
{\bf Abstract}
\end{center}
\noindent
{\small The LHCb \upgradetwo\ will fully exploit the
  flavour-physics opportunities of the HL-LHC, and study additional
  physics topics that take advantage of the forward acceptance of the LHCb
  spectrometer. The LHCb \upgradeone\ will begin operation in 2020. Consolidation will occur, and modest enhancements of the
  \upgradeone detector will be installed, in Long Shutdown 3 of the LHC
  (2025) and these are discussed here. The main \upgradetwo\ detector will be installed in long shutdown 4 of the LHC (2030) and will build on the strengths of
  the current LHCb experiment and the \upgradeone. It will operate at a luminosity up to
  $ 2 \times 10^{34}\lumunit$, ten times that of the \upgradeone detector. 
New detector components will improve the intrinsic
performance of the experiment in certain key areas. An Expression Of
Interest proposing \upgradetwo\ was submitted
in February 2017.  The physics case for the \upgradetwo\ is presented
here in more depth. \CP-violating
phases will be measured with precisions unattainable at any other
envisaged facility. The experiment will probe $b\to s \ellell$ and $b\to d
\ellell$ transitions in both muon and
electron decays in modes not accessible at \upgradeone. Minimal flavour violation will be tested with a
precision measurement of the ratio of ${\cal
  B}(\decay{\Bd}{\mumu})/{\cal B}(\decay{\Bs}{\mumu})$. Probing charm
\CP violation at the
$10^{-5}$ level may result in its long sought discovery.
Major advances in hadron spectroscopy will be possible, which will be
powerful probes of low energy QCD. \upgradetwo\ potentially will have the highest
sensitivity of all the LHC experiments on the Higgs to charm-quark couplings. Generically, the new physics mass scale
probed, for fixed
couplings, will almost double compared with the pre-HL-LHC era; this
extended reach for flavour physics is similar to that which would be achieved by the
HE-LHC proposal for the energy frontier. 
}


{\footnotesize 
\vspace*{1cm}
\centerline{\copyright~\papercopyright. \href{\paperlicenceurl}{\paperlicence}.}}
\vspace*{2mm}

\end{titlepage}

\pagestyle{empty}  


\newpage
\pagestyle{plain}
\pagenumbering{roman}
\setcounter{page}{2}
\mbox{~}

\cleardoublepage

%% file: LHCb_authorlist.tex
\centerline{\large\bf LHCb collaboration}
\begin{flushleft}
\small
{I.~Bediaga{$^{}$}}, {M.~Cruz~Torres{$^{}$}}, {J.M.~De~Miranda{$^{}$}}, {A.~Gomes{$^{a}$}}, {A.~Massafferri{$^{}$}}, {J.~Molina~Rodriguez{$^{z}$}}, {A.C.~dos~Reis{$^{}$}}, {l.~Soares~Lavra{$^{}$}}, {R.~Tourinho~Jadallah~Aoude{$^{}$}}
{\footnotesize \it
{$^{1}$}Centro Brasileiro de Pesquisas F{\'\i}sicas (CBPF), Rio de Janeiro, Brazil\\}\bigskip 

{S.~Amato{$^{}$}}, {K.~Carvalho~Akiba{$^{}$}}, {F.~Da~Cunha~Marinho{$^{}$}}, {L.~De~Paula{$^{}$}}, {F.~Ferreira~Rodrigues{$^{}$}}, {M.~Gandelman{$^{}$}}, {A.~Hicheur{$^{}$}}, {J.H.~Lopes{$^{}$}}, {I.~Nasteva{$^{}$}}, {J.M.~Otalora~Goicochea{$^{}$}}, {E.~Polycarpo{$^{}$}}, {C.~Potterat{$^{}$}}, {M.S.~Rangel{$^{}$}}, {L.~Silva~de~Oliveira{$^{}$}}, {B.~Souza~De~Paula{$^{}$}}
{\footnotesize \it
{$^{2}$}Universidade Federal do Rio de Janeiro (UFRJ), Rio de Janeiro, Brazil\\}
% ...
\bigskip 

{C.A.~Aidala{$^{}$}}
{\footnotesize \it
{$^{73}$}University of Michigan, Ann Arbor, United States, associated to Institute {$^{52}$}\\}\bigskip 

{C.L.~Da~Silva{$^{}$}}, {J.M.~Durham{$^{}$}}
{\footnotesize \it
{$^{74}$}Los Alamos National Laboratory (LANL), Los Alamos, United States, associated to Institute {$^{52}$}\\}\bigskip 

{\footnotesize \it
\bigskip 

{$^{a}$Universidade Federal do Tri{\^a}ngulo Mineiro (UFTM), Uberaba-MG, Brazil}\\

{$^{b}$Laboratoire Leprince-Ringuet, Palaiseau, France}\\

{$^{c}$P.N. Lebedev Physical Institute, Russian Academy of Science (LPI RAS), Moscow, Russia}\\

{$^{d}$Universit{\`a} di Bari, Bari, Italy}\\

{$^{e}$Universit{\`a} di Bologna, Bologna, Italy}\\

{$^{f}$Universit{\`a} di Cagliari, Cagliari, Italy}\\

{$^{g}$Universit{\`a} di Ferrara, Ferrara, Italy}\\

{$^{h}$Universit{\`a} di Genova, Genova, Italy}\\

{$^{i}$Universit{\`a} di Milano Bicocca, Milano, Italy}\\

{$^{j}$Universit{\`a} di Roma Tor Vergata, Roma, Italy}\\

{$^{k}$Universit{\`a} di Roma La Sapienza, Roma, Italy}\\

{$^{l}$AGH - University of Science and Technology, Faculty of Computer Science, Electronics and Telecommunications, Krak{\'o}w, Poland}\\

{$^{m}$LIFAELS, La Salle, Universitat Ramon Llull, Barcelona, Spain}\\

{$^{n}$Hanoi University of Science, Hanoi, Vietnam}\\

{$^{o}$Universit{\`a} di Padova, Padova, Italy}\\

{$^{p}$Universit{\`a} di Pisa, Pisa, Italy}\\

{$^{q}$Universit{\`a} degli Studi di Milano, Milano, Italy}\\

{$^{r}$Universit{\`a} di Urbino, Urbino, Italy}\\

{$^{s}$Universit{\`a} della Basilicata, Potenza, Italy}\\

{$^{t}$Scuola Normale Superiore, Pisa, Italy}\\

{$^{u}$Universit{\`a} di Modena e Reggio Emilia, Modena, Italy}\\

{$^{v}$MSU - Iligan Institute of Technology (MSU-IIT), Iligan, Philippines}\\

{$^{w}$Novosibirsk State University, Novosibirsk, Russia}\\

{$^{x}$National Research University Higher School of Economics, Moscow, Russia}\\

{$^{y}$Sezione INFN di Trieste, Trieste, Italy}\\

{$^{z}$Escuela Agr{\'\i}cola Panamericana, San Antonio de Oriente, Honduras}\\

{$^{aa}$School of Physics and Information Technology, Shaanxi Normal University (SNNU), Xi'an, China}\\

{$^{ab}$Physics and Micro Electronic College, Hunan University, Changsha City, China}\\
\medskip
$ ^{\dagger}$Deceased
}
\end{flushleft}

%% file: CONTRIBUTIONS/1_Executive_Summary/1.tex
 \section{Overview}

\begin{figure}[!htb]
\centering
\includegraphics[width=1.0\textwidth]{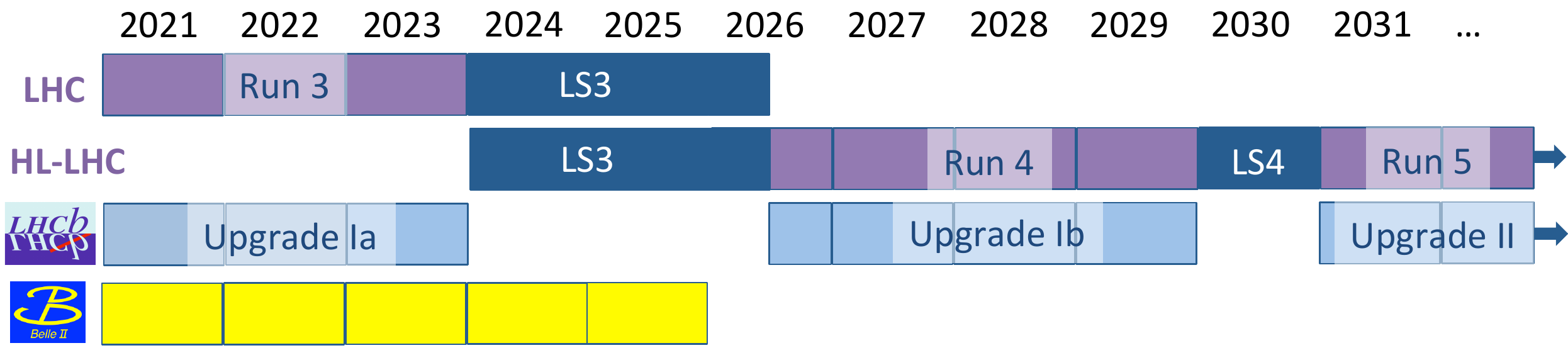}
\caption{\small Timeline of accelerator and experiment operations over the decade 2021 to
  2031. The periods of operations of the LHC and HL-LHC are
  indicated and the long shutdowns (LS). The LHCb operational periods
  are shown with gaps where the detector consolidation and upgrades
  discussed in this document occur. The running period of Belle~II,
  the other major international flavour-physics facility, is also shown.}   
\label{fig:Timeline}
\end{figure}

It is widely recognised in the particle physics community that 
the complementary approaches of the energy and intensity frontier must
both be utilised in the search for physics beyond the Standard Model. The
European Strategy for Particle Physics in 2013 emphasised the need for
flavour physics as a key element of the programme. The LHCb experiment has demonstrated emphatically that the LHC is an
ideal laboratory for quark-flavour physics.  

The LHCb \upgradetwo\ programme \cite{LHCb-PII-EoI} aims to make full use of the
 capabilities of a forward acceptance detector during the High Luminosity
 LHC (HL-LHC) operational period. Foremost in the physics programme, are the possibilities of
 the experiment in its core areas of \CP violation and rare  decays in
 flavour physics. Two chapters of the document are also dedicated 
 to its capabilities in forward and high-\pt physics and in
 spectroscopy. Opportunities in other areas of physics are described in an appendix.

 The timeline
of operations and major shutdowns of the LHC, HL-LHC, LHCb and
Belle~II are illustrated in Fig.~\ref{fig:Timeline}.
The LHCb \upgradeone\ is currently under construction and will start
data taking in 2021 after LHC Long Shutdown~2 (LS2). LHCb \upgradetwo\ will be
installed  during LS4, with operations
beginning in LHC Run 5 which is scheduled to start in 2031. This 
\upgradetwo experiment will operate at instantaneous luminosities of
up to $2 \times 10^{34} \cm^{-2} \sec^{-1}$, an order of magnitude
above \upgradeone. LHCb will accumulate
a data sample corresponding to a minimum of $300 \invfb$.
New attributes, installed in LS3 and LS4, will enhance the detector’s
capabilities to a wider range of physics signatures.  
 
Consolidation of the LHCb \upgradeone\ detector is required during
LS3. Preparatory
work for \upgradetwo\ will be performed at this time making best use of the
extended shutdown period.  These changes are known as Upgrade~Ib. 
LHCb will continue data taking at an
instantaneous luminosity of $2 \times 10^{33} \cm^{-2}
\sec^{-1}$ until LS4. 

The HL-LHC starts operations after LS3.
It is
expected that the
\belletwo\ experiment~\cite{Abe:2010gxa} will finish data taking at the
same time, around 2025. It is thus highly likely that LHCb will be the only large-scale flavour-physics
experiment operating in the HL-LHC era.

The general concepts of the LHCb Upgrade II were presented in an
Expression of Interest in 2017~\cite{LHCb-PII-EoI}. The LHCC encouraged the
collaboration to continue its studies, in particular both to elaborate
on the physics reach and also to hold further discussions with the LHC machine experts to establish
feasible running conditions. This physics-case document addresses the
former point. The HL-LHC team have prepared an independent report on
the second topic~\cite{P8HighLumi}, which is briefly summarised in
this document. The report identifies ``a range of potential
solutions for operating LHCb \upgradetwo\ at a luminosity of up to $2
\times 10^{34} \rm{cm^{-2}s^{-1}}$ and permitting the collection of
$300\invfb$ or more at IP8 during the envisaged lifetime of the LHC''. 
The physics landscape, machine considerations and detector upgrades
are summarised in Chapter~\ref{chpt:intro}.
The LHCb \upgradetwo\  has been discussed in a sequence of 
dedicated workshops from 2016--2018. The workshops have had significant
participation from the theoretical physics community and we
gratefully acknowledge their important contributions to this
document. 





 \section{Physics programme}

The LHCb \upgradetwo\ is
being designed as the flavour physics experiment for the HL-LHC era. The SM does not predict the values of the weak flavour couplings, and so all matrix elements
 must be measured experimentally. However, the unitary nature of the CKM matrix, and the
 assumptions of the SM, impose relations between the elements that are often expressed graphically
 in the complex plane as the so-called unitarity
 triangle. Discrepancies among the various measurements that can be
 made of the sides and
 angles of the unitarity triangle would be strong indications of physics beyond the SM. The
 measurements of the weak phases are discussed in Chapters~\ref{chpt:TDCPV}
and \ref{chpt:TICPV}. The angle $\gamma$ of the unitarity triangle is of particular
interest, because it can be measured extremely cleanly and the
theoretical uncertainty on its interpretation is negligible. While it
is currently one of the least well-known unitarity triangle
parameters, being measured with a
$5^\circ$ uncertainty, LHCb \upgradetwo will allow this to be determined
with an order of magnitude higher precision, or better. The precision measurement of the \Bs weak
mixing phase will be another highlight of the programme. The expected
precision on $\phi_s^{c\bar{c}s}$ after \upgradetwo\ will be $\sim 3
\mrad$. This will be at the same level as the current precision on the
indirect determination based on the CKM fit using tree-level
measurements.

A wide-ranging set of lepton universality tests in $b \to c \ellm
\neulb$ decays (charge conjugated decays are implied throughout) can be performed at \upgradetwo, exploiting the full range of
\bquark hadrons, to probe new physics models. These and other prospects in semileptonic decays, including
measurements of the unitarity triangle sides, are reviewed in
Chapter~\ref{chpt:semilept}. The copious semileptonic decays allow the
high precision search for \CP violation in \Bz and
\Bs mixing. The \upgradetwo\ dataset will allow the \CP asymmetries for
both mesons to be measured to the level of a few parts in $10^{-4}$, giving unprecedented
new physics sensitivity.

Indirect \CP violation in the charm system is predicted to be very
 small, $\order(10^{-4})$ or less in the SM. In the
 absence of new physics contributions to charm \CP violation, the \LHCb~\upgradetwo
 may well be the only facility with a realistic probability of
 observing this phenomena as it will
 be able to reach a sensitivity of $\order(10^{-5})$, through multiple
 modes. A full programme of direct \CP-violation searches in charm
 will also be performed, with complementary approaches and probing
 both new physics and Standard Model sensitive
 modes. Chapter~\ref{chpt:charm} describes the perspective in this field.

The study of rare decays is central to the LHCb programme with the
statistics of \upgradetwo providing a wealth of opportunities,
including in charm and strange decays, as described in Chapter~\ref{chpt:rare}.
A comprehensive measurement programme will be performed of observables
in a wide range of $b \to s \ellp \ellm$ and $b \to d \ellp \ellm$ transitions,
employing both muon and electron modes. The LHCb \upgradetwo\ is the
only facility with the potential to distinguish between some plausible
new physics scenarios. Another notable target of the physics programme
is the ratio of branching fractions, ${\cal
  B}(\decay{\Bd}{\mumu})/{\cal B}(\decay{\Bs}{\mumu})$, which is a
powerful observable to test minimal flavour violation and will
be probed at the $10\%$ level with $300\invfb$. 

The LHCb Upgrades, with their flexible fully software based trigger,
will be general purpose detectors in the forward
region. Chapter~\ref{chpt:qee}  describes the strong programme in forward
and high-\pt physics. The effective weak mixing angle, \ssqtwef, and
the W-mass are both fundamental parameter of the Standard Model where
LHCb's design gives advantages or complementarity to ATLAS and CMS.
The excellent vertexing abilities and large samples sizes of LHCb
\upgradetwo\ will give the experiment the important capability to
constrain the Higgs to charm
coupling, with a precision which may exceed that of the other LHC experiments.
Important results can also be obtained in the field of dark sector and
long-lived particle searches.

LHCb has had a transformative impact on the field of hadron
spectroscopy. Chapter~\ref{chpt:spect} describes measurements that can be
made on the known states to determine their nature and the
opportunities for further particles to be observed. A pentaquark
multiplet, pentaquarks containing beauty quarks and doubly charmed
tetraquarks are all amongst the states that can be searched for with a
sufficiently large dataset.


The summary section of Chapter~\ref{chpt:summary} gathers together the
estimated precision on modes which will be highlights of the \upgradetwo\ programme,
including many of the decays that have been mentioned here. 

%% file: CONTRIBUTIONS/2_Introduction/2.tex
\label{chpt:intro}
\input{CONTRIBUTIONS/2_Introduction/2.2.tex}
\input{CONTRIBUTIONS/2_Introduction/2.3.tex}
\input{CONTRIBUTIONS/2_Introduction/2.4.tex}

%% file: CONTRIBUTIONS/2_Introduction/2.2.tex
\section{Physics landscape}

The successful operation of, and exploitation of data from, the Large Hadron
Collider has transformed the landscape of high energy physics, although not in
the way that was widely anticipated a decade ago.
The discovery of the Higgs boson~\cite{Aad:2012tfa,Chatrchyan:2012xdj}
completes the Standard Model~(SM), leaving a self-consistent theory that in
principle could be valid up to the Planck scale.  
Although the Higgs mass is ``unnatural'' in the SM, requiring extreme
fine-tuning of quantum corrections, the new particles predicted in theories
that address this hierarchy problem (\eg\ supersymmetry) have not yet been
discovered.  

The SM has also survived a large number of additional tests. 
Measurements of the Higgs boson's properties, including its quantum
numbers and production and decay modes, are consistent with SM predictions.
Electroweak precision constraints on variables such as the W boson and top
quark masses and the effective weak mixing angle agree with the
self-consistency predicted in the SM.
Studies of \CP violation in the quark sector are consistent with the
predictions based on there being a sole source of matter-antimatter asymmetry,
encoded in the Cabibbo-Kobayashi-Maskawa quark mixing
matrix~\cite{Cabibbo:1963yz,Kobayashi:1973fv}; all determinations of the
properties of the so-called Unitarity Triangle fit together within the
SM. 
Searches for candidate dark matter particles, either through production in LHC
collisions or through interaction in extremely low background detectors have
yielded null results, providing stringent constraints on WIMP nuclear
cross-sections. 

Nevertheless, there remain extremely clear arguments to continue to search
for physics beyond the SM.  
There are strong reasons to believe that the hierarchy problem may be resolved
in a theoretically attractive way, at a higher energy scale than yet has been
probed by experiments. 
There is a compelling need to understand the constitution of the 95\% of the
energy density of the Universe that is not baryonic matter (68\% dark energy,
27\% dark matter).
The origin of the observed excess of matter over antimatter in the Universe also cannot
be explained in the SM, and addresses one of the most fundamental questions in
science: ``why are we here?''
The LHC and its high luminosity upgrade will continue, for about the next 20
years, to be our most powerful tool to address these major challenges to our
fundamental understanding of nature.
For this reason, the 2013 update of the European Strategy for Particle Physics
stressed that ``Europe's top priority should be the exploitation of the full
potential of the LHC,'' noting that this will ``provide further exciting
opportunities for the study of flavour physics.''

The high confidence, expressed in the European strategy statement, in the
potential for flavour physics at the HL-LHC was justified by the tremendous
achievements of the LHCb experiment during Run~1 of the LHC (2010--12).
The successful operation of LHCb during this period vindicated the concept and
design of a dedicated heavy flavour physics experiment at a hadron collider.
Among the several hundreds of publications based on the Run~1 data sample,
highlights include the first evidence (and then first observation, together
with CMS) of the very rare decay $\Bs\to\mumu$~\cite{LHCb-PAPER-2013-046,LHCb-PAPER-2014-049}, world-leading results on \CP\
violation in beauty and charm hadrons, significant improvements in precision
on Unitarity Triangle angles, and observations of new hadronic states of
matter including pentaquarks~\cite{LHCb-PAPER-2015-029}.  
LHCb has continued its successful operation during Run~2 (2015--18), during
which the increased cross-sections due to the higher collision energy and
implementation of novel online data processing strategies have led to
significant increases in the samples available for analysis.
The flexible data-collection strategies of LHCb have also allowed the physics
programme to be expanded in ways that were completely unforeseen prior to
data-taking, for example in studies of dark sectors, and by collecting data
from heavy ion and fixed target collisions.

The original LHCb detector was designed to be able to collect $8\invfb$
at an instantaneous luminosity of up to $2 \times 10^{32} \lumunit$.
By the end of Run~2 this target will have been exceeded, with the majority of
the data collected at $4 \times 10^{32} \lumunit$, and therefore in a more
intense environment with higher pile-up.  
In order to be able to continue the LHCb physics programme, a first LHCb
upgrade was approved in 2012~\cite{LHCb-TDR-012}.
The key concept of this upgrade is that the current bottleneck of the level 0
(hardware) trigger is removed.
By reading out the full detector at the maximum LHC bunch crossing rate of $30
\mhz$ and implementing all trigger decisions in software, it is possible to
increase the luminosity without suffering a compensating loss in efficiency.  
By increasing the instantaneous luminosity by a factor of five, to $2 \times
10^{33} \lumunit$, and improving the trigger efficiency for most modes by a
factor of two, the annual yields in most channels will be an order of magnitude
larger than at present. A total integrated luminosity of around $23
\invfb$ is anticipated by the end of Run~3 and $50
\invfb$ by the end of Run~4 of the LHC.   
The upgraded LHCb detector has been designed to meet these specifications, and
will withstand the occupancies with comparable performance to
the current experiment. It was already foreseen, in 2012, that further
consolidation of the detector would be required before Run~4, notably
for the inner part of the electromagnetic calorimeter.

The first LHCb upgrade will be installed during the second LHC long shutdown
(2019--20) and commence operation in 2021.
During Run~3, it will be collecting data at the same time as the \belletwo\
experiment~\cite{Abe:2010gxa} at the SuperKEKB asymmetric $\epem$ collider.
Although \belletwo\ has a superficially similar flavour physics programme to
that of LHCb, the rather different collision environments lead to the two
experiments having complementary strengths.  
For final states composed of only charged tracks, LHCb will in general have
much larger yields and lower backgrounds; \belletwo\ on the other hand tends to
have better capability for channels involving neutral particles or missing
energy. 
LHCb can study all species of beauty hadrons, and the large boost and excellent
vertexing allow $\Bs$ oscillations to be resolved; \belletwo\ can exploit the
$\epem \to \taup\taum$ production mode to study essentially all properties and
decays of the $\tau$ lepton.
\belletwo\ is expected to collect $\epem$ collision data corresponding to
$50 \invab$ at or
around the $\FourS$ resonance by 2025, when SuperKEKB will conclude operation;
there are no current plans for any subsequent upgrade.  

\begin{figure}[!tb]
\centering
\includegraphics[width=0.8\textwidth]{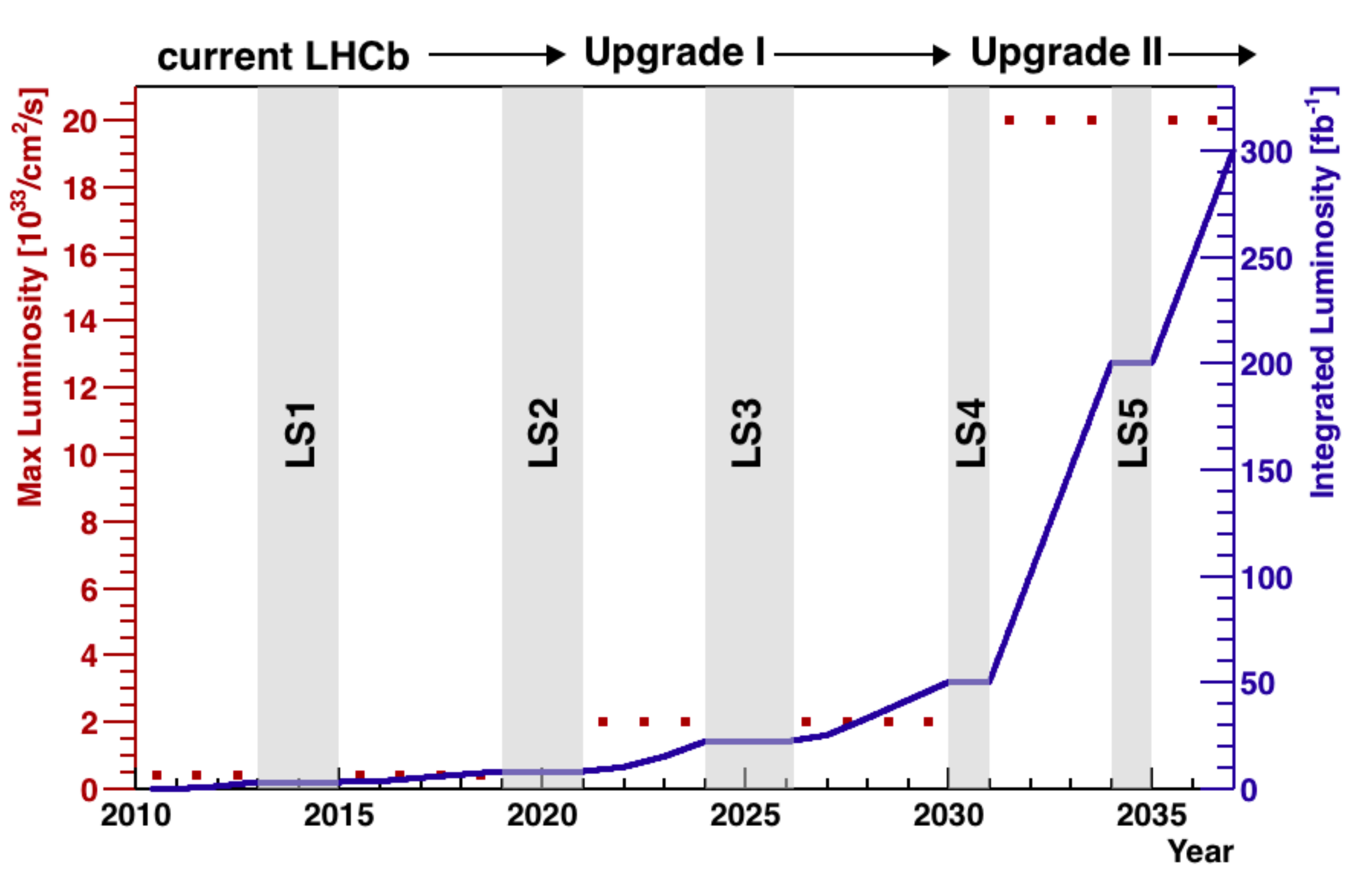}
\caption{\small Luminosity projections for the original LHCb,
  Upgrade~I, and \upgradetwo\  experiments as a function of time. The red points and the left scale indicate the
  anticipated instantaneous luminosity during each period, with the
  blue line and right scale indicating the integrated luminosity accumulated.}
\label{fig:LumiProjection}
\end{figure}

\begin{figure}[!tb]
\centering
\includegraphics[width=0.8\textwidth]{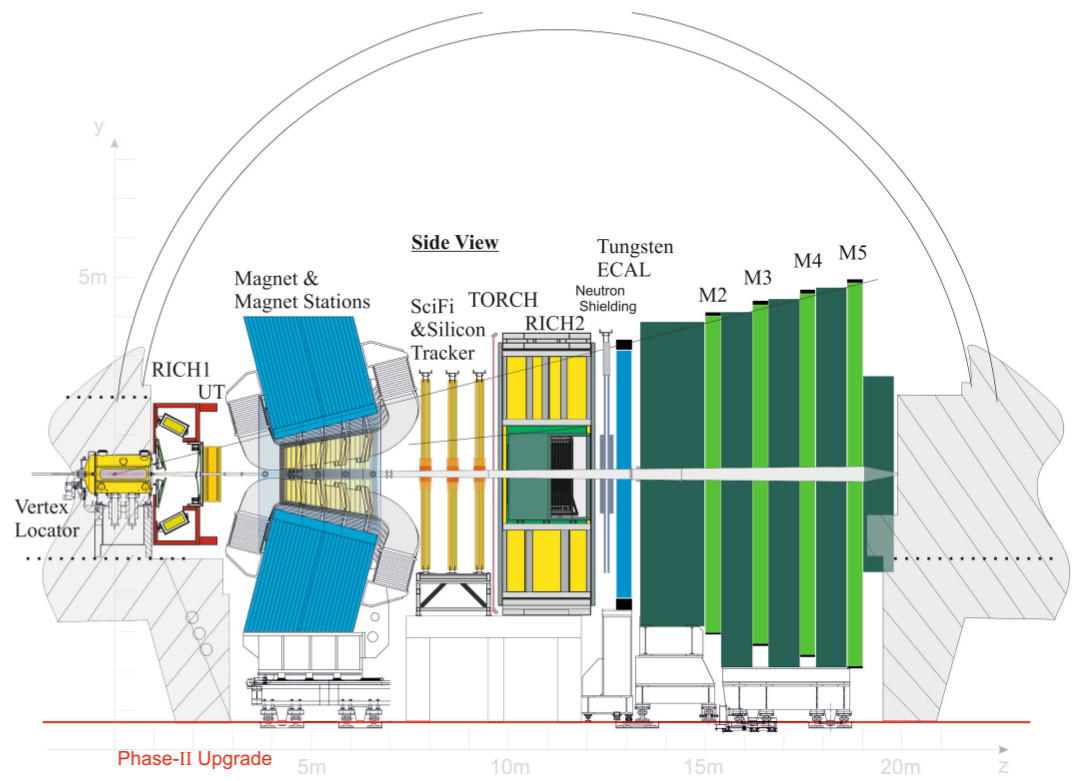}
\caption{\small Schematic side-view of the \upgradetwo detector.}
\label{fig:U2Detector}
\end{figure}

Since the approval of the HL-LHC, and the associated plan to operate the LHC
until the late 2030s, it has become clear that the integrated luminosity collected by
the first LHCb upgrade will not meet the European strategy goal of fully
exploiting the LHC potential for flavour physics.  
A further upgrade based on cutting-edge technologies could allow a further
order-of-magnitude increase in instantaneous luminosity, to achieve a total
integrated luminosity of $300 \invfb$ or more. The ultimate limit may be
higher if  the LHC inner triplet magnets can withstand a higher dose;
the HL-LHC capabilities are discussed in Section~\ref{sec:HLLHC}. The
foundations for this next generation experiment will be laid in the
third LHC long shutdown between Run~3 and Run~4,  and are referred to
as Upgrade~Ib. This will occur  coincident with the HL-LHC accelerator
upgrade, and is foreseen to consist of  consolidation and modest enhancements to the detector.
These detector enhancements
will improve the physics performance of the experiment, with running continuing in Run~4 at the same luminosity as in
Run~3. The major detector improvements of \upgradetwo will occur in
the fourth long shutdown of the LHC. These will equip the detector for
the higher luminosity data-taking of Run~5 and beyond. The integrated
luminosity projection for the experiment as a function of time is
shown in Fig.~\ref{fig:LumiProjection} and a diagram of the proposed
detector design in Fig.~\ref{fig:U2Detector}.

The data sample collected by the end of the HL-LHC period will be more than a
factor thirteen higher than that collected in the pre-HL-LHC period, and
at least a factor six higher than that at the end of Run~4. This will
lead to remarkable improvements in precision in the large number of
observables that are not expected to be limited by systematic uncertainties.
The energy scale that is probed through precision measurements scales as the
fourth root of the sample size, so the step from the pre-HL-LHC to
the post-HL-LHC, and post Belle~II, period corresponds to a factor of 1.9 or more in reach.  
Thus the impact of \upgradetwo\ in flavour physics will be comparable
to that for on-shell new particle searches of increasing the LHC
collision energy from $14 \tev$ to $27\tev$, \ie similar to the
impact of the HE-LHC~\cite{Zimmermann:2017bbr} concept at the energy frontier. 
This extrapolation assumes that the current detector performance is
maintained, and is therefore an underestimate for topics where improvement is
expected due to new detector technologies, for example the elimination
of the
vertex detector RF foil or replacement by a lower material alternative, the improved
granularity of the electromagnetic calorimeter, or the addition of low momentum hadron
identification capability through the TORCH detector.

The method of testing the SM through precision measurements in flavour physics
is fully complementary to that of searching for on-shell production of
new particles in high energy collisions.  
Mixing and decay of beauty and charm hadrons occurs through weak interactions,
mediated by gauge bosons with masses many times larger than those of the hadrons
themselves (compare, for example, the $W$ boson mass of $80 \gevcc$ with the $B$
mass of around $5 \gevcc$).
Other, as-yet unknown particles could also contribute, in which case measured
parameters such as decay rates and \CP violation asymmetries would be shifted
from the SM predictions. While direct searches for new particles are
carried out  in the \tev\ mass range at the \lhc, flavour physics
measurements can probe much higher mass scales,
considering a generic effective Lagrangian shows that mixing and \CP
violation observables in the beauty and charm system probe scales of up to 1000 to $10000 \tev$~\cite{Isidori:2010kg,*Isidori:2013ez,*Kamenik:2017znu}.

The reach of measurements of flavour physics observables is limited only by
precision, both experimental and on the SM prediction.
Rare processes, where the SM contribution is small or vanishing and as such
has low uncertainty, are therefore of special interest.
In particular, processes for which the SM contribution occurs through
loop diagrams, \ie\ flavour changing neutral currents, are often considered
golden channels for potential discoveries of physics beyond the SM.
These include $\Bd$--$\Bdb$, $\Bs$--$\Bsb$ and $\Dz$--$\Dzb$ mixing processes
as well as rare decays including $B\to\ell\ell$, $B \to X \ell \ell$ and $B
\to X \gamma$ (here $\ell$ is a lepton and $X$ is a hadronic system). 

Several anomalies recently reported by LHCb in these types of processes have
led to speculation that a discovery of physics beyond the SM may be not far
off. However, even if these anomalies are not confirmed, they serve as
a good example of the potential of flavour physics at \upgradetwo to
probe beyond the energy frontier.
The angular distribution in $\Bz \to \Kstarz\mumu$ decays shows a deviation
from the SM and branching fractions of $B \to X\mumu$ decays appear to be
consistently below the SM predictions.
Model-independent analyses indicate a consistent possible explanation of the
trends in terms of additional contributions, beyond those of the SM, to the
operators of the effective theory that describes these decays.  
Measurements of the ratios of branching fractions of $B \to X\mumu$ and $B \to
X\epem$ decays are also below the SM predictions that are close to unity and
have small uncertainty due to the assumption of lepton universality.  
Strikingly, there are also hints of lepton universality violation in
charged-current weak interactions, namely in ratios of the branching fraction
of $B \to X \tau \nu$ relative to that where the $\tau$ is replaced by a $\mu$
or $e$ lepton.  
Results from LHCb on these modes with neutrinos are consistent with, and of
comparable precision to, those from \babar\ and \belle\ (for the other modes
mentioned above, LHCb is world-leading).
None of these measurements is individually above the $5\sigma$ threshold,
so only a cautious interpretation can be made, but the consistency of the
results has led to several explicit models being proposed.
These typically introduce new particles such as leptoquarks or $Z^\prime$
bosons that may be either just within, or just out of, the reach of the LHC.
While it is too early to conclude on whether these particular models are
likely to be realised in nature, they demonstrate the discovery potential of
flavour physics, underlining the point that an LHCb \upgradetwo\ is necessary to
exploit fully the opportunities provided by the HL-LHC.

%% file: CONTRIBUTIONS/2_Introduction/2.3.tex
\section{Machine aspects for high-luminosity running at the LHCb interaction point}
\label{sec:HLLHC}

The potential for the HL-LHC to deliver the luminosity required by LHCb \upgradetwo has been studied and documented in a CERN accelerator note~\cite{P8HighLumi}. The HL-LHC baseline design \cite{Apo:2017} is compatible with LHCb running at the \upgradeone luminosity of $2 \times\,{\rm 10^{33} cm^{-2} s^{-1}}$. Running above the nominal \upgradeone luminosity requires modifications to the HL-LHC optics and layout in the LHCb Interaction Region (IR 8). 

The luminosity performance achievable at LHCb for \upgradetwo, and the impact on the integrated luminosity in ATLAS and CMS has been studied. The modifications required to the machine layout have also been investigated. These preliminary studies and beam dynamics simulations have shown no fundamental limitations to the delivery of an integrated luminosity of $\sim 50 \invfb$ per year at LHCb. They show a corresponding reduction of the integrated luminosity in ATLAS and CMS of less than $3\%$ as a result of the additional burn-off. The predictions for LHCb use the standard HL-LHC operational scenario in Refs.~\cite{Metral:2301292, MedinaMedrano:2301928} where 160 proton-proton physics collision days/year is assumed; it is foreseen by the accelerator division that this number is conservative and will increase as there will be less special runs, notably ion running, and less machine development than currently. The CERN accelerator note~\cite{P8HighLumi} concludes that ``preliminary investigations have identified a range of potential
solutions for operating LHCb \upgradetwo\ at a luminosity of up to $2
\times 10^{34} \rm{cm^{-2}s^{-1}}$ and permitting the collection of
$300\invfb$ or more at IP8 during the envisaged lifetime of the LHC''.

\subsection{Prospects for running LHCb at high luminosity}
For fixed values of the HL-LHC beam parameters (number of bunches, filling scheme, bunch population, bunch length) the luminosity delivered at  LHCb will essentially depend on the minimum ${\rm\beta^*}$ and beam crossing angles achievable at the interaction point.

The minimum $\rm{\beta^*}$ and crossing angle are constrained by available magnet strength, beam-beam effects, and aperture considerations. A possible set of HL-LHC compatible parameters have been identified and are listed in Table~\ref{TablePerf} together with the corresponding luminosity performance, under the assumption that the beam lifetime is dominated by burn-off. 

All the \upgradetwo scenarios proposed in Table~\ref{TablePerf} are based on similar layouts. Operation at high luminosity and with small $\rm{\beta^*}$ will enhance beam-beam effects, which could have the potential to reduce the dynamic aperture and therefore lower the integrated luminosities from the values given in Table~\ref{TablePerf}, and also degrade the performance at ATLAS and CMS. Detailed simulations, benefiting from the ever-increasing knowledge of the performance of the current machine, are underway to answer these questions~\cite{BeamBeam,DeMaria:Annecy}.

In the scenarios where the beams cross in the horizontal plane, the spectrometer dipole adds an internal crossing angle to the external one~\cite{Herr:2007} resulting in different net crossing angles and consequently the performance differs for the two magnet polarities. Collecting significant samples with both magnet polarities is desirable for the LHCb physics programme as it facilitates the study of some sources of systematic uncertainties in \CP-violation measurements. Injecting the beams with vertical crossing angles is not possible at LHCb because of aperture limitations from the beam screens. However, a vertical crossing plane can be implemented: a horizontal crossing angle can be used at injection and the crossing plane can be rotated before establishing collisions in LHCb. Indeed, this scheme has already been used in operation in 2012 but it adds significant operational complexity and beam dynamics constraints. The vertical crossing allows identical interaction point (IP) characteristics (Luminosity, pile-up, and size of the beam spot) for each detector magnet polarity to be achieved. However, the maximum integrated luminosity at LHCb will be achieved by running with a horizontal crossing angle and without magnet-polarity reversal. Additional scenarios based on flat optics (${\rm \beta^*_{\parallel}} < {\rm\beta^*_{x}}$) could be considered to overcome some of the aperture limitations and further increase the luminosity~\cite{DeMaria:Annecy} but they have not been studied yet. At this stage, the important message is that multiple scenarios have been identified that could deliver integrated luminosities in the range required by LHCb \upgradetwo.

\begin{table}[!tbp]
  \begin{center}
\caption{HL-LHC baseline and Luminosity Scenario at LHCb, with different levelled luminosities, crossing planes, and dipole polarities. The values provided assume standard HL-LHC beams parameters and duty cycle. See Ref.~\cite{P8HighLumi} for more details. The yearly integrated luminosity for ATLAS/CMS during \upgradeone\ LHCb operations is estimated to be 261.5 ${\rm fb^{-1}/y}$.}
 \label{TablePerf}
    \begin{tabular}{lccccccc}
\hline
      \textbf{Parameter} & \textbf{Unit} & \multicolumn{6}{c}{\textbf{Lumi scenario}}\\
\hline
\textbf{Target levelled lumi} & $10^{34}\,{\rm cm^{-2} s^{-1}}$ & \multicolumn{3}{c}{\textbf{1.0}} & \multicolumn{3}{c}{\textbf{2.0}}\\
$\beta^*$ & m  & \multicolumn{6}{c}{1.5}\\
Crossing plane &  & \multicolumn{2}{c}{H} & V & \multicolumn{2}{c}{H} & V \\
Magnet polarity &   & $-$ & $+$ & $\pm$ & $-$ & $+$ & $\pm$ \\
External x-ing angle & ${\rm \mu rad}$  & 400 & 300 & 320 & 400 & 300 & 320 \\
Full x-ing angle at IP & ${\rm \mu rad}$  & 130 & 570 & 419 & 130 & 570 & 419 \\
Virtual (Peak) luminosity & $10^{34}\,{\rm cm^{-2} s^{-1}}$  & 2.16 & 1.57 & 1.79 & 2.16 & 1.57 & 1.79 \\
Levelled pile-up & 1 & 28 &28 & 28 & 56 & 44.2 & 50.3 \\
RMS luminous region (start) & ${\rm mm}$ & 52.7 & 39.5 & 44.7 & 52.7 & 39.5 & 44.7 \\
Peak line pile-up density (start) & ${\rm mm^{-1}}$ & 0.20 & 0.28 & 0.25 & 0.41 & 0.44 & 0.44 \\  
Eff. line pile-up density (start) & ${\rm mm^{-1}}$ & 0.13 & 0.17 & 0.15 & 0.20 & 0.20 & 0.20 \\  
Fill duration & h & 8.0 & 8.0 & 8.0 & 7.7 & 8.0 & 7.9  \\
Levelling time & h & 4.7 & 3.1 & 3.6 & 0.6 & 0 & 0  \\
\hline
\textbf{Integ. lumi. at LHCb} & ${\rm fb^{-1}/y}$ & \textbf{46.3} & \textbf{40.9} & \textbf{42.5} & \textbf{61.7} & \textbf{46.2} & \textbf{51.0} \\ 
Integ. lumi. at ATLAS/CMS & ${\rm fb^{-1}/y}$ & 257.1 & 257.7 & 257.5 & 255.1 & 257.0 & 256.4 \\ 

      \hline
    \end{tabular}
    
  \end{center}
\end{table}

\subsection{Energy deposition and shielding issues}
A simulation of the LHC machine layout around LHCb, using the Fluka package~\cite{Bohlen:2014buj}, was performed in order to assess the energy deposition in the different machine components \cite{P8HighLumi}.
This study and its conclusions were already outlined in the \upgradetwo Expression of Interest in 2017 \cite{LHCb-PII-EoI}. As a large half crossing angle ($38\,{\rm \mu rad}$) was pessimistically assumed at that time, the conclusions remain valid for the new scenarios presented in this document.
In order to operate at high luminosity, additional elements will be added to the machine layout at each side of the LHCb IP, in particular: 
\begin{itemize}
\item
a TAS (Target Absorber) to protect the upstream inner triplet from quenching, and to limit its radiation dose,
\item
a TAN (Target Absorber Neutrals) to shield the recombination dipoles D2 from high-energy neutral particles,
\item
and a TCL (Target Collimator Long) to protect the cold magnets in the matching sections from collision debris.
\end{itemize}

Additional items could still be required and the cost and installation work related to raising the LHCb luminosity for \upgradetwo are being investigated. A mini-TAN will already be installed during LS2 to allow for $2 \times 10^{33}\,{\rm cm^{-2}s^{-1}}$ operation. Thanks to its effective final design and location it is possible that this could also suffice for the higher luminosity conditions of \upgradetwo, though this has still to be fully proven.

The lifetime of the triplet quadrupoles at LHCb is assumed to be identical to that of the triplets currently in place
at the high luminosity IPs. They are designed to withstand a dose 30MGy, corresponding to $300\, {\rm fb^{-1}}$ at the IP of ATLAS.
Further energy deposition studies will allow the design and the crossing angle scheme at LHCb to be optimised and consequently potentially to extend the lifetime of the quadrupoles beyond that limit. The existing triplets at ATLAS and CMS will be removed in LS3. A careful inspection of these triplets after removal will also shed light on the radiation hardness of these components and the possibility of running beyond their currently accepted lifetime. Conservatively, $300\,{\rm fb^{-1}}$ of integrated luminosity for the LHCb \upgradetwo is assumed for the physics projections reported in this document.

%% file: CONTRIBUTIONS/2_Introduction/2.4.tex
\section{Expected detector developments}
\input{CONTRIBUTIONS/2_Introduction/2.4.1.tex}
\input{CONTRIBUTIONS/2_Introduction/2.4.2.tex}
\input{CONTRIBUTIONS/2_Introduction/2.4.3.tex}

%% file: CONTRIBUTIONS/2_Introduction/2.4.1.tex
\subsection{Tracking with timing detectors}
\label{sec:tracking}

At a luminosity of $2 \times 10^{34}$\lumunit, the maximum considered, the mean number of visible proton-proton interactions
per crossing would be 56, producing around 2500 charged particles within the LHCb
acceptance. Efficient real-time reconstruction of charged particles
and interaction vertices within this environment represents a
significant challenge. To meet this challenge it is foreseen to modify
the existing spectrometer components to increase the granularity,
reduce the amount of material in the detector and to exploit the use
of precision timing \cite{LHCb-PII-EoI}. It is expected that the
performance of the \upgradeone\ detector will be
maintained or improved in terms of track-finding efficiency and
momentum resolution. 

\subsubsection{The vertex detector}

The LHCb upgrade physics programme is reliant on an efficient and precise vertex detector (VELO) that enables real time reconstruction of tracks from all  LHC bunch crossings in the software trigger system.  
The \upgradetwo\ luminosity poses significant challenges which necessitate the
construction of a new VELO with enhanced capabilities. 
Compared to \upgradeone\ there will be a further order of magnitude increase in data output rates accompanied by corresponding increases in radiation levels and occupancies. 
To cope with the large increase in pile-up, new techniques to assign correctly each \bquark hadron to the primary vertex from which it originates, and to address the challenge of real time pattern recognition, are needed.  
These challenges will be met by the development of a new 4D hybrid pixel detector with enhanced rate and timing capabilities in the ASIC and sensor.  
Improvements in the mechanical design of the \upgradetwo\ VELO will also be needed to allow for periodic module replacement. 
The mechanical design will be further optimised to minimise the material before the first measured point on a track (which is dominated by the RF foil) and to achieve a more fully integrated module design with thinned sensors and ASICs combined with a lightweight cooling solution. 
As well as improving the VELO performance, quantified by the impact parameter resolution, these changes will also be beneficial both in improving the momentum resolution of the spectrometer and reducing the impact of secondary interactions on the downstream detectors.

\begin{figure}[tb]
\centering
\includegraphics[width=0.7\textwidth]{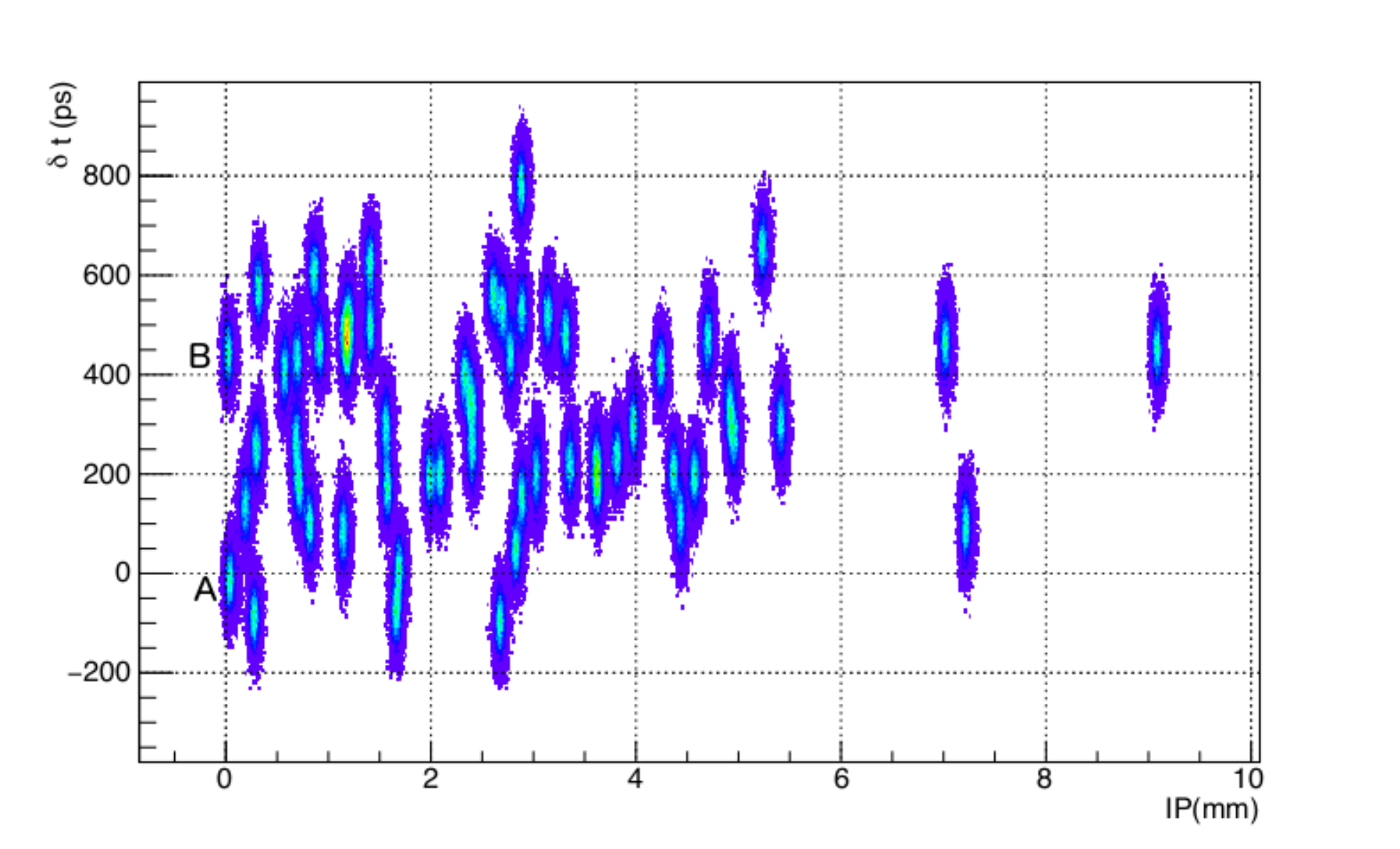}
\caption{Example event containing a $B^0 \to \pi^+ \pi^-$ candidate under Upgrade II conditions, illustrating the PV association challenge. Each PV is drawn as a 2D Gaussian distribution with the appropriate values and uncertainties for both spatial ($x$-axis) and temporal ($y$-axis) metrics used to associate the $B$ meson to a single origin PV. In this case, adding the temporal information allows the correct PV [`A', closest to $(0,0)$] to be identified where the spatial information alone would lead to the wrong choice (`B').}
\label{fig:velo_timing_example}
\end{figure}

Uniquely at the LHC, the physics programme of LHCb 
requires high efficiency to associate correctly heavy flavour decay products to the parent primary vertex (PV).  
The principle behind the use of timing information for PV association is illustrated for a single \upgradetwo\ event in Fig.~\ref{fig:velo_timing_example}. 
In this case, choosing the PV with which the \bquark-hadron candidate forms the smallest impact parameter significance ($x$-axis) leads to an incorrect assignment (labelled `B' in Fig.~\ref{fig:velo_timing_example}). 
Adding time information resolves this problem by providing an additional metric ($y$-axis) to improve the PV association performance, leading to the correct PV (`A') being assigned. 
Studies show that without timing the Upgrade II PV mis-association
levels may reach $\sim 20 \%$, while this can be reduced to $\sim5\%$
with a timing precision of $50$--$100 \ps$. Studies have also shown
that the track reconstruction efficiency and fake rate can be
addressed by decreasing the pixel pitch from the current $55 \mum$ at
\upgradetwo, particularly for the innermost region of the VELO.
The addition of timing will also have crucial benefits in track reconstruction since it allows to reduce drastically combinatorics at an early stage, saving CPU resources. 
Timing information from the VELO also provides a precise time origin
for tracks for the rest of the experiment, which will be helpful for
other subdetectors with timing such as the TORCH. 

\begin{figure}[tb]
\centering
\includegraphics[width=0.7\textwidth]{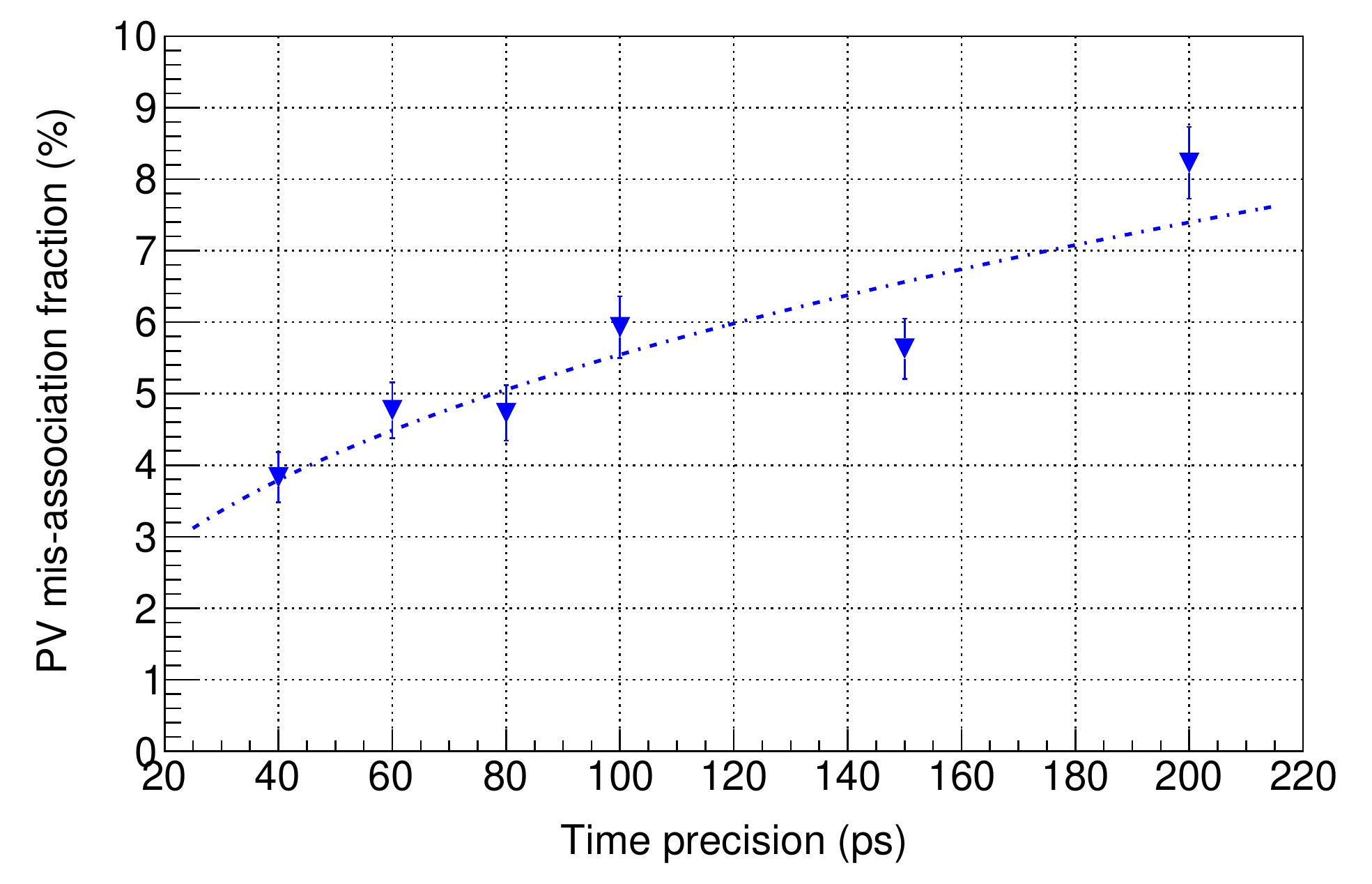}
\caption{Fraction of $B^0 \to \pi^+ \pi^-$ candidates which are associated with an incorrect primary vertex in Upgrade II conditions ($\lum = 1.5 \times 10^{34}\lumunit$), under a range of time precisions for the outer radial region ($20 < r < 35 \mm$) of the VELO as described in the text. These performance numbers should be compared with the 20\% PV mis-association fraction corresponding to a detector with no time information.}
\label{fig:velo_timing_results}
\end{figure}

The timing and rate capabilities required from the ASIC are ambitious but achievable with the foreseen R\&D timeline. 
Another possibility being considered is to have a `mixed' solution where the inner region has a smaller pitch (emphasising resolution) and the outer region has a larger pitch emphasising more precise timing. 
Studies of the performance of a possible configuration are shown in Fig.~\ref{fig:velo_timing_results}. 


\subsubsection{Downstream tracking}

Changes to the downstream tracking system are also foreseen. In \upgradeone\ this comprises a silicon strip detector located upstream (UT) and three tracking
stations located downstream of the magnet (T-stations). For \upgradeone\
the T-stations are covered by  a twelve-layer scintillating fibre
tracker (SciFi). This covers the full acceptance, corresponding to $30\ma$ per
layer.  In conjunction with the VELO, these stations provide a high precision momentum
measurement. They also measure the track directions of the charged particles as input
to the particle identification systems, notably the photon-ring
searches in the RICH detectors.  Two challenges must be met in the
design of the system for \upgradetwo. First, the higher occupancies
necessitate increased detector granularity. Second, the rate of
incorrect matching of upstream and downstream track segments needs to
be minimised. This can be achieved by optimisation of the UT,
minimisation of detector material and use of timing information.

The high occupancies in \upgradetwo\ necessitate replacing the
inner part of the T-stations with a high granularity silicon
detector, with the large area covered by scintillating fibres as shown in Fig.~\ref{fig:Tracking} (left). Based
upon occupancy studies it is proposed to cover an area of $17 \ma$ per layer with silicon \cite{LHCb-PII-EoI} for \upgradetwo. As an
intermediate step it is proposed to upgrade the inner part of the
SciFi  during Upgrade~Ib in LS3. The design of this new silicon
detector, known as the MIghTy tracker with inner and middle tracker elements,
is at an early stage with technologies based upon CMOS or HV-CMOS technology
being considered. The occupancy requirement can be met using
pixels of size $100$--$200 \mum$ in $x$ and up to $1 \cm$ in $y$.  
The number of detector layers needed is being
studied with layouts between 6 and 12 planes being considered. 
For \upgradetwo, as in the VELO, the tracking system could benefit from 
being at least partially equipped with timing information. This could
be particularly useful in helping to match upstream and downstream
track segments. 

\subsubsection{Magnet stations}

\begin{figure}[!htb]
\centering
\includegraphics[width=0.48\textwidth]{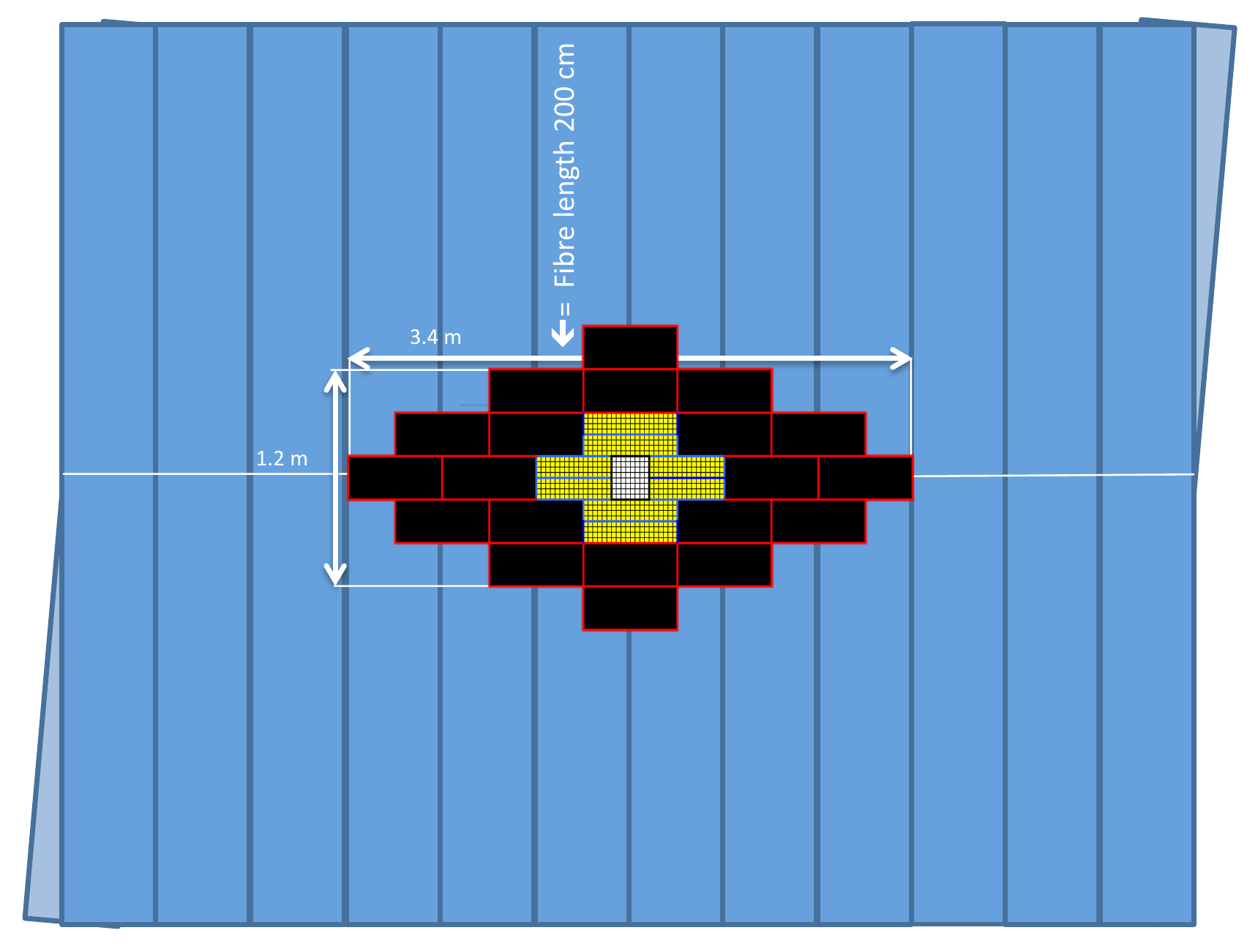}
\includegraphics[width=0.45\textwidth]{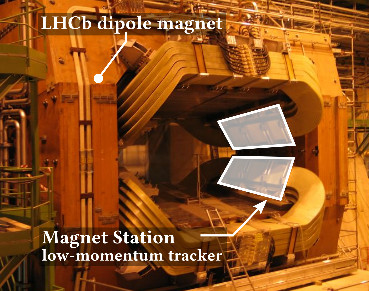}
\caption{\small (left) The proposed layout of a station of the \upgradetwo tracker with
scintillating fibres (blue), silicon middle tracker (black), and
silicon inner tracker (yellow). (right) The LHCb dipole magnet, with the white outline indicating the area to be covered by the
Magnet Station. A symmetrical module will cover the opposite face of the
magnet. }
\label{fig:Tracking}
\end{figure}

For Upgrade~Ib in LS3 it is proposed to extend the spectrometer coverage for
low momentum tracks (for example the slow pion from the decay $D^{*+}
\rightarrow D^0 \pi^+$) by instrumenting the internal surfaces of
the magnet with scintillating fibres, as shown in
Fig.~\ref{fig:Tracking} (right). A spatial resolution of the order of
a mm is sufficient in the bending-plane to obtain
the required momentum resolution. The use of a stereo arrangement of layers will be implemented
to achieve the required $y$ segmentation. Preliminary studies show that the pattern recognition
will be able to cope with the occupancy in \upgradetwo\ conditions.

%% file: CONTRIBUTIONS/2_Introduction/2.4.2.tex
\subsection{Particle identification}
\label{sec:PID}

High quality particle identification (PID) is essential for almost all precision flavour measurements.   The excellent performance of the PID subdetectors of the current experiment will be maintained for \upgradetwo, and in some cases enhanced.  A common theme of these developments will be improved granularity and, for certain subdetectors, fast timing of the order of a few tens of picoseconds, in order to associate signals with one, or a small number, of $pp$ interactions in the bunch crossing.   Here a brief overview of initial plans and early R\&D directions is presented.  A more extensive discussion can be found in the EoI~\cite{LHCb-PII-EoI}.

\subsubsection{Hadron identification: the RICH system and the TORCH}

The RICH system of \upgradetwo\ will be a natural evolution of the current detectors and those being constructed for \upgradeone.  There will be two counters, an upstream RICH~1 optimised for lower momentum tracks, and a downstream RICH~2, both occupying essentially the same footprint as now.  

In order to cope with the increased track multiplicity it will be necessary to replace the MaPMTs of \upgradeone\ with a new photodetector of higher granularity.  Several candidate technologies are under consideration, with SiPMs being a leading contender. Other possibilities include vacuum devices such as MCPs, HPDs and MaPMTs.  Fast timing is an additional desirable attribute in order to reduce the computing time required for the pattern recognition.  Active R\&D is being pursued into all of these options.

As well as reducing the occupancy it will be necessary to improve the Cherenkov angle resolution by around a factor of three in both counters with respect to the specifications of \upgradeone.  This goal can be achieved by redesigning the optics, for which a preliminary design already exists, ensuring that the response of the photodetectors is weighted towards longer wavelengths, and taking advantage of the smaller pixel size.  

There is an exciting possibility, under consideration, to enhance the low-momentum hadron-identification capabilities of the experiment  by installing a TORCH detector.  Such a detector measures time-of-flight through detecting internally reflected Cherenkov light produced in a thin ($\sim$1\,cm) quartz plane with MCP photodetectors.  A time resolution of 70\,ps per photon and an expected yield of $\sim$30 photons per track will allow for kaons to be positively identified in the region below $10 \gevc$, where currently they can only be selected by using the RICH  in `veto mode'.  Low-momentum proton identification would also become available.  These improvements would benefit flavour tagging, reconstruction of multi-body final states, physics with baryons and spectroscopy studies.  An R\&D programme has been ongoing for several years which has both developed the MCP technology and demonstrated proof-of-principle with a demonstrator module which is show in~ Fig.\ref{fig:PID}~(left).  A suitable location for the TORCH within LHCb would be upstream of RICH~2, and a half-sized prototype with this application in mind will soon be constructed.

\subsubsection{Electromagnetic calorimeter}

The electron, photon and $\pi^0$ identification provided by the current electromagnetic calorimeter (ECAL) has proved of great importance during Runs~1 and 2 of the LHC. A new and suitably designed ECAL will be essential  for obtaining high performance in certain key modes as well as ensuring a broad physics programme at \upgradetwo.

The principal challenges for the ECAL at \upgradetwo\ will be threefold.  Firstly, the radiation environment will be extremely severe, with a total dose of around 200\,Mrad foreseen for the innermost modules (although falling off rapidly with distance from the beam pipe).  Indeed, severe degradation in performance is already expected for these modules in the innermost  region of the detector by the time of LS3, which will then require replacement.  The choice of technology must therefore be very radiation hard. Secondly, the very high luminosity of $10^{34}\,{\rm cm^{-2}s^{-1}}$ operation will lead to overlapping showers and a corresponding degradation in energy resolution and shower finding efficiency.  This problem can be tackled by reducing the Moli\`{e}re radius of the converter and moving to a smaller cell size, for example 2$\times 2\,{\rm cm^{2}}$ in the inner region.  Finally, the high number of candidates in every event, for example of $\pi^0$ mesons, will lead to an unacceptably large combinatoric background.  Hence fast timing information will be essential to associate the candidates to individual $pp$ interactions in the bunch crossing.  

The baseline goal of the ECAL for \upgradetwo\ is to solve the above challenges and provide a performance that is at least as good as that of the current detector in Run~1 and Run~2 conditions.  Given that the modules of this existing detector will remain unchanged for Run~3, with only modifications to the gains and readout electronics,  it can be anticipated that its performance will degrade in Upgrade I due to the increased cluster overlap and higher combinatoric background.  This degradation will be recovered at \upgradetwo.   Other attributes of the new detector, such as excellent position resolution, and the possibility, under discussion, of accessing longitudinal-shower information, will bring additional benefits to the physics programme.

\begin{figure}[!t]
\centering
\includegraphics[width=0.3\textwidth]{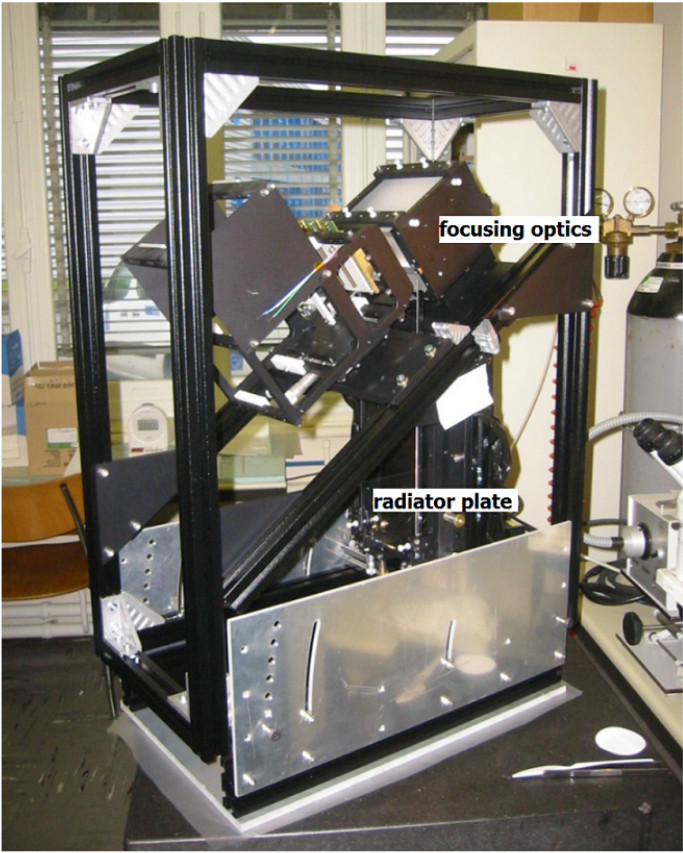}
\includegraphics[width=0.36\textwidth]{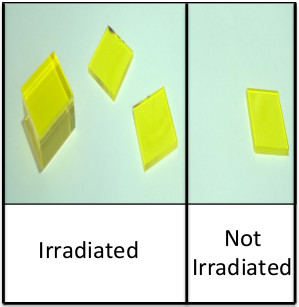}
\caption{\small (left) The small-scale TORCH module demonstrator showing the radiator and optics mechanical
vessel. (right) A promising multi-doped GAGG: Ce crystal which was recently irradiated and tested in the ECAL R\&D programme. }
\label{fig:PID}
\end{figure}

A vigorous R\&D programme has begun to investigate candidate technologies for the \upgradetwo\ ECAL.  Given the rapid variation of multiplicity and radiation flux with position it may well be the case that the final detector will make use of multiple solutions.  The mechanics and geometry of the detector allows for this possibility provided that the constituent cells can be grouped into modules of uniform dimensions  (12$\times 12\,{\rm cm^{2}})$.  One option being pursued is a homogeneous crystal calorimeter with longitudinal segmentation.  Promising materials are being investigated, see Fig.~\ref{fig:PID}~(right), which offer good radiation hardness, excellent energy resolution and very fast response.  Another possibility is a sampling calorimeter, either Shashlik or SpaCal, with a tungsten-alloy converter, with a crystal component for providing a fast-timing signal.  An alternative source of fast timing would be a preshower layer involving silicon pads.

Not all modules of the current ECAL will need to be replaced simultaneously.  It has long been known that some intervention will already be required on the innermost region in LS3 due to radiation damage.  The most important region to be overhauled is the central band across the whole detector, which encompasses the current inner region of approximate vertical dimensions 1.5\,m.

\subsubsection{The muon system}

Operating the current muon system in \upgradeone\ conditions would lead to a degraded performance due to dead-time induced inefficiencies and an increased rate of ghost hits, all coming from the higher background flux.  The problem is most acute in the innermost region of the front two stations: M2R1, M2R2, M3R1 and M3R2  (station M1, needed only for the hardware trigger, will be removed). Modifications are therefore being implemented to recover the performance.  These improvements comprise a modification to the electronics to increase the granularity, the replacement of certain detectors with new pad chambers, and additional shielding around the beam pipe.   This new system will perform well at luminosities of $2 \times 10^{33}\,{\rm cm^{-2}s^{-1}}$, but will not be adequate for the even more severe conditions of \upgradetwo.

At \upgradetwo, extra shielding will be required to suppress the flux of punchthrough to a manageable level.  This can be achieved by replacing the HCAL by up to 1.7\,m of iron, which would provide an additional four interaction lengths compared to the current situation.   (The primary role of the HCAL is to contribute to the hardware trigger, which will no longer be required in Upgrade~I and II.)   New detectors will be installed in the innermost region of all stations, with a design possessing both high granularity and high rate capabilities.  A promising solution is the micro-resistive WELL detector ($\mu$-RWELL~\cite{muRwell}), a novel MPGD with good prospects for satisfying these criteria.   In the lower flux region $\mu$-RWELL or new MWPC detectors could be installed.   R\&D is ongoing to study all of these possibilities.

%% file: CONTRIBUTIONS/2_Introduction/2.4.3.tex
\subsection{Trigger and data processing}

At an instantaneous luminosity of up to $2 \times 10^{34} \textrm{cm}^{-2} \textrm{s}^{-1}$,
the LHCb detector is expected to produce up to $400-500$~Tb of data per second, which will have to be processed
in real time and reduced by at least $4-5$ orders of magnitude before recording the remainder to permanent storage.
The ongoing evolution of radiation-hard optical links and commercial networking technology, combined with LHCb's
geometry, which places the readout cables outside the detector acceptance, is expected to allow moving this volume
of the data off the detector and into a processor farm by the time of Upgrade~II.
The essential challenge facing data processing in \upgradetwo\ is however that with up to $\sim 55$ interactions per bunch crossing, 
every bunch crossing will produce multiple heavy-flavour hadrons in the LHCb detector acceptance. For this reason,
a traditional trigger strategy of selecting bunch crossings based on generally interesting inclusive topologies (for
example a displaced vertex) will no longer be able to significantly reduce the data rate even at the earliest trigger levels. 
Instead, the \upgradetwo\ data processing will be based around pile-up suppression, in which detector
hits not associated with the individual $pp$ interaction of interest are discarded as early as possible in the processing chain.

Timing information allows for a particularly fast separation of reconstructed objects according to the $pp$ interaction that produced them, and the
availability of precise timing information in multiple subdetectors, as discussed above, is therefore crucial. 
It is equally important to fully reconstruct the charged particle trajectories and neutral particle clusters produced in the LHCb acceptance
as early as possible in the processing chain, so that the matching of specific heavy flavour signatures to $pp$ interactions can be performed in an optimal way.
LHCb has already demonstrated the ability to perform a full offline-quality detector alignment, calibration, and reconstruction 
in near-real-time in Run 2, as well as the ability to perform precision physics measurements using this real-time data processing.
Successfully scaling this problem to the much more challenging conditions of \upgradetwo\ will require LHCb to build a processor
farm out of whichever technology or technologies are most commercially viable in more than a decade from now. Given the ongoing trend
towards more and more heterogeneous computing architectures, with CPU server farms increasingly supplemented with GPU or FPGA accelerators,
it will be critical to continue and further develop collaborations with computing institutes specialising in the design of such hybrid architectures.
 R\&D has been undertaken on downstream tracking in FPGAs~\cite{LHCb-PII-EoI}, and such partnerships have already borne fruit in the context of the ongoing 
LHCb \upgradeone~\cite{DeCian:2309972,Badalov:2016bgr,Hugo:2017vxc} and will be particularly 
important in ensuring that LHCb's trigger and reconstruction algorithms are optimally tuned to take best advantage of the most affordable
architecture.

%% file: CONTRIBUTIONS/3_Time_dependent_CP_violation_measurements/3.tex

\input{CONTRIBUTIONS/3_Time_dependent_CP_violation_measurements/3.0.tex}
\input{CONTRIBUTIONS/3_Time_dependent_CP_violation_measurements/3.1.tex}

\input{CONTRIBUTIONS/3_Time_dependent_CP_violation_measurements/3.2.tex}

\input{CONTRIBUTIONS/3_Time_dependent_CP_violation_measurements/3.3.tex}

\input{CONTRIBUTIONS/3_Time_dependent_CP_violation_measurements/3.4.tex}

%% file: CONTRIBUTIONS/3_Time_dependent_CP_violation_measurements/3.0.tex
\label{chpt:TDCPV}
\label{section:TDCPV-intro}
Many of the most important observables to test the SM in the $B$ system are intrinsically related to the phenomenology of flavour oscillations: that a $\Bds$ meson can mix into its antiparticle $\Bdsb$ and vice versa.
Since this is a loop process, currently unknown physics may influence the $\Bds$--$\Bdsb$ transition amplitudes and lead to deviations from the predictions for relevant quantities assuming only SM contributions.

The mixing results in physical states, with well defined masses and lifetimes,
\begin{equation}
  \label{eq:physicalStates}
  B^0_{(\squark)\,\rm H} = p \Bds + q \Bdsb ~~ {\rm and} ~~
  B^0_{(\squark)\,\rm L} = p \Bds - q \Bdsb \, ,
\end{equation}
where $p$ and $q$ are complex coefficients that satisfy $|p|^2 + |q|^2 = 1$.
The physical states are labelled $B^0_{(\squark)\,\rm H}$ and $B^0_{(\squark)\,\rm L}$ to distinguish the heavier (H) from the lighter (L), and have mass difference $\Delta m_{\dquark\,(\squark)} = m_{B^0_{(\squark)\,\rm H}} - m_{B^0_{(\squark)\, \rm L}}$ and width difference $\Delta \Gamma_{\dquark\,(\squark)} = \Gamma_{B^0_{(\squark)\,\rm L}} - \Gamma_{B^0_{(\squark)\,\rm H}}$. 

The decay-time-dependent rates for a \B meson that is initially (at time $t=0$) known to be in the $B^0_{(\squark)}$ or $\overline{B}^0_{(\squark)}$ flavour eigenstate to decay into a final state $f$ are~\cite{HFLAV16,Gershon:2016fda}
\begin{eqnarray}
  \frac{d\Gamma_{B^0_{(\squark)} \to f}(t)}{dt} & \propto & e^{-\Gamma_{\dquark(\squark)}t} 
  \Big[
    \cosh(\frac{\Delta \Gamma_{\dquark(\squark)} t}{2}) + 
    A_{f}^{\Delta \Gamma_{\dquark(\squark)}} \sinh(\frac{\Delta \Gamma_{\dquark(\squark)} t}{2}) + \nonumber \\
    & & \hspace{30mm} C_f \cos(\Delta m_{\dquark(\squark)} t) - S_f \sin(\Delta m_{\dquark(\squark)} t)
    \Big] \, , \label{eg:tdcpv1} \\  
  \frac{d\Gamma_{\overline{B}^0_{(\squark)} \to f}(t)}{dt} & \propto & e^{-\Gamma_{\dquark(\squark)}t}
  \Big[
    \cosh(\frac{\Delta \Gamma_{\dquark(\squark)} t}{2}) +
    A_{f}^{\Delta \Gamma_{\dquark(\squark)}} \sinh(\frac{\Delta \Gamma_{\dquark(\squark)} t}{2}) - \nonumber \\
    & & \hspace{30mm} C_f \cos(\Delta m_{\dquark(\squark)} t) + S_f \sin(\Delta m_{\dquark(\squark)} t)
  \Big] \, , \label{eg:tdcpv2}
\end{eqnarray}
where 
\begin{equation}
  A^{\Delta \Gamma}_f \equiv - \frac{2\, \Re(\lambda_f)}{1 + |\lambda_f|^2} \, , ~~
  C_f \equiv \frac{1 - \left|\lambda_f\right|^2}{1 + \left|\lambda_f\right|^2} \, , ~~
  S_f \equiv \frac{2\, \Im(\lambda_f)}{1 + \left|\lambda_f\right|^2} \, .
  \label{eq:ACS-defs}
\end{equation}
The quantity $\lambda_{f}$ is defined in terms of the mixing parameters $p$ and $q$, and the amplitudes $\bar{A}_f$ ($A_f$) for the decay of a $\Bbar$ ($\B$) hadron to a final state $f$,
\begin{equation}
  \label{eq:lambda}
  \lambda_{f} = \frac{q}{p}\frac{\bar{A}_{f}}{A_{f}}\,.
\end{equation}
If $f$ is not a \CP-eigenstate, similar expressions hold for the conjugate final state $\bar{f}$, with parameters $C_{\bar{f}}$, $S_{\bar{f}}$ and $A^{\Delta \Gamma}_{\bar{f}}$. 
The decay-time-dependent asymmetry is
\begin{equation}
  \label{eg:tdcpv-asym}
  \frac{\Gamma_{\overline{B}^0_{(\squark)} \to f}(t) - \Gamma_{B^0_{(\squark)} \to f}(t)}
       {\Gamma_{\overline{B}^0_{(\squark)} \to f}(t) + \Gamma_{B^0_{(\squark)} \to f}(t)} = 
       \frac{S_f \sin(\Delta m_{\dquark(\squark)} t) - C_f \cos(\Delta m_{\dquark(\squark)} t)}
            {\cosh(\frac{\Delta \Gamma_{\dquark(\squark)} t}{2}) +
              A_{f}^{\Delta \Gamma_{\dquark(\squark)}} \sinh(\frac{\Delta \Gamma_{\dquark(\squark)} t}{2})} \, .
\end{equation}
The quantities $S_f$, $C_f$ and $A^{\Delta \Gamma}_f$ can, for certain choices of final state $f$, be related to angles of the unitarity triangle; in other cases they provide null tests of the SM.

The expressions of Eq.~(\ref{eg:tdcpv1})--Eq.~(\ref{eg:tdcpv2}) are used for all decay-time-dependent analyses of two-body \B\ meson decays.  
For more complicated final states (containing three or more pseudoscalars, or two or more vector particles) it is common to use an explicit expression for the variation of the amplitudes $\bar{A}_{f}$ and $A_{f}$ across the final state phase-space and to fit directly for the \CP-violating weak phases involved.  
It should also be noted that Eq.~(\ref{eq:physicalStates})--Eq.~(\ref{eg:tdcpv2}) assume that \CPT\ symmetry is conserved.  
Since this symmetry is fundamental to the SM, it should also be tested as precisely as possible, as can be done with \upgradetwo\ using the methods developed in Ref.~\cite{LHCb-PAPER-2016-005}.
Another common assumption in decay-time-dependent analyses is that $\left| \frac{q}{p} \right| = 1$, \ie\ absence of \CP\ violation in mixing; this will also be precisely tested with \upgradetwo\ as discussed in Sec.~\ref{sec:asl}.

\section{Measurements of mass and width differences}
\label{section:DmDG}

The values of $\Delta m_{\dquark\,(\squark)}$ and $\Delta \Gamma_{\dquark\,(\squark)}$ strongly affect the sensitivity with which the parameters $S_f$, $C_f$ and $A^{\Delta \Gamma}_f$ can be measured.
In the $\Bs$ system, the large value of $\Delta m_{\squark}$ means that excellent vertex resolution is necessary to resolve the flavour oscillations and hence to determine $S_f$ and $C_f$.
Potential improvements to the vertex resolution, as discussed in Sec.~\ref{sec:tracking}, may therefore improve the sensitivity to decay-time-dependent \CP-violation effects in the \Bs\ system beyond that from simple scaling with luminosity.

The world-leading measurements of both $\Delta m_{\dquark}$ and $\Delta m_{\squark}$ are from LHCb~\cite{LHCb-PAPER-2015-031,LHCB-PAPER-2013-006}, and can be improved further assuming that good flavour tagging performance can be maintained (see Sec.~\ref{section:FlavourTagging}).
This will not only reduce systematic uncertainties in \CP-violation measurements but will also provide a strong constraint on the length of one side of the unitarity triangle, although progress here is mainly dependent on improvements in lattice QCD calculations.

Regarding the width differences $\Delta \Gamma_{\dquark}$ and $\Delta \Gamma_{\squark}$, an important difference is that $\Delta \Gamma_{\squark}$ is relatively large while $\Delta \Gamma_{\dquark}$ has been considered negligibly small in most analyses to date.
The size of $\Delta \Gamma_{\squark}$ allows $A^{\Delta \Gamma}_f$ to be determined through either flavour-tagged or untagged analyses of \Bs\ decays (in the latter case, through the measurement of the effective lifetime), while in the \Bd\ system the $A^{\Delta \Gamma}_f$ parameters are usually unmeasurable. 
 
The decay-time-dependent angular analysis of $\Bs \to \jpsi\phi$ allows measurement of $\Delta \Gamma_{\squark}$ simultaneously with \CP-violation parameters.
Therefore, improved knowledge of $\Delta \Gamma_{\squark}$ will be obtained together with measurements of $\phi_s^{c\bar{c}s}$, as discussed in Sec.~\ref{sec:3.2.2}.
The ratio $\Delta \Gamma_{\dquark}/\Gamma_{\dquark}$, is typically measured from the difference in effective lifetimes between $\Bd$ decays to flavour-specific, namely $\jpsi \Kstarz$, and \CP-eigenstate, namely $\jpsi \KS$, final states~\cite{Gershon:2010wx}. 
Using $1\invfb$ of data, LHCb determined $\Delta \Gamma_{\dquark}/\Gamma_{\dquark} = -0.044\pm0.025\pm0.011$~\cite{LHCb-PAPER-2013-065} with this approach; ATLAS and CMS have also published competitive measurements~\cite{Aaboud:2016bro,Sirunyan:2017nbv}. 
The expected statistical uncertainty for \upgradetwo, taking into account the larger centre-of-mass energy and the increase in luminosity, is $\sigma(\Delta \Gamma_{\dquark}/\Gamma_{\dquark}) \sim 0.001$.
This can be compared with the SM prediction $\Delta \Gamma^{\rm SM}_{\dquark}/\Gamma^{\rm SM}_{\dquark}=(0.00397 \pm 0.00090)$~\cite{Artuso:2015swg}. 
Thus, if systematic uncertainties can be controlled sufficiently, it will be possible to measure a significantly non-zero value of $\Delta \Gamma_{\dquark}$ even if it is not enhanced above its SM prediction.

The challenge in controlling the systematic uncertainty is to understand precisely the decay-time acceptance difference between the two decay topologies. 
However, if the $\Bd$ vertex position is reconstructed identically, namely using only the $\jpsi$ decay products, and only $\KS$ mesons decaying inside the VELO are considered, the largest source of systematic uncertainty should cancel almost exactly and therefore not dominate the measurement. 
Similarly, the asymmetry in production rate between $\Bd$ and $\Bdb$ is expected to be precisely measured in independent control samples, and thus will not limit the achievable precision. 
In addition to the intrinsic interest in determining $\Delta \Gamma_{\dquark}$, precise knowledge of its value will benefit many other studies of $\Bd$ decay modes, since any systematic uncertainties associated with the assumption that $\Delta \Gamma_{\dquark} = 0$ can be removed.

Width differences between types of \bquark\ hadron such as $\Gamma_s - \Gamma_d$ are also of interest.
In addition to allow tests of the heavy-quark expansion framework used to make theoretical predictions of such quantities, their precise knowledge is important to control systematic uncertainties in measurements where a decay mode of one type of \bquark\ hadron is used as a control channel in studies of a decay of another.
Detailed understanding of the acceptance is necessary for such measurements, which can be achieved using topologically similar final states (see, for example, Ref.~\cite{LHCB-PAPER-2014-003}).
These measurements are therefore expected to be significantly improved with \upgradetwo.

%% file: CONTRIBUTIONS/3_Time_dependent_CP_violation_measurements/3.1.tex
\section{Flavour tagging at high pile-up}
\label{section:FlavourTagging}

In order to measure the $\Delta m$, $S_f$ and $C_f$ parameters, it is necessary to determine the initial flavour of the $\B$ meson, \ie\ to know whether it was produced as a \Bds\ or \Bdsb\ meson.
This is achieved through flavour tagging (FT) algorithms that exploit information from other particles produced in the $pp \to b\bar{b}X$ event.
At LHCb, these algorithms fall into one of two categories: those that involve measuring the flavour of the other \bquark\ hadron produced in the event (referred to as Opposite-Side tagging)~\cite{LHCb-PAPER-2011-027,LHCb-PAPER-2015-027} and those based on information from other particles associated with the hadronisation of the signal $b$ or $\bar{b}$ quark (Same-Side tagging)~\cite{LHCb-PAPER-2016-039,LHCb-PAPER-2015-056}.   
These algorithms typically combine various sources of information using multivariant techniques, and provide as output:
\begin{itemize} 
\item a discrete decision indicating whether the initial flavour is more likely to be $\B$ or $\Bbar$, or if it cannot be tagged ($q = +1, -1, 0$, respectively); 
\item an estimated probability, $\eta$, that the decision of $q$ (if non-zero) is wrong, known as the mistag fraction.
\end{itemize}
The outputs of Opposite-Side (OS) and Same-Side (SS) taggers can also be combined statistically to form a single decision.  
The flavour tagging performance at LHCb depends strongly on properties of the $pp \to b\bar{b}X$ event including the kinematics of the signal \B meson and the pile-up from other $pp$ interactions in the same bunch crossing.
Therefore the performance can differ significantly between modes, making it is necessary to calibrate the tagging output with an appropriate control sample, typically a flavour-specific \B\ decay selected with similar criteria to the signal mode.  

\begin{figure}[!tb]
\centering
\includegraphics[width=0.45\linewidth]{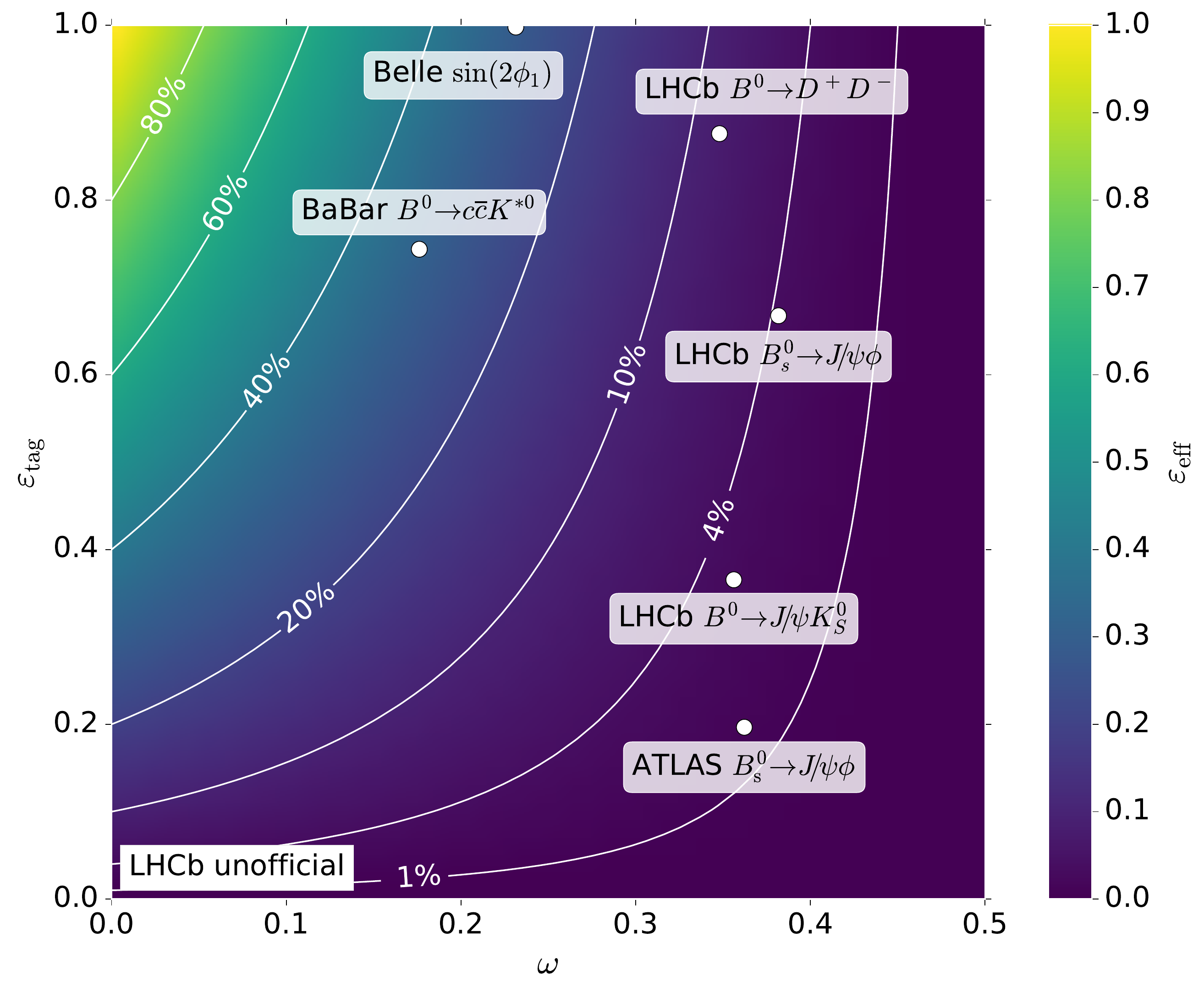}
\includegraphics[width=0.45\linewidth]{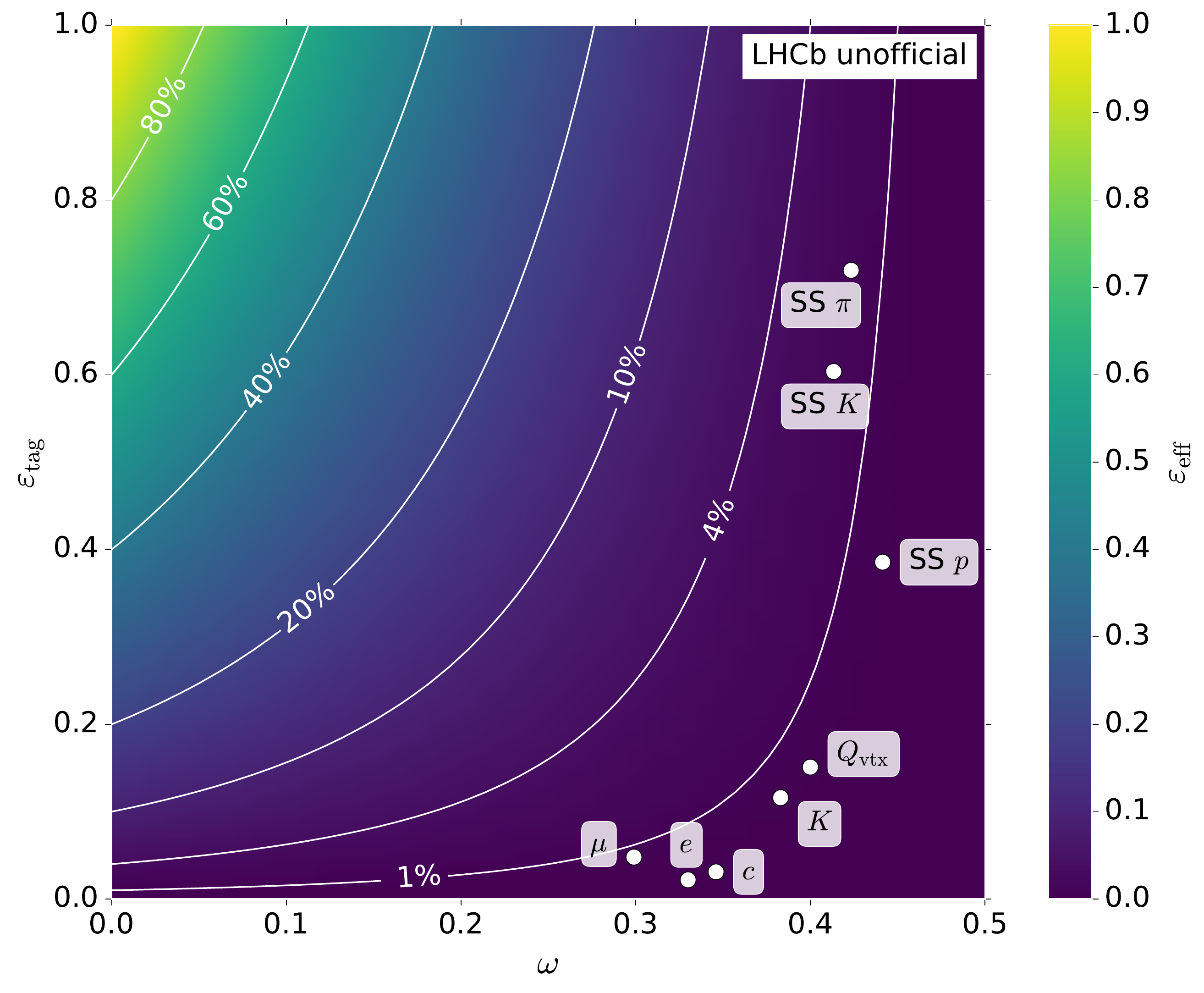}
\caption{\small 
  Effective tagging efficiency of (left) different HEP experiments and (right) LHCb flavour tagging algorithms~\cite{Heinicke:2229990}. 
  The white lines indicate contours of constant tagging power.
}
\label{fig:TagPowerPerExperiment}
\end{figure}

The sensitivity of a measured \CP asymmetry is directly related to the effective tagging efficiency $\epsilon_{\rm eff}  = \epsilon_{\rm tag} \times (1-2\omega)^{2}$, where $\epsilon_{\rm tag}$ is the fraction of candidates for which a non-zero value of $q$ is assigned and $\omega$ is the average calibrated value of $\eta$.
Figure~\ref{fig:TagPowerPerExperiment} shows the effective tagging efficiency achieved in different experiments and, at LHCb, with different channels, as well as the performance of the individual FT algorithms.

\begin{figure}[!tb]
\centering
\includegraphics[width=0.45\linewidth]{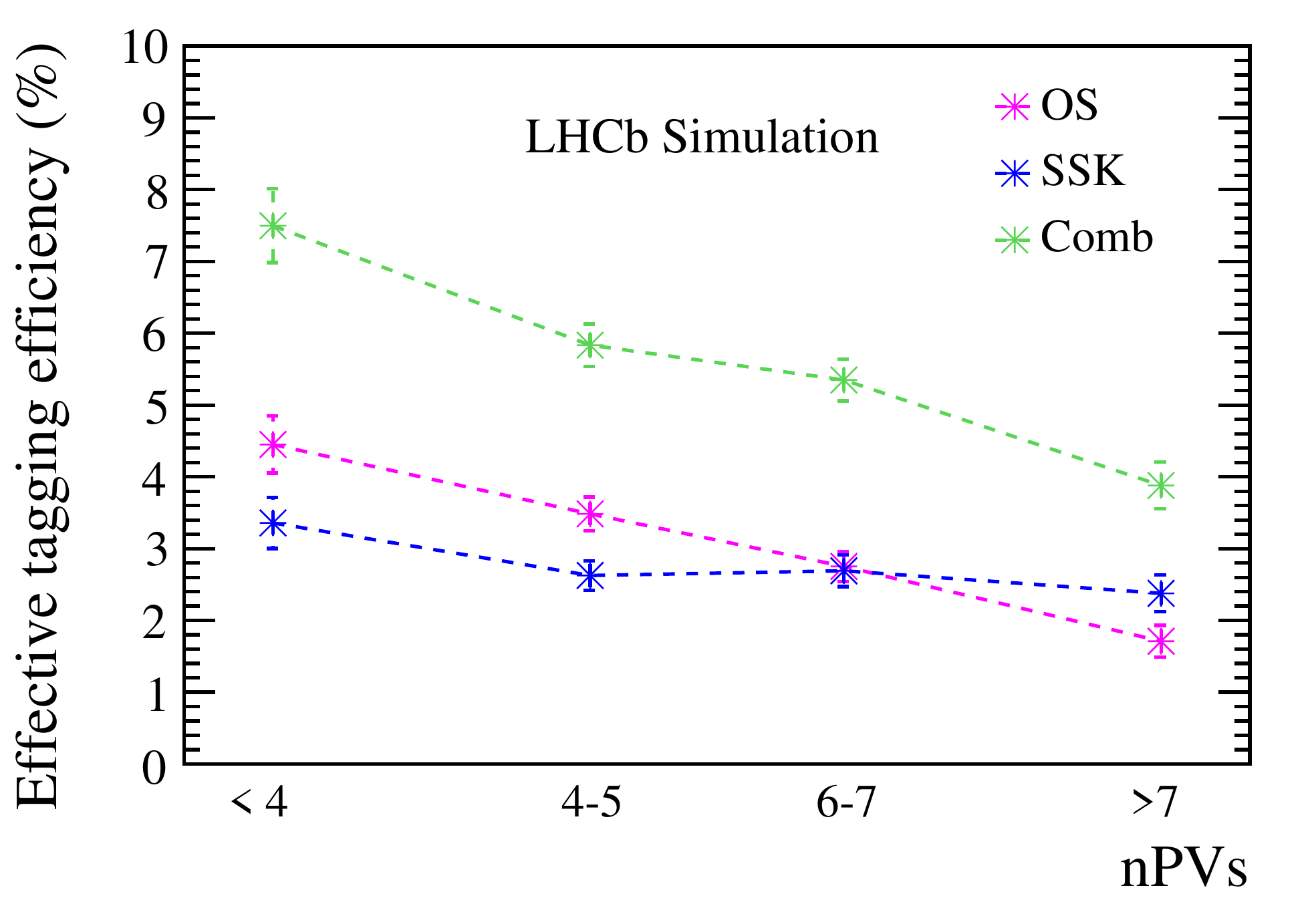}
\includegraphics[width=0.45\linewidth]{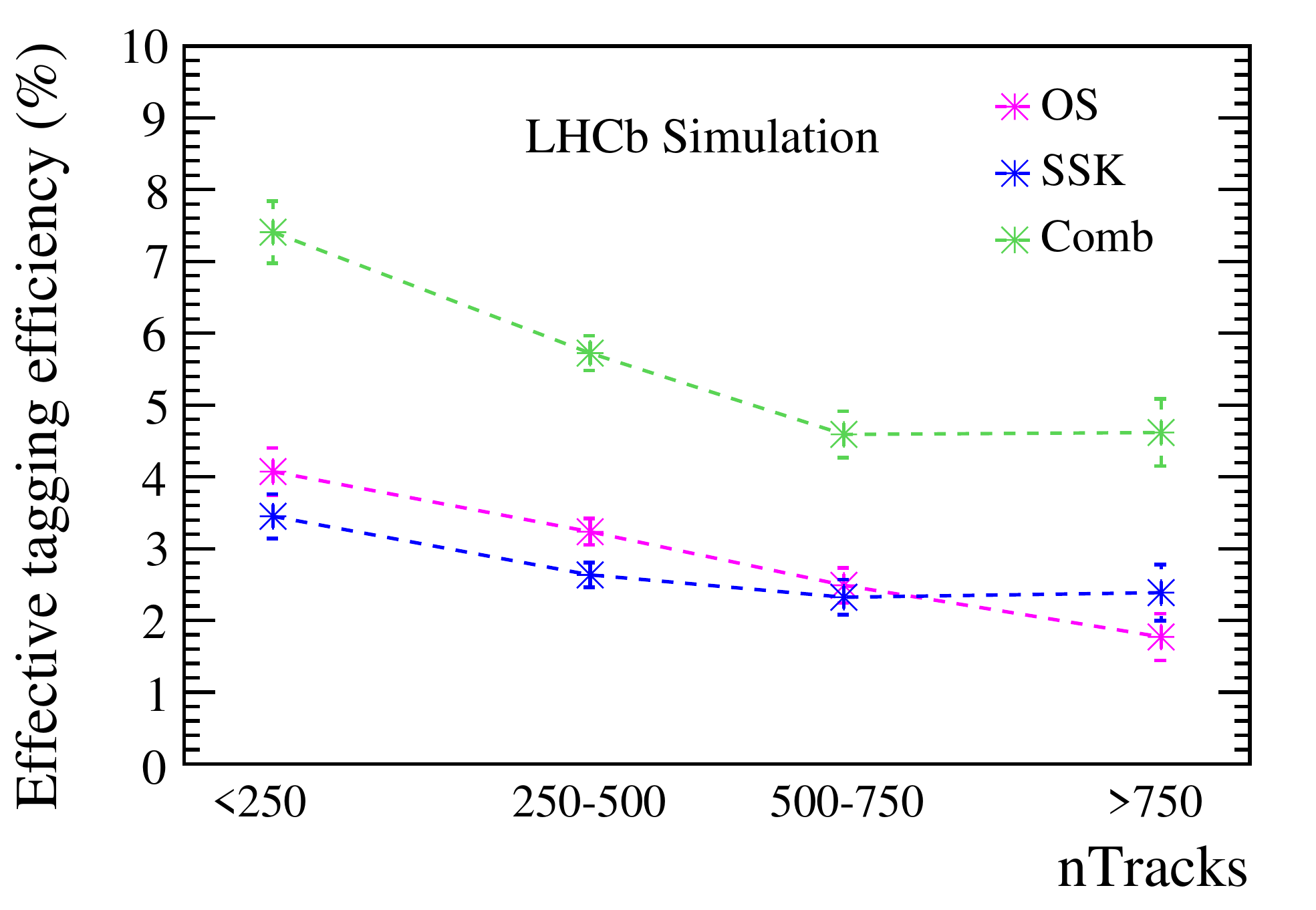}
\caption{\small 
  Effective tagging efficiency of OS and SS kaon taggers, and their combination, (left) in bins of pile-up vertices and (right) in bins of track multiplicity.
  These results are obtained from \upgradeone\ simulation of $\Bs \to \Dsm \pip$ decays. 
  The OS performances correspond to those obtained from combination of the individual OS taggers.}
\label{fig:TagPowerPerPV}
\end{figure}

The challenging environment associated with $pp$ collisions at LHCb, and the absence of the quantum correlations associated with $\PUpsilon(4S) \to B\Bbar$ decays, makes it impossible to achieve the same FT performance as at \epem\ colliders~\cite{Aubert:2009aw,Kakuno:2004cf}.
Moreover, it becomes increasingly difficult to obtain good FT performance at higher instantaneous luminosity, due to the increased pile-up, as shown in Fig.~\ref{fig:TagPowerPerPV}.
The largest degradation occurs for OS taggers, due to the increased probability to associate tracks with a different primary vertex.
Since the pile-up during \upgradetwo\ operation will be much larger than the range considered in Fig.~\ref{fig:TagPowerPerPV}, it will be critical to counter this effect through the inclusion of timing information in the detector.
As discussed in Sec.~\ref{sec:tracking}, if $50$--$100 \ps$ time resolution can be obtained in the VELO, the PV misassociation rate can be kept down to $\sim5\%$, and comparable FT performance to that achieved with the current detector can be expected.
Further improvement in FT performance may be achieved by using more sophisticated multivariate techniques, from better understanding of the hadronisation processes, and from additional information from new instrumentation in the \upgradetwo\ detector.
In particular, particle identification for low momentum tracks from the TORCH detector, and additional acceptance for tracks through magnet side stations should both help.
Therefore, it is assumed in the remainder of this section that the FT performance from existing LHCb results can be maintained for \upgradetwo, although detailed simulation studies will be necessary for a precise quantification.



%% file: CONTRIBUTIONS/3_Time_dependent_CP_violation_measurements/3.2.tex
\section{Measurements of $\phi_s$ and $\phi_d$ in theoretically clean modes}
\input{CONTRIBUTIONS/3_Time_dependent_CP_violation_measurements/3.2.2.tex}
\input{CONTRIBUTIONS/3_Time_dependent_CP_violation_measurements/3.2.1.tex}

\input{CONTRIBUTIONS/3_Time_dependent_CP_violation_measurements/3.2.3.tex}
\input{CONTRIBUTIONS/3_Time_dependent_CP_violation_measurements/3.2.4.tex}

%% file: CONTRIBUTIONS/3_Time_dependent_CP_violation_measurements/3.2.2.tex
\subsection{$\phi_{\squark}$ from $\Bs \to \jpsi \phi$ and related modes}
\label{sec:3.2.2}

\begin{figure}[!tb]
\centering
\includegraphics[width=0.51\linewidth]{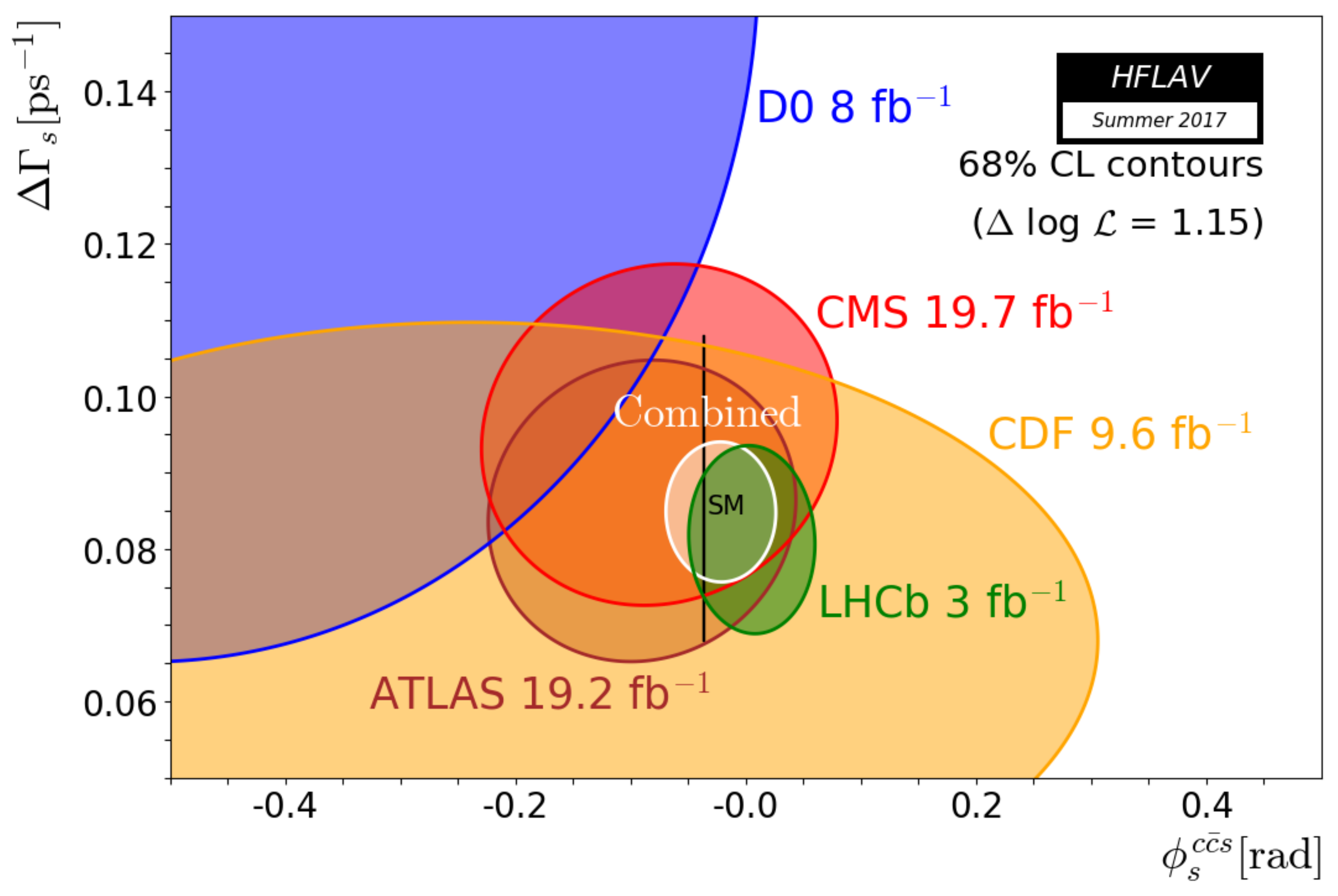}
\includegraphics[width=0.47\linewidth]{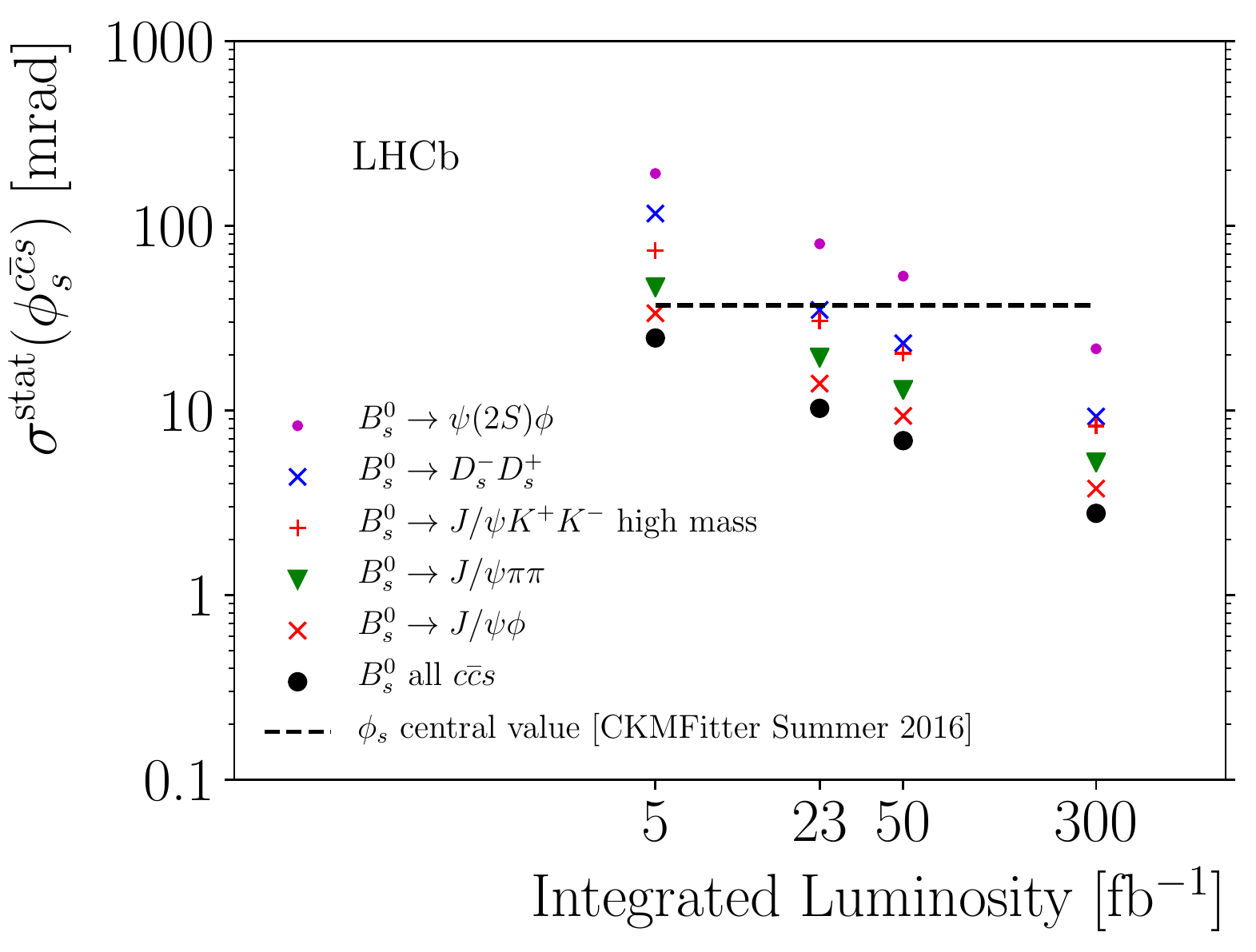}
    \caption{\small Left: Global HFLAV average of $\phi_{\squark}$ and $\Delta\Gamma_{\squark}$ from a
    variety of experiments~\cite{HFLAV16}. Right: Scaling of the statistical precision on $\phi_{\squark}$ from several
    tree-dominated \Bs meson decay modes.}
\label{fig:32:hflav}
\end{figure}

Measurements of decay-time-dependent \CP asymmetries in the \Bs system using
$b\to c\overline{c}s$ transitions are sensitive to the CKM phase $\beta_{\squark} \equiv \arg\left[ -(V^{\phantom{*}}_{ts}V_{tb}^*)/(V^{\phantom{*}}_{cs}V_{cb}^*)\right]$.
If penguin loop contributions to the decay can be neglected (see Sect.~\ref{sec:3.2.3}),
then the experimentally observable quantity is the phase, $\phi_{\squark}^{c\bar{c}s} = -2\beta_{\squark}$,
which has a precise SM prediction of $-36.4\pm 1.2$\mrad
based upon global fits to experimental data~\cite{Charles:2015gya}.
Deviations from this value would be
a clear sign of physics beyond the SM, strongly motivating the need for more
precise experimental measurements.

The single most statistically sensitive measurement $\phi_{\squark}^{c\bar{c}s}$ is given by the flavour-tagged decay-time-dependent angular analysis of the $\Bs\to \jpsi(\mu^+\mu^-)\phi(\Kp\Km)$ decay~\cite{LHCb-PAPER-2014-059}. 
This channel has a relatively high branching fraction and the presence of two muons in the final state leads to a high trigger efficiency at hadron colliders. 
Moreover, particle-identification criteria can be used in LHCb to suppress backgrounds efficiently, resulting in high sample purity (signal to background ratio of about 50 in the signal region of $\pm 20 \mevcc$ around the nominal $\Bs$ mass).
The LHCb detector has excellent time resolution ($\sim 45 \fs$) and good tagging power ($\sim 4\%$), both of which are crucial for a precision measurement. 
Angular analysis is necessary to disentangle the interfering \CP-odd and \CP-even components in the final state, which arise due to the relative angular momentum between the two vector resonances.  
In addition, there is a small ($\sim 2\%$) \CP-odd $\Kp\Km$ S-wave contribution that must be accounted for.  
To do this correctly requires detailed understanding of any variation of efficiency with angular variables and $\Kp\Km$ invariant mass.

Figure~\ref{fig:32:hflav}~(left) shows the current global average value of $\phi_{\squark}^{c\bar{c}s}$ and $\Delta\Gamma_{\squark}$, which are determined simultaneously from fits to $\Bs\to\jpsi\phi$ and, in the case of LHCb, $\Bs\to\jpsi\pip\pim$ data. 
The precision of the world average is dominated by the LHCb measurement which itself is dominated by the result using $\Bs\to\jpsi\phi$.
The averages are consistent with SM predictions~\cite{Charles:2015gya,Artuso:2015swg}, but there remains space for new physics contributions of ${\cal O}(10\%)$.
As the experimental precision improves it will be essential to have good control over possible hadronic effects~\cite{Faller:2008gt,Bhattacharya:2012ph} that could mimic the signature of beyond-the-SM physics (see Sect.~\ref{sec:3.2.3}).


Having multiple independent precision measurements is important since it allows not simply to improve the precision of the average, but also to perform a powerful consistency check of the SM.
One important way in which this can be done is by allowing independent \CP-violation effects for each polarisation state in the $\Bs\to\jpsi\phi$.
This has been done as a cross-check in the Run~1 analysis~\cite{LHCb-PAPER-2014-059}, but this strategy will become the default in \upgradetwo.
Additional information can be obtained from $\Bs\to\jpsi\Kp\Km$ decays with $\Kp\Km$ invariant mass above the $\phi(1020)$ meson, where higher spin $\Kp\Km$ resonances such as $f^\prime_2(1525)$ meson contribute~\cite{LHCb-PAPER-2017-008}.  
Among other channels, competitive precision can be obtained with $\Bs\to\jpsi\pip\pim$ decays~\cite{LHCb-PAPER-2014-019}, which have been found to be dominated by the \CP-odd component.
The $\Bs\to\Dsp\Dsm$~\cite{LHCb-PAPER-2014-051} and $\Bs\to\psi(2S)\phi$~\cite{LHCb-PAPER-2016-027} modes have also been studied with LHCb, and give less precise but still important complementary results.
Other channels, which have not been exploited yet but could be important in \upgradetwo\ if good calorimeter performance can be achieved, include $\Bs\to\jpsi\phi$ with $\jpsi \to \epem$ and $\Bs\to\jpsi\eta^{(\prime)}$ with $\etapr \to \rhoz\gamma$ or $\eta\pip\pim$, and $\eta \to \pip\pim\piz$ or $\gamma\gamma$~\cite{LHCb-PAPER-2014-056,LHCb-PAPER-2016-017}.

The scaling of the $\phi_{\squark}^{c\bar{c}s}$ precision with integrated luminosity for individual decay modes and for their combination is shown in Fig.~\ref{fig:32:hflav}~(right). 
These uncertainties are statistical only and are scaled from existing results, taking into account the gain in trigger efficiency expected for $\Bs\to\Dsp\Dsm$ after \upgradeone.
Maintaining the current performance will put stringent constraints on the design of the detector as regards momentum and vertex position resolution as well as particle identification performance.
A key ingredient is the flavour tagging that is very sensitive to event and track multiplicity, as discussed in Sec.~\ref{section:FlavourTagging}.
Systematic uncertainties are mainly based on the sizes of control samples, and are therefore expected to remain subdominant even with very large samples.
Therefore, it is expected that the small value of $-2\beta_{\squark}$ predicted in the SM can be measured to be significantly non-zero in several channels.

\begin{figure}[tb]
  \centering
  \includegraphics[width=0.50\textwidth]{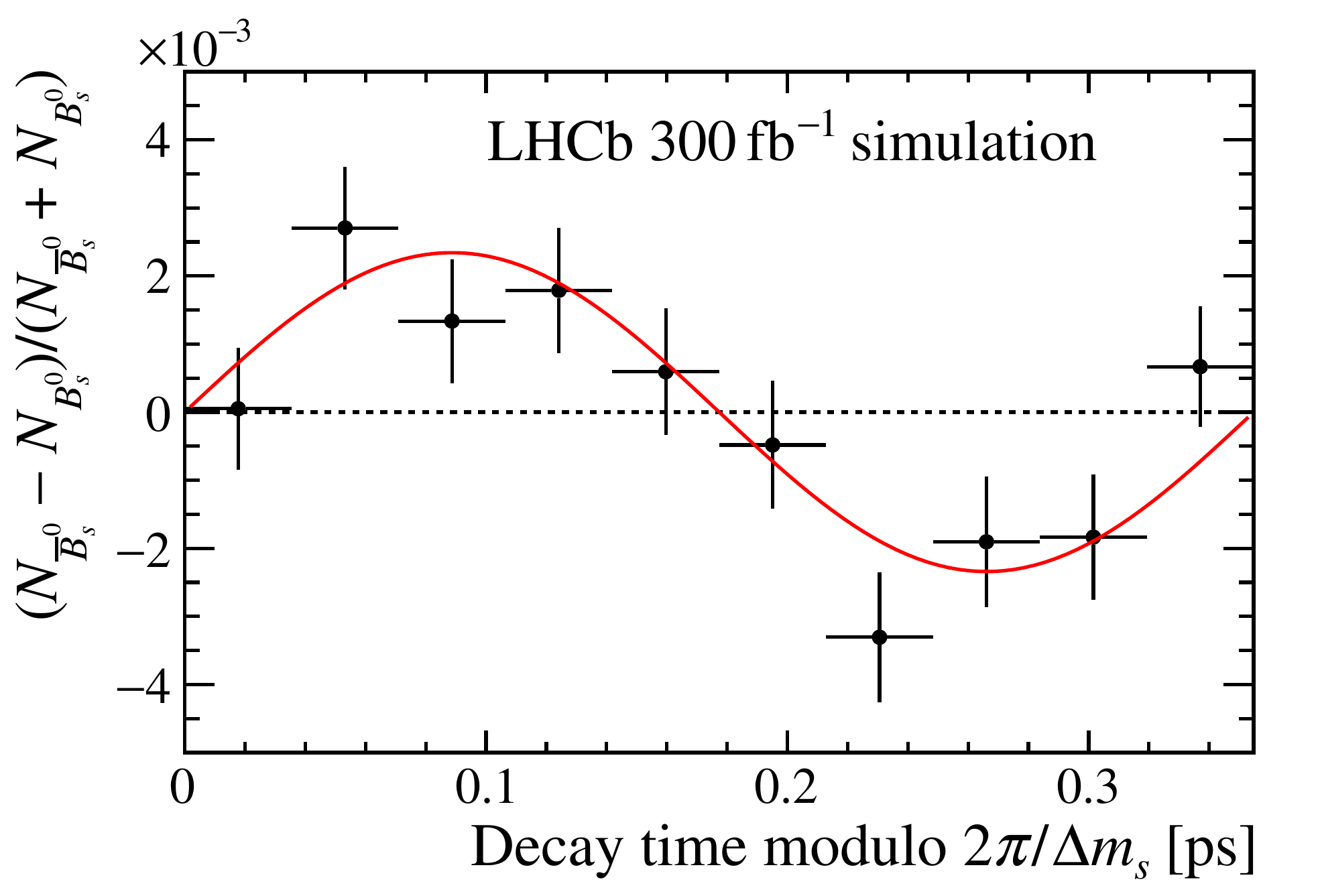}
  \includegraphics[width=0.48\textwidth]{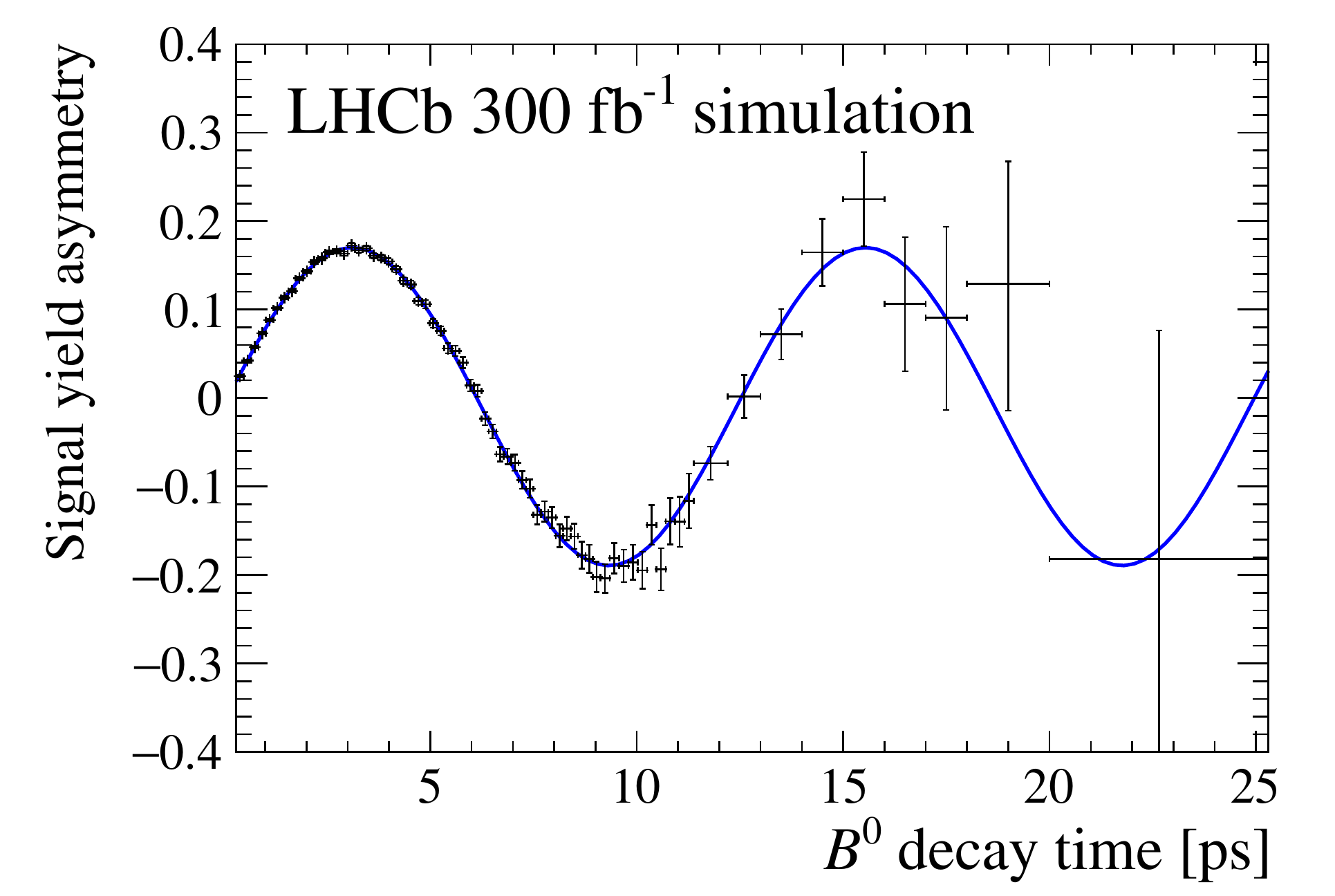}
  \caption{\small 
    Signal-yield asymmetry as a function of the $\Bds$ decay time, $(N_{\Bdsb}-N_{\Bds})/(N_{\Bdsb}+N_{\Bds})$. 
    Here, $N_{\Bdsb}$ ($N_{\Bds}$) is the number of (left) $\Bs\to\jpsi\phi$ or (right) $\Bz\to\jpsi\KS$ decays with a $\Bdsb$ ($\Bds$) flavour tag. 
    The data points are obtained from simulation with the expected sample size at $300\invfb$, and assuming the current performance of the LHCb experiment. 
    The solid curves represents the expected asymmetries for $\phi^{c\bar{c}s}_{\squark}=-36.4\mrad$~\cite{Charles:2015gya} and $\sin\phi^{c\bar{c}s}_{\dquark}=0.731$~\cite{LHCb-PAPER-2015-004}), the values used in the simulation.
    The height of the oscillation is diluted from $\sin\phi^{c\bar{c}s}_{\dquark(\squark)}$ due to mistagging, decay time resolution, and (for $\Bs\to\jpsi\phi$) the mixture of \CP-even and \CP-odd components in the final state.
  }
  \label{fig:Bx2JpsiXs_asymmetry_U2}
\end{figure}

The expected precision on $\phi_{\squark}^{c\bar{c}s}$ after \upgradetwo\ will be $\sim 4 \mrad$ from $\Bs\to\jpsi\phi$ decays alone and $\sim 3 \mrad$ from all modes combined.
This will be at the same level as the current precision on the indirect determination based on the CKM fit using tree-level measurements (this in turn is expected to improve with better measurements of other CKM matrix parameters).  
Figure~\ref{fig:Bx2JpsiXs_asymmetry_U2}(left) shows the signal-yield asymmetry as a function of the $\Bs$ decay time, folded at the frequency of $\Bs$ oscillations, for $\Bs\to\jpsi\phi$ decays from a simulated data set corresponding to $300\invfb$, and clearly shows that a visible \CP-violation effect will be observable.
The excellent precision on $\phi_{\squark}^{c\bar{c}s}$ that can be achieved with \upgradetwo\ gives exciting potential to observe deviations from the SM prediction, and in their absence will be used to impose severe constraints on possible beyond-the-SM contributions. 

%% file: CONTRIBUTIONS/3_Time_dependent_CP_violation_measurements/3.2.1.tex
\subsection{$\phi_d$ from $B^0 \to \jpsi \KS$ and related modes}
\label{sec:3.2.1}

Decay modes such as $B^0 \to \jpsi \KS$ are for the $\Bd$ system what $\Bs \to \jpsi\phi$ is to the $\Bs$ system.
In the limit that only tree-level $b \to c\bar{c}s$ amplitudes contribute, the \CP\ violation parameters of the decay-time distribution are given by $C_{\jpsi\KS} = 0$ and $S_{\jpsi\KS} = - \eta_{\CP} \sin(2\beta)$, where $\eta_{\CP}$ is the \CP eigenvalue of the final state ($-1$ for $\jpsi\KS$) and $\beta \equiv \arg\left[ -(V^{\phantom{*}}_{cd}V_{cb}^*)/(V^{\phantom{*}}_{td}V_{tb}^*) \right]$ is the angle of the unitarity triangle.
Here we denote the measured quantity as $\sin\phi_{\dquark}^{c\bar{c}s}$ for consistency with the \Bs\ case and to allow for the possibility of non-negligible penguin contributions, as discussed in the next section.
The CKM fit gives a SM prediction of $\sin2\beta =0.740\,^{+0.020}_{-0.025}$~\cite{CKMfitter2005,UTfit-UT}, consistent although in slight tension with the current world average $\sin\phi_{\dquark}^{c\bar{c}s} = 0.699 \pm 0.017$~\cite{HFLAV16}.


The world average is dominated by results from BaBar, Belle and LHCb using $\Bz\to\jpsi\KS$ decays.
The single most precise measurement is from Belle ($\sin\phi_{\dquark}^{c\bar{c}s} = 0.670 \pm 0.029 \pm 0.013$~\cite{Adachi:2012et}), with which the LHCb result has competitive uncertainty ($0.731 \pm 0.035 \pm 0.020$~\cite{LHCb-PAPER-2015-004}). With $50\invfb$ of data from \upgradeone, LHCb will reach a precision on $\sin\phi_{\dquark}^{c\bar{c}s}$ of about $0.006$ with $B^0 \to \jpsi \KS$ decays. The \belletwo\ experiment is expected to achieve a precision of about $0.005$ after accumulating $50 \invab$~\cite{Gaz:2017fgl}.  After \upgradetwo, LHCb will be able to reach a statistical precision of below $0.003$. 
Figure~\ref{fig:Bx2JpsiXs_asymmetry_U2}(right) shows the signal-yield asymmetry as a function of the $\Bz$ decay time for $\Bz\to\jpsi\KS$ decays from a simulated data set corresponding to $300\invfb$.
The impact of this measurement on the unitarity triangle fit is shown in Fig.~\ref{fig:UTprojection}.

The majority of systematic uncertainties on $\sin\phi_{\dquark}^{c\bar{c}s}$ depend on the size of control samples, and are therefore not expected to be limiting. 
At this level of precision, however, it will be necessary to understand possible biases on the result due to \CP violation in \Kz--\Kzb mixing and from the difference in the nuclear cross-sections in material between \Kz and \Kzb states, and therefore some irreducible systematic uncertainties are unavoidable. 
It is notable that the leading sources of systematic uncertainty are different between \belletwo\ and LHCb, so that having measurements from both experiments will be important.  
As for the $\phi_s^{c\bar{c}s}$ case, continued good flavour tagging performance (discussed in Sec.~\ref{section:FlavourTagging}) and improved understanding of subleading contributions to the decay amplitudes (discussed in Sec.~\ref{sec:3.2.3}) will be required.

%% file: CONTRIBUTIONS/3_Time_dependent_CP_violation_measurements/3.2.3.tex
\subsection{Control of penguin pollution}
\label{sec:3.2.3}

The $b\to c\bar{c}s$ transition involves not only tree-level contributions, but also gluonic and electroweak penguins, as well as exchange and penguin annihilation topologies, which are collectively referred to as ``penguin pollution''. 
With increasing precision in the \upgradetwo\ era, control of higher order corrections to $\phi_d^{c\bar{c}s}$ and $\phi_s^{c\bar{c}s}$ measurements will become mandatory.
Estimations of these corrections~\cite{Liu:2013nea, Frings:2015eva} and strategies to determine them directly from data using \grpsuthree\ flavour symmetry have been proposed~\cite{Fleischer:1999zi, Fleischer:2006rk, Faller:2008gt, Jung:2012mp, DeBruyn:2014oga}.
%
For example, the decay $\Bd\to\jpsi\rho^0$ proceeds through the same leading and higher order diagrams as the decay $\Bs\to\jpsi\phi$, when the quarks $s$ and $d$ are interchanged and the $\phi$ meson is replaced by a $\rho$ meson. 
But in $\Bd\to\jpsi\rho^0$ decays, the penguin pollution is not suppressed with respect to the tree amplitude, making it easier to measure and providing a good approximation of the expected contributions in the golden mode $\Bs\to\jpsi\phi$. 

These strategies have already been tested using $\Bd\to\jpsi\rho^0$~\cite{LHCb-PAPER-2014-058} and $\Bs\to\jpsi\Kstarzb$~\cite{LHCb-PAPER-2015-034} decays.
The best constraint on penguin pollution in $\phi_s^{c\bar{c}s}$ is currently obtained from the $\Bd\to\jpsi\rho^0$ channel, benefiting from the fact that $S_{\jpsi\rhoz}$ and $C_{\jpsi\rhoz}$ have been measured, giving a constraint on $\phid^{\jpsi\rhoz}$.
Following Ref.~\cite{LHCb-PAPER-2014-058}, the bias on \phis is 
\begin{equation}
\Delta\phi_{s}^{c\bar{c}s} \approx - \epsilon \left( \phid^{\jpsi\rhoz} - 2\beta \right) 
\label{eq:penguinCorrection}
\end{equation}
where $\epsilon =  \left|\Vus\right|^2 / (1 - \left|\Vus\right|^2) = 0.0534$ and $2\beta$ is mainly determined by $\Bz \to \jpsi \KS$ as discussed in Sec.~\ref{sec:3.2.1}. 
Since $2\beta$ is (and will continue to be) determined precisely, the sensitivity on $\Delta\phi_{s}^{c\bar{c}s}$ will be driven by the precision on $\phid^{\jpsi\rhoz}$.
Scaling the uncertainties obtained from Ref.~\cite{LHCb-PAPER-2014-058}, the expected statistical precision on $\phid^{\jpsi\rhoz}$ will be $\lesssim 1^\circ$ with $300 \invfb$. 
It is expected that systematic uncertainties, such as those from modelling the S-wave component in $\Bd \to \jpsi\pip\pim$ decays, can be kept under control, so that the uncertainty due to penguin pollution is expected to be $\lesssim 1.5\mrad$.
Thus, this is not expected to limit the sensitivity of the \phis measurement with $\Bs\to\jpsi\phi$.
However, if significant effects of penguin pollution become apparent it may complicate the combination of results from different modes, since a separate correction will be required for each.
For some modes such as $\Bs \to \jpsi f_0(980)$ this may be challenging, since the identification of the states in the \grpsuthree\ multiplet is not trivial; for $\Bs \to \Dsp\Dsm$ however only the U-spin subgroup is needed to relate the decay to $\Bd \to \Dp\Dm$ which can be used to control penguin pollution~\cite{Fleischer:2007zn,Jung:2014jfa,Bel:2015wha,LHCb-PAPER-2016-037}.




Similar strategies can be applied using $\Bs \to \jpsi \KS$ and $\Bd \to \jpsi \pi^0$ decays to control the penguin contributions in $\phi_d^{c\bar{c}s}$~\cite{Ciuchini:2005mg,Faller:2008zc,DeBruyn:2014oga}. 
A first analysis of $\Bs \to \jpsi \KS$ decays has been performed~\cite{LHCb-PAPER-2015-005} as a proof of concept for constraining $\Delta\phi_d^{c\bar{c}s}$ with larger datasets. 
The \CP\ violation parameters in $\Bd \to \jpsi \pi^0$ have been previously measured by BaBar and Belle~\cite{Aubert:2008bs,Lee:2007wd}, and the \belletwo\ experiment is expected to reach a sensitivity to $S_{\jpsi\piz}$ of $\sim 0.03$, which should be sufficient to keep penguin pollution under control.
LHCb can also study $\Bd \to \jpsi \pi^0$ decays, although the presence of a neutral pion in the final state makes the analysis more challenging and there is currently no public result from which to extrapolate the sensitivity.  
Improving the capabilities of the ECAL will enhance prospects for studying this mode.

It should be stressed that the methods to constrain penguin pollution rely on \grpsuthree\ symmetry, and the approximations associated with the method and inherent in Eq.~(\ref{eq:penguinCorrection}) must also be investigated.  
This can be done by studying the full set of modes related by \grpsuthree, namely $\Bds \to \jpsi \left\{ \piz, \eta, \etapr, \Kz, \Kzb \right\}$ and $\Bpm \to \jpsi \left\{ \pipm, \Kpm \right\}$ for the vector-pseudoscalar final state and $\Bds \to \jpsi \left\{ \rhoz, \omega, \phi, \Kstarz, \Kstarzb \right\}$ and $\Bpm \to \jpsi \left\{ \rhopm, \Kstarpm \right\}$ for the vector-vector final state.
Several of these modes have not yet been measured, but with the data sample of \upgradetwo\ it should be possible to measure all branching fractions and \CP\ asymmetry parameters, allowing a full theoretical treatment and more detailed understanding of subleading contributions.




%% file: CONTRIBUTIONS/3_Time_dependent_CP_violation_measurements/3.2.4.tex
\subsection{$\phi_d$ from $\Bz \to \D \pi^+ \pi^-$}

The decay $\Bzb \to \Dz \pi^+ \pi^-$ is a $b\to c\overline{u}d$ quark-level process that proceeds via tree-level processes only.
When the neutral $\D$ meson is reconstructed in a final state accessible to both \Dz\ and \Dzb\ decays, such as a \CP\ eigenstate, the direct decay can interfere with that via mixing, \ie\ $\Bzb \to \Bz \to \Dzb \pi^+ \pi^- \to \D_{\CP} \pi^+ \pi^-$, and can therefore be used to measure $\phi_d= 2\beta$ without any penguin pollution. 
In the SM, the value of $\phi_d$ is independent of the method used to determine it, so a difference between measurements from different quark-level transitions could indicate new physics effects~\cite{Grossman:1996ke,Fleischer:1996bv,London:1997zk,Ciuchini:1997zp}. 
A precise measurement of $\phi_d$ from $b\to c\overline{u}d$ transitions is therefore well motivated.

BaBar and Belle have performed measurements using $B^0 \to D^{(*)}h^0$ with both $D$ decays to \CP\ eigenstates~\cite{Abdesselam:2015gha} and to the three-body $\KS \pi^+\pi^-$ final state~\cite{Adachi:2018itz,Adachi:2018jqe}, where $h^0$ is a light neutral meson such as \piz.
The combined results are  $\sin{\phi_d} = 0.71 \pm 0.09$ and $\cos{\phi_d} = 0.91 \pm 0.25$ including all uncertainties~\cite{HFLAV16}.
While $B^0 \to Dh^0$ can also be studied at LHCb, it is more attractive to measure the same quantities using the $B^0 \to D \pi^+ \pi^-$ mode (here $D$ indicates an admixture of $\Dz$ and $\Dzb$ states).
The Dalitz-plot structure of the decay $B^0 \to \Dzb \pi^+ \pi^-$ has been previously studied~\cite{Kuzmin:2006mw,delAmoSanchez:2010ad,LHCb-PAPER-2014-070}, and the models obtained from these studies could be used in a decay-time-dependent amplitude analysis using the $D \to K^+ K^-$ and $D \to \pi^+ \pi^-$ channels to determine $\phi_d$~\cite{Charles:1998vf,Latham:2008zs}. 
In practice it will be more convenient to perform a simultaneous fit including also the $\Dzb \to K^+ \pi^-$ mode, which acts as a control mode to determine the amplitude model, flavour-tagging response and decay-time acceptance.

An estimate of the achievable sensitivity has been made using pseudoexperiments. 
The expected statistical precision for $\sin{\phi_d}$ and $\cos{\phi_d}$ are $\pm 0.06$ and $\pm 0.10$, respectively, for the LHCb Run 1 and 2 data samples combined.
Extrapolating this to $300\invfb$ gives $\pm 0.007$ for $\sin{\phi_d}$ and $\pm 0.017$ for $\cos{\phi_d}$.
Further experimental studies are needed to understand the impact of systematic uncertainties, although the use of the $\Dzb \to K^+ \pi^-$ control sample is expected to minimise effects from many potential sources of systematic bias.  
In case model uncertainties become a limiting factor, a model-independent version of the method can be used instead~\cite{Bondar:2018gpb}.
Thus, it is expected that a penguin-free measurement of $\phi_d$ can be achieved with sensitivity better than \belletwo, and comparable to that with $\Bz \to \jpsi\KS$.

A similar approach can be used to measure $\phi_s$ using $B^0_s \to \Dzb K^+ K^-$ decays~\cite{Nandi:2011uw}.
However, the expected yields are approximately 50 times smaller than for the $B^0 \to \Dzb \pi^+ \pi^-$ channel, and there is no existing amplitude model for the decay.
Therefore it is not possible to reliably estimate the sensitivity based on current results~\cite{LHCB-PAPER-2012-018}.

%% file: CONTRIBUTIONS/3_Time_dependent_CP_violation_measurements/3.3.tex
\section{Prospects with charmless $B$ decays}
\input{CONTRIBUTIONS/3_Time_dependent_CP_violation_measurements/3.3.1.tex}

\input{CONTRIBUTIONS/3_Time_dependent_CP_violation_measurements/3.3.2.tex}
\input{CONTRIBUTIONS/3_Time_dependent_CP_violation_measurements/3.3.3.tex}
\input{CONTRIBUTIONS/3_Time_dependent_CP_violation_measurements/3.3.4.tex}

\input{CONTRIBUTIONS/3_Time_dependent_CP_violation_measurements/3.3.5.tex}

%% file: CONTRIBUTIONS/3_Time_dependent_CP_violation_measurements/3.3.1.tex
\subsection{$\phi_s^{s\bar{s}s}$ from $B^0_s \to \phi\phi$}

The $\Bs\to\phi\phi$ decay is forbidden at tree level in the Standard Model (SM) and proceeds predominantly via a gluonic $\bquarkbar \to \squarkbar \ssbar$ loop (penguin) process.
Hence, this channel provides an excellent probe of physics beyond the SM that may contribute to the penguin diagram~\cite{Bartsch:2008ps,Beneke:2006hg,PhysRevD.80.114026}.
This mode is well suited for study at LHCb, as both $\phi$ mesons can be reconstructed through their decay to $\Kp\Km$.
Observables of interest include triple-product asymmetries as well as the \CP-violating phase $\phi_s^{s\bar{s}s}$ that can be measured from a polarisation-dependent 
decay-time-dependent angular analysis. 
Due to a cancellation of the small phase in $\Bs$--$\Bsb$ mixing with that in the decay amplitude, $\phi_s^{s\bar{s}s}$ is expected to be zero to a good approximation and hence different to the phase measured in modes such as $\Bs\to\jpsi\phi$.
Calculations using QCD factorisation provide a SM upper limit of $|\phisPP|< 0.02\rad$~\cite{Bartsch:2008ps,Beneke:2006hg,PhysRevD.80.114026}.
The latest LHCb result, based on $5 \invfb$ of data, is $\phisPP = -0.07 \pm 0.13 \stat \pm 0.03 \syst \rad$~\cite{LHCb-CONF-2018-001,LHCb-PAPER-2014-026}.

\begin{figure}[tb]
\begin{center}
  \includegraphics[width=0.495\textwidth]{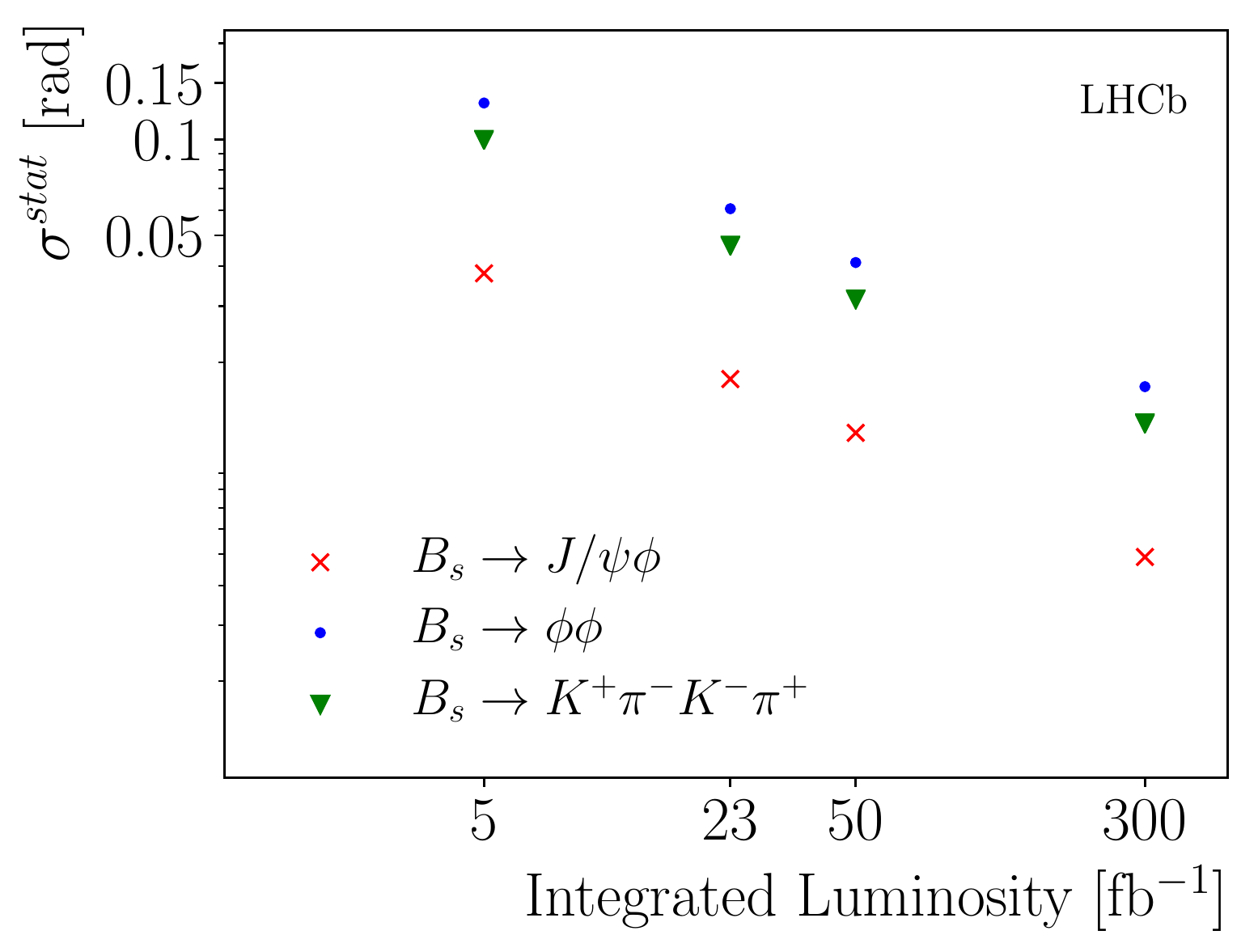}
  \caption{Comparison of $\phi_s$ statistical sensitivity from different decay modes.}
    \label{fig:td_comp}
\end{center}
\end{figure}

The measured and extrapolated statistical sensitivities for $\phisPP$ and similar \CP-violating phases measured in other decay modes are shown in Fig.~\ref{fig:td_comp}.
A statistical uncertainty on $\phisPP$ of $0.011 \rad$ can be achieved with $300 \invfb$ of data collected at \upgradetwo.
As for other measurements of \CP violation parameters through decay-time-dependent analyses, many systematic uncertainties are evaluated from control samples, and are therefore expected to scale accordingly with integrated luminosity.
Among the others, there is an important uncertainty associated with knowledge of the angular acceptance, which is determined from simulation.
This therefore relies on good agreement between data and simulation, which can be validated using control channels such as $\Bd \to \phi \Kstarz$. 
Thus the determination of $\phisPP$ is expected to remain statistically limited even with the full \upgradetwo\ data sample.

%% file: CONTRIBUTIONS/3_Time_dependent_CP_violation_measurements/3.3.2.tex
\subsection{$\phi_s^{d\bar{d}s}$ from $B^0_s \to (K^+\pi^-)(K^-\pi^+)$}

The $B_s^0\to \Kstar(892)^0\Kstarb(892)^0$ decay, which in the Standard Model (SM) is dominated by a gluonic penguin  \decay{\bquarkbar}{\dquarkbar \dquark\squarkbar} diagram, has been discussed extensively in the literature as a benchmark test for, and an excellent probe for physics beyond, the SM~\cite{Fleischer:1999zi,Fleischer:2007wg,Ciuchini:2007hx,DescotesGenon:2007qd,DescotesGenon:2011pb,Bhattacharya:2012hh,Bhattacharya:2013sga}.
A specific feature of this channel is the possibility to control the theoretical uncertainty on the SM prediction for the \CP-violating phase $\phi_s^{d\bar{d}s}$, which is expected to be close to zero, by studying the U-spin related channel $B^0\to \Kstar(892)^0\Kstarb(892)^0$.
Both decays can be studied at LHCb by reconstructing the $\Kstarz$ mesons in their decay to $\Kpm\pimp$, giving the final state $(K^+\pi^-)(K^-\pi^+)$.

LHCb has recently published the first measurement of $\phi_s^{d\bar{d}s}$~\cite{LHCb-PAPER-2017-048} using Run~1 data.
In this groundbreaking analysis, it was realised that a significant gain in sensitivity can be obtained by including the full $B_s^0\to(K^+\pi^-)(K^-\pi^+)$ phase space in the $K\pi$-mass window from 750 to $1600 \mevcc$, since the fraction of $B_s^0\to \Kstar(892)^0\Kstarb(892)^0$ in this region is only $f_{VV}=0.067 \pm 0.004 \pm 0.024$ (the other contributions are from $K\pi$ S-wave and the $K_2^*(1430)^0$ resonance).
The result, $\phi_s^{d\bar{d}s}= -0.10 \pm 0.13 \pm 0.14 \rad$, is compatible with the SM expectation.


The current result has statistical and systematic uncertainties of comparable size, but both are expected to be reducible with larger data samples.
The largest systematic uncertainty, corresponding to the treatment of the acceptance, is mostly driven by the limited size of the simulation samples --- due to the large phase space investigated in this analysis, very large simulation samples are required.
In order to produce significantly larger samples it will be necessary to exploit rapid simulation production mechanisms, since increases in available CPU power are not expected to keep pace with the size of the data samples.  
Another important systematic uncertainty due to the modelling of the $K\pi$ resonant and non-resonant components can be reduced by incorporating results of state-of-the-art studies of the $K\pi$ system, but some component of this may be irreducible.
Other systematic uncertainties are mainly based on control samples.
Therefore it is expected that the limiting systematic uncertainty will be not larger than $\sigma(\text{syst.})\sim0.03 \rad$.

The measured and extrapolated statistical sensitivities for $\phi_s^{d\bar{d}s}$ are given in \tabref{tab:BsKpiKpi_sigmastat}, both for the average over the $B_s^0\to(K^+\pi^-)(K^-\pi^+)$ system and for the exclusive $B_s^0\to \Kstar(892)^0\Kstarb(892)^0$ decay.
The sensitivities for $B_s^0\to(K^+\pi^-)(K^-\pi^+)$ are also included in Fig.~\ref{fig:td_comp}.
In the current analysis, the same weak phase is assumed for all contributions, but as the precision increases it will be possible to determine $\phi_s^{d\bar{d}s}$ separately for each, including possible polarisation dependence in the $B_s^0\to \Kstar(892)^0\Kstarb(892)^0$ decay.
The systematic uncertainty related to modelling of components is expected to be smaller when focusing on the $\Kstar(892)$ resonance, since its lineshape is well known.
Moreover, by making similar studies with the $B^0 \to (K^+\pi^-)(K^-\pi^+)$ mode, it will be possible to obtain all necessary inputs for the U-spin analysis of each component separately, leading to good control of the theoretical uncertainty on the prediction for $\phi_s^{d\bar{d}s}$.

\begin{table}[tb]
\begin{center}
\caption{Statistical sensitivity on $\phi_s^{s\bar{s}s}$ and $\phi_s^{d\bar{d}s}$.}
  \renewcommand{\arraystretch}{1.2}
 \begin{tabular}{ccccc}
 \hline
\multirow{ 2}{*}{Decay mode} & \multicolumn{4}{c}{$\sigma$(stat.) [rad]} \\
& 3~fb$^{-1}$ & 23~fb$^{-1}$ & 50~fb$^{-1}$ & 300~fb$^{-1}$ \\
\hline
$B_s^0 \to \phi\phi$ &  $0.154$  & $0.039$  & $0.026$ & $0.011$ \\
$B_s^0 \to (K^+\pi^-)(K^-\pi^+)$ (inclusive) & $0.129$  & $0.033$  & $0.022$ & $0.009$ \\
$B_s^0 \to \Kstar(892)^0\Kstarb(892)^0$ & $-$  & $0.127$  & $0.086$ & $0.035$ \\
 \hline 
\end{tabular}
\renewcommand{\arraystretch}{1.0}
 \label{tab:BsKpiKpi_sigmastat}
\end{center}
\end{table}



%% file: CONTRIBUTIONS/3_Time_dependent_CP_violation_measurements/3.3.3.tex
\subsection{$\CP$ violation in $B^0 \to \pi^+\pi^-$ and $B^0_s \to K^+K^-$}

The \BdTopipi and \BsToKK decays receive contributions from both tree-level $\bquark\to\uquark$ transitions and $\bquark\to\squark(\dquark)$ penguin diagrams.
The study of \CP violation in these decays is thus a powerful test of, and a search for physics beyond, the CKM picture of the SM~\cite{Deshpande:1994ii,He:1998rq,Fleischer:1999pa,Fleischer:2007hj,Fleischer:2010ib}.
Relying on the U-spin symmetry between the hadronic parameters entering the \BdTopipi and \BsToKK decay amplitudes, strategies have been proposed to exploit the full potential of these decays~\cite{Fleischer:1999pa,Fleischer:2007hj,Fleischer:2010ib,Ciuchini:2012gd}.
Combining the \CP asymmetries of the \BdTopipi and \BsToKK decays with the branching fractions and \CP asymmetries of the \decay{\Bd}{\piz\piz} and \decay{\Bp}{\pip\piz} decays, constraints on the CKM phases $\gamma$ and $-2\beta_s$ can be set.
From measurements of the \CP asymmetries of \BdTopipi and \BsToKK decays with a subset of the LHCb Run~1 data~\cite{LHCb-PAPER-2013-040}, an uncertainty of $0.15\rad$ in the determination of $-2\beta_s$ has been achieved, accounting also for U-spin breaking effects~\cite{LHCb-PAPER-2014-045}.
The uncertainty on $-2\beta_s$ is mainly driven by the uncertainty on the \CP asymmetry of the \BsToKK decay, with only a very mild dependence on the size of nonfactorisable U-spin breaking effects.
Information from the semileptonic \decay{\Bd}{\pim\ellp\neu} and \decay{\Bs}{\Km\ellp\neu} decays can be used to reduce the theoretical uncertainties on the determination of $-2\beta_s$ down to $0.01\rad$~\cite{Fleischer:2016jbf,Fleischer:2016ofb}.

\lhcb has measured the \CP violation parameters in \BdTopipi and \BsToKK decays (\Cpipi, \Spipi, \CKK, \SKK and \ADGKK) using the full sample of \proton\proton collisions collected during Run~1 corresponding to $3.0\invfb$ of integrated luminosity~\cite{LHCb-PAPER-2018-006}.
The results are reported in Table~\ref{tab:b2hhStatusAndProjections}, together with the projections of the statistical uncertainties to larger samples.
The scaling of statistical uncertainties assumes the same detector performances as in Run~1, in particular regarding the flavour tagging, the decay-time resolution and the particle identification performance, which are particularly important for the determination of these observables. 
The main sources of systematic uncertainties are due to limited knowledge of: the variation of the selection efficiency as a function of the \bquark meson decay time, the parameters \Gs and \DGs, and the calibration of the decay-time resolution.
The evaluation of these uncertainties is based on the study of control modes, and hence they are expected to decrease in a statistical manner as the available sample size grows.

\begin{table}[t]
	\begin{center}
	\caption{Current \lhcb measurements of \Cpipi, \Spipi, \CKK, \SKK and \ADGKK using the full sample of \proton\proton collisions collected during Run~1, where the first uncertainty is statistical and the second systematic. The projection of statistical precisions for each variables are also reported.}
\label{tab:b2hhStatusAndProjections}
	\resizebox{\columnwidth}{!}{
		\renewcommand{\arraystretch}{1.2}
		\begin{tabular}{lccccc}
			\hline 
			Data sample &  \Cpipi & \Spipi & \CKK & \SKK & \ADGKK \\
			\hline 
			Run 1 ($3\invfb$~\cite{LHCb-PAPER-2018-006}) & $-0.34 \pm 0.06 \pm 0.01$ & $-0.63 \pm 0.05 \pm 0.01$ & $\phantom{-}0.20 \pm 0.06 \pm 0.02$ & $\phantom{-}0.18 \pm 0.06 \pm 0.02$ & $-0.79 \pm 0.07 \pm 0.10$ \\
			\hline
			& \multicolumn{5}{c}{$\sigma$ (stat.)} \\
			\hline
			Run 1-3 ($23\invfb$)  & $0.015$ & $0.013$ & $0.015$ & $0.015$ & $0.018$ \\
			Run 1-6 ($300\invfb$) & $0.004$ & $0.004$ & $0.004$ & $0.004$ & $0.005$ \\
			\hline 
		\end{tabular}
		\renewcommand{\arraystretch}{1.0}
		}
	\end{center}
\end{table}

%% file: CONTRIBUTIONS/3_Time_dependent_CP_violation_measurements/3.3.4.tex
\subsection{$\alpha$ from $B$ decays to $\pi\pi$, $\rho\rho$ and $\pi^+\pi^-\pi^0$}
\label{sec:alpha}

For over a decade, it has been common for conference speakers to introduce the angle $\gamma$ as the least precisely measured of the three unitarity triangle angles.
However, the uncertainty on the world average for $\gamma$ is now around $4^\circ$, slightly below that for $\alpha$, which is $5^\circ$~\cite{HFLAV16}.
As discussed in this document, LHCb has a strong programme to continue to improve the determination of both $\beta$ and $\gamma$, and hence it is important to continue to reduce the uncertainty also on $\alpha \equiv \arg\left[ -(V^{\phantom{*}}_{td}V_{tb}^{*})/(V^{\phantom{*}}_{ud}V_{ub}^{*}) \right]$ in order to make precision tests of the SM in the CKM sector.

The method discussed in the previous subsection to determine $\gamma$ and $-2\beta_s$ from $B \to \pi\pi$ and \BsToKK decays is an extension of the well-known isospin analysis to determine $\alpha$ from $B \to \pi\pi$ decays alone~\cite{Gronau:1990ka,Bona:2007qta,Charles:2017evz} 
(the unitarity relation $\alpha + \beta + \gamma = \pi$ is used implicitly).
The main input from LHCb will be world-leading measurements of the \CP-violating parameters in \BdTopipi decay, but important input can also be obtained on the \decay{\Bu}{\pip\piz} decay, using the method pioneered in Ref.~\cite{LHCb-CONF-2015-001} for \decay{\Bu}{\Kp\piz} decays.
Clearly good performance of the electromagnetic calorimeter will be critical for such a measurement.
Progress on the \decay{\Bd}{\piz\piz} mode, which is currently limiting the precision of the $\alpha$ determination from $B \to \pi\pi$ decays, will mainly come from the \belletwo\ experiment.

The situation is quite different for the $B \to \rho\rho$ system, which currently provides the strongest constraints on $\alpha$.
Although the presence of two vector particles in the final state makes the analysis more complicated in principle~\cite{Gronau:1990ka,Dunietz:1990cj,Falk:2003uq,Beneke:2006rb}, the observed dominance of longitudinal polarisation and the smaller penguin contribution compared to $B \to \pi\pi$ lead to good sensitivity being achieved. 
The rarest of the three isospin-partner modes is the $\decay{\Bz}{\rho^0\rho^0}$ decay, which has a final state of four charged tracks following $\rho^0 \to \pip\pim$ decays, making it well suited for study at LHCb. 
Indeed, this decay was first observed by LHCb, and a time-integrated angular analysis on Run 1 data was performed~\cite{LHCB-PAPER-2015-006}.
With larger data samples it will be possible not only to improve the measurements of the branching fraction and longitudinal polarisation fraction, but to make precise determinations of the \CP-violating parameters. 
Consequently, the determination of $\alpha$ from $B \to \rho\rho$ decays will benefit from the additional information inherent in $S_{\rhoz\rhoz}$, compared to the case for the $B \to \pi\pi$ system for which $S_{\piz\piz}$ is barely measurable (\belletwo\ plans to measure $S_{\piz\piz}$ using candidates involving a $\piz$ Dalitz decay, but only rather limited sensitivity is possible).


For $\Bz \rightarrow \rho \pi$ decays, it is possible to determine $\alpha$ from a decay-time-dependent amplitude analysis of $\Bz \to \pip\pim\piz$ decays alone, without the need to include other channels in the analysis~\cite{Snyder:1993mx}.
This is possible since amplitudes for the decays $\Bz \to \rhop\pim$, $\rhom\pip$ and $\rhoz\piz$ interfere in different regions of the Dalitz plot, so that their relative phases can be measured.
First measurements have been published by BaBar and Belle~\cite{Lees:2013nwa,Kusaka:2007dv,Kusaka:2007mj}, but do not yet provide strong constraints on $\alpha$.
LHCb has not yet published any result on this channel, but it is expected that large yields will be available, and that it should be possible to control backgrounds with good understanding of $\piz$ reconstruction.  
It is worth noting that a significant proportion of the photons from neutral pions produced at the \B decay vertex convert into $\ep\en$ pairs in the LHCb detector material, with approximately half of these conversions occurring before the magnet.
Tracks from these converted photons provide additional information with which to constrain the \B decay vertex and the neutral pion momentum, resulting in improved resolution and background rejection.

An important source of systematic uncertainty for amplitude analyses, such as that for $\Bz \to \pip\pim\piz$ decays, is likely to be due to lack of knowledge of the resonant structure.
Studies of the \decay{\Bp}{\pip\pip\pim} decay will provide useful information that can help to minimise this uncertainty~\cite{Tandean:2002pe}.
Moreover, for all $\alpha$ measurements at subdegree precision, isospin-breaking effects such as electroweak penguin contributions must also be considered~\cite{Charles:2017evz}.

%% file: CONTRIBUTIONS/3_Time_dependent_CP_violation_measurements/3.3.5.tex
\subsection{\CP violation in $B^0_s \to K^0_\mathrm{S}h^+h^-$ and $B^0_s \to h^+h^-\pi^0$}
\label{sec:Kshh-hhpiz}


Charmless three-body decays of $\Bds$ mesons provide further opportunities for studies of \CP violation in penguin-dominated transitions, and are sensitive to contributions from physics beyond the SM (BSM) scenarios~\cite{Grossman:1996ke,Fleischer:1996bv,London:1997zk,Ciuchini:1997zp}.
Thus, measurements of the CKM phases \Pbeta, \betas and \Pgamma in these modes may differ from their SM benchmark values.
A good example is the determination of $\beta$ using the $\Bz \to \phi\KS$ channel, obtained from decay-time-dependent amplitude analysis of $\Bz \to \Kp\Km\KS$ decays.
Results from BaBar and Belle~\cite{Lees:2012kxa,Nakahama:2010nj} show that $S_{\phi\KS}$ is consistent with expectation, but with a large uncertainty.
Improved measurements are anticipated from \belletwo, but LHCb will also have a relevant reach 
since large yields of $\Bd \to \KS h^+h^-$ decays are available~\cite{LHCb-PAPER-2017-010}.
For the $\Bd \to \KS\pip\pim$ mode, a decay-time-integrated analysis has been performed resulting in the first observation of \CP violation in the $\Bd \to \Kstarp\pim$ channel~\cite{LHCb-PAPER-2017-033}; with more data this analysis can be updated to include decay-time-dependence and determine also \CP-violating parameters for the $\Bd \to \rhoz\KS$ and $f_0(980)\KS$ channels.

While both LHCb and \belletwo\ will study $\Bd \to \KS h^+h^-$ decays, only LHCb can make measurements of \CP-violating parameters of the counterpart \Bs decays.
Preliminary sensitivity studies of a decay-time-dependent flavour-tagged Dalitz-plot analysis of \decay{\Bs}{\KS\pip\pim} decays~\cite{Gershon:2014yma} indicate that the precision achievable on $\phi_{\squark}^{\dquark\uquarkbar\uquark}$ with the full Run 1 + Run 2 dataset is approximately $0.4\rad$. 
Extrapolation to $300\invfb$ indicates potential for a precision of around $0.07\rad$. 
Similar studies for the \decay{\Bs}{\KS\Kpm\pimp} decay mode can be performed, which requires a more complicated analysis since both $\KS\Km\pip$ and $\KS\Kp\pim$ final states are accessible to both \Bs and \Bsb decays with comparable magnitude.
Although currently the precision of such measurements are dominated by the statistical uncertainty~\cite{SilvaCoutinho:2045786}, the expected yields of more than $10^6$ signal decays for $300 \invfb$ will allow a complete understanding of the \decay{\Bs}{\KS\Kpm\pimp} phase space. 


Studies of $\Bds \to h^+ h^- \piz$ decays will provide further sensitivity.
For example, resonant contributions of the type $\Kstarpm h^\mp$ will decay to the final state $\Kpm h^\mp \piz$ in addition to $\KS \pipm h^\mp$, and therefore a combined analysis of both can provide additional information that helps to test the SM prediction~\cite{Ciuchini:2006kv,Ciuchini:2006st,Gronau:2006qn,Gronau:2007vr}.
Large yields will be available and the improved capabilities of the ECAL will allow background to be controlled. 
Two particularly important features of these decays are that background from $B \to V \gamma$ decays, where $V \to h^+h^-$, must be suppressed and that it must be possible to resolve \Bd\ and \Bs\ decays to the same final state.
Thus, it will be important to have both good $\gamma$--$\piz$ separation and good mass resolution.

%% file: CONTRIBUTIONS/3_Time_dependent_CP_violation_measurements/3.4.tex
\section{Measurements of $\gamma$ from $B^0_s \to D_s^\mp K^\pm$ and $B^0 \to D^\mp \pi^\pm$}
\label{sec:3.4}


Measurement of decay-time-dependent \CP violation in $\B^0_s\to \D_s^\mp K^\pm$ and $\B^0 \to \D^\mp \pi^\pm$ decays provide information on the combination of the CKM angles $\gamma - 2\beta_s$ and $\gamma + 2\beta$, respectively. 
Information on the angle $\gamma$ can be therefore determined by using independent determinations of the value of $\beta_{(s)}$ as input. 

Sensitivity to \CP violation, and hence to the combination of the CKM angles,  depends on the ratio between the suppressed $\bar{b} \to \bar{u}c\bar{q}$ ($q=d,s$) and the favoured $\bar{b} \to \bar{c} u\bar{q}$ decay amplitudes: $r_{D_{(s)} h} \equiv |A(B^0_{(s)} \to D^+_{(s)}h^-)/A(B^0_{(s)} \to D^-_{(s)}h^+)|$, where $h=\pi,K$.  
This ratio is sizeable ($\approx 0.4$) for $\B_s^0 \to \D_s^\mp K^\pm$ decays, and small ($\approx 0.02$) for $\B^0 \to \D^\mp \pi^\pm$ decays. 
Thus, for $\B_s^0 \to \D_s^\mp K^\pm$ decays it is possible to determine all five of the observables $C_f$, $S_f$, $S_{\bar{f}}$, $A^{\Delta\Gamma}_f$ and $A^{\Delta\Gamma}_{\bar{f}}$ in a flavour-tagged decay-time-dependent analysis ($C_f = -C_{\bar{f}}$ due to absence of \CP violation in decay in these tree-dominated modes).
In the $\B^0 \to \D^\mp \pi^\pm$ case, however, $C_f$ is indistinguishable from unity, and so only $S_f$ and $S_{\bar{f}}$ can be measured bearing in mind that $\Delta \Gamma_d$ is negligible.
In addition to $r_{D_{(s)} h}$ and the weak-phase combination $\gamma - 2\beta_s$ or $\gamma + 2\beta$, these observables depend also on the strong-phase difference between the suppressed and favoured amplitudes, denoted $\delta_{D_{(s)} h}$.


Using Run~1 data, \LHCb~has determined the $B_s^0 \to D_s^\mp K^\pm$ coefficients from a sample of about 6000 signal decays~\cite{LHCb-PAPER-2017-047}. 
From the measured parameters,  a value of $\gamma$ of $(128\,^{+17}_{-22})^\circ$ (modulo $180^\circ$) is determined, using as input $-2\beta_s = -0.030\pm0.033\rad$~\cite{HFLAV16}. 
The result is dominated by statistical uncertainties thanks to the wide use of data-driven methods to determine the decay-time acceptance and resolution, and to calibrate the flavour tagging.
The systematic uncertainties are, in decreasing order of importance, related to background from $b$-hadron decays, to uncertainty on the value of $\Delta m_s$, to the calibration of the decay-time resolution and to the flavour tagging. 
The first contribution can be significantly reduced by a tighter signal selection or by using a different fitting approach; the remaining contributions are expected to scale with the statistics accumulated due to their data-driven nature.


In the case of $\Bz \to \D^\mp \pi^\pm$ decays, the smallness of the ratio of amplitudes $r_{D \pi}$, which limits the sensitivity to $S_f$ and $S_{\bar{f}}$, is compensated for by a large signal yield.
About 480\,000 flavour-tagged signal decays are available in the Run 1 LHCb data sample~\cite{LHCb-PAPER-2018-009}. 
Analysis of this sample gives measurements of $S_f$ and $S_{\bar{f}}$ that are more precise than those from BaBar and Belle~\cite{Aubert:2006tw,Ronga:2006hv}.
Also in this case the precision is limited by the statistical uncertainty. 
The dominant sources of systematic uncertainty, such as due to knowledge of $\Delta m_d$ and of background subtraction, are expected to be reducible with larger samples.

Since $\Bz \to \D^\mp \pi^\pm$ decays there are only two observables, $S_f$ and $S_{\bar{f}}$, that depend on three unknown quantities, $r_{D \pi}$, $\delta_{D \pi}$ and $2\beta+\gamma$, external input must be used to obtain a constraint on $2\beta+\gamma$.
A common approach is to determine $r_{D\pi}$ from the branching fraction of $\Bd \to \Dsp \pim$ decays, assuming \grpsuthree\ symmetry,
\begin{equation}
r_{D\pi}= \tan \theta_c  \frac{f_{D^+}}{f_{D_s}}\sqrt{\frac{\mathcal{B}(\Bd \to \Dsp \pim)}{\mathcal{B}(\Bd \to \Dm \pip)}}\,,
\end{equation}
where $\tan\theta_c$ is the tangent of the Cabibbo angle and  $\frac{f_{D_s}}{f_{D^+}}$ is the ratio of decay constants.
Using the obtained value of $r_{D\pi}=0.0182\pm0.0012\pm0.0036$, where the second uncertainty accounts for possible nonfactorisable \grpsuthree-breaking effects~\cite{DeBruyn:2012jp}, the intervals $|\sin(2\beta + \gamma)|\in [0.77,1.00]$ and $\gamma\in [5,86]\degrees \cup [185,266]\degrees$ are obtained, at the 68\% confidence level. 
The interval of $\gamma$ is calculated using $\beta = (22.2 \pm 0.7)\degrees$~\cite{HFLAV16}. 
The uncertainties on $r_{D\pi}$ and $\beta$ have negligible impact on these intervals, as the dominant uncertainties are from the $S_f$ and $S_{\bar{f}}$ measurement. 

The expected statistical sensitivities for the \CP violation parameters in $B^0_s \to D_s^\mp K^\pm$ and $B^0 \to D^\mp \pi^\pm$ decays are shown in Table~\ref{tab:Dh_expected_uncertainties}.
These are based on scaling of yields, and as such assume that the same detector performance as achieved in Run~1 can be maintained.
In particular, the sensitivity depends strongly on the performance of the particle identification, decay time resolution and flavour tagging.  
The results can be complemented by studies of the related $B^0_s \to D_s^{*\mp} K^\pm$ and $B^0 \to D^{*\mp} \pi^\pm$ channels.
In particular, the $D^{*\mp} \pi^\pm$ mode has an all charged final state, and with a possible gain in the acceptance of slow pions from $D^*$ decays from the addition of magnet side stations, comparable precision to that for $B^0 \to D^\mp \pi^\pm$ may be possible.

 
The corresponding expected sensitivities of $\gamma$ from $\Bs \to \D_s^\mp K^\pm$ decays are about $4^\circ$, $2.5^\circ$ and $1^\circ$ after collecting 
23, 50 and 300\invfb, respectively.
It is more challenging to estimate the constraints on $\sin(2\beta +\gamma)$ and $\gamma$ from $\Bd \to D^\mp \pi^\pm$ decays, since the precision of the external value of $r_{D\pi}$ will become the dominant source of systematic uncertainty.
Theoretical advancements on understanding the nonfactorisable \grpsuthree-breaking effects are thus required.



\begin{table}[tb]
        \centering
        \caption{Expected statistical uncertainties on parameters of  $\Bs \to \D_s^\mp K^\pm$ and $\Bd \to \D^\mp \pi^\pm$ decays.}
        \begin{tabular}{lcccccccc}
		\hline
		                        & \multicolumn{4}{c}{$\Bs \to \D_s^\mp K^\pm$} & \multicolumn{4}{c}{$\Bd \to \D^\mp \pi^\pm$}  \\
		    Parameters  & Run~1 & 23\invfb & 50\invfb & 300\invfb & Run~1 & 23\invfb & 50\invfb & 300\invfb \\ 
		\hline
		$S_f$, $S_{\bar{f}}$ & 0.20 & 0.043 & 0.027 & 0.011  & 0.02 & 0.0041 & 0.0026 & 0.0010\\
		$A^{\Delta\Gamma}_f$, $A^{\Delta\Gamma}_{\bar{f}}$ & 0.28 & 0.065 & 0.039 & 0.016 & -- & -- &-- \\
		$C_f$ & 0.14 & 0.030 & 0.017 & 0.007 & -- & -- & -- \\
	  \hline
	\end{tabular}
        \label{tab:Dh_expected_uncertainties}
\end{table}

%% file: CONTRIBUTIONS/4_Time_integrated_CP_violation_measurements/4.tex
\label{sec:timeintegcpv}
\input{CONTRIBUTIONS/4_Time_integrated_CP_violation_measurements/4.0.tex}

\input{CONTRIBUTIONS/4_Time_integrated_CP_violation_measurements/4.1.tex}
\input{CONTRIBUTIONS/4_Time_integrated_CP_violation_measurements/4.2.tex}

\input{CONTRIBUTIONS/4_Time_integrated_CP_violation_measurements/4.3.tex}

\input{CONTRIBUTIONS/4_Time_integrated_CP_violation_measurements/4.4.tex}

%% file: CONTRIBUTIONS/4_Time_integrated_CP_violation_measurements/4.0.tex
\label{chpt:TICPV}
\label{section:TICPV-intro}
Measurements of decay-time-integrated \CP\ violation in the $B$ system complement decay-time-dependent measurements in testing the SM picture of quark mixing and the unitarity of the CKM matrix.
They are particularly important in giving the most precise possible tree-level 
measurement of the angle $\gamma \equiv \arg\left[ -(V_{ub}^*V_{ud}^{\phantom{*}}) / (V_{cb}^*V_{cd}^{\phantom{*}}) \right]$ of the CKM unitarity triangle, 
which is unique in not depending on a top-quark coupling.
Within the SM, the irreducible theoretical uncertainty on tree-level measurements of \g is $|\delta_\g/\g|\leq10^{-7}$~\cite{Brod:2013sga}, 
This means that with \upgradetwo\ statistics, the measurement of \g from decay-time-integrated \CP\ violation in $B \to D K$ decays 
will be the most precise {\it standard candle} of the CKM picture and the unitarity triangle apex, against which all other CKM observables can be compared. 
Crucially, \upgradetwo\ will allow for precision measurements of \g in many individual final states, particularly those with high multiplicity or neutral particles, which will have complementary experimental systematic uncertainties with the currently dominant modes. 
Such an overconstrained measurement of \g will give us particular confidence when confronting the experimental picture of \CP\ violation with the CKM expectation. 

The theoretical cleanliness of the determination of $\gamma$ in the SM is based on the fact that only tree-level processes contribute, and that it is expected that there is no new physics contribution to such amplitudes.
However, if this assumption is dropped, \upgradetwo offers the opportunity, arising from the different decay topologies in $B^0$ and $B^\pm$ decays, to probe for new physics effects within tree-level decays themselves.
It is important to underline that there is still considerable scope for new-physics effects in tree-level decays~\cite{Brod:2014bfa}, which would be signified by modifications to the tree-level Wilson coefficients $C_1$ and $C_2$.
These can be probed at the few percent level with the \upgradetwo\ dataset. 
Furthermore, with the inclusion of decay-time-dependent methods (discussed in Sec.~\ref{sec:3.4}) such as
$\Bs\to\Dsmp\Kpm$ and $\Bd\to\Dmp\pipm$ and very accurate knowledge of $\gamma$, one can obtain penguin-free measurements of $\phi_s$ and $\beta$.
\upgradetwo\ will also enable
high-precision measurements of decay-time-integrated \CP\ violation in three-body decays of $\B$ mesons and in the decays of $b$ baryons and $\B_c$ mesons.  
As with other areas of the LHCb physics programme involving baryons, \upgradetwo may be a truly unique opportunity to systematically and precisely establish the picture of \CP\ violation. 
Furthermore, many of these three-body and baryonic measurements proceed through
loop-level transitions and are therefore sensitive to NP, but a proper interpretation of any observed \CP\ violation requires a precise understanding of the hadronic phases entering
these transitions, which will only be possible by a global analysis of the possible transitions enabled by the \upgradetwo\ statistical reach.

From an experimental point of view, the main challenges and opportunities arising from \upgradetwo\ are similar and can be illustrated using $B \to D K$ decays as an example. 
These proceed via both $b \to c\bar{u}s$ and $b \to u\bar{c}s$ transitions, whose phase difference has a \CP-violating (\g) and \CP-conserving ($\delta_B$) part. 
The ratio of these amplitudes $=r_Be^{i(\delta_B\pm\gamma)}$ and the general rate equation (for \Bpm mesons) is,
\begin{equation}
\Gamma\left(\Bpm\ra D[\to f] K\right) = \left|r_{D} e^{-i\delta_{D}}+r_B e^{i(\delta_B\pm\gamma)}\right|^2 = r_{D}^2+r_B^2 +2\kappa r_{D} r_B\cos(\delta_B-\delta_{D}\pm\gamma)\,.
\label{adseq}
\end{equation}
Here, $r_{D}$ and $\delta_{D}$ are the amplitude ratio and \CP-conserving phase difference between the $\CP$-conjugate final states of the \D decay, $D\to f$ and $\D\to \bar f$. 
For two-body \B and \D decays, the coherence factor, $\kappa=1$, while
for multibody decays, $\kappa$ reduces as the interference is diluted by incoherence amongst contributing intermediate hadronic resonances. In both cases, the efficiency-corrected
time-integrated signal yields in each final state are interpreted in terms of the above physical observables, notably the $\CP$-violating phase \g. Since the bulk of the sensitivity to
\g comes from the difference in rates of the $\B$ and $\Bbar$ processes, a precise control of asymmetries in charged-particle identification and detection is crucial. Based on our knowledge of the
present detector, these systematic uncertainties will scale with integrated luminosity through the \upgradetwo\ period. An excellent vertex
resolution is equally important to separate the hadronic final states from combinatorial backgrounds, and will benefit from the expected reduction in the vertex
detector material in \upgradetwo\ , while the upgrade of the calorimeter will greatly expand LHCb's capabilities for modes with neutrals in the final state. For baryonic states,
the addition of the TORCH system may be particularly helpful in allowing for low-momentum separation of protons and kaons; this will also reduce the physics backgrounds from baryonic
decays which are often hard to model in other analyses due to their poorly known absolute branching fractions. Finally, the addition of magnet-side stations (see Sect.~\ref{sec:tracking})
may lead to important signal-yield improvements, particularly for high-multiplicity final states where there is a non-negligible probability of creating at least one low-momentum particle.


%% file: CONTRIBUTIONS/4_Time_integrated_CP_violation_measurements/4.1.tex
\section{Determination of the $\gamma$ angle from tree-level decays}
\input{CONTRIBUTIONS/4_Time_integrated_CP_violation_measurements/4.1.1.tex}

\input{CONTRIBUTIONS/4_Time_integrated_CP_violation_measurements/4.1.2.tex}
\input{CONTRIBUTIONS/4_Time_integrated_CP_violation_measurements/4.1.3.tex}
\input{CONTRIBUTIONS/4_Time_integrated_CP_violation_measurements/4.1.4.tex}

%% file: CONTRIBUTIONS/4_Time_integrated_CP_violation_measurements/4.1.1.tex
\subsection{$B \to DK$ GLW/ADS}
\label{sec.4.1.1}

The simplest $B\to DK$ topology is $\Bpm\to D\Kpm$, where the $D$ is reconstructed in a two charged-track final state.
In this case, the ratio of $b\to u\bar{c}s$ to $b\to c\bar{u}s$ amplitudes, $r_{B}^{DK} \approx 0.10$~\cite{HFLAV16}.
This is enough to generate $\mathcal{O}(10\%)$ direct \CP violation in \CP-eigenstates decays $D\to \Kp\Km, \pip\pim$ that are equally accessible from the \Dz and \Dzb.
This is an example of the GLW method~\cite{Gronau1991172} for which Eq.~\ref{adseq} simplifies with $r_D=1$ and $\delta_D=0$.
The ADS method~\cite{Atwood:1996ci} makes full use of Eq.~\ref{adseq} and is best exemplified by the case $f=\pipm\Kmp$, where $r_{D}\approx0.06$.
The similar magnitude of $r_{B}$ and $r_{D}$ gives large interference effects and high sensitivity to the phase information.

Such modes have been observed and studied during Run 1 and Run 2~\cite{LHCb-PAPER-2016-003,LHCb-PAPER-2017-021}; all ADS/GLW asymmetries are statistically limited.
The systematic uncertainties are small and arise predominantly from sources that will naturally decrease with increasing data, notably knowledge of instrumentation asymmetries. Methods for measuring and correcting for the $B$-meson production asymmetry and the \Kpm/\pipm reconstruction asymmetries are established using calibration samples; such samples will continue to be collected.
The dominant systematic uncertainties for GLW decays are due to background contributions from $\Lb \to \Lc K^-$ decays and charmless decays, while the dominant uncertainty for ADS decays arises from the $B_{s}^0 \to \Dzb \Km \pip$ background.
All will be better determined with dedicated studies as the sample size increases.

The ADS/GLW technique is being expanded to the other $\B \to D K$ decays which share the same quark-level transition.
An analysis of GLW observables in $\Bpm \to D^{*0} \Kpm$ decays has been developed for the case where the $D^{*0}$ vector meson is not fully reconstructed~\cite{LHCb-PAPER-2017-021}.
This partial reconstruction has larger background uncertainties but these will improve with more data as dedicated studies of the background are performed.
Furthermore, ADS/GLW analyses have been developed and published in quasi-two-body modes, $\Bpm \to D \Kstarpm$~\cite{LHCB-PAPER-2017-030} and $\Bz \to D K^{*0}$~\cite{LHCb-PAPER-2014-028}.
As in the case of $\Bpm \to D^{(*)0} \Kpm$ decays, these modes have no limiting systematic and they will make competitive contributions with the LHCb Upgrade datasets.

\begin{figure}[!tb]
  \centering
  \includegraphics[width=0.50\textwidth]{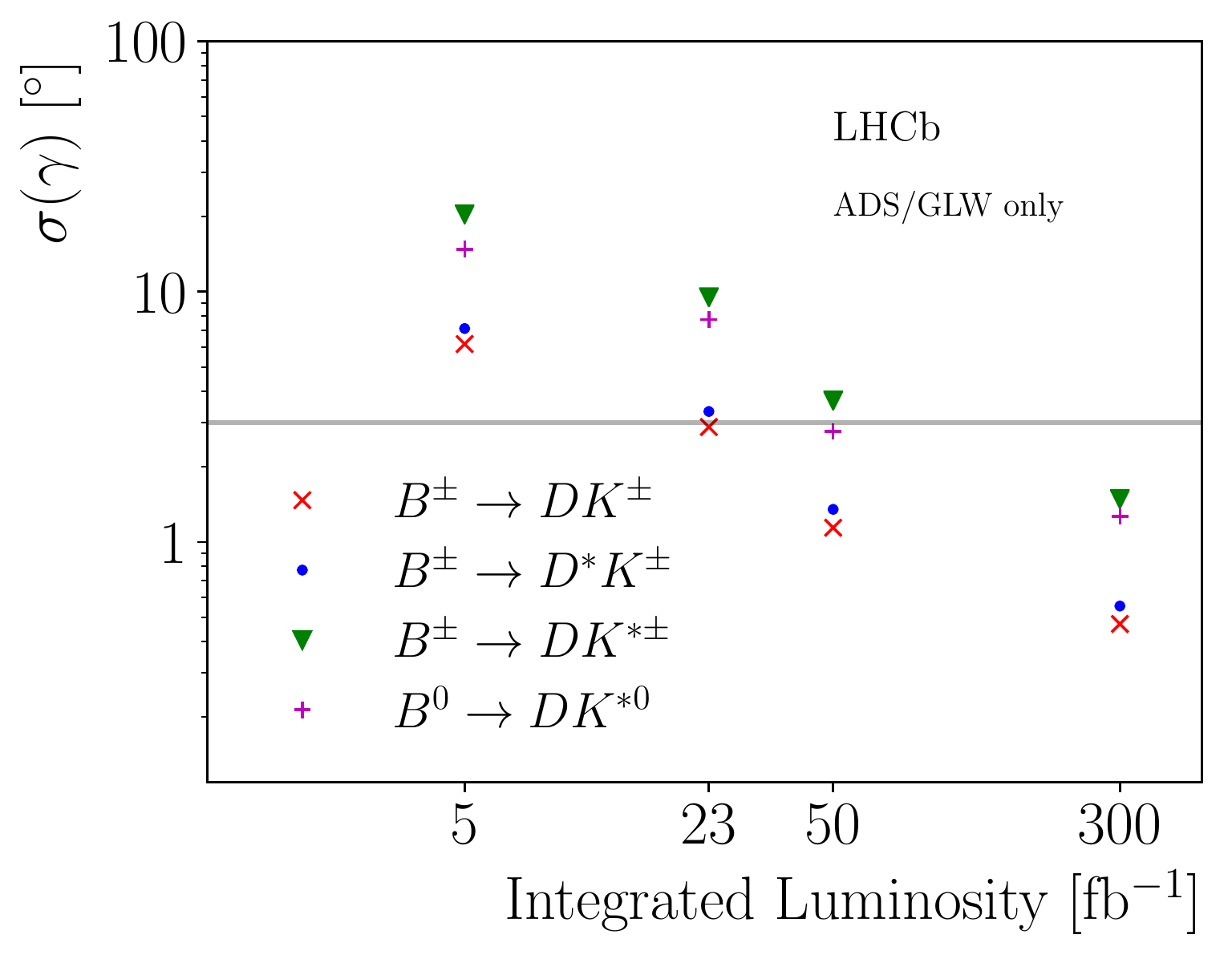}
  \caption{Extrapolation of \g sensitivity from the ADS/GLW analyses where disfavoured ambiguities are ignored. The expected Belle II $\gamma$ precision at an integrated luminosity of 50 ab$^{-1}$ is shown by the horizontal grey lines.
    \label{2body_gamma_projection}
  }
\end{figure}

Under the assumption that systematic uncertainties decrease in parallel with the statistical uncertainties, that is $\propto1/\sqrt{\mathcal{L}}$, the future precision on \g is predicted in Fig.~\ref{2body_gamma_projection}. In isolation, one ADS/GLW analysis suffers from the trigonometric ambiguities in Eq.~\ref{adseq}, but the multiple solutions are generally separated and resolved in combination with other $B\to DK$ results.
Fig.~\ref{2body_gamma_projection} uses central values and uncertainties in the published analyses of $B^\pm\to DK^\pm$, $B^\pm\to DK^{*\pm}$ and $B^0\to DK^{*0}$ decays.
For $B^\pm\to D^{*0}K^\pm$ both the partial and full reconstruction techniques are used in this study albeit with unpublished central values and uncertainties.

%% file: CONTRIBUTIONS/4_Time_integrated_CP_violation_measurements/4.1.2.tex
\subsection{$B \to DK$ GGSZ}
\label{sec:4.1.2}

\def        \DtoKspp        {\ensuremath{\D\to\KS\pip\pim}\xspace}
\def        \DtoKskk        {\ensuremath{\D\to\KS\Kp\Km}\xspace}
\def        \DtoKshh        {\ensuremath{\D\to\KS\hadron^+\hadron^-}\xspace}
\def        \BtoDpi         {\ensuremath{\Bpm\to\D\pipm}\xspace}
\def        \BtoDK          {\ensuremath{\Bpm\to\D\Kpm}\xspace}
\def        \BztoDstmuX     {\ensuremath{\Bz\to\Dstarpm\mu^\mp\nu_\mu X}\xspace}
\def        \hp            {\ensuremath{\hadron^+}\xspace}
\def        \hm            {\ensuremath{\hadron^-}\xspace}
\def        \BtoDKst       {\ensuremath{\Bz\to\PD\Kstarz}\xspace}
\def        \mpsq          {\ensuremath{m_{+}^{2}}\xspace}
\def        \mmsq          {\ensuremath{m_{-}^{2}}\xspace}
\def        \mpmsq          {\ensuremath{m_{\pm}^{2}}\xspace}
\def        \mmpsq          {\ensuremath{m_{\mp}^{2}}\xspace}
\def        \rb            {\ensuremath{r_B}\xspace}
\def        \db            {\ensuremath{\delta_B}\xspace}
\def        \g             {\ensuremath{\gamma}\xspace}
\def        \dpos          {\ensuremath{(\mpmsq,\mmpsq)\xspace}}
\def        \dbpos         {\ensuremath{(\mmpsq,\mpmsq)\xspace}}

Measurements of \g in which the $D$ meson is reconstructed using the
three-body, self-conjugate \DtoKspp and \DtoKskk final states provide powerful
input to the overall determination of \g,
as they select a single, narrow solution (see Fig.~\ref{fig:42:compare}).
Referred to as the GGSZ method~\cite{GGSZ,BONDARGGSZ}, sensitivity to \g is obtained by comparing the distributions of \DtoKshh decays across the Dalitz plane for opposite-flavour initial-state \Bp and \Bm mesons.
The partial decay rates as a function of Dalitz position depend mainly on the amplitudes of the $\Dz$ decay, with only small
deviations introduced from interference and \CP-violation.
These deviations are most conveniently probed through the measurement
of the \CP-violating observables $x_{\pm}= \rb\cos(\db\pm\g)$ and $y_{\pm}=\rb\sin(\db\pm\g)$.
For this it is necessary to have a good understanding of
the magnitudes of the \Dz and \Dzb decay amplitudes, as well as the strong-phase differences between them, $\delta_{D}$.

\begin{figure}[h]
  \centering
  \includegraphics[height=0.24\textheight]{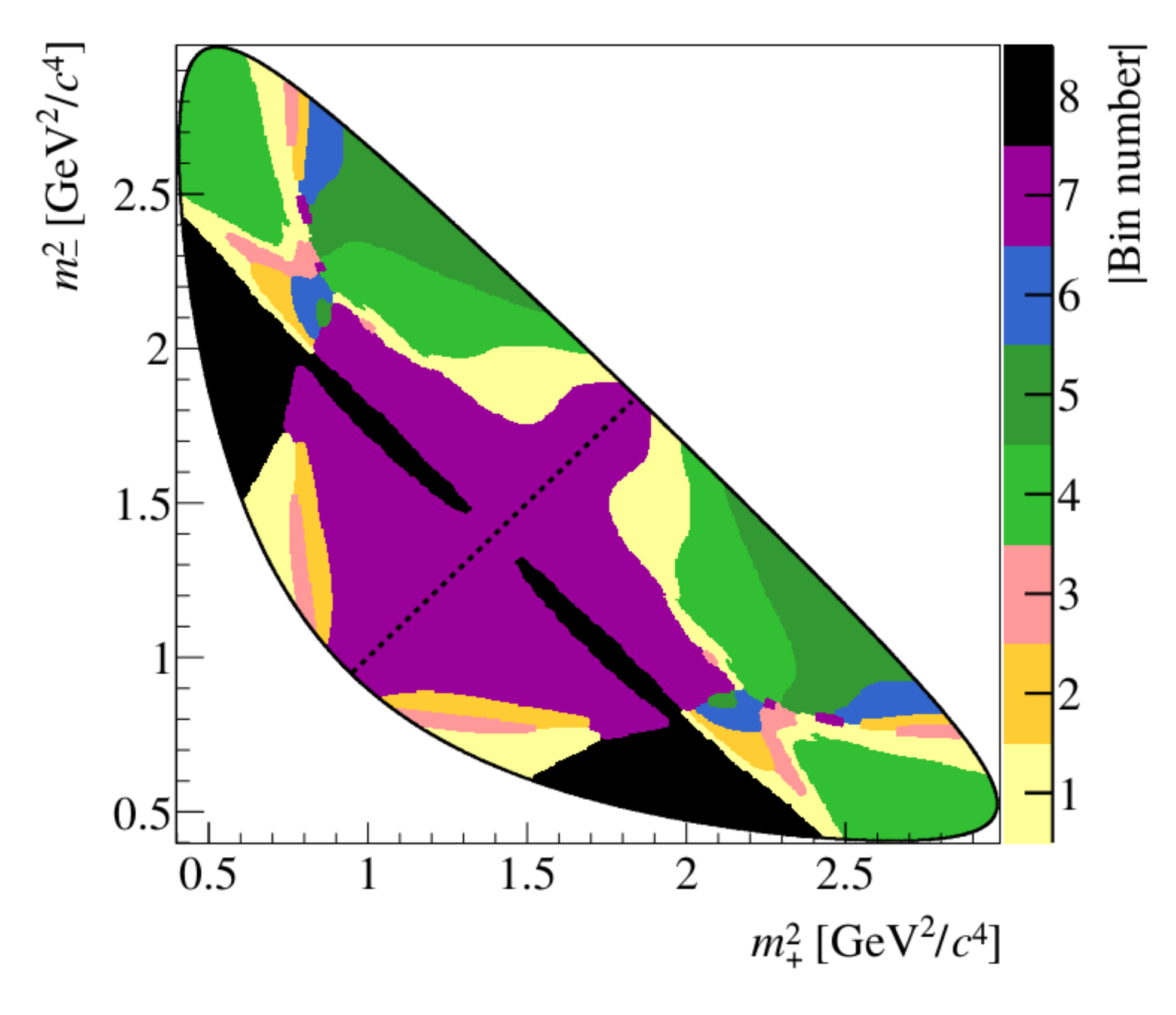}
  \includegraphics[height=0.24\textheight]{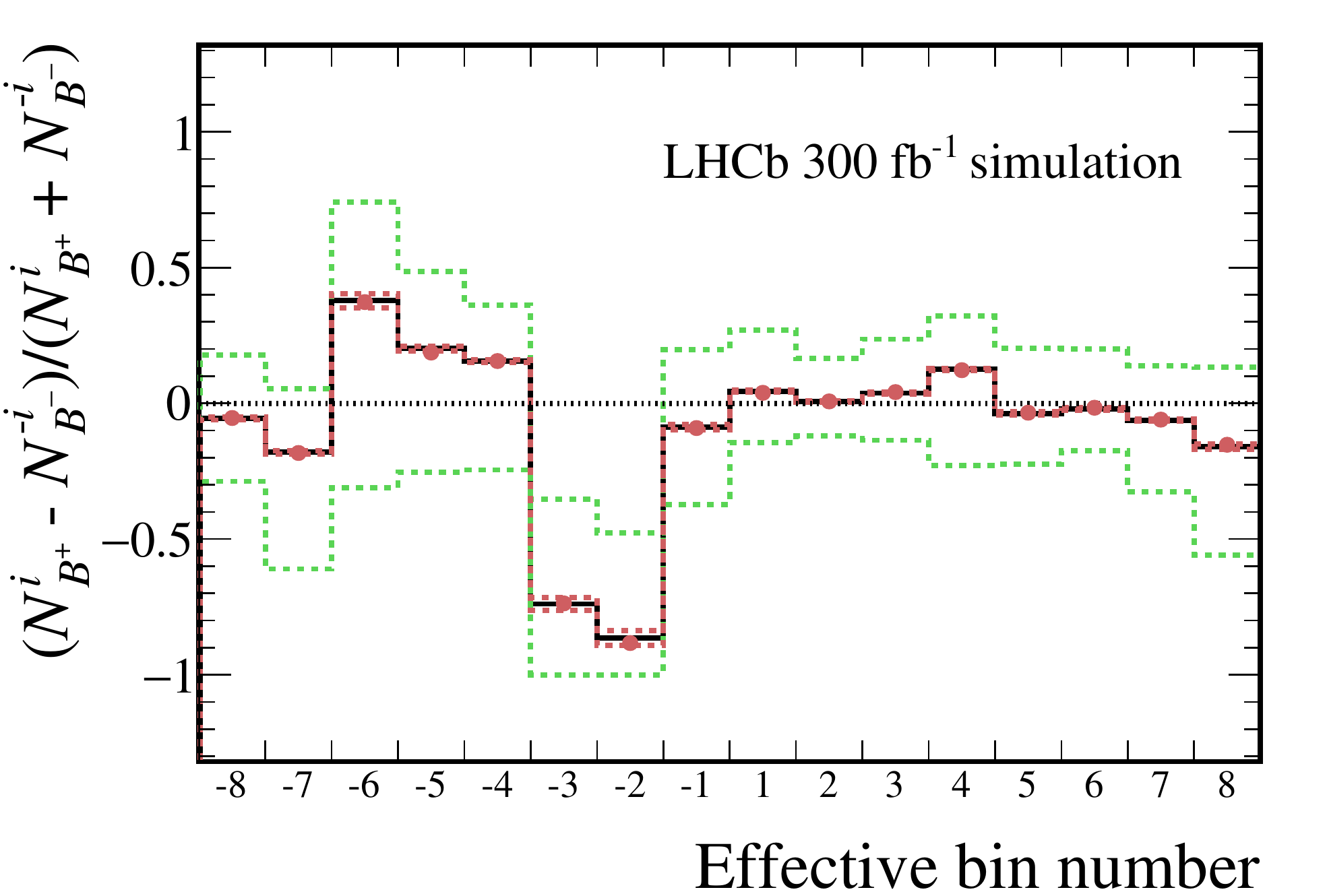}
  \caption{
    (Left) Bin definitions for \DtoKspp\ as a function of $\mmsq$ and $\mpsq$, the invariant masses squared of the $\KS h^-$ and  $\KS h^+$. Figure from Ref~\cite{LHCb-PAPER-2018-017}.
    (Right) Asymmetry between yields for $\Bpm \to D\Kpm$ decays, with \DtoKspp\ in bin $i$ (for \Bp) and $-i$ (for \Bm).
    The data points are obtained from simulation with the expected sample size at $300 \invfb$, assuming the current performance of the LHCb experiment.
    The black histogram shows the predicted asymmetry based on the current world average values of $\gamma$ and relevant hadronic parameters, the red dots show the result of a single pseudoexperiment, while the red bands show the expected uncertainties from an ensemble.
    The green bands show the corresponding uncertainties with the current LHCb data set.
    \label{fig:bins}
  }
\end{figure}

The description of $\delta_D$ has historically been
treated in two ways. One approach is to use an amplitude model determined from flavour-tagged \DtoKshh decays.
This relies on assumptions about the nature of the intermediate resonances that contribute to the \DtoKshh final state, and
choices made about these contributions to the amplitude model lead to systematic uncertainties which are not certain to scale
with luminosity in the \upgradetwo\ period.
An alternative method divides the data into bins according to the Dalitz plane coordinates, and
can then make use of direct measurements of the strong-phase differences averaged over each bin~\cite{GGSZ,BONDARGGSZ}.
One such set of binning definitions for the \DtoKspp decay mode, obtained by optimising the sensitivity to \g, is shown in Fig.~\ref{fig:bins}.

\begin{figure}[h]
  \centering
  \includegraphics[width=0.49\textwidth]{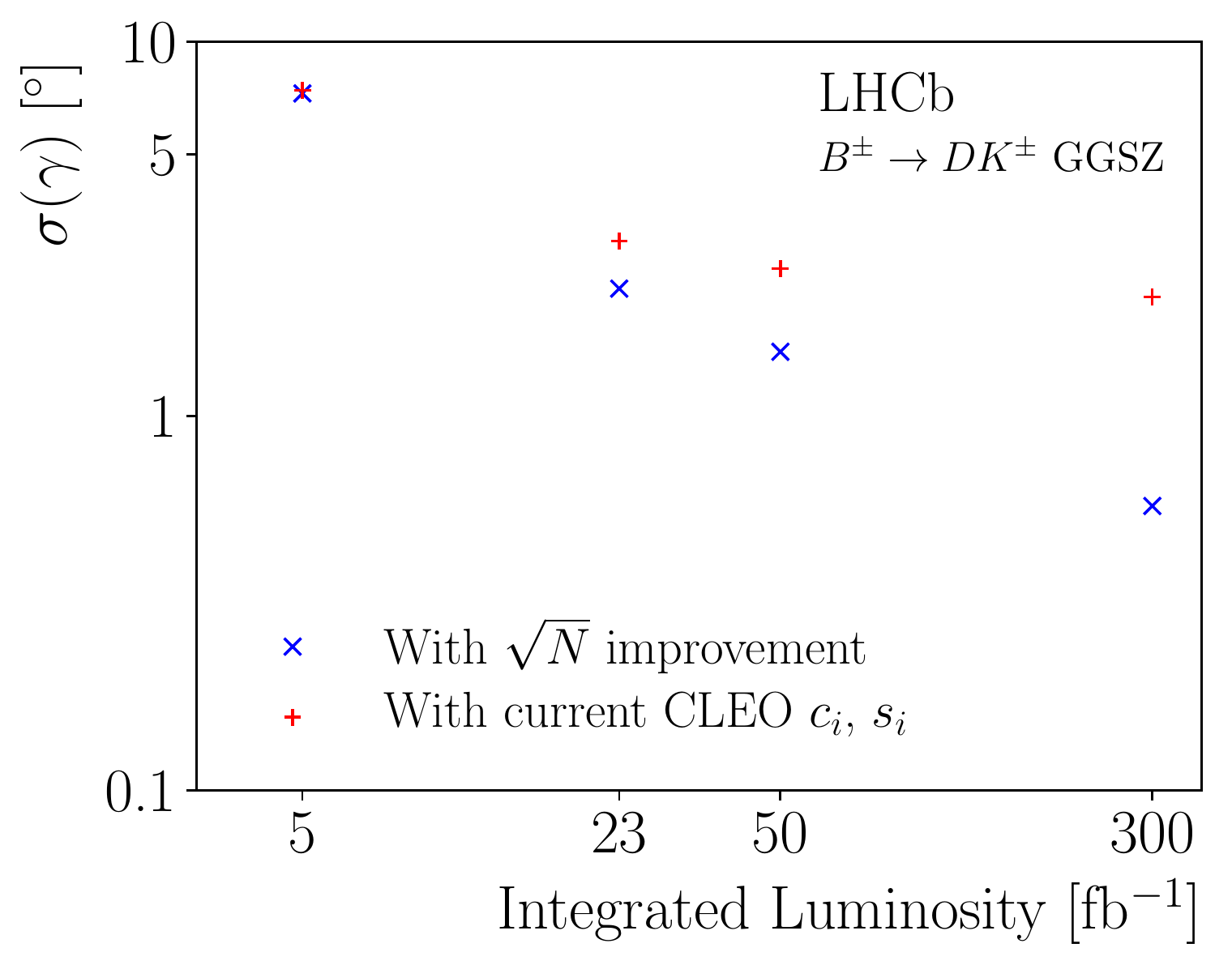}
  \caption{Expected evolution of \g sensitivity with the GGSZ method with (red crosses) current CLEO inputs and with (blue $\times$-marks) $\sqrt{N}$ improvements on their uncertainty.
    \label{fig:41:sensitivity}
  }
\end{figure}

Specifically, the values of $c_{i}$ and $s_{i}$, which correspond to amplitude-weighted average values of $\cos(\delta_D)$ and $\sin(\delta_D)$ in each Dalitz bin $i$, are required.
These can be measured in a model-independent manner by using quantum-correlated $\Dz\Dzb$ pairs from
$\psi(3770)$ decays, as was done previously with data taken at the CLEO-c experiment~\cite{Libby:2010nu}.
This allows the determination of \g to also be carried out in a completely model-independent fashion,
the price for which is a small loss in sensitivity from binning the data,
and a systematic uncertainty which depends on the precision with which the strong-phase measurements can be determined.

The model-independent method is therefore expected to be the baseline in \upgradetwo\ , and its
uncertainty is currently statistically dominated. Although systematic uncertainties
will already become significant compared to the statistical uncertainty in Run 3, studies
performed so far give confidence that the systematic uncertainties will generally scale with the
statistical reach well into the \upgradetwo\ period.


The largest systematic uncertainty is due to the precision of the
external strong-phase inputs coming from the CLEO-c data which currently contribute approximately $2\degrees$ to the overall uncertainty
on $\gamma$~\cite{LHCb-PUB-2016-025}.
The impact of this uncertainty on GGSZ measurements is estimated in Fig.~\ref{fig:41:sensitivity} which compares a $\sqrt{N}$ improvement with the
expected yield increase and the projected uncertainty if the current external information on $c_i$ and $s_i$ is not improved.
It can be seen that this starts to approach a limit originating from the fixed size of the quantum-correlated charm input from CLEO-c.
This contribution will naturally decrease with larger $\Bpm\to\Dz\Kpm$ samples, however, as the $B$ decays themselves
also have sensitivity to $c_i$ and $s_i$. Studies performed using pseudoexperiments with different size quantum correlated $D$ samples and LHCb
$B$ data suggest that the optimal sensitivity is only reached when the size of the input $D$ sample is at least as big as the overall $B$ sample.
More precise measurements with data already recorded by the BESIII experiment will be able to
reduce the external contribution to the uncertainty by around 50\% but analysis of future larger datasets with BESIII and at \lhcb will be vital
in order to avoid the external input compromising the ultimate sensitivity to \g.

The current second largest source of systematic uncertainty
comes from the knowledge of the distribution of $D$ decays
over the Dalitz plane in the flavour-specific $\B$ final state, with reconstruction and efficiency effects incorporated.
These are determined with a flavour-specific control decay mode \BztoDstmuX, where the \Dstarm decays
to \Dzb\pim and $X$ represents any unreconstructed particles.
The ultimate systematic uncertainty is particularly sensitive to data-simulation agreement and size of simulated samples because
of a need to model unavoidable differences in the signal and control modes.
Fast simulation techniques which are being deployed and further
developed at LHCb will therefore be crucial for keeping up with the large data samples,
while a fully software-based trigger will allow for a better alignment of the signal and control channel selections compared to today.
Uncertainties from sources such as low mass backgrounds can be expected to remain subdominant
with higher statistics, as further studies
will give better understanding to their rates and shapes.
Some additional complications will present themselves as the yields will eventually be high enough
that it will become necessary to take into account effects induced from asymmetries in the
\KS system, and eventually through mixing in the $\Dz$ system. However these are tractable problems,
and studies have already been done to understand when these effects will become important.

%% file: CONTRIBUTIONS/4_Time_integrated_CP_violation_measurements/4.1.3.tex
\subsection{Prospects with modes involving neutrals}

There are many prospects for adding orthogonal information on \g by applying the ADS/GLW and GGSZ techniques to modes that have an additional \piz\ meson.
The first use of \piz mesons in a \g analysis occurred with the Run 1 ADS/GLW analysis of $\Bpm \to D\Kpm$ decays with $D \to K\pi \piz,\ KK\piz, \pi\pi\piz$ 
final states~\cite{LHCb-PAPER-2015-014}. The $K\pi\piz$ and $\pi\pi\piz$ modes have branching fractions 3 and 10 times larger than their two-body equivalents. 
However, the net \piz reconstruction and selection efficiency in these decays is low, around 3\% with the current calorimeter. 
Also the analysis is complicated by a combinatorial background arising from random $\piz$ association, which will certainly worsen in \upgradeone\ . 
Improvements to the \upgradetwo\ calorimeter granularity and energy 
resolution will therefore be crucial in achieving the ultimate sensitivity with these modes especially by improving the \piz mass resolution.
Enhancements in the reconstruction of \piz\ mesons reconstructed from a single cluster due to the small spatial separation of their decay photons in the calorimeter, and the ability to separate such candidates from single photons, will also be important.

An important mode under development for the upgrade era is $\Bpm \to D^{*0} \Kpm$ decays, with $D^{*0}\to\Dz\piz$ and $D^{*0}\to\Dz\gamma$ decays. 
These twin modes provide an excellent sensitivity to \g as an exact phase difference between the two $D^{*0}$ modes can be exploited~\cite{Bondar:2004bi}. 
For this case, the efficient distinction of \piz and \g calorimeter objects is critical as the two $D^{*0}$ modes exhibit opposite \CP asymmetries. 
The initial studies show small, but clean signals with potential sensitivity to \g comparable to the partial reconstruction technique described in Sec.~\ref{sec.4.1.1}. 
As long as the fully and partially reconstructed datasets are kept statistically independent, the final sensitivity from $\Bpm \to D^{*0} \Kpm$ decays will be around $0.5^\circ$ as seen in Fig.~\ref{2body_gamma_projection}.

A GGSZ-like analysis of $\Bpm\to\D[ \to K_{s}^{0} \pi\pi\piz]\Kpm$ decays has recently been proposed for \belletwo, where a sensitivity approaching that of 
the $D \to K_{s}^{0} \pi\pi$ GGSZ analysis is expected~\cite{K:2017qxf}. With improved \piz efficiency, LHCb \upgradetwo can exploit this mode competitively. 
Lastly, higher \piz efficiency will merit the analysis of $B^\pm\to DK^{*\pm}[\to \Kpm\piz]$ decays. The \piz reconstruction efficiency is typically a factor 
3-4 lower than that of the \KS so the $\KS\pipm$ mode is preferred. 
However the $B^\pm\to D\Kpm\piz$ Dalitz analysis for \g should share many advantages of the isospin-conjugate decays $\Bd\to D \Kp\pim$ analysis (discussed next) 
but with reduced \Bs feed down and large asymmetries in the ADS-like region of the Dalitz space.

%% file: CONTRIBUTIONS/4_Time_integrated_CP_violation_measurements/4.1.4.tex
\subsection{Prospects with high-multiplicity modes}
\label{sec:4.1.4}

A variety of high-multiplicity $B$ and $D$ modes are already being established and will play an important role in a future determination of \g.
The most developed multibody $B$ decay channel is $\Bd\to D \Kp\pim$ decays, where the \D meson is found in an ADS/GLW-like ($K\pi,\ KK,\ \pi\pi,\ K3\pi,\ ... $) or GGSZ-like ($\KS\pi\pi$,\ $\KS KK,\ ...$) final states. 
These modes are less abundant than the equivalent \Bpm modes but the fact that both the $b\to c\bar{u}s$ and $b\to u\bar{c}s$ transitions proceed by colour-suppressed amplitudes means the GLW asymmetries can be very large. Furthermore, understanding the pattern of asymmetry across the \Bd Dalitz plane quashes the trigonometric ambiguities. 
This multibody technique will eventually surpass the quasi-two-body analysis introduced in Sec.~\ref{sec.4.1.1}. 
This analysis has been established with the $KK$ and $\pi\pi$ modes using Run 1 data in Ref.~\cite{LHCb-PAPER-2015-059} after the development of the $\Bd\to\Dzb\Kp\pim$ amplitude model~\cite{LHCB-PAPER-2015-017}. 
Although the statistical sensitivity to \CP violation using Run 1 data alone was not significant, the method remains promising for future analysis given the high value of $r_B$ in this mode ($\sim0.25$).
The extension to $\Bd\to D[\to\KS\pip\pim] \Kp\pim$ decays is of particular importance as it allows for a so-called ‘double Dalitz’ model-independent analysis to be performed~\cite{Gershon:2016fda,Gershon:2008pe}. 
Extrapolating yields from a Run 1 $\Bd\to\Dz\Kstarz$ analysis to a dataset corresponding to an integrated luminosity of 23 \invfb, sensitivity studies indicate that a precision on $\gamma$ of $3^\circ$ can be expected~\cite{Craik:2017dpc}, thus by extension, a sub-degree precision is expected for \upgradetwo.
Another high-multiplicity mode that holds promise is $B^\pm \to D \Kpm \pip\pim$, which has been studied with Run 1 data with two-body \D decays~\cite{LHCB-PAPER-2015-020}, though due to the unknown Dalitz structure of the $B$ decay a much larger dataset is needed to fully exploit this mode. 

\begin{figure}[h]
\begin{center}
\includegraphics[width=0.5\textwidth]{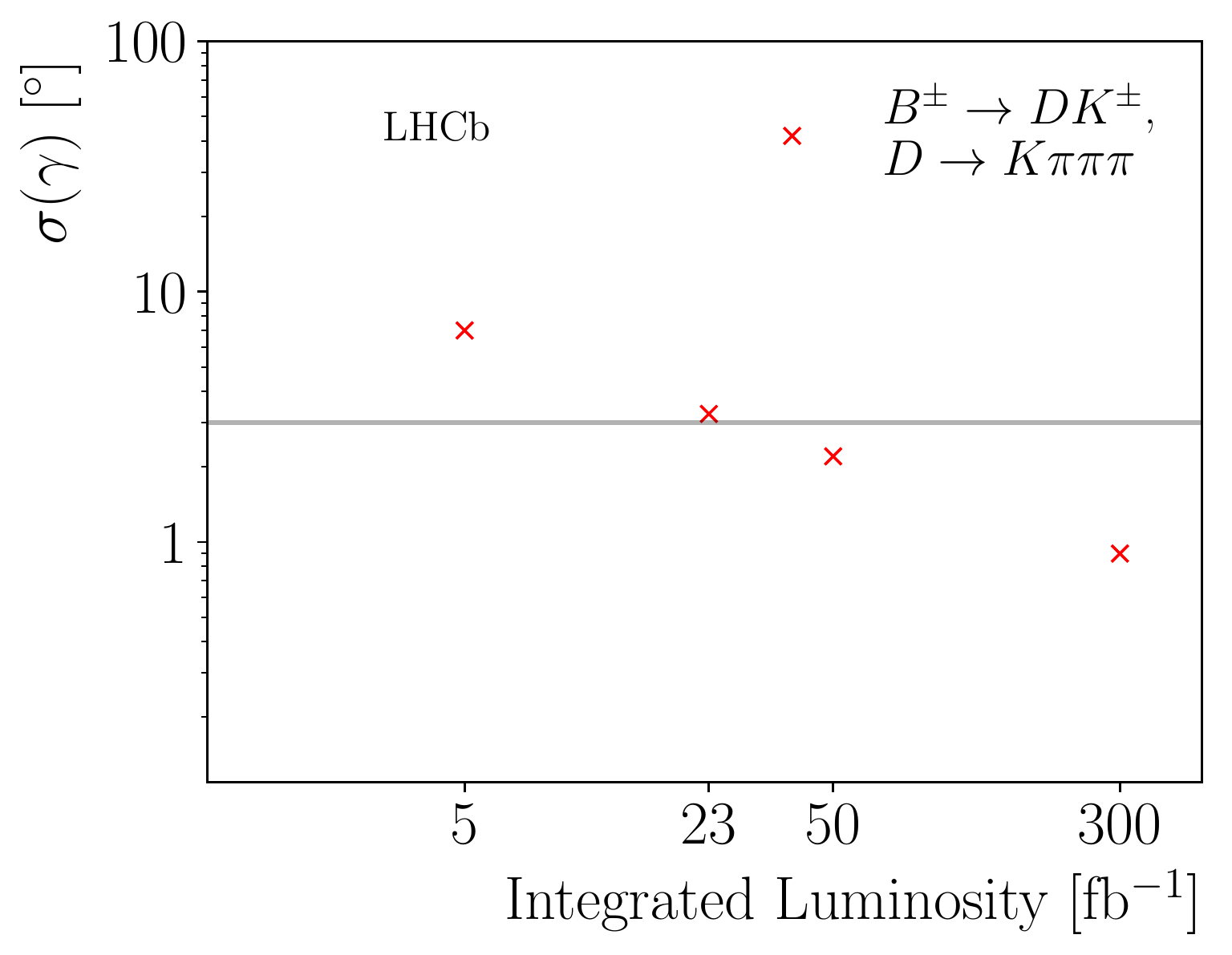}
    \captionof{figure}{Projected sensitivity for a binned analysis of $\Bpm \to D[\to\pi K\pi\pi] \Kpm$ decays. 
    \label{4body_gamma_projection}}
  \end{center}
\end{figure}

Another exciting extension to the standard ADS technique uses $\Bpm \to D[\to\pi K\pi\pi] \Kpm$ decays where the five-dimensional (5D) Dalitz volume of the $D$ decay is split into bins. 
In each bin, ADS/GLW relations derived from in Eq.~\ref{adseq} can be measured. Excellent sensitivity to \g is achieved as long as the $D$-system parameters ($r_D,\,\delta_D,\,\kappa$) in each bin are known. 
An optimal binning scheme is soon to be defined from the amplitude analysis of $D^0\to\pi K\pi\pi$ and $D^0\to K\pi\pi\pi$ decays~\cite{LHCb-PAPER-2017-040}.
The expected sensitivity is shown in Fig.~\ref{4body_gamma_projection}. An additional irreducible uncertainty from the \D system measurements of $<1^\circ$ is expected.

Other multibody \D decays are under development with reciprocal charm-system measurements underway: a $\Bpm \to D[\to4\pi] \Kpm$ analysis can build on the \D-system knowledge gathered in Ref.~\cite{Harnew:2017tlp}; an ADS analysis of $\Bpm \to D[\to\KS K\pi] \Kpm$ decays has been demonstrated with Run 1~\cite{LHCb-PAPER-2013-068}; and work on $\Bpm \to D[\to KK\pi\pi] \Kpm$ decays~\cite{Rademacker:2006zx} is envisaged.

%% file: CONTRIBUTIONS/4_Time_integrated_CP_violation_measurements/4.2.tex
\section{Ultimate $\gamma$ sensitivity and considerations on external inputs to the analyses}
\newcommand{\rbk}{\ensuremath{r_{B}^{DK}}\xspace}

\begin{figure}[tb]
\centering
\includegraphics[width=0.48\textwidth]{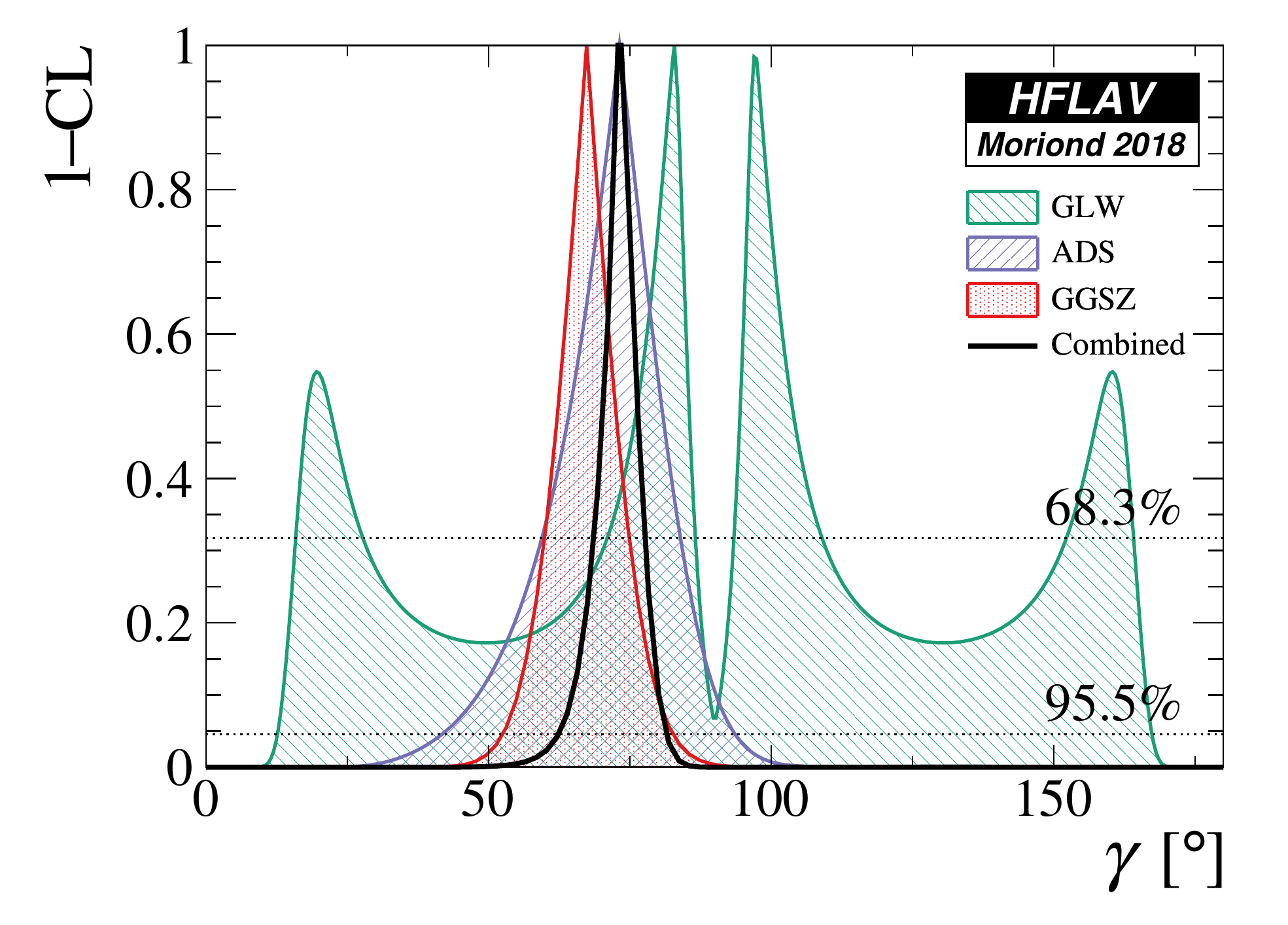}
\includegraphics[width=0.48\textwidth]{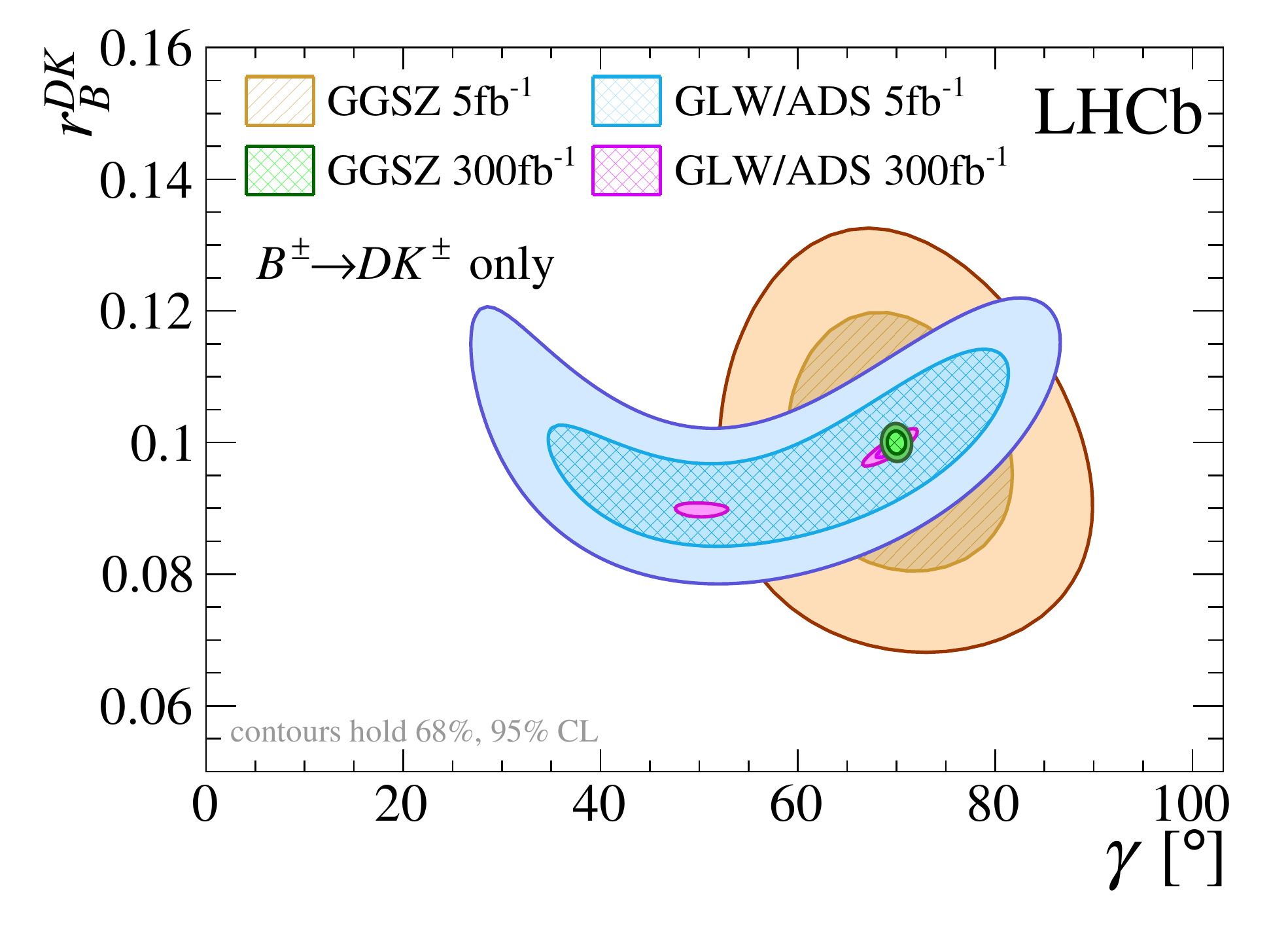}
  \caption{\small Left: Comparison of different methods used to determine the angle $\gamma$, reproduced from Ref.~\cite{HFLAV16}. Right: Comparison
  between the current LHCb 3-body GGSZ and 2-body GLW/ADS measurements alongside their future projections with 300\invfb in the plane of \g \vs
   \rbk (note the curtailed $y$-axis for \rbk). The scan is produced
  using a pseudo-experiment, centred at $\g=70\degrees$, $\rbk=0.1$, with $\Bpm\to\D\Kpm$ decays only.}
\label{fig:42:compare}
\end{figure}

The best sensitivity to $\gamma$ is achieved through a combination of measurements, as highlighted in Fig.~\ref{fig:42:compare} (left), where the contributions from three different methods are shown.
The GLW method provides four narrow solutions for $\gamma$ whilst the ADS and GGSZ methods have poorer sensitivity but, importantly, give a single solution.
Combining this information makes it possible to pick out the most consistent narrow solution. Furthermore, the various methods are complementary because their dominant sources
of systematic uncertainty have different origins.
A detailed description of the LHCb combination strategy is given in Ref.~\cite{LHCb-PAPER-2016-032}. The latest LHCb $\gamma$ combination, which dominates the world average, gives $\gamma = (74.0^{+5.0}_{-5.8})^\circ$~\cite{LHCb-CONF-2018-002}.

\begin{figure}[!htb]
\centering
\includegraphics[scale=0.5]{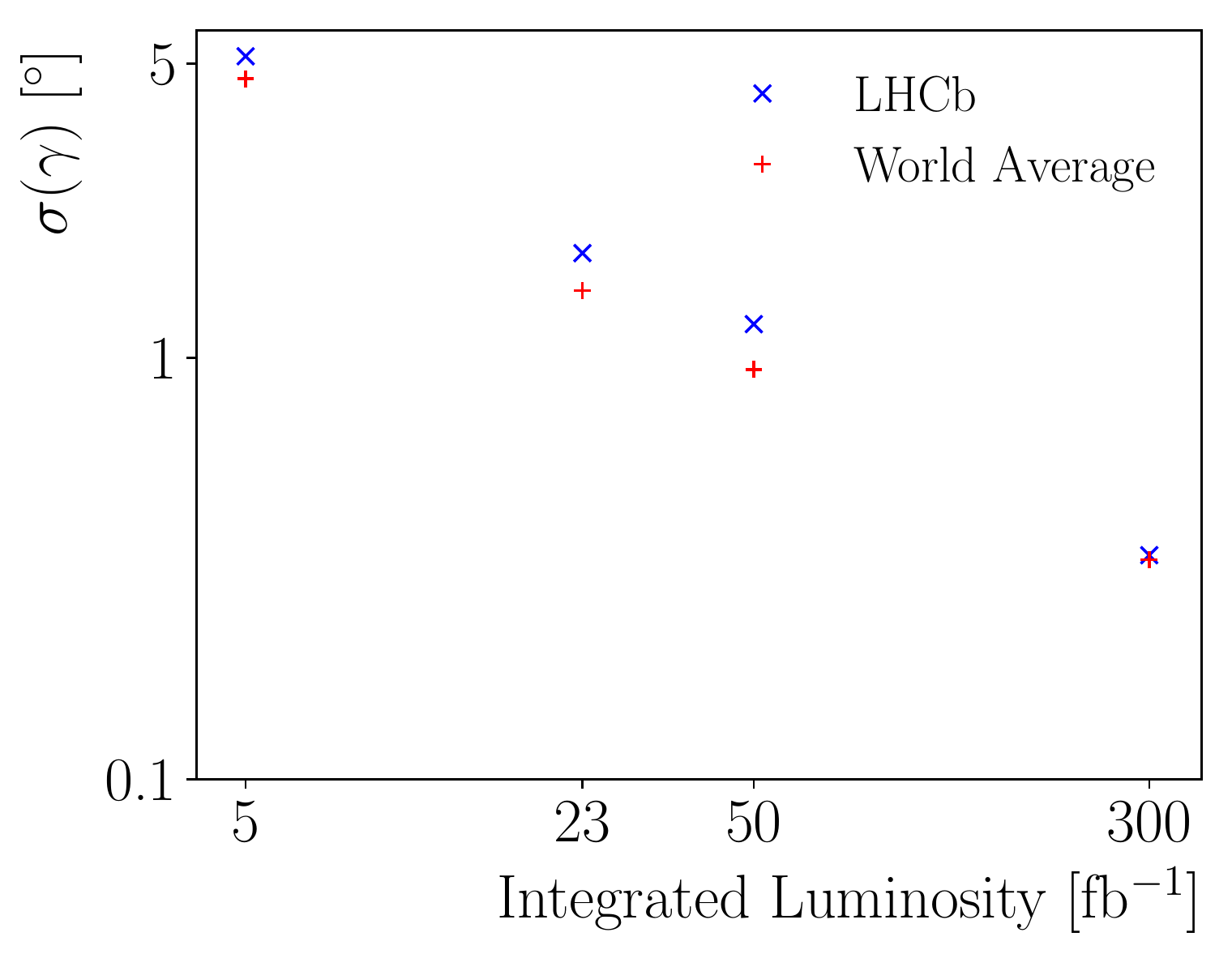}
  \caption{\small Projected sensitivity for the LHCb $\gamma$ combination with the currently used strategies and the world average using projections from \belletwo in addition.}
\label{fig:42:proj}
\end{figure}

Projections for the expected precision of the LHCb $\gamma$ combination are shown in Fig~\ref{fig:42:proj}, estimating the uncertainty on $\gamma$ to be $1.5^\circ$ and $0.35^\circ$ with 23~\invfb and 300~\invfb data samples, respectively.
The LHCb projections assume that the statistical uncertainty scales with the data sample size
and  include the effect of the increased centre-of-mass energy, increased trigger performance and increased integrated luminosity. Most of the systematic uncertainties are driven by the size of
 the data samples and corresponding simulation samples.
The overall sensitivity is predominantly driven by the GLW modes which provide
the narrowest solutions for \g. The dominant systematic uncertainty for these modes will depend on knowledge of both the shape and rate of the background
from $\Lb\to\Lc\Km$ as well as uncertainties arising from particle identification calibration and instrumental charge asymmetries.
In order to obtain the best possible precision on \g it will be necessary to ascertain the correlation of these uncertainties between
the different GLW modes. The ultimate sensitivity to these modes from \belletwo\ will be considerably less than that at \lhcb, however detailed analysis from
both experiments will provide an important cross check. For decays with neutrals in the final state, particularly the \CP-odd GLW mode with \Dz\to\KS\piz,
the \lhcb detector has considerable disadvantages over \belletwo. 
However these would be mitigated by an improved electromagnetic calorimeter for the \lhcb \upgradeone{b} and \upgradetwo.
The GGSZ modes are a powerful way to unambiguously resolve the multiple solutions of the GLW method and furthermore offer considerable standalone
sensitivity to \g.
Accurate knowledge of the selection efficiency across the Dalitz plane
is vital for these modes and contributes a considerable systematic uncertainty. This will naturally reduce with larger datasets as it is obtained
via semi-leptonic $\Bd\to\Dstarp\mun\neum X$ control modes but requires large simulation samples.
More precise measurement of important external parameters, particularly $c_i$ and $s_i$ from BESIII, will be required to reduce the uncertainty associated with the model independent GGSZ method.
The uncertainties of inputs from charm threshold data collected by CLEO-c will begin to limit the sensitivity by the end of Run 2, so it is essential to work together with BESIII to provide updated measurements for the suite of charm decays and \DtoKshh in particular.
Provided that the charm inputs are improved sub-degree level precision on $\gamma$ is attainable. Understanding the correlations between different
$B$ decay modes that all use these external parameters will be vitally important as they are likely to contribute one of the largest overall systematic
uncertainties in the combination.
A comparison between the current LHCb GGSZ and GLW/ADS measurements~\cite{LHCb-PAPER-2018-017,LHCb-PAPER-2017-021} with their future projections
at 300\invfb is shown in Fig.~\ref{fig:42:compare} (right). The order of magnitude increase in precision is very apparent and the importance of the
combination clear, given the multiple ambiguous solutions for GLW/ADS measurements is not resolved with increased luminosity.

The GGSZ modes are considered the \emph{golden modes} at \belletwo and drive the overall uncertainty on \g which is expected to reach
$1.5^\circ$ with a data sample of $50~{\rm ab}^{-1}$. This is comparable to the sensitivity that the \lhcb \g combination will
achieve with a data sample
corresponding to approximately $23\invfb$. Subsequently input from \belletwo will still contribute towards the world average by the end
of \lhcb's \upgradeone but \lhcb will dominate \g measurements with \upgradetwo (300\invfb) contributing entirely towards a world average precision of
$\sim0.35\degrees$.
The impact of this measurement on the unitarity triangle fit is shown in Fig.~\ref{fig:UTprojection}.
It should be emphasised that this projection includes only the currently used strategies, and does not include improvements from other approaches.
A comparison between the projected uncertainties for \lhcb and the world average as a function of integrated luminosity is shown in Fig.~\ref{fig:42:proj}.

%% file: CONTRIBUTIONS/4_Time_integrated_CP_violation_measurements/4.3.tex
\section{Amplitude analysis of $B^+ \to h^+h^+h^-$ decays}
\label{sec:Bu2hhh}

\begin{figure}[tb]
\begin{center}
\begin{overpic}[width=0.48\linewidth]{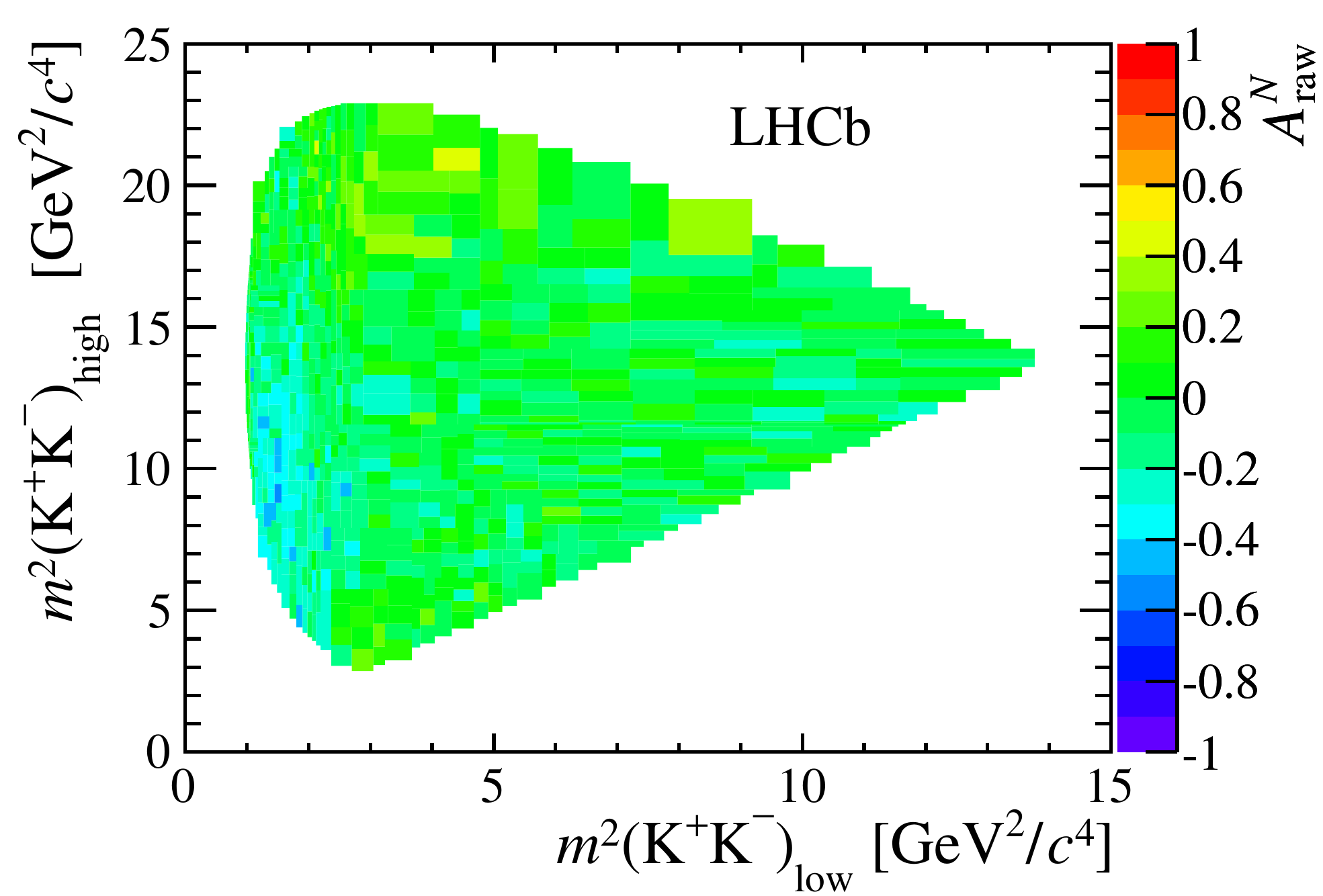}
 \put(73,58){\bf{(a)}}
\end{overpic}\hskip 0.04\textwidth
\begin{overpic}[width=0.48\linewidth]{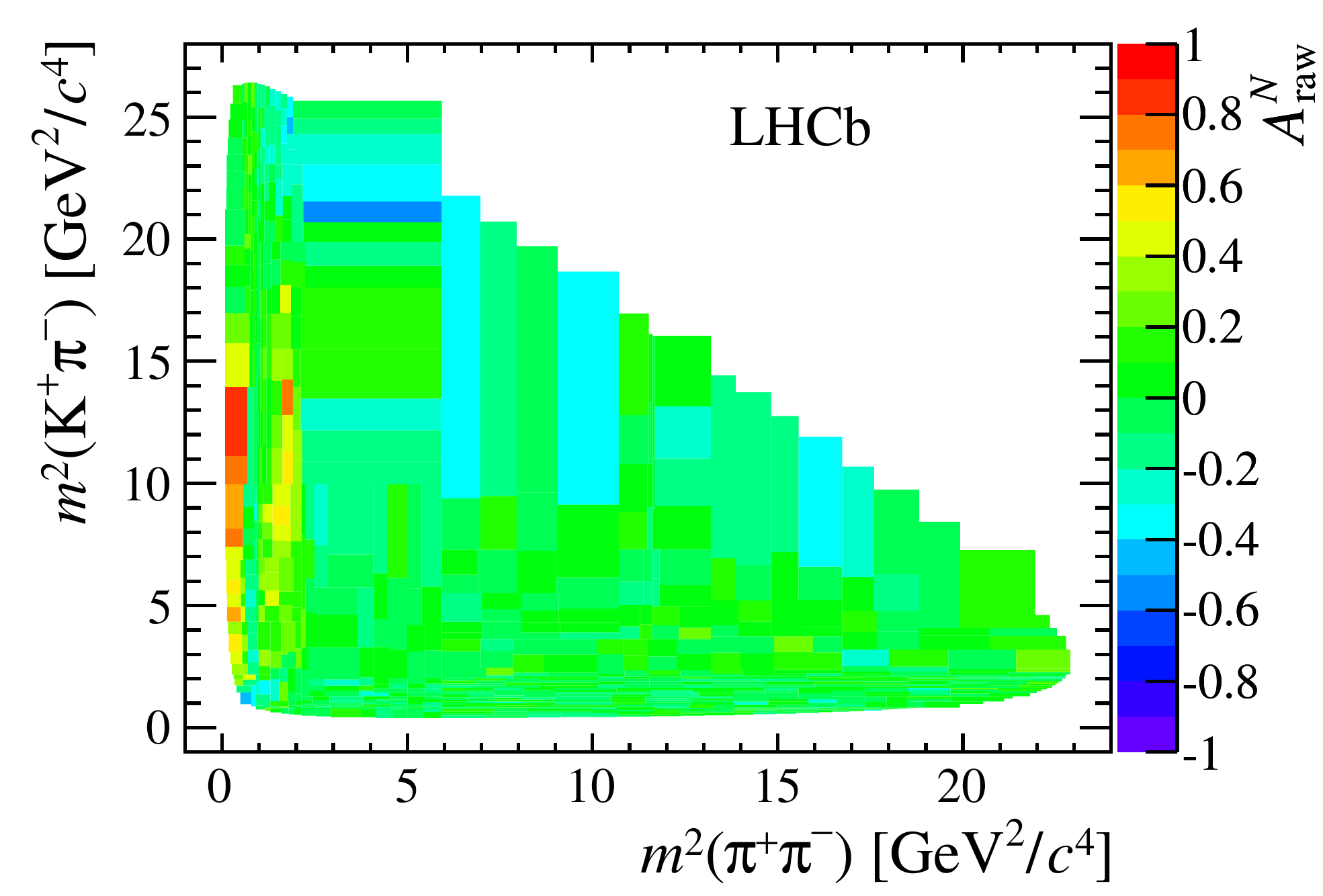}
 \put(73,58){\bf{(b)}}
\end{overpic}
\begin{overpic}[width=0.48\linewidth]{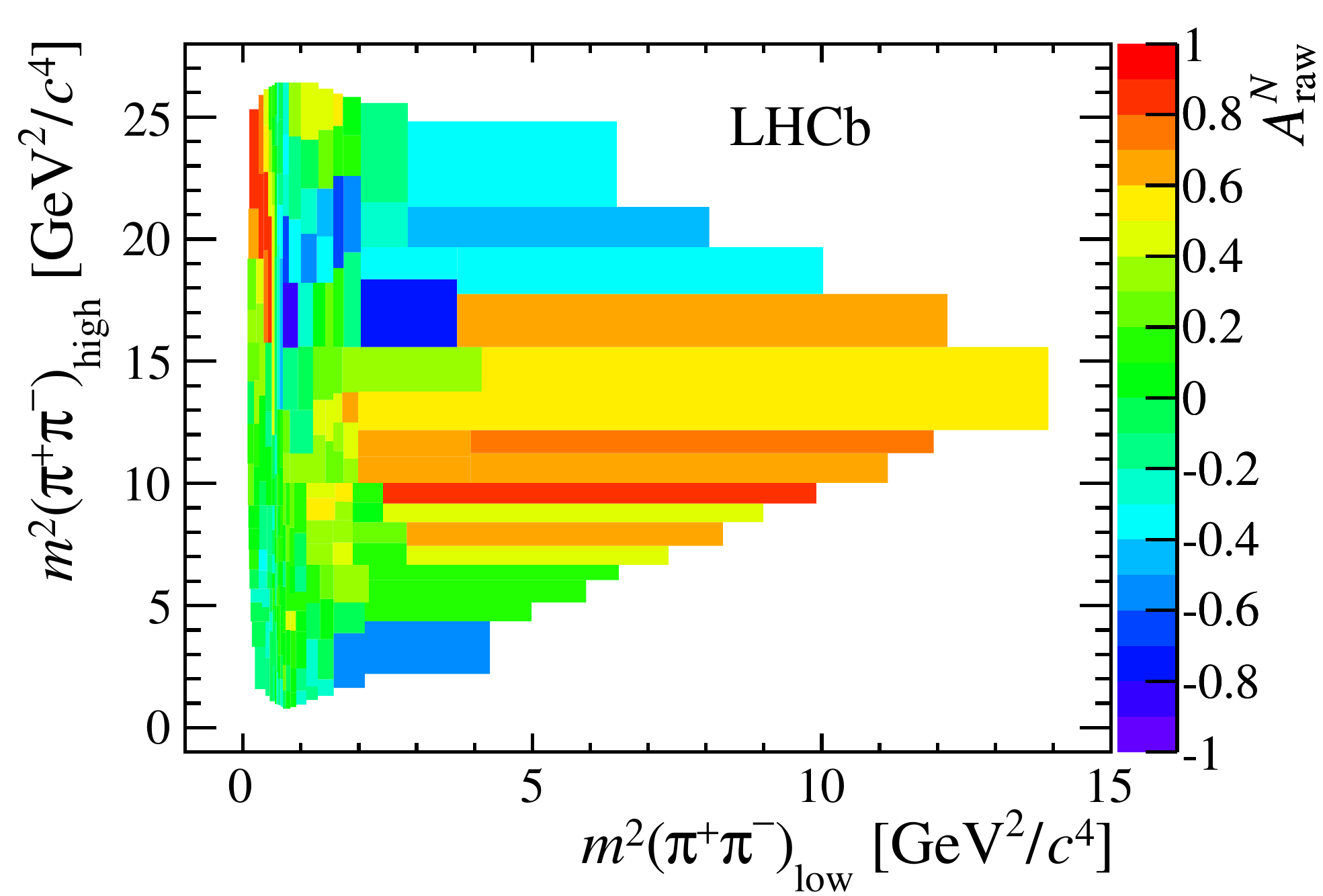}
 \put(73,58){\bf{(c)}}
\end{overpic}\hskip 0.04\textwidth
\begin{overpic}[width=0.48\linewidth]{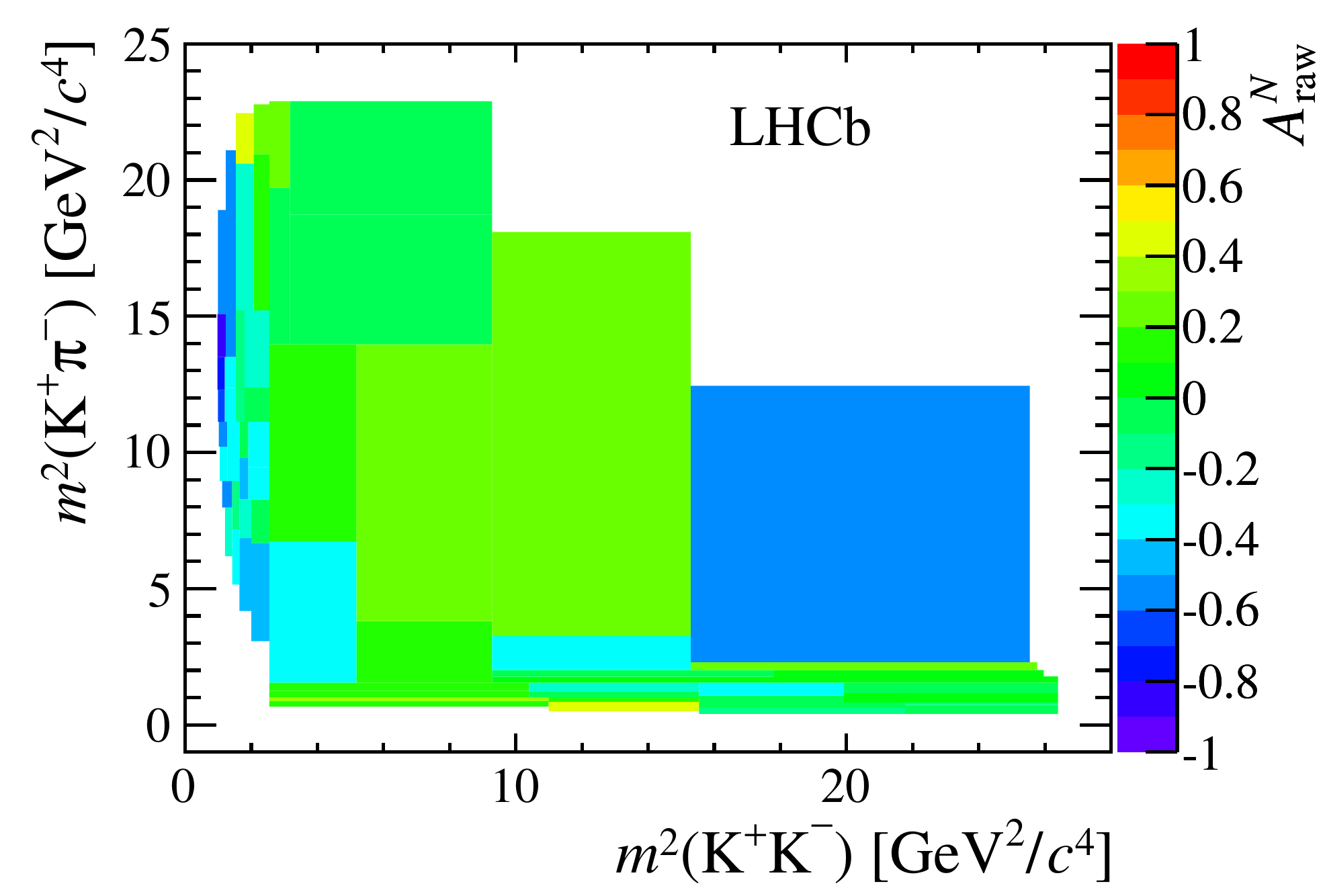}
 \put(73,58){\bf{(d)}}
\end{overpic}
\caption{
Asymmetry in the number of background-subtracted and efficiency-corrected
candidates as a function of position in phase space for
(a)~$\Bp\to\Kp\Kp\Km$,
(b)~$\Bp\to\Kp\pip\pim$,
(c)~$\Bp\to\pip\pip\pim$, and
(d)~$\Bp\to\Kp\pip\Km$ decays~\cite{LHCb-PAPER-2014-044}.
}
\label{fig:Bu2hhh-ACP}
\end{center}
\end{figure}

The decays of charged \B mesons to three-body final states containing
charged pions and kaons have elicited much recent interest due to the
observation~\cite{LHCb-PAPER-2014-044} of extremely large \CP-violating
asymmetries.
While the phase-space integrated asymmetries are of order a few percent,
the asymmetries as a function of the position in phase space are
considerably larger, even approaching $\pm 1$ in some areas, as illustrated
in Fig.~\ref{fig:Bu2hhh-ACP}.
The pattern in which the \CP asymmetries vary is also intriguing.
While some of the asymmetries appear to be associated with well established
resonances, such as the $\rho(770)^0$ meson, there are others in regions of the
phase space where there are no such structures.
It may be that additional strong phase differences are being generated by
rescattering processes, the effects of which should be particularly apparent at
the opening of thresholds.
Furthermore, it is possible that new physics processes could be
contributing and perhaps enhancing the size of the asymmetries.
It is vital therefore to perform amplitude analyses to try to disentangle
the different contributions and hence to better understand the mechanisms
behind these asymmetries.
Such amplitude analyses are currently ongoing at LHCb and have previously
been performed for a subset of modes, albeit with much smaller event
samples, at the \B factories~\cite{Garmash:2004wa,Garmash:2005rv,Aubert:2008bj,Aubert:2009av,Lees:2012kxa}.

Extrapolating the observed yields in Ref.~\cite{LHCb-PAPER-2014-044} to the data sample expected to be collected at the end
of \upgradetwo\ gives signal yields of between 1.2 and 36 million depending on the final state.
These can be compared to an extrapolation of the yields obtained by the
\B factories~\cite{Aubert:2008bj,Hsu:2017kir} to the expected 50\invab
sample to be collected by \belletwo, which gives approximately 47,000 to
660,000 events for the same range of modes. 
The far larger yields, combined with the much better signal to background
ratios, mean that LHCb will continue to dominate the precision in these modes.
Even with such large data samples, model-independent analyses of the \CP
asymmetries as a function of phase space can, with appropriate reduction of
the bin sizes, avoid becoming systematically limited.
The results of these analyses can be compared with predictions from
theoretical models, for example as developed in Ref.~\cite{Klein:2017xti}.
Such theoretical models for charmless hadronic \B decays are in their relative
infancy at present and it is hard to predict how far they can develop on
the timescale of the \upgradetwo.
However, it is clear that significant advances will be needed in order for
these asymmetry measurements to be related back to CKM parameters.

There are numerous methods that allow the extraction of 
CKM information by combining amplitude measurements made in
different decay channels, in particular the ${B^0_{(s)} \to \KS h^+h^-}$,
${B^0_{(s)} \to h^+h^-\piz}$, and ${B^+ \to h^+h^+h^-}$ families.
Some such methods exploit the interference between various $\Kstar\pion$ or
$\rho\kaon$ contributions, which can be related using isospin
symmetry~\cite{Ciuchini:2006kv,Ciuchini:2006st,Gronau:2006qn,Gronau:2007vr,Bediaga:2006jk,Charles:2017ptc}.
Constructing such relations enables the determination of the angle
$\gamma$, up to corrections for contributions from electroweak penguins.
These methods are particularly promising when using decays such as
${\Bs \to \KS \pip \pim}$ and ${\Bs \to \Km \pip \piz}$, since here the
electroweak penguin contributions are expected to be negligible.
Other methods use the whole Dalitz plot amplitude and relate numerous
decays via flavour symmetries~\cite{ReyLeLorier:2011ww,Bhattacharya:2013cla,Bhattacharya:2015uua}
or exploit interference between the charmless decays and those that proceed
via intermediate charmonium states such as
$\Bp\to\chiczero\pip$~\cite{Bediaga:1998ma,Blanco:2000gw}.

Amplitude analyses cannot benefit from cancellations of systematic
effects to the same degree as the binned measurements of asymmetries, and are more likely to become systematically limited.
On the other hand, the extremely large ${B^+ \to h^+h^+h^-}$ signal samples
will allow new information to be extracted.
In particular, by performing coupled-channel analyses of these decay modes,
contributions from $\pip\pim\leftrightarrow\Kp\Km$ rescattering processes
can be better understood and constrained.
The development of such new, improved models will also be of enormous
benefit to analyses of many other decay modes, such as the closely related
$\B \to \KS h^+ h^-$ and $\B \to h^+ h^- \piz$ families mentioned above and
discussed in more detail in Secs.~\ref{sec:alpha} and~\ref{sec:Kshh-hhpiz}.
This will help to reduce the corresponding uncertainty on the CKM phases
that can be determined from those channels.

The class of very rare hadronic decays such as \decay{\Bp}{\Kp\Kp\pim} and
\decay{\Bp}{\pip\pip\Km}~\cite{LHCb-PAPER-2016-023} proceed via diagrams
that are extremely suppressed in the SM and hence have
predicted branching fractions of order $10^{-14}$--$10^{-11}$~\cite{Huitu:1998vn}.
However, many BSM scenarios predict significant enhancements to these
decays~\cite{Fajfer:2006av,Fajfer:2000ny} to levels of between
$10^{-9}$--$10^{-6}$.
Preliminary studies indicate that sensitivities better than $10^{-9}$ can be
achieved in an \upgradetwo scenario assuming similar detector performance to
the present one can be maintained.
Of particular importance is the use of particle identification criteria to
suppress the backgrounds from the favoured decays such as
\decay{\Bp}{\Kp\pip\pim} that have branching fractions of order $10^{-5}$,
and the addition of TORCH may play a particularly important role in this.

%% file: CONTRIBUTIONS/4_Time_integrated_CP_violation_measurements/4.4.tex
\section{\CP violation in \bquark-baryon decays}

In contrast with the study of \CP violation in beauty-meson decays, the sector of beauty baryons remains
almost unexplored. Previous to the \lhc era, only a measurement of direct \CP asymmetries
in \decay{\Lb}{\proton\Km} and \decay{\Lb}{\proton\pim} decays was available with $\order(0.1)$
precision~\cite{Aaltonen:2014vra}. Thanks to the large production cross-section of beauty baryons in \proton\proton collisions at the \lhc, the \lhcb experiment is the only experiment capable of expanding our knowledge
in this sector, as these decays are not accessible at the \ep\en KEK collider. Hence, even though a handful
of \CP asymmetries of \Lb decays have been measured so far by
\lhcb~\cite{LHCb-PAPER-2013-061,LHCb-PAPER-2016-004,LHCb-PAPER-2016-059,LHCb-PAPER-2017-034},
the landscape of \CP violation in the decays of beauty baryons is expected to change rapidly in the
next few years. The first observation of \CP violation in a baryon decay is already within the reach of \lhcb 
with the data collected during the Run 2 of the \lhc, considering that a first evidence for \CP violation
in baryon decays has been reported in \decay{\Lb}{\proton\pim\pip\pim} decays~\cite{LHCb-PAPER-2016-030}. 

The unprecedented number of beauty baryons available with the data sample expected to be collected in the \lhcb \upgradetwo phase, will allow a precision measurement programme of \CP violation observables in \bquark-baryon decays to be pursued, analogously to \bquark-meson decays.
A very interesting example is the study of decays governed by \decay{\bquark}{\uquark} and \decay{\bquark}{\cquark} tree-level transitions, like the decays \decay{\Lb}{\Dz\Lambdares} and \decay{\Lb}{\Dz\proton\Km}. These decays can be used to measure the angle $\gamma$ of the unitarity triangle~\cite{Dunietz:1992ti,Fayyazuddin:1998aa,Giri:2001ju} in a similar way to what can be done with \decay{\Bd}{\D\Kp\pim} decays. The \lhcb experiment reported the first observation of the \decay{\Lb}{\Dz(\Km\pip)\proton\Km}, based on a signal yield of $163 \pm 18$ using a sample of \proton\proton collisions corresponding to 1\invfb of integrated luminosity at a centre-of-mass energy of 7\tev~\cite{LHCb-PAPER-2013-056}. Extrapolating to \upgradetwo\ approximately 95000 signal decays are expected. However, extrapolating the sensitivity to $\gamma$ is not easy, since it strongly depends on the values of the hadronic parameters involved in the process. In addition, even though the determination of $\gamma$ from the analysis of these decays is expected to be theoretically very clean, the possible polarisation of \Lb baryons produced in \proton\proton collisions has to be taken into account and might represent a limiting factor for high-precision measurements.

Another very interesting sector is that of beauty baryons decaying to final states without a charm quark. These decays receive relevant contributions from \decay{\bquark}{\dquark,\,\squark} loop-level transitions, where new physics beyond the SM may appear. Also in this case, similar quantities to those measured with \B-meson decays are available. For example, statistical precisions of $\order(10^{-3})$ and $\order(10^{-2})$ are expected for the \CP asymmetries of \decay{\Lb}{\proton h^-} and \decay{\Lb}{\Lz h^+h^-} decays (with $h = \kaon,\,\pion$), respectively. Very large signal yields are also expected in several multibody final states of \Lb and \Xib decays: about $10^6$ \decay{\Lb}{\proton\pim\pip\pim} and \decay{\Lb}{\proton\Km\Kp\Km} decays, and about $10^5$ \decay{\Xibz}{\proton\Km\pip\Km} decays~\cite{LHCb-PAPER-2016-030,LHCb-PAPER-2018-001}.
Such a signal yield will allow very precise measurements of \CP-violating quantities to be made over the phase space of these decays, characterised by a rich set of resonances. Unfortunately, as for the charmless decays of \B mesons, the interpretation of these quantities in terms of CKM parameters is still unclear from the theoretical point of view. Hence, more theoretical work is crucial to exploit the full potential of beauty baryons.

Experimentally, the main issues are the determination of particle-antiparticle production asymmetries and detection asymmetries that could mimic \CP-violation effects. This task is generally more difficult for heavy baryons, with respect to \B mesons, since methods used for measuring meson production asymmetries~\cite{LHCb-PAPER-2016-062} cannot be applied. In addition, different interactions of baryons and antibaryons with the detector material are difficult to calibrate. Nonetheless, several quantities can be measured in \bquark-baryon decays that are sensitive to different manifestations of \CP violation and are largely unaffected by experimental effects. A few examples are the difference of \CP-violating asymmetries of particles decaying to a similar final state, $\Delta A_{\CP}$~\cite{LHCb-PAPER-2017-044}, triple-product asymmetries (TPA)~\cite{LHCb-PAPER-2016-030} and energy-test (ET)~\cite{LHCb-PAPER-2014-054}. It is important to note that TPA and ET are important tools for discovery of \CP violation in multibody decays, while an amplitude analysis is required  to study the source of \CP violation.

%% file: CONTRIBUTIONS/5_Measurements_of_unitarity_triangle_sides_and_semileptonic_decays/5.tex
\label{chpt:semilept}

\input{CONTRIBUTIONS/5_Measurements_of_unitarity_triangle_sides_and_semileptonic_decays/5.0.tex}
\input{CONTRIBUTIONS/5_Measurements_of_unitarity_triangle_sides_and_semileptonic_decays/5.1.tex}
\input{CONTRIBUTIONS/5_Measurements_of_unitarity_triangle_sides_and_semileptonic_decays/5.3.tex}

\input{CONTRIBUTIONS/5_Measurements_of_unitarity_triangle_sides_and_semileptonic_decays/5.4.tex}

%% file: CONTRIBUTIONS/5_Measurements_of_unitarity_triangle_sides_and_semileptonic_decays/5.0.tex
\def\vub {\ensuremath{|V_{\uquark\bquark}|}\xspace}
\def\vcb {\ensuremath{|V_{\cquark\bquark}|}\xspace}
\def\qq {\ensuremath{q^{2}}\xspace}
\def\PMuNu{\decay{\Lb}{p \mun \neumb}}
\def\LcMuNu{\decay{\Lb}{\Lc \mun \neumb}}
\def\KMuNu{\decay{\Bs}{\Km \mup \neum}}
\def\DsMuNu{\decay{\Bs}{\Dsm \mup \neum}}
\def\Bq      {{\ensuremath{\B^0_\quark}}\xspace}
\def\Bqb     {{\ensuremath{\Bbar{}^0_\quark}}\xspace}
\def\dbtaunu {\decay{\B}{D^{(*)} \tau \nu}}
\def\dbmunu {\decay{\B}{D^{(*)} \mu \nu}}
\def\dbenu {\decay{\B}{D^{(*)} e \nu}}
\def\dmunu {\decay{\B}{D \mu \nu}}
\def\denu {\decay{\B}{D e \nu}}
\def\dsttaunu {\decay{\B}{D^{*} \tau \nu}}
\def\dstztaunu {\decay{\B}{D^{*0} \tau \nu}}
\def\dstptaunu {\decay{\B_{d}}{D^{*-} \tau^{+} \nu}}
\def\dstmunu {\decay{\B}{D^{*} \mu \nu}}
\def\dstenu {\decay{\B}{D^{*} e \nu}}
\def\dstpmunu {\decay{\B}{D^{*+} \mu \nu}}
\def\dstzmunu {\decay{\B}{D^{*0} \mu \nu}}
\def\dstlnu {\decay{\B}{D^{*} \ell \nu}}
\def\dblnu {\decay{\B}{D^{(*)} \ell \nu}}
\def\RDst {\ensuremath{\mathcal{R}(\Dstar)}\xspace}
\def\RDstp {\ensuremath{\mathcal{R}(\Dstarp)}\xspace}
\def\RD {\ensuremath{\mathcal{R}(D)}\xspace}
\def\RDb {\ensuremath{\mathcal{R}(D^{(*)})}\xspace}

Tree-level charged-current semileptonic decays of the type $b \to c \ell\nu$ and $b \to u \ell\nu$ play
a critical role in the search for new physics through quark flavour mixing.
That is because these decays can provide a theoretically clean determination of the CKM matrix elements
\vub and \vcb, which are essential ingredients in global fits that assume the unitarity of the CKM matrix.
These measurements have had rather a long and puzzling history,
in which those made with inclusive and exclusive final states have differed by over three standard deviations.
Given these controversies it is clear that precision measurements with the full suite of $b$ hadron species,
only accessible to LHCb, will be crucial to provide the maximum confidence if a discrepancy appears in the
unitarity triangle fits.

Semileptonic decays, being flavour specific, provide a unique probe of \Bq, where $q = s,d$, meson mixing phenomena.
In particular, \CP violation in \Bq meson mixing can be expressed through the semileptonic asymmetries
\begin{equation}
\label{eq:5.3.1}
  a_{\rm sl}^{q} = \frac{\Gamma (\Bqb \to f) - \Gamma (\Bq \to \bar{f})}{\Gamma (\Bqb \to f) + \Gamma (\Bq \to \bar{f})} \approx \frac{\Delta\Gamma_q}{\Delta M_q}\tan\phi_{12}^q,
\end{equation}
where $f$ is a flavour-specific final state that is only accessible through the decay of the $B_q^0$ state.
Mixing is required to mediate the transition $\Bqb \to \Bq \to f$, and its conjugate.
Semileptonic decays of the type $\Bq \to D_q^- \mu^+ \nu_{\mu} X$ are well suited because (i) they are immune to any unknown \CP violation in decay and (ii) they have large branching ratios.
These asymmetries are precisely predicted to be tiny in the SM~\cite{Lenz:2006hd,Artuso:2015swg}, while being highly sensitive to new mixing amplitudes in NP scenarios.

In the SM, the \dbmunu and \dbtaunu decays differ only through the effect of the lepton mass on the available phase space.
The ratio of their decay rates,
\begin{equation}
\RDb \equiv \frac{\dbtaunu}{\dbmunu},
\end{equation}
is theoretically extremely clean, since most form factor uncertainties cancel, while being sensitive to a range of NP models with 
preferential couplings to the third generation.
At the time of writing, the world averages for these quantities are in $\sim4\sigma$ tension with the SM expectation~\cite{HFLAV16}. 
This situation will be clarified with more precise measurements in the relatively near future.
If the discrepancy with the SM expectation persists, new observables with a range of different decay modes are needed to distinguish between different NP explanations.

%% file: CONTRIBUTIONS/5_Measurements_of_unitarity_triangle_sides_and_semileptonic_decays/5.1.tex
\def\vub {\ensuremath{|V_{\uquark\bquark}|}\xspace}
\def\vcb {\ensuremath{|V_{\cquark\bquark}|}\xspace}
\def\qq {\ensuremath{q^{2}}\xspace}
\def\PMuNu{\decay{\Lb}{p \mun \neumb}}
\def\LcMuNu{\decay{\Lb}{\Lc \mun \neumb}}
\def\KMuNu{\decay{\Bs}{\Km \mup \neum}}
\def\DsMuNu{\decay{\Bs}{\Dsm \mup \neum}}

\section{\label{sec:VubVcb}Determination of \vub and \vcb}

LHCb is well suited to measuring ratios of $b\to u\ell\nu$ to $b \to c\ell\nu$ decays rates,
in which the unknown $b$ production cross sections, and to some extent also efficiency corrections, cancel.
LHCb reported a first study of the decays \PMuNu and \LcMuNu with Run~1 data, which resulted in a determination
of \vub/\vcb~\cite{LHCb-PAPER-2015-013}, exploiting precise lattice QCD calculations of the decay form factors~\cite{Detmold:2015aaa,Flynn:2015mha}.

LHCb Upgrade~II presents an exciting opportunity for new measurements of this type.
An excellent example is the analogous analysis with \KMuNu and \DsMuNu decays.
The relatively large spectator $s$ quark mass allows the form factors of these decays to 
be computed with lattice QCD to higher precision than decays of other $b$ species.
There are also good prospects to extend the approach of~\cite{LHCb-PAPER-2015-013}, with a single $q^2$ bin,
to perform a differential measurement in many fine bins of $q^2$~\cite{Ciezarek:2016lqu},
which clearly demands substantially larger sample sizes.
Furthermore, there are several reasons to expect that, compared to the study of \Lb decays~\cite{LHCb-PAPER-2015-013}, far larger integrated luminosities are required for the
ultimate precision with \Bs decays. 
Firstly the \KMuNu signal rate is roughly one order of magnitude smaller compared to \PMuNu.
Secondly the \KMuNu decay is subject to backgrounds from all $b$ hadron species whereas \PMuNu is primarily contaminated
by other \Lb decays.
Thirdly the \KMuNu decay rate is further suppressed with respect to \PMuNu at the higher $q^2$ values at which the lattice QCD uncertainties are smallest.

\begin{figure}[!tb]
\centering
\includegraphics[width=0.6\linewidth]{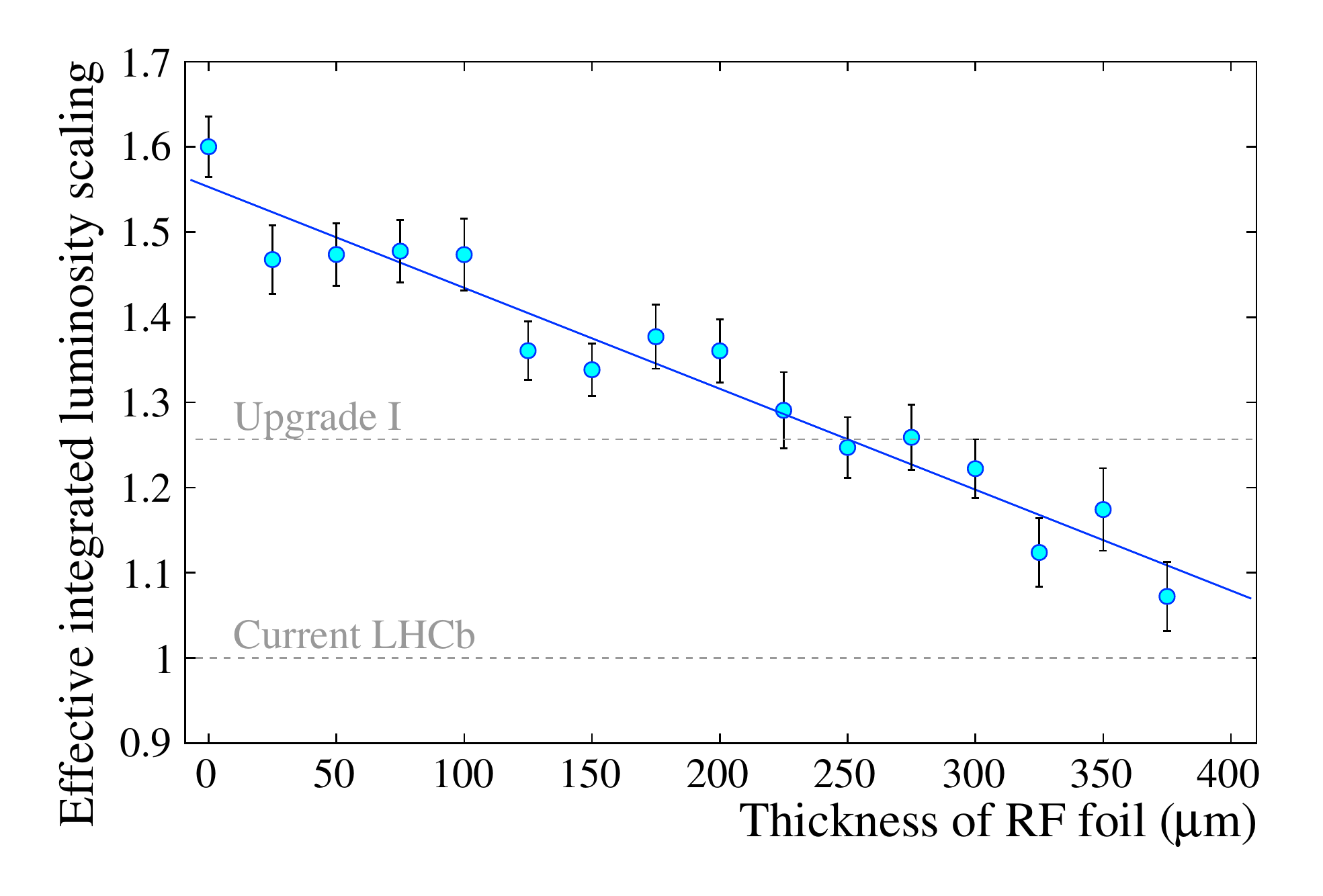} 
\caption{ The factor by which the effective integrated luminosity would be increased in an analysis of \KMuNu decays, as a function of the thickness of the RF foil.
This factor is defined with respect to the vertex resolution of the current detector.
It is determined using pseudoexperiments in which the \KMuNu signal yield is determined through a likelihood fit to the corrected mass distribution.
The other variables (primarily the first hit radius and resolution) affecting the vertex resolution are fixed to the Upgrade I values,
which should be possible to maintain for Upgrade II as discussed in Sect.~\ref{sec:tracking}.
The improvement in these variables, with respect to the current detector, is the reason that the blue trend line would cross the lower horizontal line
at a larger foil thickness than that of the real (current) detector (while crossing the upper horizontal line at the Upgrade I thickness of 250 \microns).
}
\label{fig:CorrectedMass}
\end{figure} 

Upgrade~II will also include several potential gains in detector performance which are highly relevant to the reconstruction
of decays like \KMuNu.
The key variable which distinguishes the signal from background processes is the {\em corrected mass},
which depends on the reconstructed line-of-flight between the PV and the \Bs decay vertex.
It is the resolution on this direction which dominates the corrected mass resolution.
The removal, or further thinning, of the RF foil is therefore a really appealing prospect, since this would reduce
the multiple scattering contribution to the corrected mass resolution.
Fig.~\ref{fig:CorrectedMass} shows the potential gain in effective integrated luminosity that can be achieved through a reduction of the RF foil thickness.
Hence, in the limit the RF foil could be removed entirely, then there would be almost a 1.6 increase in effective integrated luminosity. 
This analysis only considers the effect of the improved corrected mass resolution, while further improvements
are expected in the selection efficiency, purity and $q^2$ resolution.

The dominant systematic uncertainty in the \Lb analysis~\cite{LHCb-PAPER-2015-013} (material budget
and its effect on the charged hadron reconstruction efficiency) can be tightly constrained 
with new methods, the performance of which will be greatly enhanced by any reduction in the RF foil, as discussed in Sect.~\ref{sec:asl}. 
The lattice QCD form factors are most precise at large $q^2$ values, which correspond to low momentum kaons
that are not efficiently identified with the RICH approach of the current detector. 
The low momentum PID performance of the proposed TORCH detector would greatly enhance 
our capabilities with \KMuNu decays at high $q^2$.

The combination of these detector improvements with the Upgrade II dataset is expected to reduce the experimental systematic uncertainties to the 0.5\% level on \vub/\vcb. 
External branching ratio uncertainties will also be greatly reduced at the BESIII experiment. 
For example $\mathcal{B}(D_{s}^{+}\to K^{+} K^{-} \pi^{+})$ will be determined to the 1\% level, translating to a $\sim 0.5\%$ uncertainty on \vub/\vcb.
This combined with the differential shape information of the signals and continued improvements to the lattice 
calculations will lead to a \vub/\vcb measurement uncertainty of less than 1\% with the Upgrade II dataset.




Upgrade II will allow several currently inaccessible decays, not least those of the rarely produced \Bc mesons,
to be studied in detailed.
A prime example is the decay \decay{\Bc}{\Dz\mup\neum}, which is potentially
very clean from a theory point of view, due to the heavy charm spectator. 
Approximately 30,000  reconstructed decays can be expected with the 300\invfb Upgrade II dataset, which could lead to a competitive measurement of \vub
in this currently unexplored system.
Purely leptonic decays such as $B^{+} \to \mu^{+}\mu^{-}\mu^{+}\nu_{\mu}$ 
will also become competitive, with similar signal yields, and can provide information on the 
$B$-meson light cone distribution amplitude which is a crucial input to the widely used theoretical tool of QCD factorisation.



The standalone determination of $|V_{cb}|$ will also be increasingly important as other measurements get more precise,
as it will become the limiting factor in many SM predictions such as the branching fraction of $\Bs\to\mumu$~\cite{Buras:2012ru}. 
The current uncertainty is inflated due to the disagreement between measurements from inclusive and exclusive final states and currently appears to 
critically depend on the parameterisation used to fit the form factors~\cite{Grinstein:2017nlq,Bigi:2017njr}. 
LHCb has already performed a measurement of the differential rate of the decay $\Lb \to \Lc \mu \nu$, allowing a determination of the form factors
of that decay~\cite{LHCb-PAPER-2017-016}. A first determination of the absolute value of $|V_{cb}|$, exploiting theoretical predictions for the ratios of semileptonic
decay widths between different $b$ hadron species~\cite{Bigi:2011gf}, is in progress.
The LHCb Upgrade II dataset would provide large samples of exclusive $b \to c \ell\nu$ decays, with the full range of $b$ hadron species,
with which incredibly precise shape measurements could be performed, as a crucial ingredient to reach the ultimate precision of \vcb.

%% file: CONTRIBUTIONS/5_Measurements_of_unitarity_triangle_sides_and_semileptonic_decays/5.3.tex
\section{Semileptonic asymmetries $a^d_{\rm sl}$ and $a^s_{\rm sl}$}
\label{sec:asl}

\def\Bq      {{\ensuremath{\B^0_\quark}}\xspace}
\def\Bqb     {{\ensuremath{\Bbar{}^0_\quark}}\xspace}


The observable defined in Eq.~\ref{eq:5.3.1} would require flavour tagging to be measurable.
Including the effect of an unknown production asymmetry, $a_p$, the time-dependent {\em untagged} asymmetry is defined as: 
\begin{equation}
\label{eq:5.3.2}
  A_{\rm sl}^{q}(t) \equiv \frac{N - \bar{N}}{N + \bar{N}} = \frac{a_{\rm sl}^{q}}{2} - \left[ a_{p} + \frac{a_{\rm sl}^{q}}{2} \right] \cdot \left[\frac{\cos \Delta M_q t }{\cosh \Delta \Gamma_q t /2}\right],
\end{equation}
where $N$ and $\bar{N}$ denote the number of observed decays to the $f$ and $\bar{f}$ final states, respectively.
A decay-time-dependent fit can disentangle the $B_d^0-\bar{B}_d^0$ production asymmetry from $a_{\rm sl}^d$~\cite{LHCb-PAPER-2014-053}.
In the $B_s^0$ case the {\em time-integrated} asymmetry is employed~\cite{LHCb-PAPER-2016-013}.
Owing to the large value of $\Delta M_s$ the term containing $a_p$ is suppressed to a calculable correction of a few $\times 10^{-4}$,
after integrating over decay time.
These approaches have been applied in the measurements with the Run 2 dataset that are listed in Table~\ref{tab:5.3.1}.
These are already the world's best single measurements of these parameters.
However these measurements are far from any uncertainty floor in the SM predictions, while being sensitive to a vast number of anomalous NP contributions to $\Gamma_{12}^{q}$ and $M_{12}^{q}$.

The following briefly reviews the dominant sources of uncertainty on the current LHCb measurements, and considers strategies to 
fully exploit the potential of the \upgradetwo.
All uncertainties are as evaluated on $a_{\rm sl}^{q}$ (\ie all sources of raw asymmetry, and their uncertainties, are scaled by  
a factor of two as expected from Eq.~\ref{eq:5.3.2}).

\begin{figure}[!tb]
\centering
\includegraphics[width=0.6\linewidth]{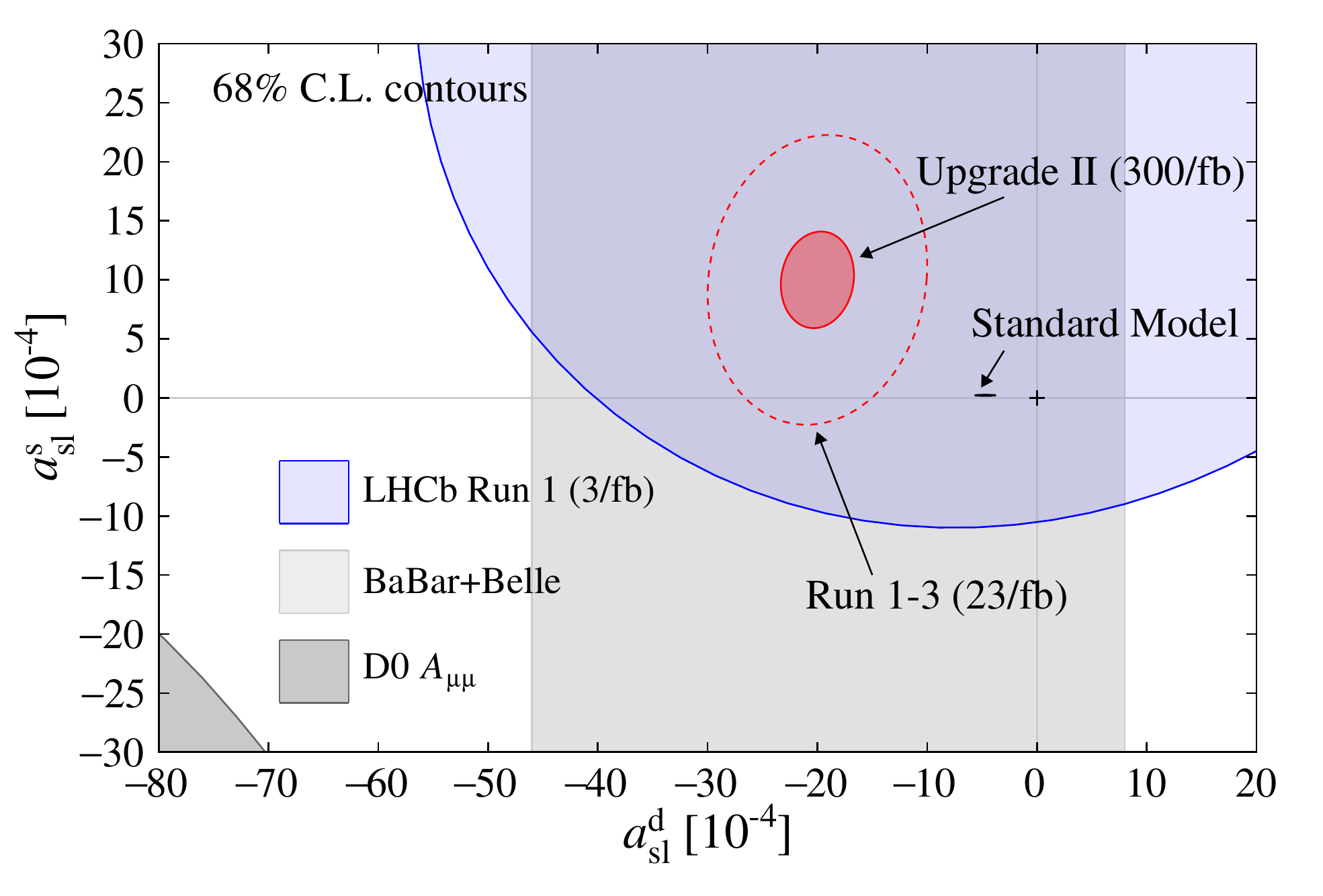}
\caption{
Current and future landscape for the semileptonic asymmetries.
The grey vertical band indicates the current \B-Factory average for $a_{\rm sl}^d$~\cite{HFLAV16}.
The blue ellipse represents the current LHCb Run 1 measurements~\cite{LHCb-PAPER-2016-013,LHCb-PAPER-2014-053}.
The red ellipse, which is arbitrarily centred, delineates the \upgradetwo projected precision.
The black ellipse shows the SM prediction, the uncertainty of which is barely visible with these axes.
\label{fig:5.3}}
\end{figure}

\begin{description}
\item[Statistical precision:] 
The statistical uncertainty on $a^s_{\rm sl}$ straightforwardly extrapolates to $2.1 \times 10^{-4}$.
In the case of $a^d_{\rm sl}$ it should be noted that stringent fiducial cuts and weights were imposed on the signal sample
to control certain tracking asymmetries that were not well known at the time. By the time of the subsequent $a^s_{\rm sl}$ measurement,
a new method with $J/\psi \to \mu^+\mu^-$ decays had been developed, and others are in the pipeline. 
A factor of two further increase in yields is therefore assumed,
which extrapolates to an uncertainty of $1.1 \times 10^{-4}$.
\item[Detection asymmetries:] The single largest contributor is the $K^-\pi^+$ asymmetry in $a^d_{\rm sl}$.
This asymmetry was determined with a single method using $D^+$ 
decays to the $K^-\pi^+\pi^+$ and $\KS\pi^+$ final states,
with a precision of around $2.0 \times 10^{-3}$~\cite{LHCb-PAPER-2014-053}.
Thanks to trigger improvements, a factor of two increase in the effective yield of the
most limiting $\KS\pi^+$ final state~\cite{Davis:2284097} can be anticipated. 
This will extrapolate to an uncertainty of $1.1 \times 10^{-4}$. 
With improvements in the reconstruction of downstream tracks in the first HLT stage, we may also be able to exploit 
$\KS\pi^+$ final states with $\KS$ decays downstream of the VELO.
Further methods have since been proposed using partially reconstructed $D^{*+}$ decays and $D^0 \to K^-K^+$ decays.
The partial reconstruction method will be greatly improved by the reduction of material before the first VELO measurement point.
The systematic uncertainties in these approaches will be controlled by using ultra high statistics fast simulations to track the 
kinematic dependencies in the asymmetries. We target an uncertainty of $1.0 \times 10^{-4}$ including systematic uncertainties.
The equivalent $K^+K^-$ asymmetry in the $a^s_{\rm sl}$ measurement will be smaller and more precisely controlled.
The $\mu^+\pi^-$ asymmetry will be controlled by a combination of $J/\psi \to \mu^+\mu^-$ decays, partially reconstructed $D^{*+}$ decays,
$D^0 \to h^-h^+$ decays, and high statistics fast simulations. 
\item[Background asymmetries:] These measurements are challenging because the \hbox{$\Bq \to D_q^- \mu^+ \nu_{\mu} X$} final states can be fed by
the decays of other $b$ hadron species. This dilutes the relation between the raw asymmetry and $a_{\rm sl}^q$ and leads to a cocktail of 
production asymmetry corrections. We assume that these backgrounds can be statistically subtracted by extending the signal fits to include the
$D_q^- \mu^+$ corrected mass dimension. It is assumed that the background asymmetry uncertainties can be controlled to the $1.0 \times 10^{-4}$ level.
\item[Other considerations:] We must assess the impact of having unequal sample sizes in the two polarities. This can be partially compensated for by
assigning weights to one polarity~\cite{Vesterinen:1642153}. 
We note that the choice of crossing angles should be carefully considered~\cite{Dufour:2304546}.
While we do not account for them in the current estimation, we utilise other $D^{+}_{q}$ decay modes to better align the detection asymmetries between $a_{\rm sl}^s$ and $a_{\rm sl}^d$.
For example, $D^+ \to K^-K^+\pi^+$ and $D^+ \to \KS \pi^+$ decays can be used, taking advantage of possible improvements in the trigger efficiency for \KS decays in Upgrade II.
While the former decay is singly Cabibbo suppressed, its \CP asymmetry could be measured using promptly produced $D^+$ mesons.
\end{description}
In summary the Upgrade II dataset should allow both asymmetries to be measured to the level of a few parts in $10^{-4}$,
as listed in Table~\ref{tab:5.3.1}, which will give unprecedented new physics sensitivity, and is still far from saturating the {\em current}
theory uncertainties in the SM predictions.
Figure~\ref{fig:5.3} shows the prospective \upgradetwo measurement, arbitrarily centred at a value that differs from the SM
prediction at the $10^{-3}$ level. 




\begin{table}\caption{\label{tab:5.3.1}Current theoretical and experimental determinations of the semileptonic asymmetries $a^d_{\rm sl}$ and $a^s_{\rm sl}$.}
\centering


\begin{tabular}{lcc}
\hline
Sample ($\mathcal{L}$) & $\delta a^s_{\rm sl} [10^{-4}]$ & $\delta a^d_{\rm sl} [10^{-4}]$ \\
\hline
Run 1 (3 \invfb)~\cite{LHCb-PAPER-2016-013,LHCb-PAPER-2014-053} & $33$  & $36$\\
Run 1-3 (23 \invfb) &  $10$ &  $8$ \\
Run 1-3 (50 \invfb) &  $7$ &  $5$ \\
Run 1-5 (300 \invfb) &  $3$ &  $2$ \\
\hline
Current theory~\cite{Lenz:2006hd,Artuso:2015swg} & $0.03$ & $0.6$\\
\hline
\end{tabular}

\end{table}


%% file: CONTRIBUTIONS/5_Measurements_of_unitarity_triangle_sides_and_semileptonic_decays/5.4.tex
\section{Lepton-flavour universality tests with $b \to c l \nu$ transitions}
\input{CONTRIBUTIONS/5_Measurements_of_unitarity_triangle_sides_and_semileptonic_decays/5.4.1.tex}

\input{CONTRIBUTIONS/5_Measurements_of_unitarity_triangle_sides_and_semileptonic_decays/5.4.2.tex}

%% file: CONTRIBUTIONS/5_Measurements_of_unitarity_triangle_sides_and_semileptonic_decays/5.4.1.tex
\subsection{$R(D)$ and $R(D^\ast)$ with muonic and hadronic $\tau$ decays}

\def\dbtaunu {\decay{\B}{D^{(*)} \tau \nu}}
\def\dbmunu {\decay{\B}{D^{(*)} \mu \nu}}
\def\dbenu {\decay{\B}{D^{(*)} e \nu}}
\def\dmunu {\decay{\B}{D \mu \nu}}
\def\denu {\decay{\B}{D e \nu}}
\def\dsttaunu {\decay{\B}{D^{*} \tau \nu}}
\def\dstztaunu {\decay{\B}{D^{*0} \tau \nu}}
\def\dstptaunu {\decay{\B_{d}}{D^{*-} \tau^{+} \nu}}
\def\dstmunu {\decay{\B}{D^{*} \mu \nu}}
\def\dstenu {\decay{\B}{D^{*} e \nu}}
\def\dstpmunu {\decay{\B}{D^{*+} \mu \nu}}
\def\dstzmunu {\decay{\B}{D^{*0} \mu \nu}}
\def\dstlnu {\decay{\B}{D^{*} \ell \nu}}
\def\dblnu {\decay{\B}{D^{(*)} \ell \nu}}
\def\RDst {\ensuremath{\mathcal{R}(\Dstar)}\xspace}
\def\RDstp {\ensuremath{\mathcal{R}(\Dstarp)}\xspace}
\def\RD {\ensuremath{\mathcal{R}(D)}\xspace}
\def\RDb {\ensuremath{\mathcal{R}(D^{(*)})}\xspace}


\begin{figure}[!tb]
\centering
\includegraphics[width=0.7\linewidth]{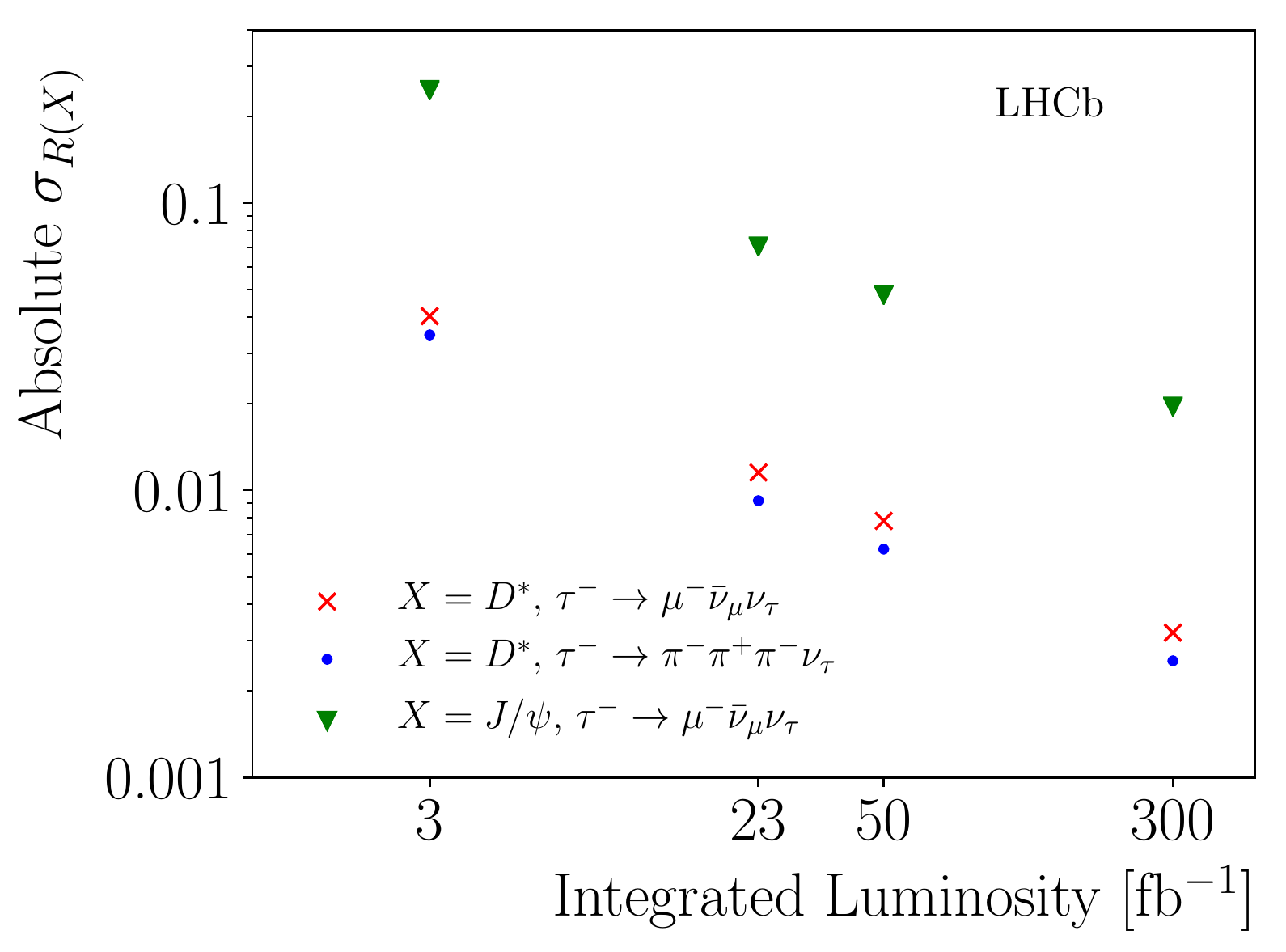}
\caption{
The projected absolute uncertainties on $\mathcal{R}(D^{\ast})$ and $\mathcal{R}(J/\psi)$ (see Sect.~\ref{subsec:OtherBhadronRX}) from the current sensitivities (at 3\invfb) to 23\invfb, 50\invfb, and 300\invfb.
}
\label{fig:RXproj}
\end{figure} 

LHCb has made measurements of \RDb using both muonic ($\tau^{+} \to \mup \nu \nu$) and hadronic ($\tau^{+} \to \pip \pim \pip \nu$) decays of the tau lepton~\cite{LHCb-PAPER-2015-025,LHCb-PAPER-2017-017,LHCb-PAPER-2017-027}.
Due to the presence of multiple neutrinos these decays are extremely challenging to measure.
The measurements rely on isolation techniques to suppress partially reconstructed backgrounds, 
\B meson flight information to constrain the kinematics of the unreconstructed neutrinos, and a multidimensional template fit to determine the signal yield. 
Figure~\ref{fig:RXproj} shows how the absolute uncertainties on the LHCb muonic and hadronic $\mathcal{R}(D^{\ast})$ measurements are projected to evolve with respect to the current status.
The major uncertainties are the statistical uncertainty from the fit, the uncertainties on the background modelling and the limited size of simulated samples.
A major effort is already underway to commission fast simulation tools.
The background modelling is driven by a strategy of dedicated control samples in the data, and so this uncertainty will continue to improve with larger data samples.
From Run 3 onward it is assumed that, taking advantage of the full software HLT, the hadronic analysis can normalise directly to the $B^0 \to D^{\ast -}\mu^+\nu_{\mu}$  decay,
thus eliminating the uncertainty from external measurements of $\mathcal{B}(B^0 \to D^{\ast -} \pi^+\pi^-\pi^+)$.
It is assumed that all other sources of systematic uncertainty will scale as $\sqrt{\mathcal{L}}$.
With these assumptions, an absolute uncertainty on $\mathcal{R}(D^{\ast})$ of $0.003$ will be achievable for the muonic and hadronic modes with the 300\invfb Upgrade II dataset.

\begin{figure}[!tb]
\centering
\includegraphics[width=0.45\linewidth]{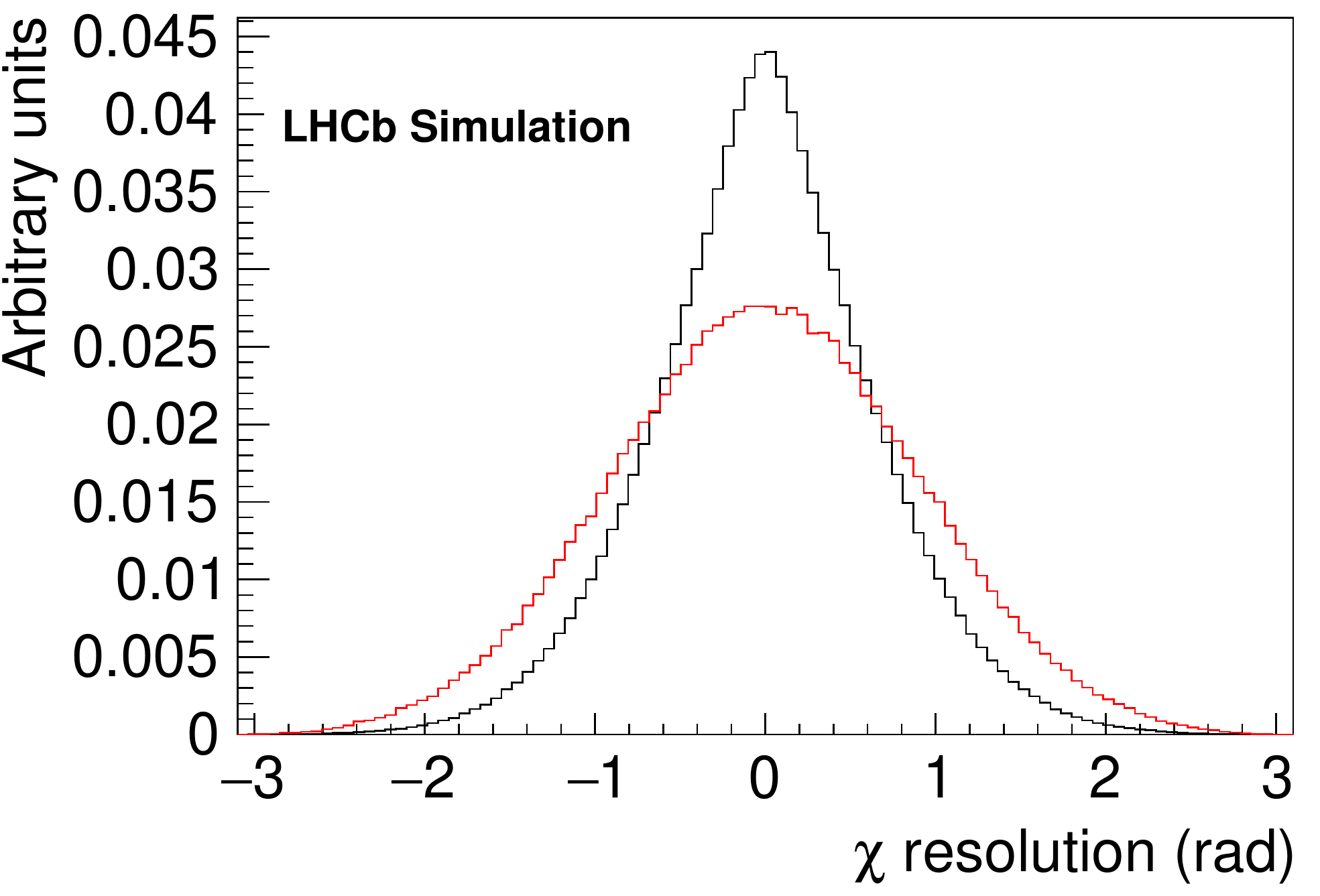}
\includegraphics[width=0.45\linewidth]{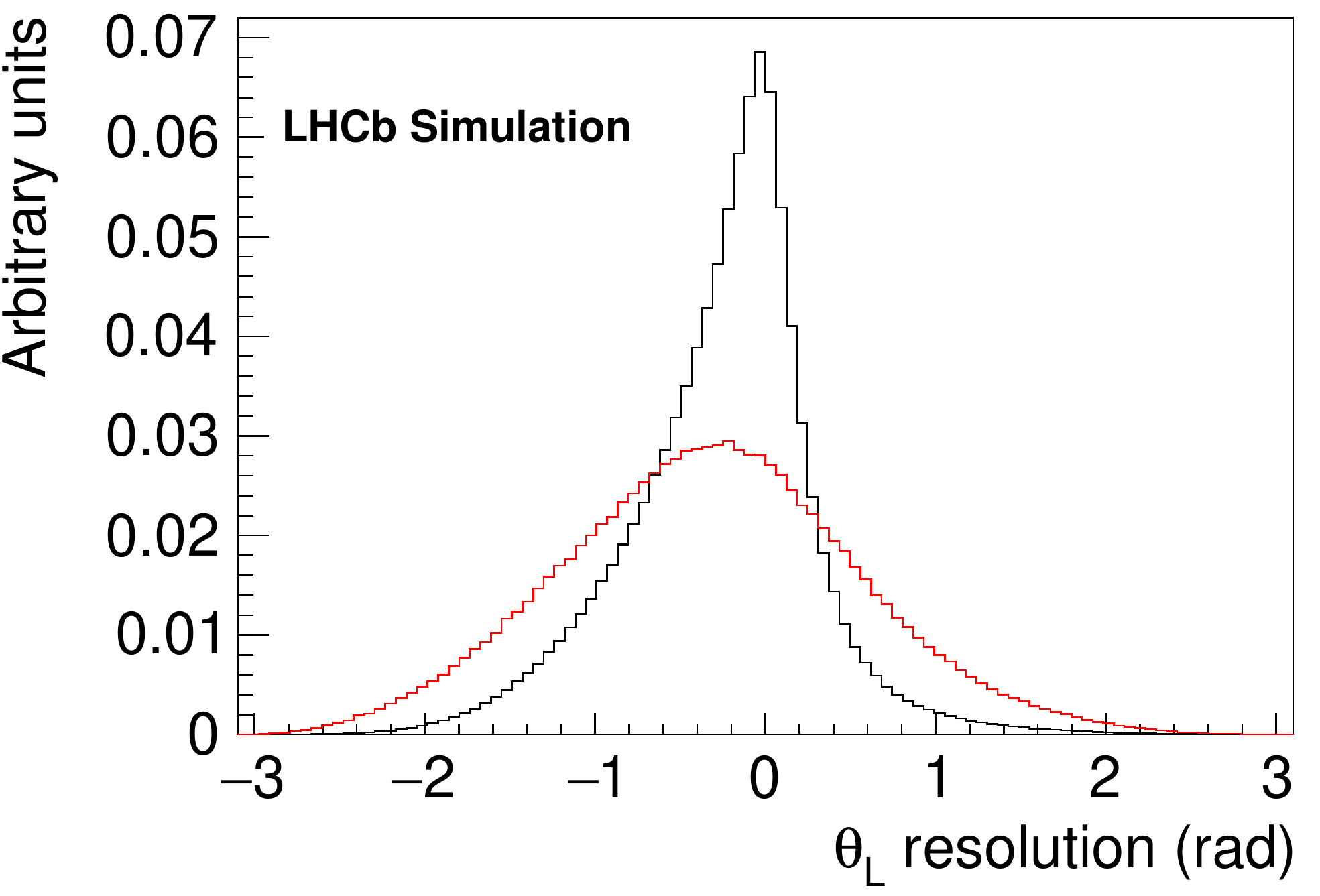}
\includegraphics[width=0.45\linewidth]{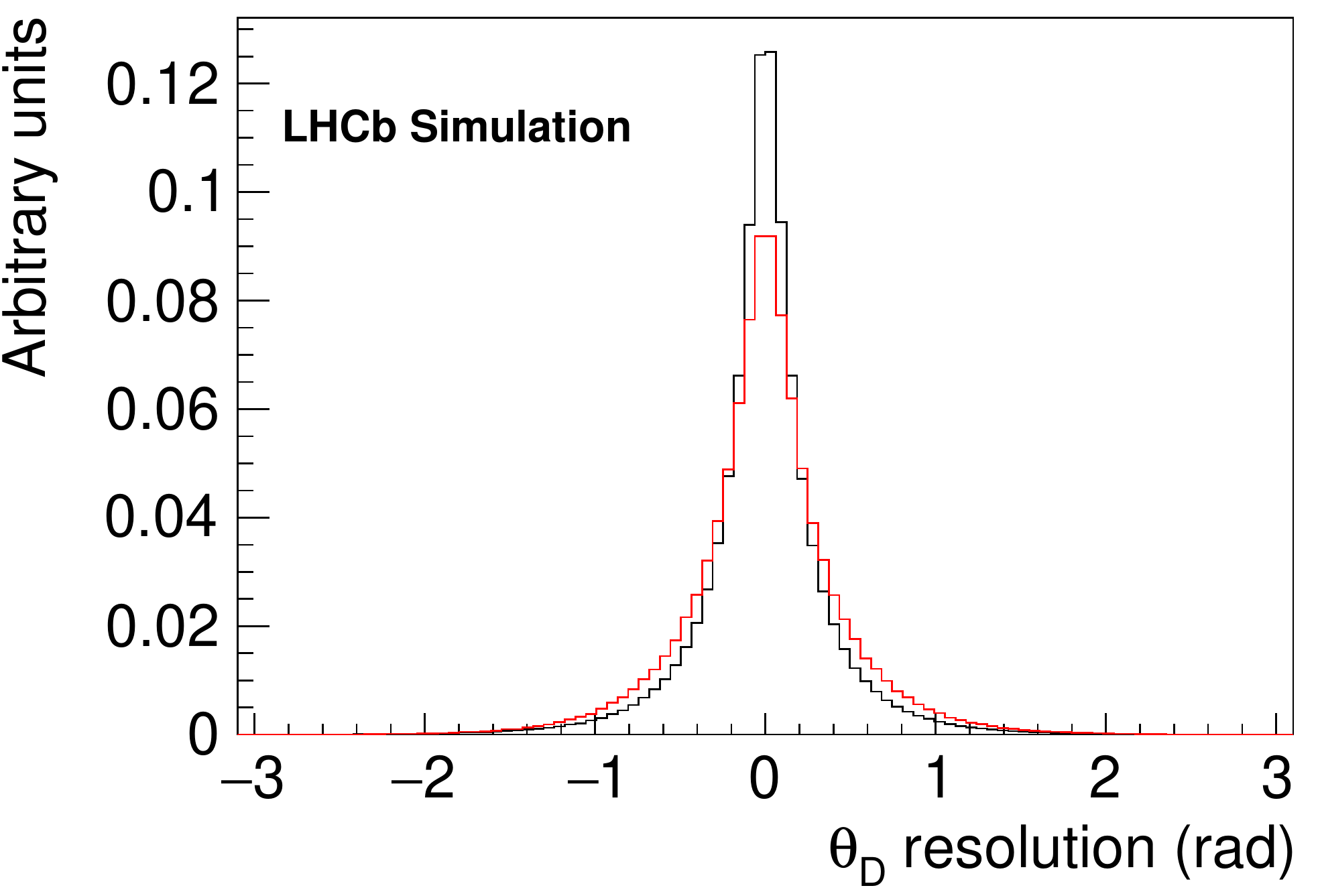}
\caption{
Angular resolution for simulated \dstmunu (black) and \dsttaunu  (red) decays, with $\tau^{+} \to \mup \nu \nu$. 
This demonstrates our ability to resolve the full angular distribution, with some level of statistical dilution.
}
\label{fig:RDAngles}
\end{figure}

On the timescale of \upgradetwo, interest will shift toward new observables  beyond the branching fraction ratio~\cite{Becirevic:2016hea}.
The kinematics of the \dsttaunu decays is fully described by the dilepton mass, and three angles which are denoted $\chi$, $\theta_L$ and $\theta_D$. 
LHCb is capable of resolving these three angles, as can be seen in~\figref{fig:RDAngles}.
However, the broad resolutions demand very large samples to extract the underlying physics.
The decay distributions within this kinematic space are governed by the underlying spin structure, and precise measurements of these distributions will allow the different helicity amplitudes to be disentangled.
This can be used both to constrain the spin structure of any potential new physics contribution, and to measure the hadronic parameters
governing the \dsttaunu decay, serving as an essential baseline for SM and non-SM studies.
The helicity-suppressed amplitude which presently dominates the theoretical uncertainty on \RDb is too strongly suppressed in the 
\dbmunu decays to be measurable, however this can be accessed in the \dbtaunu decay directly.
If any potential new physics contributions are assumed not to contribute via the helicity-suppressed amplitude then the combined measurements of \dbmunu and \dbtaunu decays will allow for a fully data-driven prediction for
\RDb under the assumption of lepton universality, eliminating the need for any theory input relating to hadronic form factors.
However, these measurements have yet to be demonstrated with existing data.
This exciting programme of differential measurements needs to be developed on Run~1 and 2 data before any statement is made about the precise sensitivity, 
but it offers unparalleled potential to fully characterise both the SM and non-SM contributions to the $b \to c \tau \nu$ transition.

%% file: CONTRIBUTIONS/5_Measurements_of_unitarity_triangle_sides_and_semileptonic_decays/5.4.2.tex
\subsection{Prospects with other \bquark hadrons}
\label{subsec:OtherBhadronRX}


\def\RDst {\ensuremath{\mathcal{R}(\Dstar)}\xspace}
\def\RDorDst {\ensuremath{\mathcal{R}(D^{(*)})}\xspace}
\def\RDsteq {\ensuremath{\mathcal{R}(\Dstar)\equiv\frac{\mathcal{B}\left(\decay{\B}{\Dstarb\taup\neu}\right)}{\mathcal{B}\left(\decay{\B}{\Dstarb\ellp\neu}\right)}}\xspace}
\def\Req {\ensuremath{\mathcal{R}(D^{*})\equiv\frac{\mathcal{B}\left(\decay{\B}{\Db^{(*)}\taup\neu}\right)}
         {\mathcal{B}\left(\decay{\B}{\Db^{(*)}\ellp\neu}\right)}}\xspace}
\def\RD {\mathcal{R}(D)}
\def\threepim {\ensuremath{\pim \pip \pim}\xspace}
\def\threepip {\ensuremath{\pip \pim \pip}\xspace}

As measurements in \RDst become more statistically precise, it will become increasingly more
important to provide supplementary measurements in other \bquark-hadron species with different background 
structure and different sources of systematic uncertainties. For example, the \decay{\Bsb}{\Dsp \taum\neub} and \decay{\Bsb}{\Dssp\taum\neub} decays
will allow supplementary 
measurements at high yields, and do not suffer as badly from cross-feed backgrounds from other 
mesons, unlike, for example, \decay{\Bzb }{ \Dstarp \taum\neub}, where the \Bp and \Bs both contribute to 
the $\Dstarp \mu X$ or $\Dstarp \threepim X$ final states. 
Furthermore, the comparison of decays with different spins of the $b$ and $c$ hadrons can enhance the sensitivity 
to different NP scenarios~\cite{Azatov:2018knx,LHCb-PAPER-2017-016}.
No published measurements exist for the $\Bs$ case yet, but based on known relative efficiencies and assuming 
the statistical power of this mode tracks \RDorDst, we expect less than $6\%$ relative uncertainty 
after Run~3, and $2.5\%$ with the \upgradetwo data, where limiting systematic uncertainties 
are currently expected to arise from corrections to 
simulated pointing and vertex resolutions, from knowledge of particle identification efficiencies, 
and from knowledge of the backgrounds from random combinations of charm and muons. It is conceivable 
that new techniques and control samples could further increase the precision of these measurements. 

Methods 
are currently under development for separating the \decay{\Bsb}{\Dssp\ell^{-}\neub} and \decay{\Bsb}{\Dsp\ell^{-}\neub} modes, and given the
relative slow pion (\decay{\Dstarp }{ \Dz\pip}) and soft photon (\decay{\Dssp }{ \Dsp\gamma}) efficiencies, the precision in \decay{\Bsb}{ \Dsp \tau \nu} decays can be expected to exceed that in
\decay{\Bsb }{ \Dssp \tau \nu}, the reverse of the situation for $\RDorDst$.
An upgraded ECAL would extend the breadth and sensitivity of $\mathcal{R}(D^{*(*)+}_{s})$ measurements possible
in the \upgradetwo\ scenario above and beyond the possible benefits of improved neutral isolation in 
$\RD$ or $\mathcal{R}(\Dsp)$ measurements.

Of particular interest are the semitauonic decays of
\bquark baryons and of \Bcp mesons.
The former provides probes of entirely new Lorentz structures of NP operators
which pseudoscalar to pseudoscalar or vector transitions simply do not access.
The value of probing this supplementary space of couplings has already been demonstrated by
LHCb with its Run~1
measurement of $|\Vub|$ via the decay \decay{\Lb}{\proton\mun\neub}, which places strong constraints on 
right-handed currents sometimes invoked to explain the inclusive-exclusive tensions in that quantity. By the end of Run~3, it is expected that the relative uncertainty for $\mathcal{R}(\Lc)$ will reach below $4\%$, and $2.5\%$ by the end of \upgradetwo. 
A further exciting prospect is lepton universality tests with $b \to u\ell \nu$ decays, with $\ell=\mu,\tau$, which have been beyond experimental reach thus far.
For example the decay $B^+ \to p\bar{p}\ell\nu$ offers a clean experimental signature. 
Our capabilities with this decay could benefit from the enhanced low momentum proton identification with the TORCH subdetector.

Meanwhile, the \decay{\Bcp }{ \jpsi \taum \neub} decay
is an entirely unique state among the flavoured mesons as the bound state of two distinct 
flavors of heavy quark, and, through its abundant decays to charmonium final states, provides a 
highly efficient signature for triggering and reconstruction at high instantaneous luminosities. 
Measurements of \decay{\Bcp }{ \jpsi \ell \neub} decays involve a trade-off between the approximately 100 times 
smaller production cross-section for \Bcp verses the extremely efficient \decay{\jpsi}{\mup\mun} signature in the LHCb 
trigger. For illustration, in Run~1, LHCb reconstructed 
and selected 19\,000 \decay{\Bcp }{ \jpsi \mun \neub} decays, compared with 360\,000 \decay{\Bzb}{\Dstarp\mun\neub} decays. This resulted in a measurement of $\mathcal{R}(\jpsi)=0.71\pm 0.17\pm 0.18$ \cite{LHCb-PAPER-2017-035}. As a result of the smaller production cross-section, the muonic measurements have large backgrounds from \decay{h}{ \mu} misidentification from the relatively abundant \decay{\B }{ \jpsi X_h} decays, where $X_h$ is any collection of hadrons, and so they
are very sensitive to the performance of the muon system and PID algorithms in the future. Here it is assumed that 
it will be possible to achieve similar performance to Run 1 in the upgraded system. 

To project the sensitivity for \decay{\Bcp }{ \jpsi \taum \neub} based on Ref.~\cite{LHCb-PAPER-2017-035}, it is assumed that 
all the systematic uncertainties can be reduced with the size of the input data except for those that were assumed not to
scale with data for the previous predictions. For these, we assume that they can be reduced up until they reach 
the same absolute size as the corresponding systematic uncertainties in the Run~1 muonic $\RDst$ analysis. In addition, it is
assumed that sometime in the 2020s lattice QCD calculations of the form factors for this process will allow 
the systematic uncertainty due to signal form factors to be reduced by an additional factor of two. This results in
a projected absolute uncertainty for the muonic mode of $0.07$ at the end of Run~3 
and $0.02$ by the end of \upgradetwo,  as can be seen in Fig.~\ref{fig:RXproj}.
Measurements in the hadronic mode can be expected to reach similar sensitivities.

%% file: CONTRIBUTIONS/6_CP_violation_and_mixing_in_charm/6.tex
\label{chpt:charm}

The mixing of neutral \D mesons probes flavour changing neutral currents between
up-type quarks, which could be affected by beyond-the-SM physics in
fundamentally different ways to the down-type quarks that make up the \Kz,
\Bz and \Bs systems.

A fully software trigger will allow \LHCb\ to take full advantage of
the enormous production rate of charm hadrons at the HL-LHC, and collect
the largest sample, by far, of charm decays ever recorded: two orders
or magnitude more than Belle~II in the key channels discussed here. The
flavour-tagged sample will be further boosted for \upgradetwo by the addition of
magnet stations (see Section \ref{sec:dtokpipi}). 

\CP violation in mixing-related phenomena are predicted to be very
 small, $\order(10^{-4})$ or less in the SM \cite{Bobrowski:2010xg, Silvestrini:2015kqa}: there are thus
 excellent prospects for observing new physics contributions beyond
 the SM level. In the
 absence of new physics contributions to charm \CP violation, the \LHCb~\upgradetwo
 may well be the only facility with a realistic probability of
 observing these phenomena as it will
 be able to reach a sensitivity of $\order(10^{-5})$. The \LHCb\ programme
 is also robust, as strong sensitivity is obtained in a  range of
 modes with complementarity in their dominant systematic uncertainties. The use of
 the doubly Cabibbo-suppressed \decay{\Dz}{K^+\pi^-}
 (Sect.~\ref{sec:dtokpipi}) and \decay{\Dz}{\Kp\pim\pim\pip} decays
 (Sect.~\ref{sec:dtok3pi}), \decay{D^0}{\KS \pi^+ \pi^-}
 (Sect.~\ref{sec:dtokshh}) and \agamma in Cabibbo suppressed
 \decay{\Dz}{ \hadron^+ \hadron^-} \CP eigenstates
 (Sect.~\ref{sec:AGamma}) are all discussed
 here. Sect.~\ref{sec:charmindirectsummary} and Fig.~\ref{fig:charm_indirect}
 summarise the remarkable impact that \upgradetwo will have on charm indirect \CP violation.
 
 Effects of \CP violation in decay are less cleanly predicted, and
 could be as large as $\order(10^{-3})$~\cite{Bigi:2016jii} in charm, close to the
current levels of sensitivity.  If current hints \cite{LHCb-PAPER-2016-044} are confirmed,
 or if other signals emerge with future data, it will be possible to launch a comprehensive survey of such effects in
charm decays. The higher yields of the two-body \decay{\Dz}{\hadron^+\hadron^-} 
 (Sect.~\ref{sec:dtopipi}) modes, and the ability to probe the interfering
 structures in the phase space of multi-body decays, such as 
 \decay{D_{(s)}^+}{h^+h^+h^-} (Sect.~\ref{sec:dtopipi}) and \decay{\Dz}{h^+h^-h^+h^-}(Sect.~\ref{sec:dto4h}), provide
 interestingly complementary programmes. The potential for \LHCb\ to study neutral modes (Sect.~\ref{sec:charmneutral}), and perform a thorough exploration of the largely
 unstudied charm-baryon sector (Sect.~\ref{sec:charmbaryon}) are also discussed.

\input{CONTRIBUTIONS/6_CP_violation_and_mixing_in_charm/6.1.tex}
\input{CONTRIBUTIONS/6_CP_violation_and_mixing_in_charm/6.2.tex}

%% file: CONTRIBUTIONS/6_CP_violation_and_mixing_in_charm/6.1.tex
\newcommand{\secVIpix}{./CONTRIBUTIONS/6_CP_violation_and_mixing_in_charm/figs}
\section{Neutral $D$-meson mixing and indirect $C\!P$ violation}
\label{sec:sec61}
\newcommand{\inlineComplexmix}{\ensuremath{\nicefrac{q}{p}}}
\newcommand{\inlineInvcomplexmix}{\ensuremath{\nicefrac{p}{q}}}
\newcommand{\inlineRmix}{\ensuremath{\left | \inlineComplexmix \right |}}
\newcommand{\gp}{\ensuremath{g_{+}}}
\newcommand{\gm}{\ensuremath{g_{-}}}
\newcommand{\phimix}{\ensuremath{\phi_{\mathit{mix}}}}

The precision measurement of the mixing parameters and above all the
discovery of \CP violation in the \Dz\ system, and the precise
determination of its parameters, will be the key goals of \upgradetwo.

The neutral \D mesons mix to form mass eigenstates
$\ket{\D_1}=p\ket{\Dz} + q \ket{\Dzb}$ and
$\ket{\D_2} = p\ket{\Dz} - q\ket{\Dzb}$ with masses $M_1$, $M_2$ and
widths $\Gamma_1$, $\Gamma_2$, respectively. The complex-valued
parameters $q$ and $p$ are normalised such that $|p|^2+|q|^2=1$. We
use the phase convention $\CP\ket{\Dz} = +\ket{\Dzb}$ implying that, in
the limit of \CP\ symmetry with $p=q$, $\D_1$ is the \CP-even eigenstate 
and $\D_2$ is the \CP-odd state. 
We define the dimensionless mixing parameters, $x = (M_1 - M_2)/\Gamma$
and $y = (\Gamma_1 - \Gamma_2)/(2\Gamma)$, where
$\Gamma=(\Gamma_1+\Gamma_2)/2$ is the average width. The deviation of
$|q/p|$ from unity parameterises \CP violation in mixing. The relative
phase $\phi=\arg(q\overline{A}_f/(p A_f))$ between $q/p$ and the ratio
of \Dzb and \Dz decay amplitudes to a common final state $f$,
$\overline{A}_f/A_f$, is sensitive to \CP violation in the
interference between mixing and decay. 
Within the SM $\phi$ is approximately independent of the decay-mode.

%
%
\input{CONTRIBUTIONS/6_CP_violation_and_mixing_in_charm/6.1.1.tex}

\input{CONTRIBUTIONS/6_CP_violation_and_mixing_in_charm/6.1.2.tex}

\input{CONTRIBUTIONS/6_CP_violation_and_mixing_in_charm/6.1.3.tex}

\input{CONTRIBUTIONS/6_CP_violation_and_mixing_in_charm/6.1.4.tex}

\subsection{Summary of indirect \CP violation}
\label{sec:charmindirectsummary}
\begin{figure}
  \centering
  \includegraphics[width=1.0\textwidth]{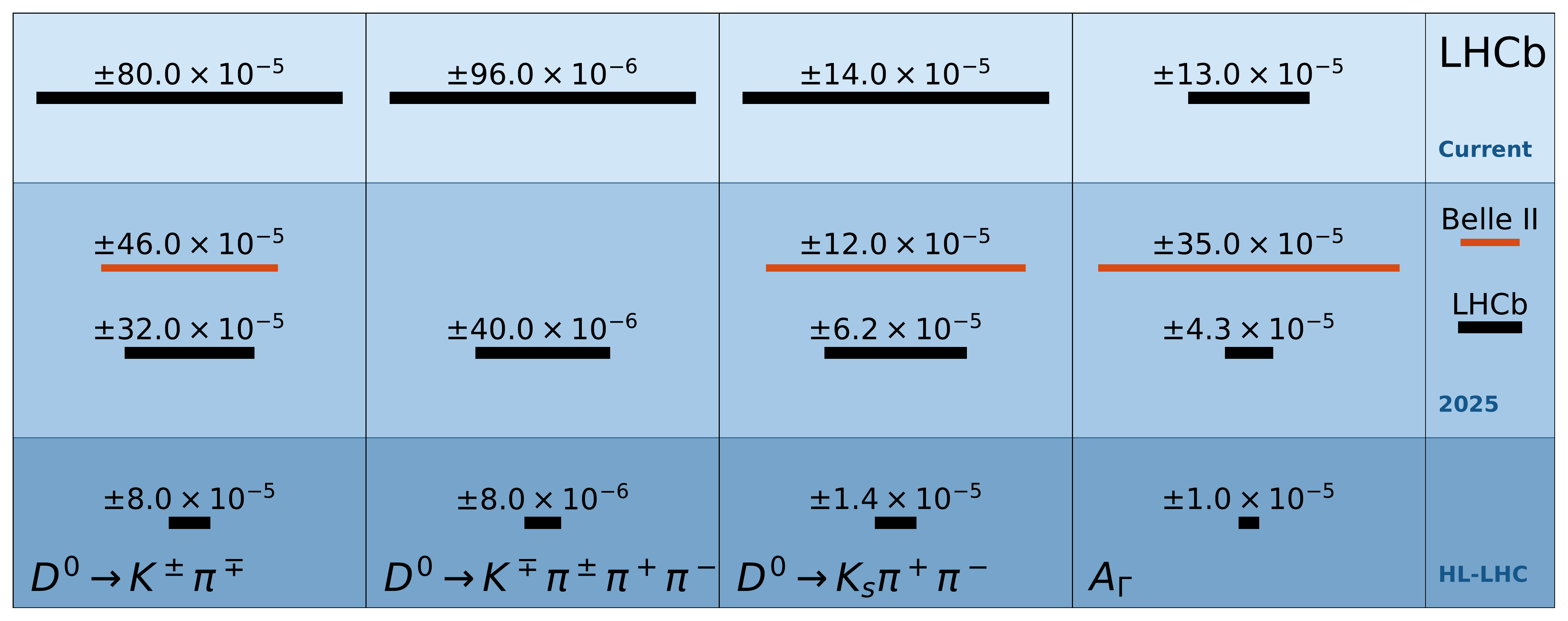}
  \caption{The predicted constraints on the indirect \CP violation
    asymmetry in
    charm from the decay channels indicated in the labels at the
    bottom of the columns. Predictions are shown in LS2
    (2020) from LHCb, LS3 (2025) from LHCb, at the end of Belle~II (2025), and at the end
    of the HL-LHC LHCb \upgradetwo\ programme. } 
\label{fig:charm_indirect_channels}
\end{figure}

A comparison of the potential precision of the analyses presented in
the previous sections is shown in
Fig.~\ref{fig:charm_indirect_channels}, and compared with the expected
precision from Belle~II. The expected LHCb constraints on $\phi$ are
translated into asymmetry constraints ($A_{\CP}^{\rm ind.} \approx x \sin(\phi)$) by multiplying by the current
HFLAV average of $x$ and neglecting the uncertainty on this. This
comparison assumes that $x$ is non-zero, $\phi$ not overly large, and $x$ will be determined
comparatively well in the future. This comparison neglects the additional constraining power from $|q/p|$.

\begin{figure}
  \centering
  \includegraphics[width=0.6\textwidth]{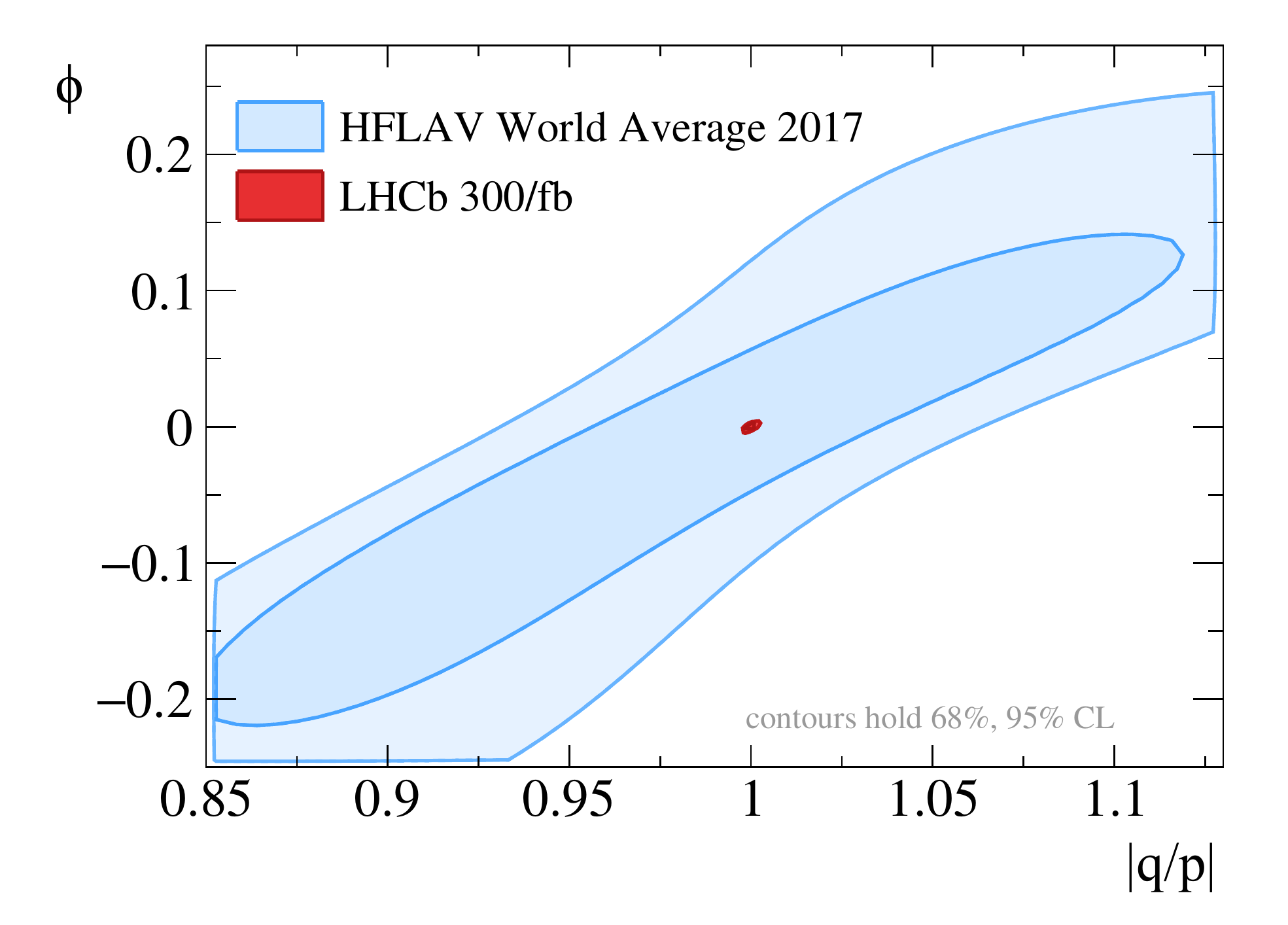}
  \caption{The estimated constraints for LHCb \upgradetwo\ on $\phi$, $|q/p|$ from the combination of the analyses in the previous section (red) compared to the current world-average precision~\cite{HFLAV16} (light blue).} 
\label{fig:charm_indirect}
\end{figure}

The analyses presented in the previous sections are also combined to
establish the sensitivity to the \CP-violating parameters $|q/p|$ and
$\phi$. The combination is performed using the method described in Ref.~\cite{LHCb-PAPER-2016-032}.
At an integrated luminosity of $300\invfb$ the sensitivity to $|q/p|$ is expected to be $0.001$ and that to $\phi$ to be $0.1^\circ$.
This remarkable sensitivity is contrasted with the HFLAV world average
as of 2017 in Fig.~\ref{fig:charm_indirect}. We thus conclude that the
LHCb \upgradetwo will have impressive power to characterise new physics
contributions to \CP violation and is the only foreseen facility with
a strong potential of probing the Standard Model contribution.

%% file: CONTRIBUTIONS/6_CP_violation_and_mixing_in_charm/6.1.1.tex
\subsection{Measurements with \decay{\Dz}{\Kmp\pipm}}
\label{sec:dtokpipi}

The mixing and \CP-violation parameters in \Dz--\Dzb oscillations can be accessed through the comparison of the decay-time-dependent ratio of \decay{\Dz}{K^+\pi^-} to \decay{\Dz}{K^-\pi^+} rates with the corresponding ratio for the charge-conjugate processes.

\begin{figure}
  \centering
  \includegraphics[width=0.6\textwidth]{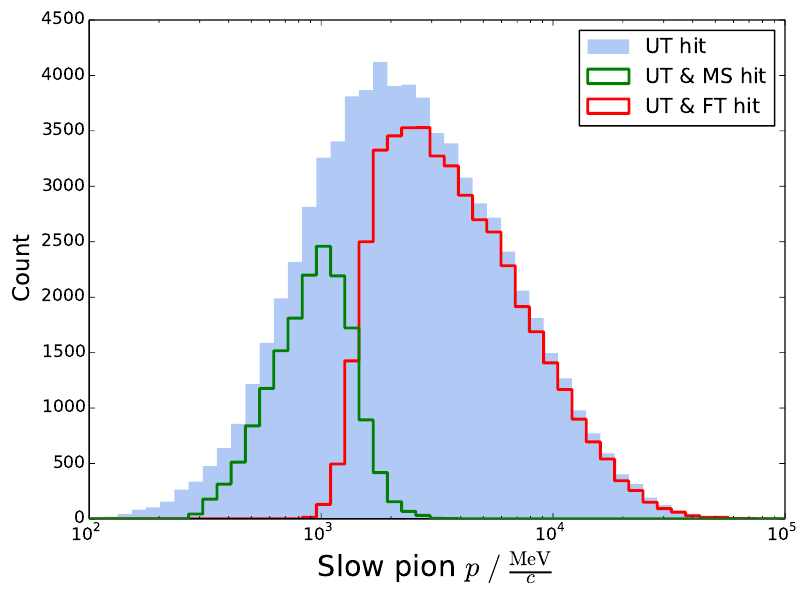}
  \caption{The momentum spectra in \LHCb\ simulation of slow pions from the decay \decay{\Dstarp}{\Dz\pi^+} that leave hits in the UT. The slow pions observed in the SciFi in \upgradeone and those that will be added by the Magnet Stations (MS) for \upgradetwo are indicated.} \label{fig:charm_ms}
\end{figure}

The neutral \D-meson flavour at production can be determined from the charge of the low-momentum pion (slow pion) produced in the flavour-conserving strong-interaction decay \decay{\Dstarp}{\Dz\pi^+}. This flavour-tagging technique is used throughout many measurements in this chapter. These low-momentum pions are strongly deflected by the magnetic field in the experiment. The addition of the Magnet Stations (see Sect.~\ref{sec:tracking}) will allow the momentum and charge of previously lost pions to be determined. As illustrated in Fig.~\ref{fig:charm_ms} the flavour-tagged charm sample can be increased by 40\% in size by the inclusion of Magnet Station information.

The \decay{\Dstarp}{\Dz(\to K^-\pi^+)\pi^+} process, which is dominated by a Cabibbo-favoured amplitude, is denoted as right-sign (RS).
Wrong-sign (WS) decays, \decay{\Dstarp}{\Dz(\to K^+\pi^-)\pi^+}, arise from the doubly Cabibbo-suppressed \decay{\Dz}{K^+\pi^-} decay and the Cabibbo-favoured \decay{\Dzb}{K^+\pi^-} decay that follows \Dz--\Dzb oscillation.
Since the mixing parameters are small, $|x|,|y|\ll1$, the \CP-averaged decay-time-dependent ratio of WS-to-RS rates is approximated as~\cite{Blaylock:1995ay,Bianco:2003vb,Burdman:2003rs}
\begin{equation}
R_{K\pi}(t) \approx R^{K\pi}_D + \sqrt{R^{K\pi}_D}\ y_{K\pi}'\ (\Gamma t)+\frac{x_{K\pi}'^2 +y_{K\pi}'^2}{4}\left(\Gamma t\right)^2,
\end{equation}
where $t$ is the $D$ proper decay time. The parameter $R_D^{K\pi}$ is the ratio
of suppressed-to-favoured decay rates at $t=0$. The parameters $x_{K\pi}'$
and $y_{K\pi}'$ depend linearly on the mixing parameters,
$x_{K\pi}' \equiv x\cos\delta_{K\pi}+y\sin\delta_{K\pi}$ and
$y_{K\pi}' \equiv y\cos\delta_{K\pi}-x\sin\delta_{K\pi}$, through the
strong-phase difference $\delta_{K\pi}$ between the suppressed and
favoured amplitudes. 
If \CP violation occurs,
the decay-rate ratios $R^+(t)$ and $R^-(t)$ of mesons produced as \Dz
and \Dzb, respectively, are functions of independent sets of
parameters, $R_D^{K\pi\,\pm}$, $(x_{K\pi}'^{\pm})^2$ and $y_{K\pi}'^\pm$.  The parameters
$R^{K\pi\,+}_D$ and $R^{K\pi\,-}_D$ differ if the ratio between the suppressed and
favoured decay amplitudes is not \CP symmetric, indicating direct \CP
violation, which is commonly described by the asymmetry
$A_D^{K\pi}=(R_D^{K\pi\,+}-R_D^{K\pi\,-})/(R_D^{K\pi\,+} + R_D^{K\pi\,-})$.  Violation of \CP symmetry either
in mixing or in the interference between mixing and decay amplitudes
generates differences between
$(x_{K\pi}'^\pm)^2 = |q/p|^{\pm2}(x_{K\pi}'\cos\phi \pm y_{K\pi}'\sin\phi)^2$ and
$y_{K\pi}'^\pm = |q/p|^{\pm 1}(y_{K\pi}'\cos\phi \mp x_{K\pi}'\sin\phi)$.  
The \CP-violation parameters ($|q/p|$ and $\phi$) can
be determined in the analysis without the need of external inputs.

The latest measurement from LHCb~\cite{LHCb-PAPER-2017-046} uses Run~1
and early Run~2 (2015--2016) data, corresponding to a total sample of
about $\mathcal{L}=5\invfb$ of integrated luminosity.  Assuming \CP
conservation, the mixing parameters are measured to be
$x_{K\pi}'^2=(3.9 \pm 2.7) \times10^{-5}$,
$y_{K\pi}'= (5.28 \pm 0.52) \times 10^{-3}$, and
$R_D^{K\pi} = (3.454 \pm 0.031)\times10^{-3}$.  Studying \Dz and \Dzb decays
separately shows no evidence for \CP violation and provides the
current most stringent bounds on the parameters $A_D^{K\pi}$ and $|q/p|$ from
a single measurement, $A_D^{K\pi} =(-0.1\pm9.1)\times10^{-3}$ and
$1.00< |q/p| <1.35$ at the $68.3\%$ confidence level.  

\begin{table}[tb]
\centering
\caption{Extrapolated signal yields, and statistical precision on the mixing and \CP-violation parameters, from the analysis of promptly produced WS \decay{\Dstarp}{\Dz(\to K^+\pi^-)\pi^+} decays.
Signal yields of promptly produced RS \decay{\Dstarp}{\Dz(\to K^-\pi^+)\pi^+} decays are typically $250$ times larger.\label{tab:WS-D2Kpi-yields}}
\begin{tabular}{lcccccc} 
\hline
Sample ($\mathcal{L}$)  & Yield ($\times10^6$) & $\sigma(x_{K\pi}'^2)$ & $\sigma(y_{K\pi}')$ & $\sigma(A_D)$ & $\sigma(|q/p|)$ & $\sigma(\phi)$ \\
\hline
Run 1--2 (9\invfb)   &  1.8   & $1.5\times10^{-5}$ & $2.9\times10^{-4}$ & 0.51\% & 0.12 & 10\degrees \\
Run 1--3 (23\invfb)  &   10   & $6.4\times10^{-6}$ & $1.2\times10^{-4}$ & 0.22\% & 0.05 &  4\degrees \\
Run 1--4 (50\invfb)  &   25   & $3.9\times10^{-6}$ & $7.6\times10^{-5}$ & 0.14\% & 0.03 &  3\degrees \\
Run 1--5 (300\invfb) &  170   & $1.5\times10^{-6}$ & $2.9\times10^{-5}$ & 0.05\% & 0.01 &  1\degrees \\
\hline
\end{tabular}
\end{table}

In Table~\ref{tab:WS-D2Kpi-yields} the signal yields and the statistical precision from Ref.~\cite{LHCb-PAPER-2017-046} are extrapolated to the end of Run 2
and to the end of \upgradetwo, assuming that the central values of the measurements stay the same. This assumption is particularly important for the \CP-violation parameters, as their precision may depend on the measured values.

Systematic uncertainties are estimated using control samples of data and none of them are foreseen to have irreducible contributions that exceed the ultimate statistical precision, if the detector performance (particularly in terms of vertexing/tracking and particle identification capabilities) is kept at least in line with what is currently achieved at LHCb.

%% file: CONTRIBUTIONS/6_CP_violation_and_mixing_in_charm/6.1.2.tex
\subsection{Measurements with \decay{\Dz}{\Kmp\pipm\pip\pim}}
%
 \label{sec:dtok3pi}
%
Like \decay{\Dz}{\Km\pip} and \decay{\Dz}{\Kp\pim}, the decays
\decay{\Dz}{\Km\pip\pim\pip} and \decay{\Dz}{\Kp\pim\pim\pip} are a
pair of RS and WS decays with high sensitivity to charm mixing.
However, the rich amplitude structure across the five dimensional phase
space that describes these decays offers unique opportunities (and
challenges) in these four-body modes.

In a phase-space integrated analysis using $3\mathrm{fb}^{-1}$ of data,
LHCb made the first observation of mixing in this decay mode, and
measured the quantities $R_D^{K3\pi} = (3.21 \pm 0.014)\cdot 10^{-3}$,
as well as
$R^{K3\pi}_{\mathrm{coher}}y'_{K3\pi} = (0.3\pm 1.8)\cdot 10^{-3}$,
$\frac{1}{4}(x^2 + y^2) = (4.8\pm 1.8) \cdot
10^{-5}$~\cite{LHCb-PAPER-2015-057}; the parameter
$R^{K3\pi}_{\mathrm{coher}}$ 
is the coherence factor, which takes into account the effect of integrating
over the entire four-body phase
space~\cite{Atwood:coherenceFactor,Evans:2016tlp}.

The unique power of multibody decays lies to a large extent in the
fact that the strong-phase difference between the interfering \Dz and
\Dzb amplitudes varies across phase space. This can only be fully
exploited through analyses of phase space distributions (either in
bins or unbinned), rather than a phase-space-integrated approach. Such
a ``phase space resolved'' approach allows a direct measurement of
$x_{K\pi\pi\pi}'$ and $y_{K\pi\pi\pi}'$ (rather than only to
$x'^2$ and $y'$ as in the 2-body case), and
crucially provides high sensitivity to the \CP-violating variables $\phi$ and
$|q/p|$.

However, the same phase variations that make multibody decays so
powerful, are also a major challenge, as they need to be known in
order to cleanly determine the mixing and \CP-violation parameters of
interest. In principle, the relevant phases can be inferred from an
amplitude model such as that obtained from $3\mathrm{fb}^{-1}$ of LHCb
data~\cite{LHCb-PAPER-2017-040}. Such models may introduce unacceptable theoretical uncertainties for
the precision era of \lhcb \upgradetwo  unless there are significant innovations
in the theoretical description of four-body amplitudes.
Model-independent approaches will also be applied. These use 
quantum-correlated events at the charm threshold to infer
the required phase information in a model-unbiased way. BESIII is
working closely with \lhcb~\cite{Malde:2223391} to provide the necessary model-independent
input for \decay{\Dz}{\Kp\pim\pim\pip} across different regions of
phase space for measurements of the $\gamma$ angle (see Chapter~\ref{sec:timeintegcpv}) as well as charm mixing and
\CP-violation measurements.

Simulation sensitivity studies with model-dependent approaches give a
useful indication of the precision that can be achieved. A recent such
study~\cite{Muller:2297069} uses LHCb's latest
\decay{\Dz}{\Kp\pim\pim\pip} amplitude
model~\cite{LHCb-PAPER-2017-040}.  Table~\ref{tab:WS-D2K3pi} gives the
yields and sensitivities scaled from this study,  illustrating the
impressive sensitivity of this decay mode.
The study is based on
promptly produced \Dstarp mesons decaying in the flavour-conserving
decay mode \decay{\Dstarp}{\Dz\pip}.  Several systematic uncertainties
require improvements in the analysis method in order to scale with
increasing sample sizes. However, given the huge potential of this
channel, sufficient effort is expected to be dedicated to this challenge, such
that adequate methods can be developed, and the necessary input from
threshold measurements is both generated at BESIII and exploited
optimally at \lhcb. Indeed, once these are in place, this channel has
the potential for probing $\order(10^{-5})$ \CP violation, assuming the
current world average value of $x$. 
%
\begin{table}[tb]
\centering
\caption{Extrapolated signal yields, and sensitivity to the mixing and
  \CP-violation parameters, from the analysis of
  \decay{\Dz}{\Kp\pim\pim\pip} decays (statistical uncertainties, only).\label{tab:WS-D2K3pi}}
\begin{tabular}{lccccc} 
\hline
Sample ($\mathcal{L}$)  & Yield ($\times10^6$) & $\sigma(x_{K\pi\pi\pi}')$ & $\sigma(y_{K\pi\pi\pi}')$ & $\sigma(|q/p|)$ & $\sigma(\phi)$ \\
\hline
Run 1--2 (9\invfb)       &  0.22                & $2.3\times10^{-4}$       & $2.3\times10^{-4}$       & 0.020           & 1.2\degrees \\
Run 1--3 (23\invfb)      &  1.29                & $0.9\times10^{-4}$
                                                                           & $0.9\times10^{-4}$       & 0.008           & 0.5\degrees \\
Run 1--4 (50\invfb)      &  3.36                & $0.6\times10^{-4}$       & $0.6\times10^{-4}$       & 0.005           & 0.3\degrees \\
Run 1--5 (300\invfb)     & 22.5\phantom{00}     & $0.2\times10^{-4}$       & $0.2\times10^{-4}$       & 0.002           & 0.1\degrees \\
\hline
\end{tabular}
\end{table}

%% file: CONTRIBUTIONS/6_CP_violation_and_mixing_in_charm/6.1.3.tex
\subsection{Amplitude analysis of \decay{\Dz}{\KS h^+ h^-}}\label{Sec:6.1.3}

\label{sec:dtokshh}

The self-conjugate decay \decay{D^0}{\KS \pi^+ \pi^-} includes both the Cabibbo-favoured and doubly Cabibbo-suppressed as well as $\CP$-eigenstate processes reconstructed in the same final state.
This allows the relative strong phase between these contributions to be determined from the data, and in turn enables both the mixing parameters $x$ and $y$ and the \CP-violation parameters $|q/p|$ and $\phi$ to be directly measured without need for external input.
As a result, this channel provides the dominant constraint on the parameter $x$ in the global fits.

The mixing and CPV parameters modulate the time-dependence of the complex amplitudes, which also depend on the two-dimensional phase-space of the decay.
As such, the measurement relies on both a precise understanding of the detector acceptance effects as a function of phase-space and decay time, and on an accurate description of the evolution of the underlying decay amplitudes over the Dalitz plane. As discussed in the previous section both model-dependent and model-independent approaches using quantum-correlated $D\bar{D}$ pairs from $\psi(3770)$ decays can be applied.


Previous measurements from the CLEO~\cite{Asner:2005sz}, BaBar~\cite{delAmoSanchez:2010xz}, and Belle~\cite{Peng:2014oda} collaborations have used the model-dependent approach, with the Belle measurement having the best precision to date, $x = (0.56 ^{+0.20}_{-0.23})\%$, $y = (0.30 ^{+0.16}_{-0.17})\%$ (assuming \CP symmetry), and $|q/p| = 0.90 ^{+0.18}_{-0.16}$, $\phi = (-6 \pm 12)\degrees$.
The one published LHCb result was based on 1~fb$^{-1}$ of Run 1 data~\cite{LHCb-PAPER-2015-042}, and used a model-independent approach with strong phases taken from a CLEO measurement~\cite{Libby:2010nu} to determine $x = (-0.86 \pm 0.56)\%$, $y = (0.03 \pm 0.48)\%$.
This analysis used around $2\times10^5$ \decay{D^{*+}}{D^0 \pi^+}, \decay{D^0}{\KS \pi^+ \pi^-} decays from 2011, which suffered from low \KS trigger efficiencies that were significantly increased for 2012 and beyond, and will benefit further from software trigger innovations in the Upgrade era.

At LHCb these decays can be reconstructed either through semileptonic decays such as \decay{B^-}{D^0 \mu^- \bar{\nu_{\mu}}} (where the muon charge is used to tag the initial $D^0$ flavour), or through prompt charm production (where the charge of the slow pion in the decay \decay{D^{*+}}{D^0 \pi^+} tags the initial flavour).
The two channels have complementary properties and both will be important components of future mixing and \CP violation analyses at LHCb.

The prompt charm yields are significantly larger than for the semileptonic channel, due to the increased production cross-section; however, triggering signal candidates is much more efficient, and introduces fewer non-uniformities in acceptance, for the semileptonic channel.
The estimated future yields are presented in Table~\ref{tab:D0KSPiPi_yields}.
Also shown are projected statistical precisions on the four mixing and \CP-violating parameters, which have been extrapolated from complete analyses of the Run 1 data for both SL and prompt cases. 

\begin{table}[tb]
\begin{center}
\caption{Extrapolated signal yields, and statistical precision on the mixing and \CP violation parameters, for the analysis of the decay \decay{D^0}{\KS \pi^+ \pi^-}.
Candidates tagged by semileptonic $B$ decay (SL) and those from prompt charm meson production are shown separately.}
  \renewcommand{\arraystretch}{1.2}
 \begin{tabular}{lcccccc} 
 \hline
 Sample (lumi $\mathcal{L}$)              & Tag    & Yield & $\sigma(x)$ & $\sigma(y)$ & $\sigma(|q/p|)$ & $\sigma(\phi)$ \\ \hline
\multirow{2}{*}{Run 1--2 (9~fb$^{-1})$}   & SL     & 10M   & 0.07\%      & 0.05\%      & 0.07            & 4.6\degrees    \\
                                          & Prompt & 36M   & 0.05\%      & 0.05\%      & 0.04            & 1.8\degrees    \\ 
\multirow{2}{*}{Run 1--3 (23~fb$^{-1})$}  & SL     & 33M   & 0.036\%     & 0.030\%     & 0.036           & 2.5\degrees    \\
                                          & Prompt & 200M  & 0.020\%     & 0.020\%     & 0.017           & 0.77\degrees   \\ 
\multirow{2}{*}{Run 1--4 (50~fb$^{-1})$} & SL     & 78M   & 0.024\%     & 0.019\%     & 0.024           & 1.7\degrees    \\
                                         & Prompt & 520M  & 0.012\%     & 0.013\%     & 0.011           & 0.48\degrees   \\
\multirow{2}{*}{Run 1--5 (300~fb$^{-1})$} & SL     & 490M  & 0.009\%     & 0.008\%     & 0.009           & 0.69\degrees   \\
                                          & Prompt & 3500M & 0.005\%     & 0.005\%     & 0.004           & 0.18\degrees   \\
 \hline
\end{tabular}
\renewcommand{\arraystretch}{1.0}
 \label{tab:D0KSPiPi_yields}
\end{center}
\end{table}

The dominant systematic uncertainties on mixing parameters in this channel come from two main sources.
Firstly, the precision with which the non-uniformities in detector-acceptance versus decay time and as a function of phase space can be determined; secondly, the knowledge of the strong-phase evolution across the Dalitz plane.
For the Run 2 analysis, both contributions are significantly lower than the statistical precision, while in the longer term new approaches will be necessary to further reduce these systematic uncertainties.
Trigger and event selection techniques should be adapted to emphasise uniform acceptance, even if this comes at the cost of significant efficiency loss.
New techniques, such as the bin-flip method~\cite{DiCanto:2018tsd}, 
can further reduce dependence on the non-uniform acceptance, although at the cost of 
degraded statistical precision on the mixing and \CP-violation parameters.
In the model-dependent approach many of the model systematic uncertainties may reduce or vanish with increased integrated luminosity, as currently fixed parameters are incorporated into the data fit, and the data become increasingly capable of rejecting unsuitable models provided that there is suitable evolution in the model descriptions.
For the model-independent approach, the uncertainty from external inputs (currently from CLEO-c, later from BESIII) will also reduce with luminosity as the LHCb data starts to provide constraining power.
There are no systematic uncertainties which are known to have irreducible contributions that exceed the ultimate statistical precision.

For the \CP violation parameters additional sources of systematic uncertainty come from the knowledge of detector-induced asymmetries.
In particular, there is a known asymmetry between \Kz and \Kzb in their interactions with material.
The limitation here will be the precision with which the material traversed by each $\KS$ meson can be determined.
%
%
The \upgradetwo detector will be constructed to minimise material, and to allow precise evaluation of the remaining contributions. This channel thus has comparable power on \CP violating parameters, but with a simpler two-dimensional phase space and complementary detector systematic uncertainties, as the four-body decay discussed in the previous section.

%% file: CONTRIBUTIONS/6_CP_violation_and_mixing_in_charm/6.1.4.tex
\subsection{Measurement of $A_\Gamma$}
\label{sec:AGamma}

A further sensitive probe of indirect \CP violation in the charm sector is given by the time evolution of decays of \Dz mesons into \CP eigenstates.
Due to the smallness of the charm mixing parameters, the \CP asymmetry as a function of decay time can be written as
\begin{equation}
\begin{aligned}
A_\CP(t) &= \frac{\Gamma(\decay{\Dz(t)}{ f}) - %
 	 			\Gamma(\decay{\Dzb(t)}{ f})}%
				{\Gamma(\decay{\Dz(t)}{ f}) + %
 				\Gamma(\decay{\Dzb(t)}{ f})} = A_\CP^\mathrm{dir} + \frac{t}{\tau_D} A_\CP^\mathrm{indir} +
                \order\left(\left(\frac{t}{\tau_D}\right)^2\right)\simeq A_\CP^\mathrm{dir} - \frac{t}{\tau_D} \agamma, \label{eq:acp_time} 
\end{aligned}
\end{equation}  
where $\Gamma (\Dz(t) \to f)$ and $\Gamma (\Dzb(t) \to f)$ indicate
the time-dependent decay rates of an initial \Dz or \Dzb decaying to a final state $f$ at decay time $t$,
$\tau_D = 1/\Gamma$ is the average lifetime of the \Dz meson, 
and $A_\CP^\mathrm{dir}$ is the asymmetry related to direct \CP violation.
The parameter \agamma is related to indirect \CP violation ($\simeq -A_\CP^\mathrm{indir}$) and is defined as
\begin{align}
\agamma &\equiv \frac{\hat{\Gamma}(\decay{\Dz}{ \hadron^+ \hadron^-)} - %
 			\hat{\Gamma}(\decay{\Dzb}{ \hadron^+ \hadron^-)}}%
 			{\hat{\Gamma}(\decay{\Dz}{ \hadron^+ \hadron^-)}+%
 			\hat{\Gamma}(\decay{\Dzb}{ \hadron^+ \hadron^-)}} \label{eq:agamma_def} 
 \end{align}
 where $\hat{\Gamma}$, the effective lifetime, is defined by
 \begin{equation}
 1/\hat{\Gamma} = \frac{\int t\, \Gamma(t) \, {\rm d}t}{\int \Gamma(t) \, {\rm d}t}.
 \end{equation}
Neglecting contributions from subleading amplitudes
, \agamma is independent of the final state $f$. 
It is therefore possible to measure indirect \CP violation by either measuring the two lifetimes separately, or by the time evolution of the asymmetry.

Neglecting \CP violation in mixing, it can be shown that $\agamma \approx - x \sin \phi $, implying that $|\agamma| <  |x| \lsim 5\times 10^{-3}$~\cite{HFLAV16},thus  setting the scale of the size of the effect that needs to be measured. While this means that the quantity to be measured is necessarily small, the large yields available in the Cabibbo-suppressed modes $f=\pi^+\pi^-$ or $f=K^+K^-$, and tagging from the \Dstarpm decay, allow precise measurements provided the systematic uncertainties can be controlled with a high degree of precision.
Tagging based on semileptonic decays of a parent beauty hadron is also possible and has been used in a published LHCb measurement~\cite{LHCb-PAPER-2014-069}, but contributes significantly lower yields. 

Most potential systematic effects are essentially constant in $t$ and
therefore cause little uncertainties in the observed decay time
evolution of the asymmetry.  However, second-order effects and
detector--induced correlation between momentum and proper decay time
are sufficient to produce spurious asymmetries, that must be
appropriately corrected. In addition, contamination from secondary
decays is a first-order effect in time that must be suppressed, and
its residual bias accounted for.  Both corrections are dependent on
the availability of a large number of favoured $\Dz\to\Km\pip$  decays
as calibration, and can be expected to scale with statistics;
collection of this sample with the same trigger as for the signal
modes is therefore a crucial tool for performing this measurement with high precision in the future.

The Run~1 LHCb measurement of this quantity gave consistent results in the two $h^+h^-$ modes, averaging  $\agamma= (- 0.13 \pm 0.28 \pm 0.10)\times 10^{-3}$ \cite{LHCb-PAPER-2016-063}, which is still statistically dominated. For the reasons mentioned above, this precision is at the threshold of becoming physically interesting, making it a high-priority target to pursue with more data.
It seems highly likely that LHCb \upgradetwo\ will be the only
experiment built in the foreseeable future that will be able to do this.

Table~\ref{tab:Agamma_yields} shows expected yields and precisions attainable in  \LHCb~\upgradetwo, under the same assumptions on efficiencies adopted in the previous sections; this must include provisions for acquiring and storing $5\times 10^{10}$ Cabibbo-favoured decays. The ultimate combined precision is $1\times 10^{-5}$.

\begin{table}[tb]
\begin{center}
\caption{Extrapolated signal yields, and statistical precision on
  indirect \CP violation from \agamma. 
}
  \renewcommand{\arraystretch}{1.2}
 \begin{tabular}{lc|cc|cc} 
 \hline
 Sample ($\mathcal{L}$)   & Tag   & Yield $K^+K^-$ & $\sigma(\agamma)$ & Yield  $\pi^+\pi^-$  & $\sigma(\agamma)$ \\ \hline
Run 1--2 (9~fb$^{-1})$    & Prompt   & 60M    & 0.013\% & 18M & 0.024\%       \\ 
Run 1--3 (23~fb$^{-1})$  & Prompt   & 310M   & 0.0056\% &92M& 0.0104 \%       \\ 
Run 1--4 (50~fb$^{-1})$  & Prompt   & 793M   & 0.0035\% &236M& 0.0065 \%       \\ 
Run 1--5 (300~fb$^{-1})$ & Prompt  & 5.3G   & 0.0014\% & 1.6G & 0.0025 \%         \\
 \hline
\end{tabular}
\renewcommand{\arraystretch}{1.0}
 \label{tab:Agamma_yields}
\end{center}
\end{table}

%% file: CONTRIBUTIONS/6_CP_violation_and_mixing_in_charm/6.2.tex
\section{Direct $C\!P$ violation}

\LHCb\ has a comprehensive programme of searches for direct \CP
violation in charm. Neutral \D mesons, where
indirect $\CP$ violation can also contribute, and charged \Dspm
meson decays, where only direct effects can occur, are both studied.
The higher statistics of two-body decays and the
interfering amplitudes of multi-body decays are both probed. Four-body
decays and the new field of baryon decays, allow the complementary
studies of $P$-even and $P$-odd \CP violation contributions.  
Singly and doubly Cabibbo-suppressed decays are both studied and
channels with new physics sensitive loop processes or those with
exchange diagrams where larger SM contributions are expected. The
precision study of modes containing neutral particles will be opened up
by the proposed calorimeter of \upgradetwo.
 
Direct \CP violation effects in the charm system could be larger than
those in indirect \CP violation, and \upgradetwo will be able to 
characterise the direct \CP sources. Alternatively \CP violation effects may be
very small and \upgradetwo will be needed to probe them. In either scenario
the experiment will have a strong programme in this field.


\label{sec:charmdirectcpv}
\input{CONTRIBUTIONS/6_CP_violation_and_mixing_in_charm/6.2.1.tex}

\input{CONTRIBUTIONS/6_CP_violation_and_mixing_in_charm/6.2.2.tex}
\input{CONTRIBUTIONS/6_CP_violation_and_mixing_in_charm/6.2.3.tex}
\input{CONTRIBUTIONS/6_CP_violation_and_mixing_in_charm/6.2.4.tex}
\input{CONTRIBUTIONS/6_CP_violation_and_mixing_in_charm/6.2.5.tex}

%% file: CONTRIBUTIONS/6_CP_violation_and_mixing_in_charm/6.2.1.tex
\subsection{Measurement of $A_\CP$ in \decay{\Dz}{\Kp\Km} and \decay{\Dz}{\pip\pim}  and \CP violation in other two-body modes}
\label{sec:dtopipi}


The singly Cabibbo-suppressed $D^0 \rightarrow \Km\Kp$ and $D^0 \rightarrow \pim\pip$ decays discussed in Sect.~\ref{sec:AGamma} for indirect \CP violation studies, also play a critical role in the measurement of time-integrated direct \CP violation.
The amount of \CP violation in these decays is expected to be below the percent level~\cite{Feldmann:2012js, Bhattacharya:2012ah, Pirtskhalava:2011va,
Brod:2012ud, Cheng:2012xb, Muller:2015rna,Golden:1989qx,Li:2012cfa}, but large theoretical uncertainties due to long-distance interactions prevent precise SM predictions. In the presence of physics beyond the SM, the expected \CP asymmetries could be enhanced~\cite{Giudice:2012qq}, although an observation near the current experimental limits would be consistent with the SM expectation. 
The direct \CP violation is associated with the breaking of \CP symmetry in the decay amplitude.  It is measured through the time-integrated \CP asymmetry in the $h^-h^+$ decay rates

\begin{equation}
A_{\CP}(\Dz\to h^-h^+)\equiv \frac{\Gamma(D^0\rightarrow h^-h^+)-\Gamma(\Dzb\rightarrow h^-h^+)}{\Gamma(D^0\rightarrow h^-h^+)+\Gamma(\Dzb\rightarrow h^-h^+)}.
\end{equation}
The sensitivity to direct \CP violation is enhanced through a measurement of the difference in \CP asymmetries between \Dz\to\Km\Kp and \Dz\to\pim\pip decays, $\Delta A_{\CP}=A_{\CP}(\Km\Kp)-A_{\CP}(\pim\pip)$, in which detector asymmetries largely cancel.

The individual asymmetries $A_{\CP}(\Km\Kp)$ and $A_{\CP}(\pim\pip)$ can also be measured. 
A measurement of the time-integrated \CP asymmetry in $D^0 \rightarrow \Km\Kp$ has been performed at \LHCb\ with 3\invfb 
collected at centre-of-mass energies of 7 and 8\tev. The flavour of the charm meson at production is determined from the charge of the pion in
$\Dstarp\to\Dz\pi^+$ decays, or via the charge of the muon in semileptonic \bquark-hadron decays ($\Bb\to\Dz\mun\neumb X$). 
The analysis strategy so far relies on the $D^+\rightarrow K^-\pi^+\pi^- $, $D^+\rightarrow K_s^0\pi^+$ and $\Dstarp\to\Dz (\to K^-\pi^+)\pi^+$ decays as control samples~\cite{LHCb-PAPER-2016-035}. In this case, due to the weighting procedures aiming to fully cancel the production and reconstruction asymmetries, the effective prompt signal yield for $A_{\CP}(\Km\Kp)$ is reduced. The expected signal yields and the corresponding statistical precision in \upgradetwo are summarised in Table~\ref{tab:expectedyields}.



\begin{table}[t]
\caption{\small Extrapolated signal yields and statistical precision on direct \CP violation observables for the promptly produced samples.}
\centering
\begin{tabular}{l c  c  c  c  c }
\hline
Sample ($\mathcal{L}$) & Tag & Yield & Yield  & $\sigma(\Delta A_{\CP})$ &  $\sigma(A_{\CP}(hh))$ \\
 &  & \Dz\to\Km\Kp & \Dz\to\pim\pip & [$\%$] &  [\%] \\
\hline
Run 1--2 (9 \invfb)  & Prompt      &  52M &   17M & $0.03$  & $0.07$ \\   
Run 1--3 (23 \invfb) & Prompt      & 280M &   94M & $0.013$  & $0.03$ \\   
Run 1--4 (50 \invfb) & Prompt     &  1G  &  305M & $0.007$ & $0.015$ \\ 
Run 1--5  (300 \invfb) & Prompt    & 4.9G &  1.6G & $0.003$ & $0.007$ \\ 
\hline
\end{tabular}
\label{tab:expectedyields}
\end{table}
 
\begin{figure}
  \centering
  \includegraphics[width=0.7\textwidth]{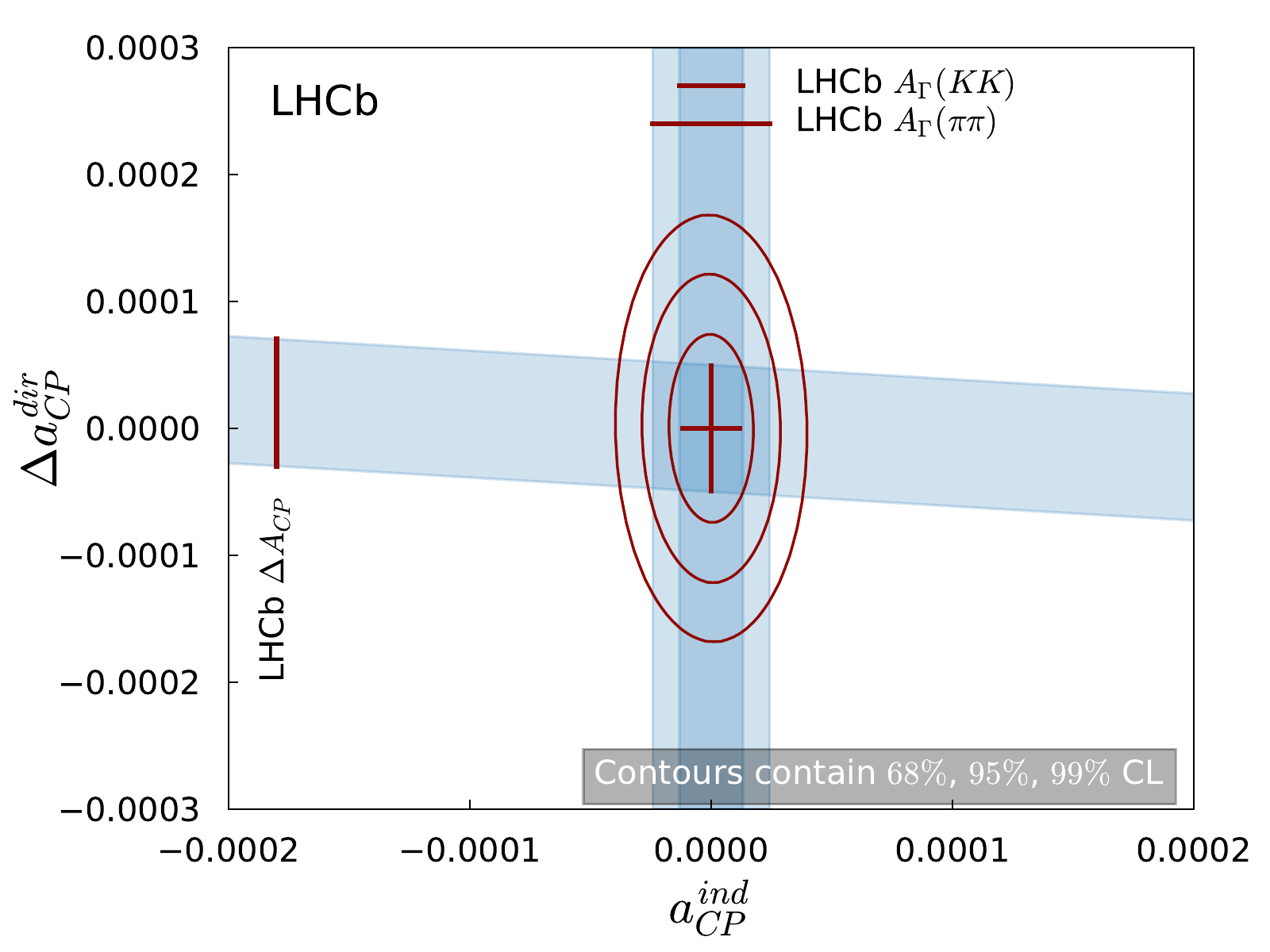}
  \caption{The estimated constraints for LHCb \upgradetwo\ on indirect and direct charm \CP violation from the analysis of two-body \CP eigenstates. The current world-average precision~\cite{HFLAV16} is $\pm2.6 \times 10^{-4}$ for indirect and $\pm 18 \times 10^{-4}$ for direct \CP violation and thus larger than the full scale of this plot.} 
\label{fig:agammadeltaacp}
\end{figure}

The $\Delta A_{\CP}$ observable is robust against systematic uncertainties. The main sources of systematic uncertainties are inaccuracies in the fit model, the weighting procedure, the contamination of the prompt sample with secondary $\Dz$ mesons and the presence of peaking backgrounds. 
There are no systematic uncertainties which are expected to have irreducible contributions which exceed the ultimate statistical precision.  This channel is already entering the upper range of the physically interesting sensitivities, and will likely continue to provide the world's best sensitivity to direct \CP violation in charm in \upgradetwo. The power of these two-body \CP eigenstates at LHCb \upgradetwo is illustrated in Fig.~\ref{fig:agammadeltaacp}, which shows the indirect (see Sect.~\ref{sec:AGamma}) and direct \CP constraints that will come from these modes.

There are a significant number of other two-body modes of strong physics interest where \upgradetwo will also make important contributions. These include the decay modes $\Dz\to\KS\KS$ (0.28\%), 
$\Dz\to\KS\Kstarzb$ $(0.6\times10^{-4})$, $\Dz\to\KS\Kstarz$ $(0.8\times10^{-4})$, $\Dsp\to\KS\pip$ $(3.2\times 10^{-4})$, $\Dp\to\KS\Kp$ $(1.2\times 10^{-4})$, $\Dp\to \phi \pip$ $(6\times 10^{-5})$, $\Dp \to \eta'\pip$ $(3.2\times 10^{-5})$, $\Dsp \to \eta'\pip$ $(3.2\times 10^{-4})$, where the projected statistical only \CP asymmetry sensitivities are given in parentheses. The first three modes mentioned are notable as they receive sizeable contributions from exchange amplitudes at tree-level and could have a relatively enhanced contribution from penguin annihilation diagrams which are sensitive to New Physics. Consequently some authors have highlighted them as potential \CP violation discovery channels~\cite{Nierste:2015zra,Nierste:2017cua}.


%% file: CONTRIBUTIONS/6_CP_violation_and_mixing_in_charm/6.2.2.tex
\subsection{Measurements with $D^+ \to h^+h^-h^+$ and $D_s^+ \to h^+h^-h^+$ decays}

Searches for direct \CP violation in the phase space of Cabibbo-suppressed (CS) $D^+\to h_1^+h_2^-h_3^+$ 
decays, hereafter referred to as $D\to 3h$, are complementary to that of $D^{(0,+)}\to h_1h_2$ 
($h_i = \pi, K$). In charged $D$ systems, only \CP violation in the decay is possible. The main observable 
is the \CP asymmetry, which, in the case of two-body decays, is a single number. In contrast,  $D\to 3h$, 
decays allow the study of the distribution of the \CP asymmetry across a two-dimensional phase space 
(usually represented by the Dalitz plot). 

The $D\to 3h$ decays proceed 
mostly through intermediate resonant states. With a relatively small phase space, the resonances are spread 
over the entire Dalitz plot. The different resonant amplitudes overlap and interfere, providing the strong phase 
difference needed for direct \CP violation to occur. This is a unique feature of multi-body decays.
In some regions of the phase space the difference in strong phase  
between the various resonant amplitudes is very large, which may cause relatively large and localised \CP asymmetries.
The difference in strong phases often change sign across the Dalitz plot, causing eventual \CP
asymmetries to change sign as well, reducing the sensitivity of phase-space-integrated measurements. 

The most appealing aspect of this study is that one does not need to rely on models for the resonant 
structure of $D\to 3h$ decays. The \CP violation may be observed in a direct, model-independent comparison 
between $D^+$ and $D^-$ Dalitz plots. In both $K^-K^+\pi^+$ and $\pi^-\pi^+\pi^+$ spectra the $D^+_s$ signals
can be used as control channels to investigate possible phase-space-dependent asymmetries in detection efficiency.
This is the most sensitive technique for a first observation of \CP violation in charged $D$ mesons.

In the SM, small \CP asymmetries in charged $D$ mesons are expected in CS decays, but are highly suppressed in
doubly-Cabibbo-suppressed (DCS) and Cabibbo-favoured decays. New mechanisms of \CP violation would
require subleading amplitudes involving new particles --- a charged Higgs, for instance. 
This mechanism would generate \CP violation in Cabibbo-favoured decays as well, but the observation
of the \CP asymmetries would be obscured by the vastly dominant SM amplitudes. In DCS decays, however, the
SM amplitudes are suppressed by a factor $\sim \tan^4 \theta$, increasing the possibility of an observation. 
Therefore, DCS decays,  such as $D^+ \to K^-K^+K^+$ and $D^+ \to \pi^-K^+\pi^+$,
offer an unique opportunity to search for new sources of \CP violation in an almost ``background free'' environment.

The estimated  signal yields in future upgrades  are summarised in Table~\ref{tab:Dhhh-yields}. The yields are based on
an extrapolation of the Run 2 yields per unit luminosity, made under the same assumptions of Section 6.1. The
estimated sensitivities to observation of \CP violation, using the decay $D^+ \to \pi^-\pi^+\pi^+$ as example, are presented 
in Table~\ref{tab:Dhhh-sensitivity}.

\begin{table}[tb]
\centering
\caption{Extrapolated signal yields, in units of 10$^6$, of the Cabibbo-suppressed decays $D^+ \to K^-K^+\pi^+$, $D^+ \to \pi^-\pi^+\pi^+$,
and of the doubly Cabibbo-suppressed decays $D^+ \to K^-K^+K^+$, $D^+ \to \pi^-K^+\pi^+$. \label{tab:Dhhh-yields}}
\begin{tabular}{lcccc}
\hline 
Sample ($\mathcal{L}$)  & $D^+ \to K^-K^+\pi^+$  & $D^+ \to \pi^-\pi^+\pi^+$ &  $D^+ \to K^-K^+K^+$ & $D^+ \to \pi^-K^+\pi^+$ \\
\hline
Run 1--2 (9\invfb)     &   200       & 100     & 14       & 8   \\
Run 1--4 (23\invfb)   &   1,000    & 500  & 70     & 40   \\
Run 1--4 (50\invfb)   &   2,600    & 1,300  & 182     & 104   \\
Run 1--6 (300\invfb) &  17,420   &  8,710 & 1,219  &  697  \\
\hline 
\end{tabular}
\end{table}

\begin{table}[bt]
\centering
\caption{Sensitivities to \CP-violation scenarios for $D^+ \to \pi^-\pi^+\pi^+$ decays. Simulated $D^+$ and $D^-$
Dalitz plots are generated with relative changes in the phase of the $R\pi^{\pm}$ amplitude, $R=\rho^0(770)$, 
$f_0(500)$ or $f_2(1270)$. The values of the phase differences are given in degrees and correspond to 
a 5$\sigma$  \CP-violation effect. Simulations are performed with $3\invfb$ and extrapolated to the 
expected integrated luminosities.}
\begin{tabular}{lc c c c}
\hline
  resonant channel           & $9\invfb$   & $23\invfb$     & $50\invfb$     & $300\invfb$\\
\hline
$f_0(500)\pi$                    & 0.30 &  0.13   &  0.083   &  0.032 \\
$\rho^0(770\pi$                & 0.50 &  0.22   &  0.14     &  0.054\\
$f_2(1270)\pi$                  & 1.0  &  0.45   &   0.28    & 0.11\\
\hline
\end{tabular}
\label{tab:Dhhh-sensitivity}
\end{table}

%% file: CONTRIBUTIONS/6_CP_violation_and_mixing_in_charm/6.2.3.tex
\subsection{Measurements with \decay{\Dz}{h^+h^-h^+h^-} decays}
\label{sec:dto4h}

Standard-Model \CP violation could be observed in the Cabibbo-suppressed \decay{\Dz}{\pip\pim\pip\pim} and \decay{\Dz}{\Kp\Km\pip\pim} decays, while New Physics is needed to justify any observation in the doubly Cabibbo-suppressed \decay{\Dz}{\Kp\pim\pip\pim} decays.

Many techniques can be adopted to search for \CP violation, all of them exploiting the rich resonant structure of the decays.
Those that have been so far used at \lhcb are based on ${\widehat{T}}$-odd asymmetries and the energy test, while studies are ongoing to measure model-dependent \CP asymmetries in the decay amplitudes.

The study of ${\widehat{T}}$-odd asymmetries exploits potential $P$-odd \CP violation from the interference of the different amplitude structures in the decay, as described in Ref.~\cite{Durieux:2015zwa}.  This uses a triple product $C_T=\vec{p}_A\cdot (\vec{p}_B\times \vec{p}_C)$ constructed from the momenta of three of the final state particles $\vec{p}_A,\,\vec{p}_B,\,\vec{p}_C$.
\lhcb has studied ${\widehat{T}}$-odd asymmetries using data corresponding to 3\invfb from the Run~1 dataset, obtaining a sensitivity of $2.9\times10^{-3}$ with very small systematic uncertainties~\cite{LHCB-PAPER-2014-046}.
The peculiarity of this measurement is the absence of instrumental asymmetries, since it is given by the difference of two asymmetries measured separately on \Dz and \Dzb decays,
\begin{align}
  A_T = \frac{(\Gamma(\Dz,C_T>0)-\Gamma(\Dz,C_T<0))}{(\Gamma(\Dz,C_T>0)+\Gamma(\Dz,C_T<0))}\quad \bar{A}_T = \frac{(\Gamma(\Dzb,\bar{C}_T>0)-\Gamma(\Dzb,\bar{C}_T<0))}{(\Gamma(\Dzb,\bar{C}_T>0)+\Gamma(\Dzb,\bar{C}_T<0))}
\end{align}
and $a_{\CP}=(A_T-\bar{A}_T)/2$.
One can therefore expect the measurement to scale with luminosity to reach a sensitivity down to $2.9\times10^{-5}$ ($9.4\times10^{-5}$) for $\Dz\to\pip\pim\pip\pim$ ($\Dz\to\Kp\Km\pip\pim$) decays, as detailed in Table~\ref{tab:CS-D24h-yields}.
\begin{table}[tb]
\centering
\caption{\small Extrapolated signal yields, and statistical precision on ${\widehat{T}}$-odd \CP-violation observables.\label{tab:CS-D24h-yields}}
\begin{tabular}{lcccc}
\hline
  & \multicolumn{2}{c}{$\Dz\to\pip\pim\pip\pim$} & \multicolumn{2}{c}{$\Dz\to\Kp\Km\pip\pim$} \\
Sample ($\mathcal{L}$)  & Yield ($\times10^6$) & $\sigma(a_{\CP}^{\widehat{T}\text{-odd}})$  & Yield ($\times10^6$) & $\sigma(a_{\CP}^{\widehat{T}\text{-odd}})$ \\
\hline
Run 1--2 (9\invfb)   & 13.5   & $2.4\times10^{-4}$ &  4.7   & $5.4\times10^{-4}$ \\
Run 1--3 (23\invfb)  &   69   & $1.1\times10^{-4}$ &   12   & $3.4\times10^{-4}$ \\
Run 1--4 (50\invfb)  &  150   & $7.5\times10^{-5}$ &   26   & $2.3\times10^{-4}$ \\
Run 1--5 (300\invfb) &  900   & $2.9\times10^{-5}$ &  156   & $9.4\times10^{-5}$ \\
\hline
\end{tabular}
\end{table}

The energy test method is an unbinned model-independent two-sample test that assesses the p-value for the hypothesis that two distributions are sampled from the same underlying distribution.
Applied to the phase-space distribution of \PD and \Db decays, this gives a probability for the hypothesis of \CP symmetry.
The energy test is described in detail in Ref.~\cite{Parkes:2016yie,Barter:2018xbc} and has been applied at \lhcb to \decay{\Dz}{\pip\pim\piz}~\cite{LHCb-PAPER-2014-054} and \decay{\Dz}{\pip\pim\pip\pim} decays~\cite{LHCb-PAPER-2016-044}.
In the latter, two variants were applied, one testing for $P$-even and the other for $P$-odd \CP violation, with the latter giving an intriguing $2.7\sigma$ deviation from the no-\CP violation hypothesis.

The method is insensitive to global asymmetries; however, it is expected that it will become sensitive to variations in phase space of production and detection asymmetries.
These can be controlled in data by application of the method to Cabibbo-favoured decays such as \decay{\Dz}{\Km\pip\pip\pim}.
Assuming scaling with the square-root of the ratio of sample sizes, the same p-values can be expected for the \CP asymmetries given in Table~\ref{tab:charm_623_energy_test}.

\begin{figure}[t]
  \begin{center}
    \includegraphics[width=0.45\textwidth]{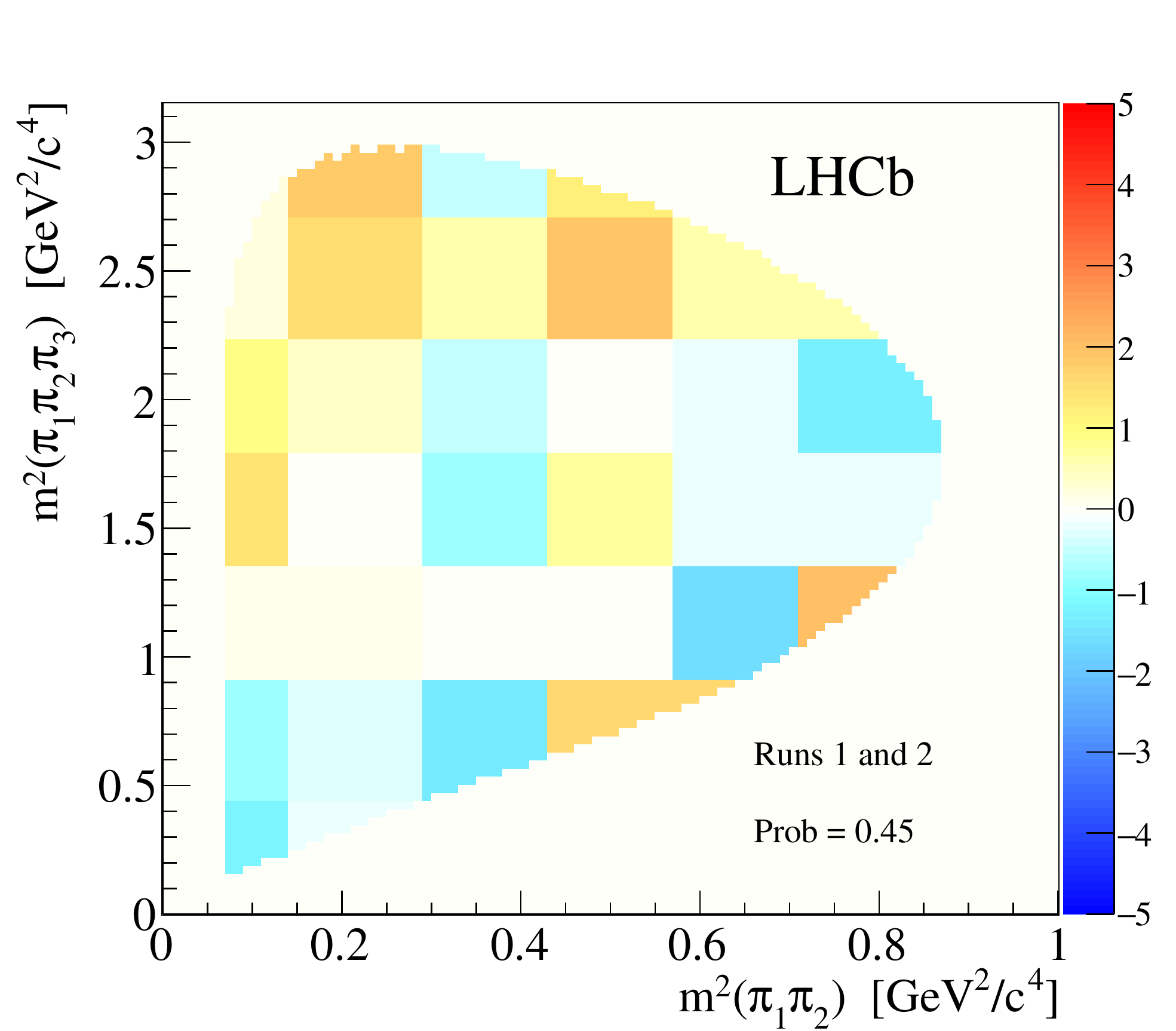}
    \includegraphics[width=0.45\textwidth]{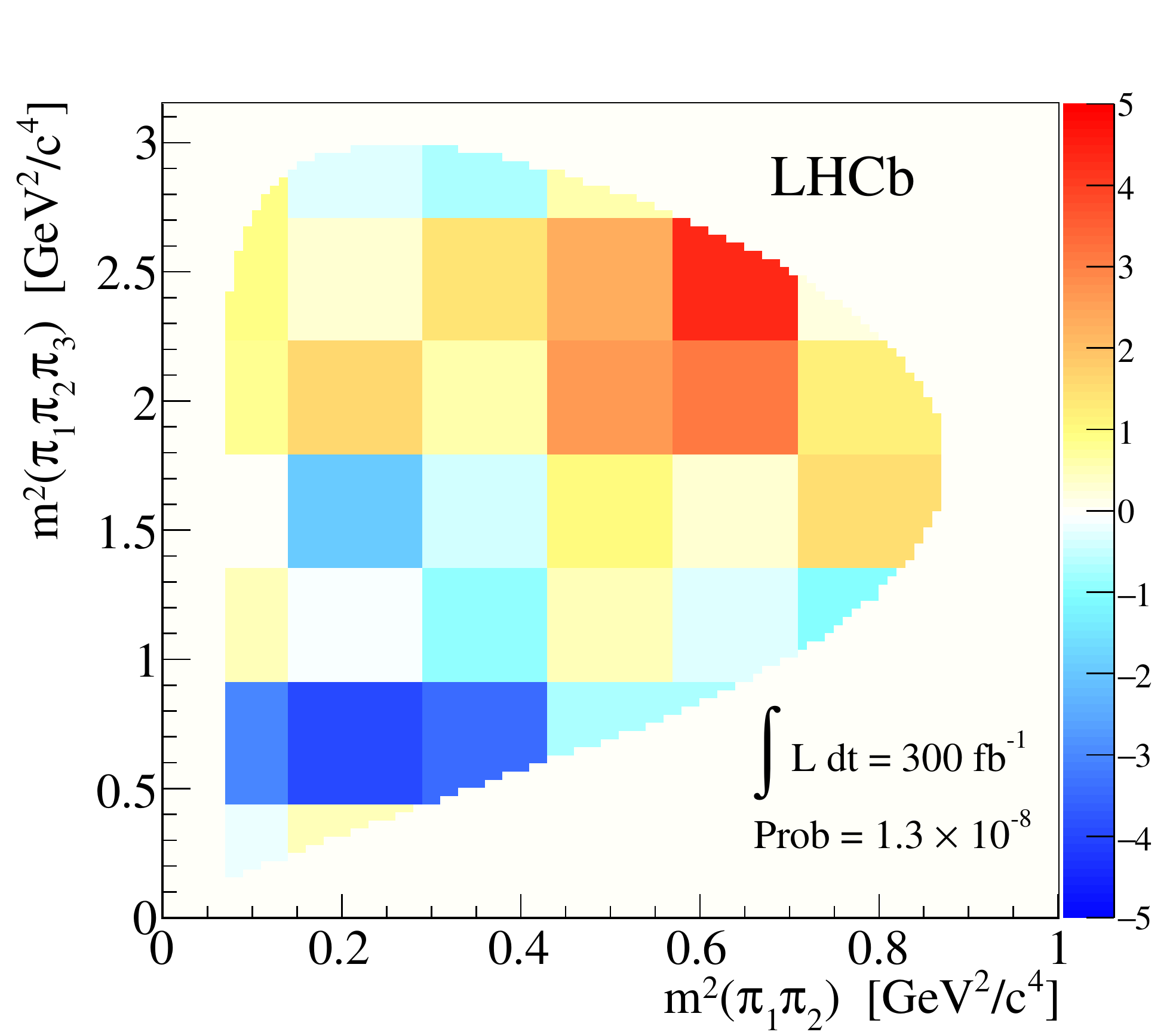}
  \end{center}
  \caption{The binned $S_\CP$ test~\cite{Parkes:2016yie} applied to simplified simulation for the $\Dz\to\pip\pim\pip\pim$ decay. The left hand plot shows the significance of the (normalised) asymmetry in each bin for the event yield expected following Runs 1 and 2, with the right hand plot using the expected event yield corresponding to 300\invfb. The contributions from the four main intermediate resonances associated with the $\Dz$ meson decay (as determined using CLEO-c data~\cite{dArgent:2016rbp}) are simulated; \CP-violation is introduced to the matter and anti-matter samples in the test by varying the relative yield of the $\Dz\to\rho^0\rho^0$ contribution in the samples being compared by $\pm 0.4\%$. Large bin significances and a significant $\chi^2$ value are only observed in the dataset corresponding to 300\invfb; in order to observe this \CP-violation the \upgradetwo dataset is required.}
  \label{fig:charm_scp_d4h}
\end{figure}

The increased sensitivity of the \upgradetwo dataset is demonstrated in Fig.~\ref{fig:charm_scp_d4h}, where the binned $S_{\CP}$ test~\cite{Parkes:2016yie} is applied to data from a simplified simulation. Inserting \CP violation in the same resonances as indicated by the current $2.7\sigma$ anomaly, it is illustrated that in the simulated scenario the \upgradetwo dataset is required to observe and study the \CP violation contribution.

\begin{table}[bt]
\centering
\caption{Overview of sensitivities to various \CP-violation scenarios for \decay{\Dz}{\pip\pim\pip\pim} decays as extrapolated from Ref.~\cite{LHCb-PAPER-2016-044}. The relative changes in magnitude and phase of the amplitude of the resonance $R$ to which sensitivity is expected are given in $\%$ and $^\circ$, respectively. The $P$-wave $\rho^0(770)$ is a $P$-odd component. The phase change in this resonance is tested with the $P$-odd \CP-violation test. Results for all other scenarios are given with the standard $P$-even test.}
\begin{tabular}{lc c c c}
\hline
$R$ (partial wave)        & $9\invfb$   & $23\invfb$ & $50\invfb$ & $300\invfb$\\
\hline
\decay{a_1}{\rhoz\pi} (S) & $1.4\%$     & $0.6\%$     & $0.4\%$  & $0.17\%$\\
\decay{a_1}{\rhoz\pi} (S) & $0.8^\circ$ & $0.35^\circ$  & $0.24^\circ$ & $0.10^\circ$\\
\rhoz\rhoz (D)            & $1.4\%$     & $0.6\%$     & $0.4\%$  & $0.17\%$\\
\hline
\rhoz\rhoz (P)            & $0.8^\circ$ & $0.35^\circ$  & $0.24^\circ$ & $0.10^\circ$\\
\hline
\end{tabular}
\label{tab:charm_623_energy_test}
\end{table}

%% file: CONTRIBUTIONS/6_CP_violation_and_mixing_in_charm/6.2.4.tex
\subsection{Prospects with neutral final states}
\label{sec:charmneutral}

Radiative charm decays are discussed  in Sect.~\ref{sec:radCharm}, with decays containing neutral hadrons discussed here. Charm decays with neutrals in the final state can help to shed light on the SM or beyond-SM origin of possible \CP-violation signals by testing correlations between \CP asymmetries measured in various flavour-\grpsuthree related decays \cite{Bhattacharya:2012ah,Pirtskhalava:2011va,Feldmann:2012js}. They are, however, particularly challenging in hadronic collisions, where the calorimeter background for low energy clusters is high, while the trigger retention rate needs to be kept low to allow for affordable rates. 

Nevertheless, good performances are achieved when considering decays, such as \decay{\Dz}{\pi^+\pi^-\pi^0}, with at least two charged particles in the final states that can help to identify the displaced decay vertex of the charm meson. In only 2\invfb of data collected during 2012, LHCb has reconstructed about 660,000 \decay{\Dz}{\pi^+\pi^-\pi^0} decays~\cite{LHCb-PAPER-2014-054}, \ie about five times larger than that from the full BaBar data set~\cite{TheBABAR:2016gom}, with comparable purity. Preliminary estimates for Run 2 data, give about 500\,000 signal decays per \invfb, making future \CP-violation searches in this channel very promising. Similarly, large samples of \decay{\D_{(\squark)}^+}{\eta^{(\prime)}\pi^+} decays, with \decay{\eta^{(\prime)}}{\pi^+\pi^-\gamma}, or \decay{\Dp}{\pi^+\pi^0} decays, with \decay{\pi^0}{e^+e^-\gamma}, are already possible with the current detector. The \decay{\D_{(\squark)}^+}{\eta^{\prime}\pi^{+}} mode, as an example, has been used by LHCb during Run 1 to perform the most precise measurement of \CP asymmetries in these channels to date, with uncertainties below the 1\% level~\cite{LHCb-PAPER-2016-041}.

More challenging final states consisting only of neutral particles, such as $\pi^0 \pi^0$ or $\eta \eta$, can still be reconstructed with \decay{\pi^0}{\gamma\gamma} or \decay{\eta}{\gamma\gamma} candidates made of photons which, after interacting with the detector material, have converted into an $e^+e^-$ pair. Such conversions must occur before the tracking system to have electron tracks reconstructed. Although the reconstruction efficiency of these ``early'' converted photons in the current detector reaches only a few percent of the calorimetric photon efficiency, their purity is much higher. This approach may become interesting only with the large data sets that are expected to be collected by the end of \upgradetwo. The $\eta$ decays can also be reconstructed through the $\pi^+\pi^-\gamma$ final state.

Unlike the \Dz decays which are usually tagged with a soft pion from \Dstarp decays, there is no easy tagging of the \Dp modes, which thus often suffer from a high combinatorial background. Employing a $\pi^0$ tag using \decay{\Dstarp}{\Dp\pi^0} decays could facilitate future studies of \Dp decays, in particular those with challenging and/or high multiplicity final states.

The study of these modes may be challenging in \upgradeone due to the cluster pile-up at higher luminosities and radiation damage of the current calorimeter. A central element of the \upgradetwo detector is the proposed new calorimeter system (see Sect.~\ref{sec:PID}). The proposed increase in granularity would improve the efficiency for \decay{\pi^0}{\gamma\gamma} decays and, in particular, could make the $\pi^0$ tag feasible.

%% file: CONTRIBUTIONS/6_CP_violation_and_mixing_in_charm/6.2.5.tex
\subsection{Prospects with charmed baryons}\label{Sec:6.2.5}
\label{sec:charmbaryon}

The first charmed baryon, \Lc, was discovered over 40 years ago, but
its properties are
still poorly understood.
Even less is known about the heavier states such as the 
$\PXi_{\cquark}$ and $\POmega_{\cquark}$ baryons.
This is due to a historical lack of large experimental datasets, which has in 
turn disfavoured theoretical study~\cite{Bigi:2012ev}.
Given the large \Lc fiducial cross-section measured by \lhcb at $\sqs = 
7\tev$~\cite{LHCb-PAPER-2012-041}, and the recent discovery of the 
doubly charmed $\PXi_{\cquark\cquark}^{++}$ baryon~\cite{LHCb-PAPER-2017-018} 
there is now the experimental possibility to execute a broad charmed baryon 
programme.
Along with the evidence for \CP violation in beauty baryon 
decays~\cite{LHCb-PAPER-2016-030}, there is also theoretical interest in 
understanding if similar behaviour exists in the charm sector.

In general, charmed baryon decays offer a rich laboratory within which to study 
matter-antimatter asymmetries.
The most easily accessible modes for \lhcb are multibody decays containing one 
proton and several kaons or pions, such as the Cabibbo-favoured 
$\decay{\Lc}{\proton\Km\pip}$ decay and the singly Cabibbo-suppressed 
$\decay{\Lc}{\proton\Km\Kp}$ and $\proton\pim\pip$ decays.
Unlike the analogous final states of the spinless neutral and charged \PD mesons, such 
charmed baryon final states have at least five degrees of freedom, allowing for 
a complex variation of the strength of \CP violation across the phase space.
A complete description of such a space is experimentally challenging.
This is compounded by the relatively short lifetime of the \Lc baryon with 
respect to those of the \PD mesons, which decreases the power of selections 
based on the displacement of the charm vertex, increasing the prompt 
combinatorial background, which mainly comprises low-momentum pions and kaons.
Selections of charm baryons then heavily rely on good discrimination between 
proton, kaon, and pion hypotheses in the reconstruction of charged tracks, and 
on a precise secondary vertex reconstruction.
The latter is also necessary when reconstructing charm baryons originating from 
semileptonic \bquark-hadron decays, which has the advantage of both a simple 
trigger path (a high-\pt, displaced muon) and a cleaner experimental signature 
due to the large \bquark-hadron lifetime.
The presence of the proton in the final state, whilst a useful handle for 
selections, poses experimental challenges for \CP-violation measurements in 
addition to those present for measurements with \PD mesons, as the 
proton-antiproton detection asymmetry must be accounted for.
This has not yet been measured at \lhcb due to the lack of a suitable control 
mode.

The most precise measurement of \CP violation in charm baryons was made 
recently by the \lhcb collaboration using Run 1 data, corresponding to 
$3$\invfb of integrated luminosity~\cite{LHCb-PAPER-2017-044}.
The difference between the phase-space-averaged \CP asymmetries in 
$\decay{\Lc}{\proton\Km\Kp}$ and $\decay{\Lc}{\proton\pim\pip}$ decays was 
found to be consistent with \CP symmetry to a statistical precision of 
$0.9\,\%$.
This difference is largely insensitive to the proton detection asymmetry, but 
masks like-sign \CP asymmetries between the two modes.
Further studies must then gain a precise understanding of the proton detection 
asymmetry, in addition to measuring the variation of \CP violation across the 
decay phase space.

Although there is little literature on the subject, the magnitude of direct
\CP violation in charm baryon decays is expected to be similar to that for 
charmed mesons~\cite{Bigi:2012ev}, and so the first step in furthering our 
understanding is to reach a precision of $0.5\times10^{-4}$, which can be met 
given the $300$\invfb of integrated luminosity collected by the end of Run 5.
The acquisition of more data is vital in enabling studies of states heavier 
than the \Lc baryon, as their alternate compositions may permit considerably different 
dynamics.

The \Xiccpp baryon was discovered using the $\Lc\Km\pip\pip$ final state with 
data taken in 2016 corresponding to an integrated luminosity of $1.7$\invfb.
A signal yield of $313 \pm 33$ was determined, which can be 
extrapolated to around $100,000$ such decays obtainable with data corresponding to
$300$\invfb.
This will allow asymmetry measurements with a precision of $0.4$\,\%.

Consequently the high statistics of \upgradetwo offers the possibility
to open up the detailed study of this new research area.

%% file: CONTRIBUTIONS/7_Rare_and_radiative_decays/7.tex
\label{chpt:rare}

The precision on measurements with rare decays of beauty and charm hadrons, or $\tau$ leptons, is dominated by the limited available data samples. This will still be the case after the \upgradeone data taking period. For example, to reach a 
precision $\sim 10\%$ on the ratio $R={\cal B} (\Bdmm)/{\cal B}(\Bsmm)$ will provide strong constraints (or evidence) for new flavour structures 
beyond the SM. Moreover, the \upgradetwo yields will allow meaningful measurements of new observables, namely the effective lifetime ($\tau_{\mu\mu}^{\rm eff}$) and the time-dependent \CP\ asymmetry 
of \Bsmm decays, and will allow for improved searches for other (semi-)leptonic \B and \D decays.
If the current indications of 
lepton non-universality NP contributions to $\bquark\to\squark \ellell$ processes are confirmed with more data, \upgradetwo will be needed to be able to 
discriminate between potential NP models, by increasing the precision, but more importantly, by expanding the range of complementary decay modes studied (including the less abundant 
$\bquark\to\dquark \ellell$ processes). In this case, a more performant calorimeter system with improved granularity, as discussed in this document, should be able to improve the efficiency for bremsstrahlung recovery in  $\bquark\to\squark e^+ e^-$ and therefore provide a significant improvement in sensitivity. 
A better calorimeter performance should also help to improve the precision of the \upgradetwo in the measurements of radiative beauty and charm decays.
In the following sections  a more detailed description of the potential of the \upgradetwo in rare and radiative decays is given.
 
\input{CONTRIBUTIONS/7_Rare_and_radiative_decays/7.1.tex}
\input{CONTRIBUTIONS/7_Rare_and_radiative_decays/7.2.tex}

\input{CONTRIBUTIONS/7_Rare_and_radiative_decays/7.3.tex}

\input{CONTRIBUTIONS/7_Rare_and_radiative_decays/7.4.tex}

\input{CONTRIBUTIONS/7_Rare_and_radiative_decays/7.5.tex}

\input{CONTRIBUTIONS/7_Rare_and_radiative_decays/7.6.tex}

%% file: CONTRIBUTIONS/7_Rare_and_radiative_decays/7.1.tex
\section{Leptonic $B$ decays}\label{sec:B2ll}
\input{CONTRIBUTIONS/7_Rare_and_radiative_decays/7.1.1.tex}

\input{CONTRIBUTIONS/7_Rare_and_radiative_decays/7.1.3.tex}

%% file: CONTRIBUTIONS/7_Rare_and_radiative_decays/7.1.1.tex
\subsection{Measurements with $B \to \mumu$ decays}
\input{CONTRIBUTIONS/7_Rare_and_radiative_decays/7.1.1.1.tex}

\input{CONTRIBUTIONS/7_Rare_and_radiative_decays/7.1.1.2.tex}

%% file: CONTRIBUTIONS/7_Rare_and_radiative_decays/7.1.1.1.tex
\subsubsection{Branching fractions}

The decay \Bsmm represents one of the most theoretically clean and sensitive probes to reveal effects of New Physics beyond the SM unreachable by direct searches.
This decay is both loop- and helicity-suppressed and is therefore very rare, with an expected branching fraction of ${\cal B}(\decay{\Bs}{\mumu})=(3.66\pm0.23) \times 10^{-9}$ in the SM~\cite{Bobeth:2013uxa}. 
The related mode \Bdmm is further suppressed by the ratio of CKM matrix elements $|\Vtd|^2/|\Vts|^2$, resulting in a SM prediction ${\cal B}(\decay{\Bd}{\mumu})=(1.06\pm0.09) \times 10^{-10}$~\cite{Bobeth:2013uxa}. These decays are particularly sensitive probes for new scalar or pseudoscalar contributions.

\begin{figure}[!tb]
\centering
\includegraphics[scale=0.37]{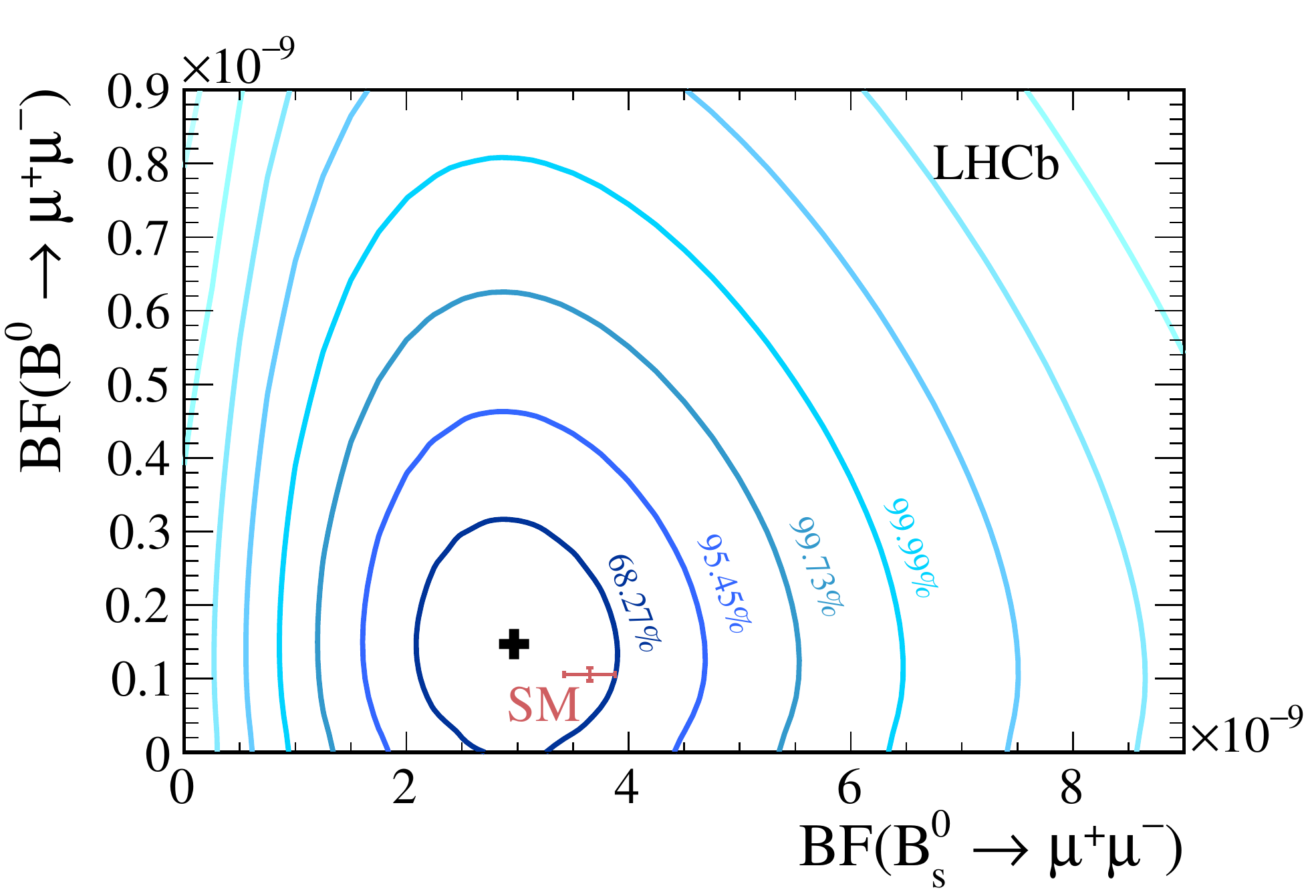}
\includegraphics[scale=0.4]{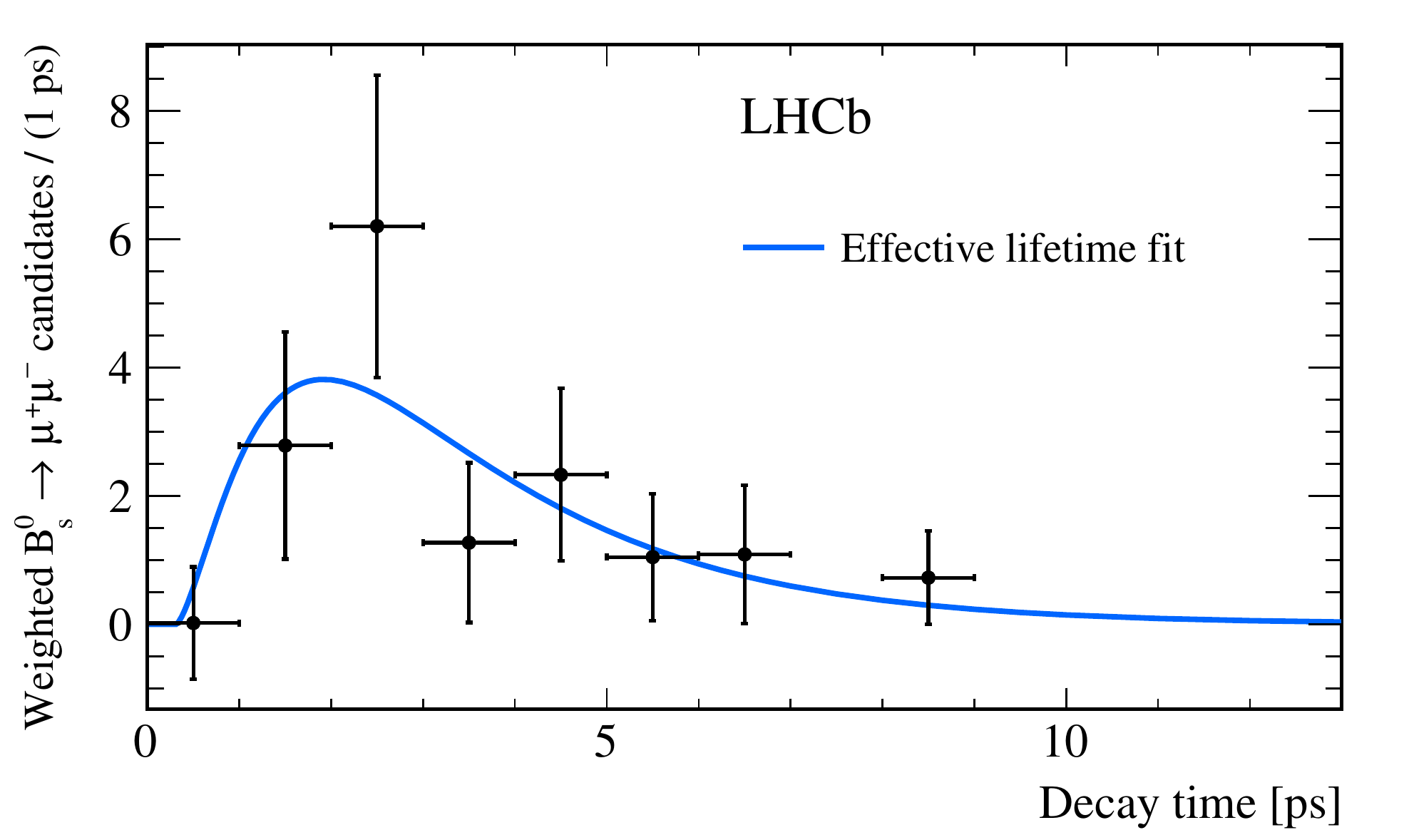}
\caption{(Left) A two-dimensional representation of the LHCb branching fraction measurements for \Bdmm and \Bsmm. The central values are indicated with the black plus sign. The profile likelihood contours for 1,2,3\dots $\sigma$ are shown as blue contours. The Standard Model value is shown as the red cross labelled SM. (Right) Background-subtracted \Bsmm decay-time distribution with the fit result superimposed.}
\label{fig:btomumu}
\end{figure}

In 2017 the LHCb collaboration reported the first observation by a single experiment of the decay $\decay{\Bs}{\mumu}$~\cite{LHCb-PAPER-2017-001} and measured ${\cal B}(\decay{\Bs}{\mumu})=\left(3.0\pm 0.6^{+0.3}_{-0.2}\right)\times 10^{-9}$ and ${\cal B}(\decay{\Bd}{\mumu})=\left(1.5^{+1.2 +0.2}_{-1.0 -0.1}\right) \times 10^{-10}$ using data collected in $pp$ collisions corresponding to a total integrated luminosity of 4.4\invfb (Fig.~\ref{fig:btomumu}, left). Both measurements are compatible with the SM predictions but affected by large statistical uncertainties. The analysis exploits the excellent performance of the LHCb detector in terms of muon identification and vertex and mass resolution. Compared to the previous analysis~\cite{LHCb-PAPER-2013-046}, the contamination of combinatorial and \BTohh backgrounds were reduced by more than a factor two thanks to the development of improved isolation variables and the optimisation of the muon identification requirements.
This has led to a reduction of the statistical uncertainty of about 30\% for ${\cal B}(\decay{\Bs}{\mumu})$ for a given integrated luminosity. 

While the uncertainty of ${\cal B}(\decay{\Bd}{\mumu})$ will be dominated by the statistical uncertainty, the projected uncertainty of ${\cal B}(\decay{\Bs}{\mumu})$ with a 300\invfb dataset depends on the assumptions made for the systematic uncertainties. The current systematic uncertainty is dominated by the 5.8\% relative uncertainty associated with the $b$-quark fragmentation probability ratio ($f_s/f_d$)~\cite{fsfd} followed by approximately 3\% from the branching fractions of the normalisation modes, 2\% from particle identification and 2\% from track reconstruction.
The projected relative statistical uncertainty of the current analysis on a dataset of 300\invfb is 1.8\%. 
At the end of the Upgrade II data taking period,
a conservative assumption for the systematic uncertainty on ${\cal B}(\decay{\Bs}{\mumu})$ is about 4\%, which
would imply an
uncertainty of ${\cal B}(\decay{\Bs}{\mumu})$ to be approximately $0.30\times 10^{-9}$ with 23\invfb and $0.16\times 10^{-9}$ with $300\invfb$. The increased precision will be able to cover larger part of the unconstrained parameter space of MSSM models. 
The ratio of branching fractions, ${\cal B}(\decay{\Bd}{\mumu})/{\cal B}(\decay{\Bs}{\mumu})$, is a powerful observable to test minimal flavour violation. The relative uncertainty of this ratio is expected to remain limited only by statistics and decrease from $90\%$ for the current measurement to about $34\%$ with $23\invfb$ and $10\%$ with $300\invfb$. 
All estimates use the quoted SM predictions as central values for the branching ratios and similar detector and analysis performance as in Ref.~\cite{LHCb-PAPER-2017-001}. 

The estimated experimental uncertainties at 300\invfb are close to the uncertainty of the current SM prediction from theory, which is dominated by the uncertainty of the $\Bs$ decay constant, determined from lattice QCD calculations, and the CKM matrix elements. Both are expected to improve in precision in the future.

%% file: CONTRIBUTIONS/7_Rare_and_radiative_decays/7.1.1.2.tex
\subsubsection{Effective lifetime and $C\!P$ violation}

With a 300\invfb dataset, precise measurements of additional observables are possible, namely the effective lifetime ($\tau_{\mu\mu}^{\rm eff}$) and the time-dependent \CP\ asymmetry of \Bsmm decays. Both quantities are sensitive to possible new contributions from the scalar and pseudo-scalar sector in a way complementary to the branching ratio measurement\cite{DeBruyn:2012wk}. 

The effective lifetime is related to the mean \Bs lifetime $\tau_{B_s}$ through the relation 
\begin{equation}
\tau_{\mu\mu}^{\rm eff}=\frac{\tau_{B_s}}{1-y^2_s}\frac{1+2A^{\mu\mu}_{\Delta\Gamma}y_s+y^2_s}{1+A^{\mu\mu}_{\Delta\Gamma}y_s}\,, 
\end{equation}
where $y_s=\tau_{B_s}\Delta\Gamma_s/2$ and $\Delta\Gamma_s=\Gamma_{B^0_{sL}}-\Gamma_{B^0_{sH}}$. The parameter $A^{\mu\mu}_{\Delta\Gamma}$ is equal to 1 in the SM, where only the heavy mass eigenstate decays to \mumu, but can take any value between $-1$ and $+1$ in scenarios beyond the SM.
LHCb has performed the first measurement of the \Bsmm effective lifetime by fitting the signal decay-time distribution using a dataset of 4.4\invfb, 
resulting in $\tau_{\mu\mu}^{\rm eff}=2.04\pm 0.44\pm 0.05 \ps$~\cite{LHCb-PAPER-2017-001} (Fig.~\ref{fig:btomumu}, right).
The relative uncertainty on $\tau_{\mu\mu}^{\rm eff}$ is expected to decrease to approximately 8\%  with 23\invfb and 2\% with 300\invfb of data.
While the current experimental uncertainty is larger than $\tau_{B^0_{sH}}-\tau_{B^0_{sL}}$,
a 2\% uncertainty on $\tau_{\mu\mu}^{\rm eff}$ allow stringent constraints to be set on $A^{\mu\mu}_{\Delta\Gamma}$ and in particular would allow the degeneracy between any possible contribution from new scalar and pseudoscalar mediators to be broken.

Assuming a tagging power of about 3.7\%\cite{LHCb-PAPER-2014-059},
a dataset of 300\invfb allows a pure sample of more than 100 flavour-tagged \Bsmm decays (effective yield) to be reconstructed and their time-dependent \CP asymmetry to be measured. 
From the relation  
\begin{equation}
\frac{\Gamma(B^0_s(t)\to\mu^+\mu^-)-\Gamma(\bar{B^0_s}\to\mu^+\mu^-)}{\Gamma(B^0_s(t)\to\mu^+\mu^-)+\Gamma(\bar{B^0_s}\to\mu^+\mu^-)}=\frac{S_{\mu\mu}\sin(\Delta M_st)}{\cosh(y_st/\tau_{B_s})+A^{\mu\mu}_{\Delta\Gamma}\sinh(y_st/\tau_{B_s})}\,, 
\end{equation}
where $t$ is the signal proper time and $\Delta M_s$ is the mass difference of the heavy and light \Bs mass eigenstates, $S_{\mu\mu}$ can be measured with an uncertainty of about 0.2. On the other hand, the signal yield expected in a 23\invfb dataset is too low to allow a meaningful constraint to be set on $S_{\mu\mu}$. A nonzero value for $S_{\mu\mu}$ would automatically indicate evidence of \CP-violating phases beyond the SM.

%% file: CONTRIBUTIONS/7_Rare_and_radiative_decays/7.1.3.tex
\subsection{Search for $B \to \tautau$ decays}

LHCb has performed a search for the $B\to\tautau$ decays using a data sample corresponding to an integrated luminosity of $3\,\invfb$ \cite{LHCb-PAPER-2017-003}.
Assuming no contribution from $\Bd\to\tautau$ decays, an upper limit is set on the $\Bs\to\tautau$ branching fraction of $6.8 \times 10^{-3}$ at 95\% CL.
This is the first direct limit on $\BR(\Bs\to\tautau)$.
If instead no contribution from $\Bs\to\tautau$ decays is assumed, the limit is $\BR(\Bd\to\tautau)< 2.1 \times 10^{-3}$ at 95\% CL.
This is the world's best limit on $\BR(\Bd\to\tautau)$ and constitutes a factor 2.6 improvement with respect to the \babar result \cite{Aubert:2005qw}.

New Physics models proposed to explain the tensions currently observed in $b\to s\ell^+\ell^-$ and $b\to c\ell^+\nu_{\ell}$ transitions allow for a potential enhancement of the $B\to\tautau$ branching fractions by several orders of magnitude \cite{Dighe:2012df, Alonso:2015sja,Cline:2015lqp,Becirevic:2016yqi,Becirevic:2016oho} with respect to their SM predictions, $\BR(\Bd\to\tautau) = (2.22 \pm 0.19)\times 10^{-8}$ and $\BR(\Bs\to\tautau) = (7.73 \pm 0.49)\times 10^{-7}$ \cite{Bobeth:2013uxa}.
With potential enhancements of the branching fractions close to the current best upper limits 
it therefore remains interesting to pursue a search for the $B\to\tautau$ decays, even though, based on the current experimental precision, a measurement of the SM value will remain out of reach for LHCb.
Naively scaling our obtained limits with the collected luminosity, and taking into account the factor two increase in the $b\bar{b}$ cross-section 
when going from 7--8$\tev$ to 14$\tev$ 
we expect to improve the limit on the $\Bs\to\tautau$ branching fraction by a factor 5 to $1.3\times 10^{-3}$ by the end of \upgradeone and by another factor 2.6 to $5\times 10^{-4}$ by the end of \upgradetwo.

The understanding of the intermediate resonance structure of the $\tau^-\to\pi^-\pi^+\pi^-\nu_{\tau}$ decay, which is exploited in the current analysis to define a region with enhanced signal sensitivity, forms a potential systematic limitation in the $B\to\tautau$ searches.
Here, we most likely will have to rely on \belletwo to provide improved input measurements.


%% file: CONTRIBUTIONS/7_Rare_and_radiative_decays/7.2.tex
\section{Lepton-flavour, lepton-number and baryon-number violating decays}
\input{CONTRIBUTIONS/7_Rare_and_radiative_decays/7.2.1.tex}

\input{CONTRIBUTIONS/7_Rare_and_radiative_decays/7.2.2.tex}

\input{CONTRIBUTIONS/7_Rare_and_radiative_decays/7.2.3.tex}
\input{CONTRIBUTIONS/7_Rare_and_radiative_decays/7.2.4.tex}

\input{CONTRIBUTIONS/7_Rare_and_radiative_decays/7.2.5.tex}

%% file: CONTRIBUTIONS/7_Rare_and_radiative_decays/7.2.1.tex
\subsection{Search for $\B \to e^\pm \mu^\mp$ decays}

   \def\epm    {{\ensuremath{e^\pm}}\xspace}
   \def\emp    {{\ensuremath{e^\mp}}\xspace}
   \def\mupm   {{\ensuremath{\mu^\pm}}\xspace}
   \def\mump   {{\ensuremath{\mu^\mp}}\xspace}
    
   \def\BToEMu    {\decay{\B}{\epm\mump}}
   \def\BdToEMu   {\decay{\Bd}{\epm\mump}}
   \def\BsToEMu   {\decay{\Bs}{\epm\mump}}

   The lepton-flavour-violating (LFV) process \BToEMu is forbidden within the SM, however recent hints of
   lepton-flavour non-universality in \B-meson decays~\cite{LHCb-PAPER-2014-024,LHCb-PAPER-2017-013}
   suggest the presence of a mechanism at higher energies able to differentiate the flavour of charged leptons.
   In relation to the anomalies observed, several models~\cite{Crivellin:2015era,Becirevic:2016yqi,Varzielas} have been
   developed where the branching fractions of the \BToEMu can be as large as $\mathcal{O}(10^{-11})$.

   The LHCb collaboration has recently published~\cite{LHCb-PAPER-2017-031} the world's best limits on the branching fractions of the \BsToEMu
   and \BdToEMu decays using the first 3\invfb collected in 2011 and 2012 at 7 and 8\tev respectively.
   The decay-time acceptance of the \BsToEMu decays can be affected by the relative contribution of the two \Bs mass eigenstates
   to the total decay amplitude, due to their large lifetime difference. 
   Therefore, the upper limit on the branching fraction of \BsToEMu decays is evaluated in two extreme hypotheses:
   where the amplitude is completely dominated by the heavy eigenstate or by the light eigenstate.
   The results are  $\BF(\BsToEMu) < 6.3\,(5.4) \times 10^{-9}$ and $\BF(\BsToEMu) < 7.2\,(6.0) \times 10^{-9}$
   at $95\%\,(90\%)$ CL, respectively. The limit for the branching fraction of the \Bd mode is 
   $\BF(\BdToEMu) < 1.3\,(1.0) \times 10^{-9}$ at $95\%\,(90\%)$ CL.

   Assuming similar performances in background rejection and signal retention as in the current analysis, at the end of the \upgradeone data taking period the LHCb
   experiment will be able to probe branching fractions of \BsToEMu and \BdToEMu decays down to $8\times 10^{-10}$
   and $2\times 10^{-10}$, respectively. The additional samples accumulated during the \upgradetwo data taking period will push down
   these limits to $3\times 10^{-10}$ and $9\times 10^{-11}$ respectively, close to the interesting region
   where NP effects may appear. The improvement at \upgradetwo\ over
   \upgradeone\ in electron  reconstruction will be very important 
   in attaining, or exceeding, this goal.

%% file: CONTRIBUTIONS/7_Rare_and_radiative_decays/7.2.2.tex
\subsection{Search for $\B \to \tau^\pm\mu^\mp$ decays}
    \newcommand{\subdecay}[2]{\ensuremath{#1(\!\to #2)}\xspace}
    \def\mupm   {{\ensuremath{\mu^\pm}}\xspace}
    \def\mump   {{\ensuremath{\mu^\mp}}\xspace}
    \def\taupm   {{\ensuremath{\tau^\pm}}\xspace}
    \def\taump   {{\ensuremath{\tau^\mp}}\xspace}
    \def\tauppp{\decay{\taupm}{\pi^\pm\pi^\mp\pi^\pm\nu}}
    \def\tauppppi0{\decay{\taupm}{\pi^\pm\pi^\mp\pi^\pm\piz\nu}}
    \def\BToTauMu    {\decay{\B}{\taupm\mump}}
    \def\BdToTauMu   {\decay{\Bd}{\taupm\mump}}
    \def\BsToTauMu   {\decay{\Bs}{\taupm\mump}}
    \def\BdToTauMupm   {\decay{\Bd}{\taupm\mump}}
    \def\BsToTauMupm   {\decay{\Bs}{\taupm\mump}}
    \def\BdsToTauMu   {\decay{\B^0_{\left( s\right) }}{\taupm\mump}}
    \def\BdsToTauMuSS {\decay{\B^0_{\left( s\right) }}{\taupm\mupm}}
    \def\Btaumu{\decay{\B}{\subdecay{\taupm}{\pipm\pimp\pipm\nu}\mump}}
    \def\Bstotaumu{\decay{\Bs}{\subdecay{\taupm}{\pipm\pimp\pipm\nu}\mump}}
    \def\Bdtotaumu{\decay{\Bd}{\subdecay{\taupm}{\pipm\pimp\pipm\nu}\mump}}
    \def\Bdstaumu{\decay{\B^0_{\left( s\right) }}{\subdecay{\taupm}{\pipm\pimp\pipm\nu}\mump}}
    \def\Bdtaumu{\decay{\Bd}{\subdecay{\taupm}{\pipm\pimp\pipm\nu}\mump}}
    \def\Bstaumu{\decay{\Bs}{\subdecay{\taupm}{\pipm\pimp\pipm\nu}\mump}}
    \def\Bstaumupi0{\decay{\Bs}{\subdecay{\taupm}{\pipm\pimp\pipm\piz\nu}\mump}}
    \def\Bdtaumupi0{\decay{\Bd}{\subdecay{\taupm}{\pipm\pimp\pipm\piz\nu}\mump}}
    \def\BToDPi    {\decay{\Bd}{\D^\pm\pi^\mp}}

    A wide variety of Beyond Standard Model (BSM) scenarios predict
    the occurrence of lepton-flavour violating \B decays, with the highest rates where third generation leptons are involved, such as \BdsToTauMu. 
For different models, not excluded by the current experimental constraints, the expected branching fraction for the LFV process \BsToTauMu is in the range $10^{-9} - 10^{-4}$~\cite{Boubaa:2012xj,Parry:2006mv,Harnik:2012pb,Dedes:2002rh,Lee:2016dcb,Guetta:1998id,Becirevic:2016oho,Smirnov:2018ske,Crivellin:2015mqa,Becirevic:2016zri,Hiller:2016kry,Dighe:2012df,Alonso:2015sja}.

On the experimental side, an upper limit on the \BdToTauMu channel has been already set by $\babar$: $\BR\left( \BdToTauMu\right)< 2.2\times 10^{-5} $ at $90\%$~CL~\cite{Aubert:2008cu}.
The first search on the \BsToTauMu channel is in progress in LHCb and the results are expected soon on data recorded in 2011 and 2012
using the \tauppp and \tauppppi0 decay modes.
Given the presence of a neutrino that escapes detection this kind of analysis is much more complicated than those investigating electron or muon final states. 
A specific reconstruction technique is used in order to infer the energy of the $\nu$, taking advantage of the known $\tau$ decay position given by the $3\pi$ reconstructed vertex. 
This way, the complete kinematics of the process can be solved up to a two-fold ambiguity. LHCb expects to reach sensitivities of a few times $10^{-5}$ with the Run~1 and 2 data sets. 
Extrapolating the current measurements to the \upgradetwo LHCb could reach  $\BR\left( \BdToTauMu\right)< 3\times 10^{-6} $ at $90\%$~CL. The mass reconstruction technique depends heavily on the resolution of the primary and the $\tau$ decay vertices, hence improvement in the tracking system in \upgradetwo, including a removal or reduction in material of the VELO RF foil, will be very valuable.


%% file: CONTRIBUTIONS/7_Rare_and_radiative_decays/7.2.3.tex
\subsection{Search for $B \to Ke\mu$  and $B \to K\tau\mu$ decays}

If New Physics allows for charged lepton-flavour violation then the branching fractions of $B\to K \ell \ell^{'}$  or $\Lb \to \Lz \ell \ell^{'}$ will be enhanced with respect to their purely leptonic counterparts, since the helicity suppression is lower. 
Furthermore, if observed, they would allow the measurement of more observables with respect to the lepton-flavour violating decays discussed in the previous sections, thanks to their multi-body final states and, in the case of $\Lb$, to the non-zero initial spin. 

In many generic new physics models with lepton flavour universality violation, charged lepton-flavour violating decays of $b$-hadrons can be linked with the anomalies recently measured in $b\rightarrow s\ell\ell$ decays~\cite{LHCb-PAPER-2014-024,LHCb-PAPER-2017-013,LHCb-PAPER-2015-051}. 
%
%
The current  limits set by the \bfactories on the branching fractions of $B \to Ke \mu$ and $B \to K\tau \mu$ decays are $< 13\times 10^{-8}$~\cite{Aubert:2006vb} and $< 4.8\times 10^{-5}$~\cite{Lees:2012zz}  at 90\% confidence level, respectively.  

At LHCb, searches for $\decay{\Bu}{\Kp e^\pm\mu^\mp}$, $\decay{\Bd}{\Kstarz\tau^\pm\mu^\mp}$, $\decay{\Bu}{\Kp\tau^\pm\mu^\mp}$ and $\decay{\Lb}{\Lz e^\pm\mu^\mp}$ are ongoing. These searches are complementary, as charged lepton-flavour violation couplings among different families are expected to be different.
The analyses involving $\tau$ leptons reconstruct candidates via the $\decay{\taum}{\pim\pim\pip\nu_{\tau}}$ channel,
which allows the reconstruction of the $\tau$ decay vertex.\footnote{It should be noted that searches for $\decay{\Bu}{\Kp\tau^\pm\mu^\mp}$ from $B_{s2}^*$ without $\tau$ reconstruction can give complementary information.} 
All these decays contain at least one muon, which is used to efficiently trigger on the event. 
Usually, since these decays involve combinations of leptons that are not allowed in the Standard Model, the backgrounds can be kept well under control,  leaving very clean samples only polluted by candidates formed by the random combinations of tracks. This combinatorial effect is higher for the channel with a $\tau$ lepton in the final state decaying into three charged pions.  The other relevant background  comes from chains of semileptonic decays, where two or more neutrinos are emitted and therefore combinations of leptons of different flavours are possible. These decays have typically a low reconstructed invariant mass, due to the energy carried away by the neutrinos, and so they do not significantly pollute the signal region. 

The expected upper limits at LHCb using the first $9 \invfb$ of data taken are
$\mathcal{O}(10^{-9})$ and $\mathcal{O}(10^{-6})$ for the $\decay{\Bu}{\Kp e^\pm\mu^\mp}$ and $\decay{\Bd}{\Kstarz\tau^\pm\mu^\mp}$ decays respectively, at the 90\% confidence level.
The limit for $\decay{\Bu}{\Kp\tau^\pm\mu^\mp}$ is expected to be similar to $\decay{\Bd}{\Kstarz\tau^\pm\mu^\mp}$.
The sensitivity of these analyses scales almost linearly with integrated luminosity  for $\decay{\Bu}{\Kp e^\pm\mu^\mp}$, 
and with the square root of the integrated luminosity for $\decay{\Bd}{\Kstarz\tau^\pm\mu^\mp}$. 
In both cases, the expected limits using the \upgradetwo data are in the region of interest of the models currently developed for explaining the $B$ anomalies, so they will provide strong constraints on the New Physics scenarios with charged lepton-flavour violation. 

%% file: CONTRIBUTIONS/7_Rare_and_radiative_decays/7.2.4.tex
\subsection{Search for $\taup \to \mumu\mup$ decays}
    \def\mupm   {{\ensuremath{\mu^\pm}}\xspace}
    \def\mump   {{\ensuremath{\mu^\mp}}\xspace}
    \def\taupm   {{\ensuremath{\tau^\pm}}\xspace}
    \def\taump   {{\ensuremath{\tau^\mp}}\xspace}
    \def\taummm{\decay{\taupm}{\mupm \mu^+ \mu^-}}
    \def\Dsetamunu{\decay{\Ds}{\subdecay{\eta}{\mu^+\mu^-\gamma}\mu^+\nu_{\mu}}}

An important test of the SM is the search for the  lepton-flavour-violating process \taummm. 
Within the SM with zero neutrino masses this process is strictly forbidden. Depending on the mechanism of neutrino mass generation, 
many theories~\cite{Babu:2002et,Brignole:2003iv,Paradisi:2005fk,Hays:2017ekz} beyond the SM predict this branching ratio to be in the region $(10^{-9}-10^{-8})$. 
The current experimental limit~\cite{LHCb-PAPER-2014-052,Hayasaka:2010np,Lees:2010ez}, $\cal{B} (\taummm) <$ $1.2 \times 10^{-8}$, 
is a combination of the results from LHCb and the \bfactories and reaches the starting point of this range. 
\belletwo will probe this interesting region of sensitivity when accumulating up to $50~\rm{ab}^{-1}$. 
The LHC proton collisions at 13 TeV produces $\tau$ leptons, primarily in the decay of heavy flavour hadrons, with a cross-section five orders of magnitude 
larger than at \belletwo. This compensates for the higher background levels and lower integrated luminosity,
and means that during the \upgradetwo LHCb would also be able to probe down to ${\cal O}(10^{-9})$, and independently confirm any \belletwo discovery or 
significantly improve the combined limit. In addition, the proposed improvements to the LHCb calorimeter during the \upgradetwo  will be helpful in suppressing 
backgrounds such as \Dsetamunu, enabling LHCb to make best use of its statistical power.

%% file: CONTRIBUTIONS/7_Rare_and_radiative_decays/7.2.5.tex
\subsection{Search for lepton-number and baryon-number violating decays}

%
%

In the Standard Model, baryon and lepton number conservation are
accidental, low-temperature symmetries that can be violated through the
Bell-Jackiw (chiral) anomaly~\cite{tHooft:1976rip}.
Despite this, neither baryon number violation (BNV) nor lepton number violation (LNV) has been
observed experimentally, with BNV in particular being constrained by
stringent lower limits
on the mean lifetimes of protons and of bound neutrons~\cite{PDG2018}.
Physics beyond the SM may introduce additional sources of LNV and BNV,
so that searches for these effects are sensitive to new physics and can
constrain---or potentially reveal---its presence.

Experimentally, since these are null searches, sensitivity is assumed
to scale linearly with integrated luminosity ${\cal L}$ when the background
is negligible 
and as $\sqrt{{\cal L}}$ if the background is significant. LHCb has already published searches in certain channels, and others are
in progress:
\begin{itemize}
  \item Searches for LNV in various $B$-meson decays of the form
    $B \to X \mu^+ \mu^+$, where $X$ is a system of one or more hadrons.
    The principal motivation is the sensitivity to contributions from Majorana neutrinos~\cite{Atre:2009rg},
    which may be on-shell or off-shell, depending on the decay mode.
    The published results consist of
      searches for $\decay{\Bp}{\Km \mu^+ \mu^+}$, $\decay{\Bp}{\pim \mu^+ \mu^+}$ and $\decay{\Bu}{D^+_{(s)}\mun\mun}$~\cite{LHCb-PAPER-2011-009, LHCb-PAPER-2011-038, LHCb-PAPER-2013-064}.
      A limit of 
    $\mathcal{B}(B^+ \to \pi^- \mu^+ \mu^+) < 4 \times 10^{-9}$ is set at the 95\% confidence level,
    along with more detailed limits as a function of the Majorana neutrino mass.
    Since the combinatorial background was found to be low but not negligible with the Run~1 data,
    we estimate that the limit can be improved by a factor of ten with the full \upgradetwo dataset.
  \item Search for BNV in \Xibz oscillations~\cite{LHCb-PAPER-2017-023}.
    Six-fermion, flavour-diagonal operators, involving two fermions from each
    generation, could give rise to BNV/LNV without violating the nucleon stability
    limit~\cite{Smith:2011rp,Durieux:2012gj}.
    Since the \Xibz~($bsd$) baryon has one valence quark from each generation, it could
    couple directly to such an operator and oscillate to a \Xibzbar baryon.
    The published search used the Run1 data and set a lower limit on the oscillation
    period of 80\ps. Since events are tagged by decays of
    the \XibPrimeMinus and \XibStarMinus resonances, with the former being particularly
    clean, and since the analysis also uses the decay-time distribution of events,
    the sensitivity is expected to scale linearly. 
    Although the decay mode used in the published analysis is hadronic
    ($\Xibz \to \Xicp \pim$), future work could also benefit from the
    lower-purity but higher-yield semileptonic mode $\Xibz \to \Xicp \mun \neumb$.
  \item $\Lc \to \antiproton \mu^+ \mu^+$.
    This channel has previously been investigated at the $e^+ e^-$ \bfactories.
    The current upper limit, obtained by \babar~\cite{Lees:2011hb}, is
    $\mathcal{B}(\Lc \to \antiproton \mup \mup) < 9.4 \times 10^{-6}$
    at the 90\% confidence level.
    With Run~1 and~2 data alone, it should be possible to reduce this
    to $1 \times 10^{-6}$. Further progress depends on the background
    level, but an additional factor of 5--10 with the full \upgradetwo statistics
    is likely.
  \item $\Lc \to \mu^+ \mu^- \mu^+$.
    Experimentally, this is a particularly promising decay mode:
    the final state with three muons is very clean, and there are
    no known sources of peaking background.
    This search could be added for little extra effort to the
    $\taum \to \mup \mun \mun$ search described in the
    preceding section.
\end{itemize}

The discussion in this section is certainly not exhaustive; the list of possible few-body decay
modes that violate BNV or LNV is long. In order to make best use of the LHCb
data, it will be important to have input from theory to identify those modes
that could best evade the bounds from nucleon stability.

%% file: CONTRIBUTIONS/7_Rare_and_radiative_decays/7.3.tex
\section{Flavour-changing $b \to s\ellell$ and $b \to d\ellell$ transitions}
\label{sec:penguins}

\input{CONTRIBUTIONS/7_Rare_and_radiative_decays/7.3.1.tex}
\input{CONTRIBUTIONS/7_Rare_and_radiative_decays/7.3.2.tex}
\input{CONTRIBUTIONS/7_Rare_and_radiative_decays/7.3.3.tex}

\input{CONTRIBUTIONS/7_Rare_and_radiative_decays/7.3.4.tex}
\input{CONTRIBUTIONS/7_Rare_and_radiative_decays/7.3.5.tex}

\input{CONTRIBUTIONS/7_Rare_and_radiative_decays/7.3.6.tex}

%% file: CONTRIBUTIONS/7_Rare_and_radiative_decays/7.3.1.tex
\subsection{Introduction} 
\label{sec:penguins:introduction}

Recent LHCb measurements of rare semileptonic decays show discrepancies with
respect to SM predictions. 
None of these deviations is by itself significant enough to be considered as evidence for NP but global fits~\cite{Capdevila:2017bsm,Altmannshofer:2017yso,Neshatpour:2017qvi} show that they can be interpreted in a consistent picture, with an ${\cal O}(1)$ NP contribution to the vector (and potentially axial-vector) coupling strength of the decays. 
Regardless of whether these discrepancies are confirmed with additional data, the fact that ${\cal O}(1)$ NP contributions are still allowed  demonstrates the importance of making precise measurements of $\bquark\to\squark \ellell$ and $\bquark\to\dquark \ellell$ processes. 
The NP contribution can be associated with new particles at mass scales well above the LHC energy reach, \eg by a multi-\tev-scale $Z'$ boson or a leptoquark.
A precise determination of the effective couplings, through measurements of different $\bquark\to\squark \ellell$ and  $\bquark\to\dquark \ellell$ processes, is therefore critical to understand or constrain the structure of any NP model. 

In the rest of this section, a number of benchmark NP scenarios
are considered (see Table~\ref{tab:penguins:NPscenarios}). 
Scenarios~I and II are inspired by the current discrepancies. 
The first scenario is that which best explains the present rare semileptonic decay data. 
The second scenario best explains the rare semileptonic measurements if a purely left-handed coupling to
quarks and leptons is required for NP. 
This requirement is theoretically well motivated and arises in models designed to simultaneously explain the discrepancies seen in both tree-level semitauonic and loop-level semileptonic decays.
The third and fourth scenarios assume that the current discrepancies are not confirmed but there is instead a small contribution from right-handed currents that would not be visible with the current level of experimental precision. 
These scenarios will serve to illustrate the power of the large \upgradetwo data set to distinguish between different NP models.
This power relies critically on the ability to exploit multiple related decay channels.

\begin{table}[!tb]
   \centering
\caption{
    Benchmark NP scenarios. The first scenario is inspired by the present discrepancies in the
    rare decays, including the angular distributions of the decay \decay{\Bz}{\Kstarz\mumu} and the measurements of the branching fraction ratios $R_K$ and $R_{K^*}$. 
    The second
    scenario is inspired by the possibility of explaining the
    rare decays discrepancies and those measured in the observables
    $R(D^{(*)})$. The third and fourth scenarios assume a small 
    right-handed chirality coupling. The Wilson coefficients ($C_i$) are discussed in Sec.~\ref{sec:penguins:framework}} 
            \label{tab:penguins:NPscenarios}
   \begin{tabular}{crrrr}
       \hline
            scenario & $C_9^{\rm NP}$ & $C_{10}^{\rm NP}$ & $C_{9}^{\prime}$ & $C_{10}^{\prime}$ \\
       \hline
       I & $-1.4$ & 0 & 0 & 0 \\
       II & $-0.7$ & 0.7 & 0 & 0\\
       III & 0 & 0& 0.3 & 0.3\\
       IV & 0 & 0& 0.3 & $-0.3$\\
       \hline
   \end{tabular}
\end{table}

%% file: CONTRIBUTIONS/7_Rare_and_radiative_decays/7.3.2.tex
\subsection{Theoretical framework}
\label{sec:penguins:framework}

\begin{figure}[tb!]
\centering
\includegraphics[width=0.76\linewidth]{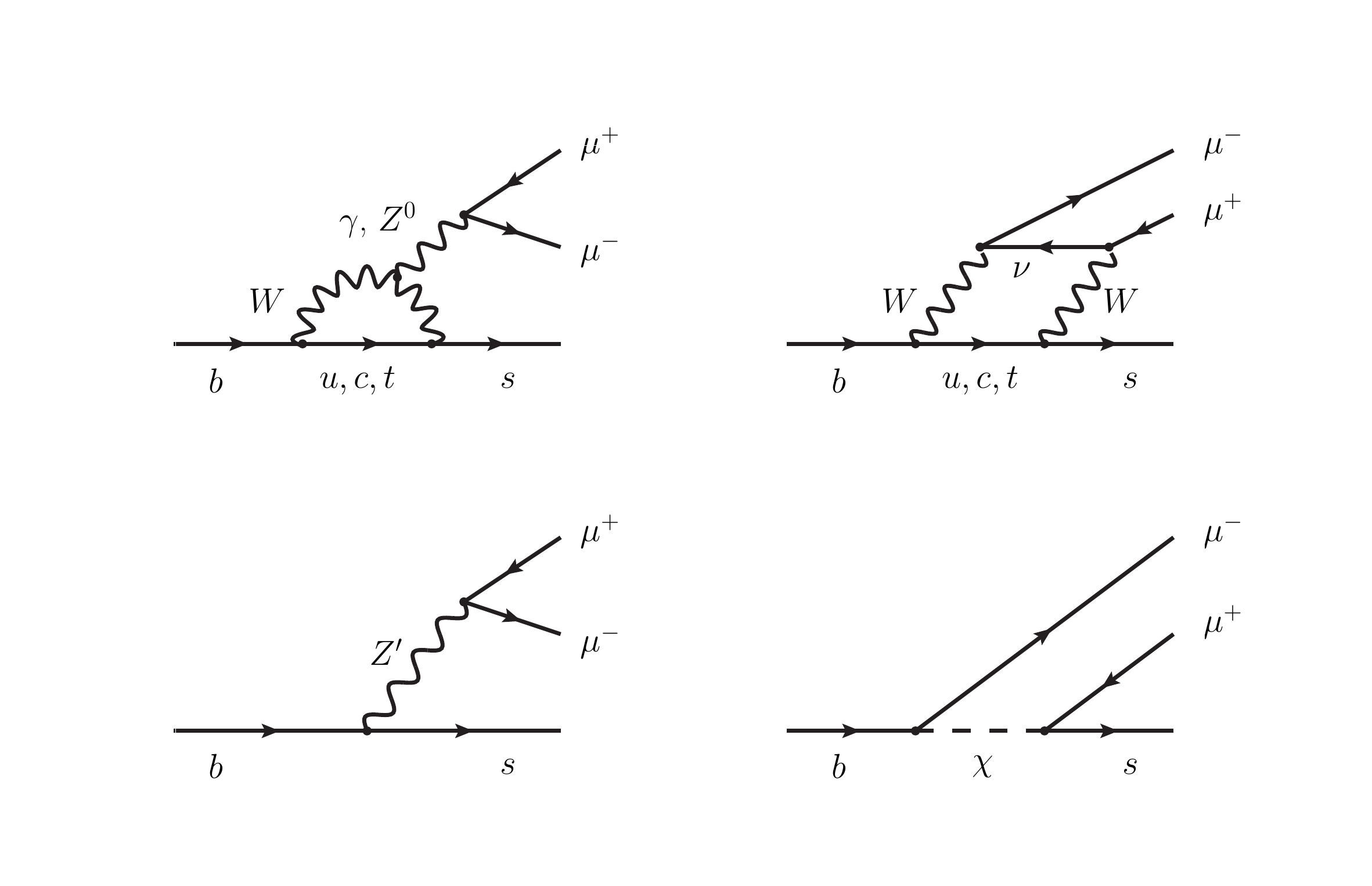}
\caption{
  Feynman diagrams representing (top) SM contributions to
  $b\to s \mumu$ transitions and  (bottom) possible contributions in 
  NP models. The bottom left diagram shows the contribution of a
  $Z^{\prime}$ boson, while the bottom right shows that of a leptoquark.
  }
  \label{fig:penguins:diagrams}
\end{figure}

Flavour-changing neutral-current decays involving $b\to s \ellell$ and $b\to d \ellell$  transitions are suppressed by the GIM
mechanism in the SM and are therefore promising places to search for effects of NP. 
New particles that arise in extensions of the SM can contribute  to the amplitude of these decays with a similar strength to the SM processes.
Feynman diagrams for SM and possible NP extensions are shown in Fig.~\ref{fig:penguins:diagrams}. 

The $\bquark \to \squark$ and $\bquark\to\dquark$ FCNC processes are described by the effective Hamiltonian,
\begin{equation}
{\cal H}_{\rm eff} =-\frac{4G_F}{\sqrt{2}}V_{tb}V_{tq}^*\left[
    \frac{\alpha_e}{4\pi}\sum_i {\cal C}_i(\mu) {\cal
      O}_i(\mu) \right], 
\end{equation}
where $G_F$ is the Fermi constant, $V_{ij}$ are elements of the CKM matrix and $\alpha_e$ is the fine structure constant~\cite{Buchalla:1995vs}.
The $C_{i}$ coefficients are the so-called Wilson coefficients that encode short-distance contributions to the decay amplitude at scales above $\mu$. 
The Wilson coefficients are universal and apply to both the $\bquark \to \squark$ and $\bquark\to\dquark$ processes.
Finally, the ${\cal O}_{i}$ are local operators with different Lorentz structures.
The most important operators for $\bquark \to \squark \ellell$ and $\bquark \to \dquark \ellell$ decays are
\begin{align}
\begin{split}
{\cal O}_{7}^{(\prime)} &= \frac{m_b}{e}\left(  \overline{s}  \sigma^{\mu\nu} P_{\rm R(L)} b \right) F_{\mu\nu}~,  \\
{\cal O}_{9}^{(\prime)} &= \left( \overline{s}P_{\rm L(R)}b \right)\left(\overline{\ell}\gamma^{\mu}\ell\right)~, \\
{\cal O}_{10}^{(\prime)} &= \left( \overline{s} P_{\rm L(R)}b\right)\left( \overline{\ell} \gamma^{\mu}\gamma^5\ell\right)~,
\end{split}
\end{align}
where $m_b$ is the \bquark-quark mass and $P_{\rm L, R}$ are the left- and right-handed projection operators. 
The coefficients $C_9$ and $C_{10}$ correspond to the coupling to a vector and an axial-vector leptonic current, respectively. 
The V--A structure of the weak interaction leads to the expectation that $C_9 \approx -C_{10}$ and $C^{\prime}_{9,10} \approx 0$. 
Hadronic matrix elements are parametrised in terms of form
factors that can be determined with non-perturbative methods, such as QCD sum rules or lattice QCD. 
Contributions from new particles can modify the Wilson coefficients, leading to observable effects in the rate and angular distribution of the decays or introducing new sources of \CP violation. 

Different regions of dilepton mass squared (\qsq) and decays with different final-state hadrons provide sensitivity to different combinations of the Wilson coefficients. 
New physics contributions at  energy scales above  $m_b$ can modify the Wilson coefficients from their SM-values, or introduce entirely new Lorentz structures. 
Decomposing possible NP contributions in terms of effective operators with a well defined Lorentz structure allows the model-independent correlations between different observables and decays to be exploited. 
The low-\qsq region of the decay \decay{\Bz}{\Kstarz\mumu} (and other vector meson final-states) is  particularly sensitive to NP effects, due to the large interference effects between  $C_{7}^{(\prime)}$ and $C_{9}^{(\prime)}$.
The large number of different $\bquark\to\squark \ellell$ and $\bquark\to\dquark \ellell$ decay modes that will be accessible in \upgradetwo will improve the sensitivity to the Wilson coefficients and will allow the consistency between different measurements to be tested comprehensively. 

The interpretation of observables in terms of Wilson
coefficients requires knowledge of QCD matrix elements in the non-perturbative regime and these are difficult to calculate. 
For example, long-distance contributions involving intermediate charm-loops have generated much debate about the interpretation of global fits for the Wilson coefficients~\cite{Jager:2014rwa,Ciuchini:2015qxb,Capdevila:2017ert}.  
Fortunately, the correlation with different observables allows theoretical uncertainties to be reduced to a significant extent, 
either by constraining the matrix elements from the data, or by forming ratios where the matrix elements cancel such as in lepton universality tests.

%% file: CONTRIBUTIONS/7_Rare_and_radiative_decays/7.3.3.tex
\subsection{Branching fractions and angular observables in  $b \to s\ellell$ transitions}
\label{sec:penguin:btosll}

The branching fractions of a number of exclusive $\bquark\to\squark \ellell$ processes have been studied using the LHCb Run~1 data set. 
The experimental measurements are systematically lower than their corresponding SM predictions. 
The largest discrepancy appears for $\BF(\decay{\Bs}{\phi\mumu})$ which, in the  $1<q^2<6\gevgevcccc$ region, is more than $3\,\sigma$ from the SM predictions~\cite{LHCb-PAPER-2015-023}.
For both the \upgradeone and \upgradetwo datasets,
the precision of the measurement of the branching fractions will be limited by the knowledge of the \decay{\B}{\jpsi X} decay modes that are used to normalise the observed signals. 
The knowledge of these 
branching fractions will be improved by the \belletwo collaboration but will inevitably limit the precision of the absolute branching fractions of rare $\bquark\to\squark \ellell$ processes. 
The comparison between the predicted and measured branching fractions will in any case be limited by the theoretical knowledge of the form factors, even if in the future these are determined parameterically using the data.
A better comparison between theory and experiment can be achieved by studying isospin and \CP asymmetries, which with the \upgradetwo data set will be experimentally probed with percent level precision. 
The \upgradetwo data set will also enable new decay modes to be studied, for example higher-spin \Kstar states and modes with larger numbers of decay products.  

It is also possible to reduce theoretical and experimental uncertainties by comparing regions of angular phase-space of $\bquark\to\squark \ellell$ decays. 
The angular distribution of \decay{\B}{V \ellell} decays, where $V$ is a vector meson, can be expressed in terms of eight \qsq-dependent angular coefficients that depend on the Wilson coefficients and the form-factors. 
Measurements of angular observables in \decay{\Bz}{\Kstarz\mumu} decays show a
discrepancy with respect to SM predictions~\cite{LHCb-PAPER-2013-037,LHCb-PAPER-2015-051,LHCb-PAPER-2014-006,LHCb-PAPER-2013-019,Aubert:2006vb,Lees:2015ymt,Wei:2009zv,Aaltonen:2011ja,Chatrchyan:2013cda,Khachatryan:2015isa,Aaboud:2018krd,Sirunyan:2017dhj,Beneke:2004dp,Egede:2015kha,Kruger:2005ep,Lyon:2014hpa,Hurth:2013ssa,Beaujean:2013soa,Altmannshofer:2013foa,Descotes-Genon:2013wba}. 
This discrepancy is largest in the form-factor-independent observable $P'_{5}$~\cite{LHCb-PAPER-2015-051}. 
The decay \decay{\Bs}{\phi\mumu} can also be described by the same angular formalism as the \decay{\Bz}{\Kstarz\mumu} decay.
However, in this case the \Bs and the \Bsb mesons decay to a common final state and it is not possible to determine the full set of observables   without tagging the initial flavour of the \Bs. 
This is discussed further in Sec.~\ref{sec:penguin:timedependent}. 

\begin{figure}[!tb]
\centering
\includegraphics[width=0.45\linewidth]{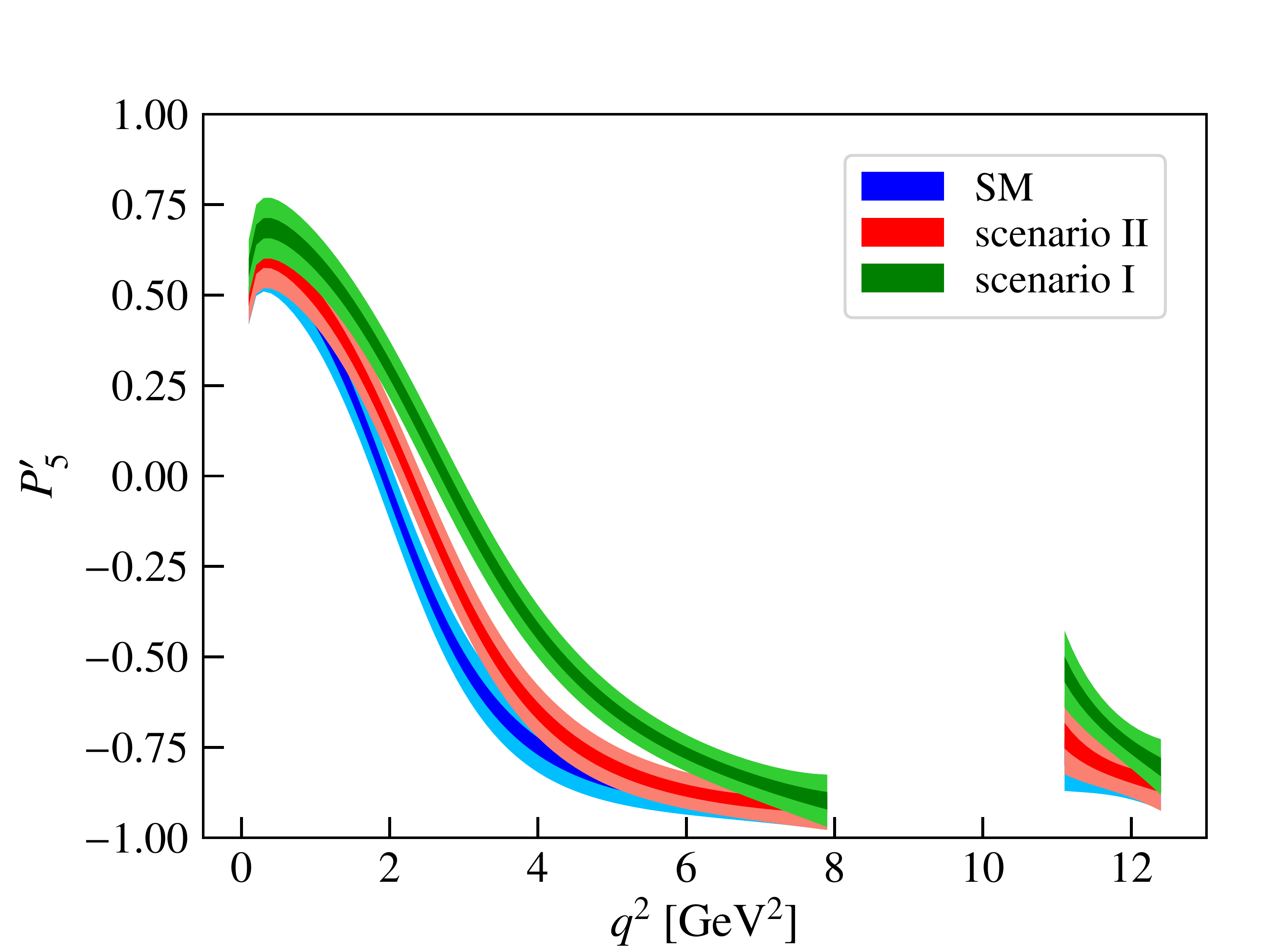}
\includegraphics[width=0.45\linewidth]{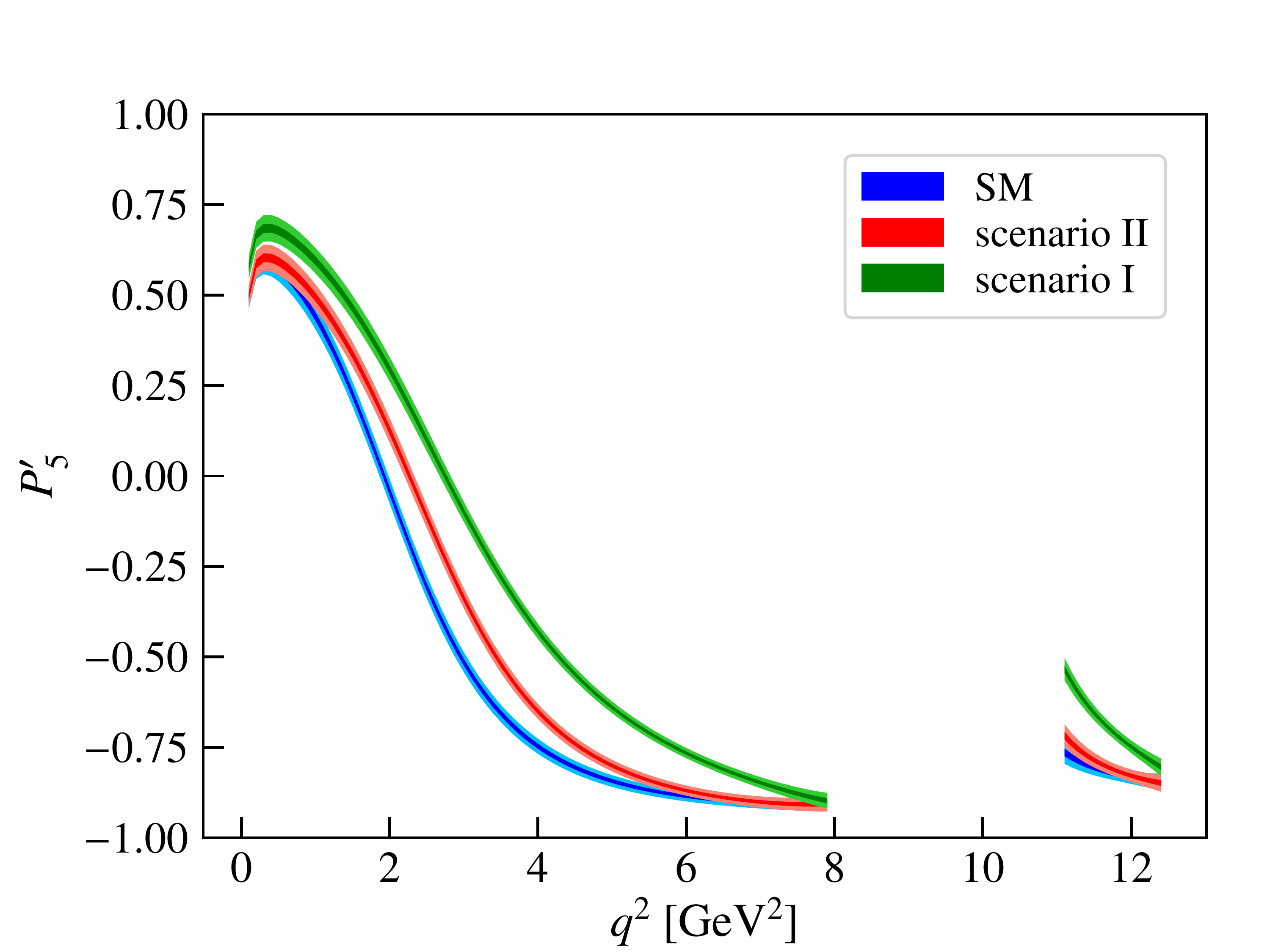}
\caption{Experimental sensitivity to the $P_5^{\prime}$ angular observable in the SM,
  Scenarios~I and II for (left) the Runs~1--3 and (right) the \upgradetwo data sets. 
  The sensitivity is computed assuming that the charm-loop contribution is determined from the data.
}
\label{fig:penguin:P5primeUnbinned}
\end{figure}

\begin{figure}[!tb]
\centering
\includegraphics[width=0.4\linewidth]{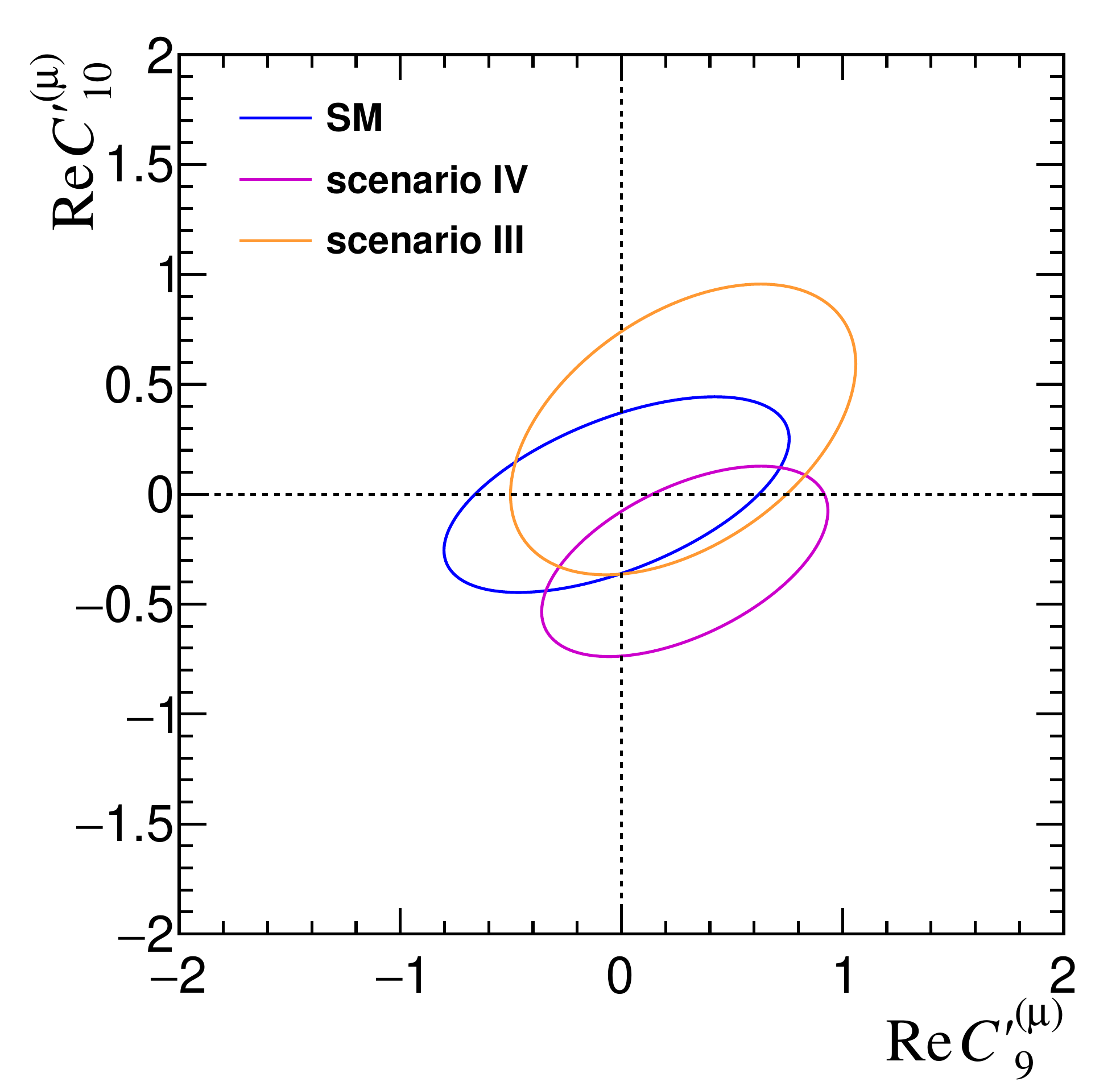}
\includegraphics[width=0.4\linewidth]{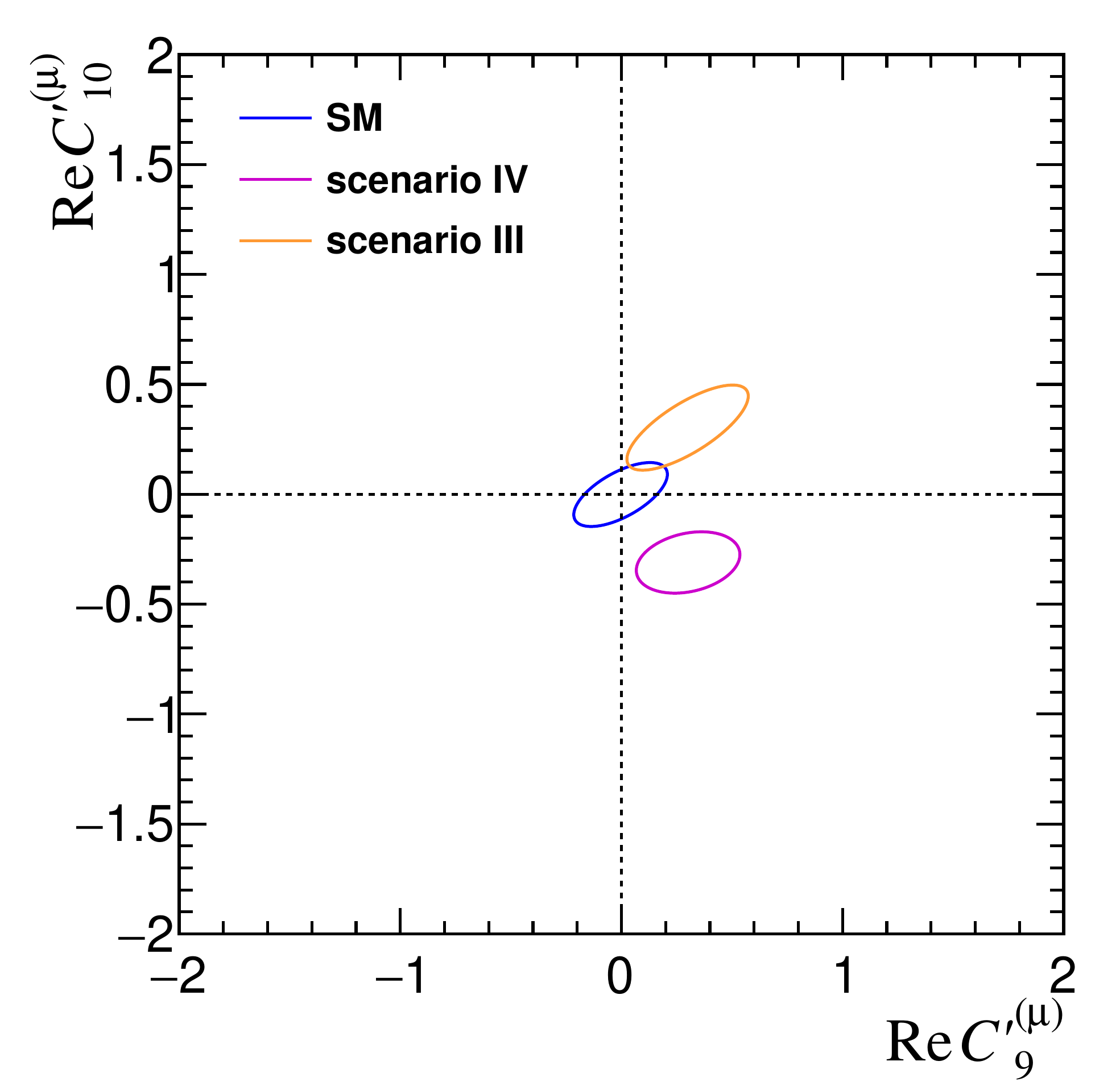}
\caption{Expected sensitivity for the Wilson coefficients $C_9^{\prime}$ and
  $C_{10}^{\prime}$ from the analysis of the decay \decay{\Bz}{\Kstarz\mumu}. 
  The ellipses correspond to 3\,$\sigma$ contours for
  the SM, Scenario~III and Scenario~IV for (left) the Runs~1--3 and (right) the \upgradetwo data sets. 
  }
  \label{fig:penguin:WCKstmm}
\end{figure}

With the large data set that will be collected with \upgradetwo,  
corresponding to around 440\,000 fully reconstructed \decay{\Bz}{\Kstarz\mumu} decays, it will
be possible to make a precise determination of the angular observables in narrow
bins of \qsq or using a \qsq-unbinned approach~\cite{Hurth:2017sqw,Chrzaszcz:2018yza}. 
The expected precision of an unbinned determination of $P_5^{\prime}$ in the SM and in Scenarios~I and II is illustrated in Fig.~\ref{fig:penguin:P5primeUnbinned}. 
\upgradetwo will enable these scenarios to be clearly separated from the SM and from each other.
By combining information from all of the angular observables in the decay, it will also be possible to distinguish models with much smaller NP contributions. 
Figure~\ref{fig:penguin:WCKstmm} shows the expected $3\,\sigma$ sensitivity for the Wilson
coefficients for $C_{9,10}^{\prime}$ for the SM, Scenario~III and Scenario~IV.  
These scenarios are also clearly distinguishable with the precision that will be available with the \upgradetwo data set.

The major challenge for \decay{\B}{V \ellell} decays is to disentangle NP effects from
SM contributions. 
With a large data set it will be possible to probe the SM contributions, under the premise that a genuine NP contribution is expected to have no $q^2$ dependence, while \eg a 
charm loop contribution is expected to grow approaching the pole of
the charmonia resonances.  
A measurement using Breit-Wigner functions to parametrise the
resonances, and their interference with the
short-distance contributions to the decay, is proposed in Ref.~\cite{Blake:2017fyh}. 
A similar technique has already been applied to the Run~1 data for the \decay{\Bp}{\Kp\mumu} decay~\cite{LHCb-PAPER-2016-045}. 
An alternative approach using additional phenomenological inputs has also been proposed~\cite{Bobeth:2017vxj}.
A precise knowledge of the charm loop contribution and a parametric determination of the form factors,
 will come from a combination of phenomenological and experimental
methods and will allow $C_9$ and
$C_{10}$ to be determined with great precision in $\bquark\to\squark\mumu$ transitions.

%% file: CONTRIBUTIONS/7_Rare_and_radiative_decays/7.3.4.tex
\subsection{Branching fractions and angular observables in $b \to d\ellell$ transitions}

The \upgradetwo data set will provide a unique opportunity to make precise measurements of $\bquark\to\dquark \ellell$ processes. 
Using the Run~1 and 2 data sets, LHCb data have been used to observe the decays \decay{\Bp}{\pip\mumu}~\cite{LHCb-PAPER-2012-020,LHCb-PAPER-2015-035} and \decay{\Lb}{p\pim\mumu}~\cite{LHCb-PAPER-2016-049} and find evidence for the decays \decay{\Bz}{\pip\pim\mumu} (in a $\pip\pim$ mass region that  is expected to be dominated by \decay{\Bz}{\rhoz\mumu}) and \decay{\Bs}{\Kstarzb\mumu}~\cite{LHCb-PAPER-2018-004} with branching fractions at the $\mathcal{O}(10^{-8})$ level.
The existing data samples comprise $\mathcal{O}(10)$ decays in these decay modes. 
The \upgradetwo will provide samples of thousands, or tens of thousands of such decays.
The ability to measure the properties of these processes depends heavily on the PID performance of the LHCb subdetectors. 
In the case of the \decay{\Bs}{\Kstarzb\mumu} decay, excellent mass resolution is also critical to separate \Bs and \Bz decays. 

The ratio of branching fractions between the CKM-suppressed $b \to d\ellell$ transitions and their CKM-favoured  $\bquark \to \squark \ellell$ counterparts, together with theoretical input on the ratio of the relevant form factors, enables the ratio of CKM elements  $|V_{td}|/|V_{ts}|$ to be determined. The precision on $|V_{td}|/|V_{ts}|$ from such decays is presently dominated by the statistical uncertainty on the experimental measurements of \decay{\Bp}{\pip\mumu}, and is significantly worse than the determination from mixing measurements. The theoretical uncertainty is at the level of 4\% and is expected to improve with further progress on the form-factors from lattice QCD. Around 17\,000 \decay{\Bp}{\pip\mumu} decays are expected in the full 300\invfb dataset, allowing an experimental precision better than 2\%.

The current set of measurements of $\bquark\to\squark \ellell$ processes have demonstrated the importance of angular measurements in the precision determination of Wilson coefficients. With the \upgradetwo dataset, where a sample of 4300  \decay{\Bs}{\Kstarzb\mumu} decays is expected,  it will be possible to make a full angular analysis of a $\bquark\to\dquark \ellell$ transition. The \decay{\Bs}{\Kstarzb\mumu} decay is both self-tagging and has a final state involving only charged particles. The \upgradetwo data set will allow the angular observables in this decay to be measured with better precision than the existing measurements of the \decay{\Bz}{\Kstarz\mumu} angular distribution. 

The \upgradetwo dataset will also give substantial numbers of \decay{\B^{0,+}}{\rho^{0,+}\mumu}  and \decay{\Lb}{N\mumu} decays. Although the  \decay{\Bz}{\rhoz\mumu} decay does not give the flavour of the initial \B meson, untagged measurements will give sensitivity to a subset of the interesting angular observables. 
Analysis of the \decay{\Lb}{N\mumu} decay will require statistical separation of  overlapping \decay{N}{p\pim} resonances with different $J^P$ by performing an amplitude analysis of the final-state particles. 

The combination of information from $\BF(\decay{\Bz}{\mumu})$, the differential branching fraction of the \decay{\Bp}{\pip\mumu} decay, and angular measurements, notably of \decay{\Bs}{\Kstarzb\mumu}, will indicate whether NP effects are present in $\bquark\to\dquark$ transitions at the level of 20\% of the SM amplitude with more than $5\sigma$ significance. 

%% file: CONTRIBUTIONS/7_Rare_and_radiative_decays/7.3.5.tex
\subsection{Lepton-flavour universality tests}
\label{sec:penguin:LFUtest}

In the SM, the electroweak couplings of leptons are flavour independent, or lepton ``universal''.
The ratios of branching fractions measured with different lepton families are therefore free from hadronic corrections and such quantities can be precisely predicted.
For example, the ratios 
\begin{align}
R_{X} = \int \frac{ \deriv\Gamma(\decay{\B}{X\mumu}) }{\deriv\qsq} \, \deriv\qsq \Bigg/ \int \frac{ \deriv\Gamma(\decay{\B}{X\epem}) }{\deriv\qsq} \, \deriv\qsq \, ,
\end{align}
between \B decays to final states with muons and electrons, where $X$ is a hadron containing a \squark or a \dquark quark, are predicted to be very close to unity in the SM~\cite{Hiller:2003js,Bobeth:2007dw,Bouchard:2013mia}.
The uncertainties from QED corrections are found to be at the percent level~\cite{Bordone:2016gaq}.

\begin{figure}[!t] 
\centering
\includegraphics[width=0.75\textwidth]{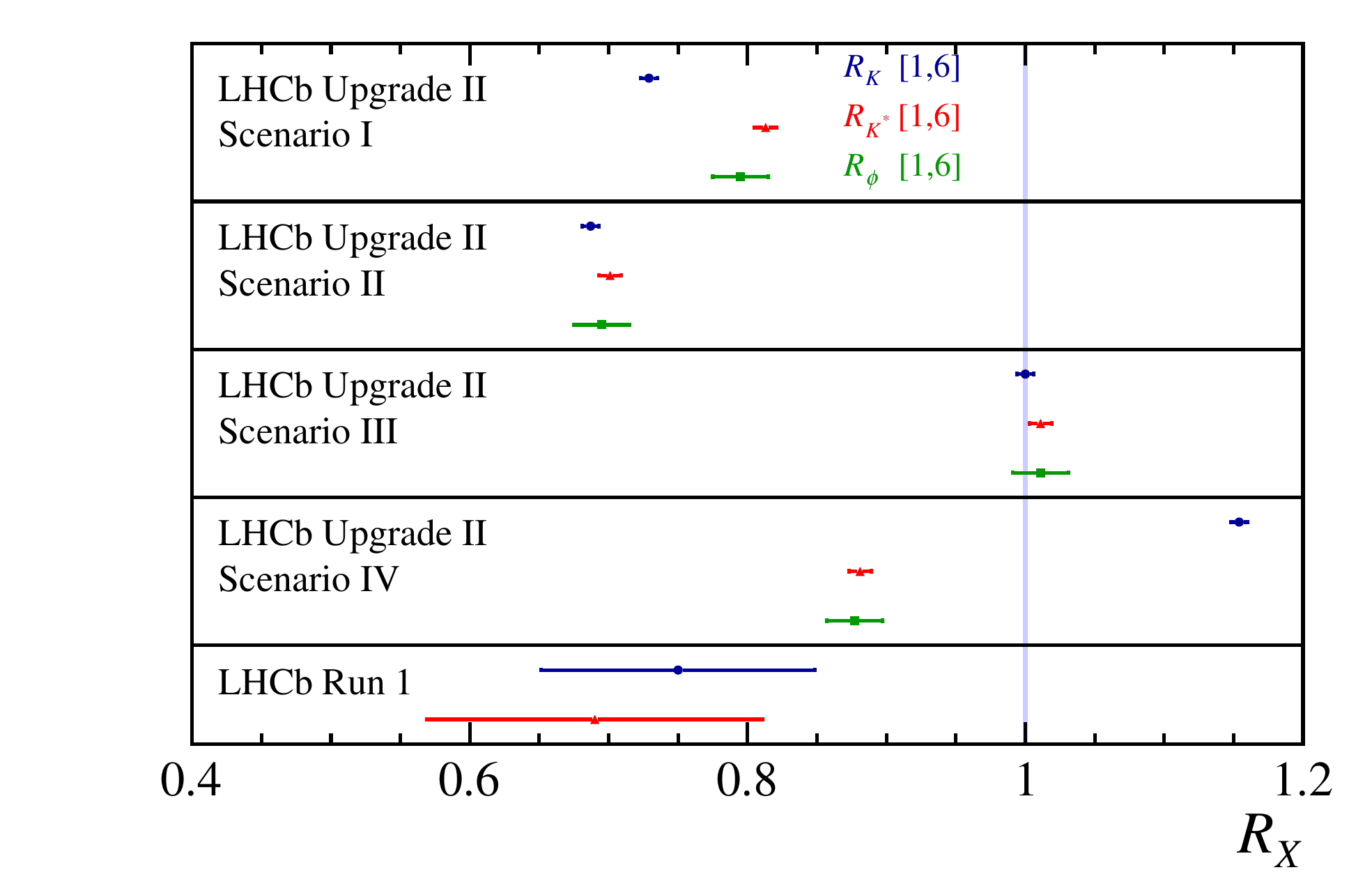}
\caption{
Projected sensitivity for the $R_{K}$, $R_{K^*}$ and $R_{\phi}$ measurements in different NP scenarios with the Upgrade~II data set. 
The existing Run~1 measurements of $R_{K}$ and $R_{K^*}$ are shown for comparison. 
}
\label{fig:penguin:RXscenarios}
\end{figure}

The Run~1 \lhcb data have been used to perform the most precise measurements of $R_{K}$ and $R_{\Kstar}$ to-date~\cite{LHCb-PAPER-2017-013,LHCb-PAPER-2014-024} (see Fig.~\ref{fig:penguin:RXscenarios}).
The measurements show some tension with the SM, all deviating from predictions at the level of 2.1--2.6 standard deviations. 
Assuming the current detector performance, approximately 46\,000 \decay{\Bp}{\Kp\epem} and 20\,000 \decay{\Bz}{\Kstarz\epem} candidates are expected in the range $1.1 < \qsq < 6.0\gevgevcccc$ in the \upgradetwo data set.
The ultimate precision on $R_{K}$ and $R_{\Kstar}$ will be better than 1\%.
The importance of the \upgradetwo data set in distinguishing between different NP scenarios is highlighted in Fig.~\ref{fig:penguin:RXscenarios}. 
With this data set all four NP scenarios could be distinguished at more than 5\,$\sigma$ significance.

The \upgradetwo data set will also enable the measurement of other $R_{X}$ ratios \eg $R_{\phi}, R_{p\kaon}$ and the ratios in CKM suppressed decays. 
For example, with 300\invfb, it will be possible to determine $R_{\pi}=\BR(\decay{\Bp}{\pip\mumu})/\BR(\decay{\Bp}{\pip\epem})$ with a few percent statistical precision.
A summary of the expected performance for a number of different $R_X$ ratios is indicated in Table~\ref{tab:penguin:LU_extrapolations}.

\begin{table}[!tb]
\caption{
Estimated yields of $\bquark\to\squark\ep\en$ and $\bquark\to\dquark\ep\en$ processes and the statistical uncertainty on $R_{X}$ in the range $1.1<\qsq<6.0\gevgevcccc$ extrapolated from the Run~1 data. 
A linear dependence of the \bbbar production cross section on the $pp$ centre-of-mass energy and unchanged Run~1 detector performance are assumed. 
Where modes have yet to be observed, a scaled estimate from the corresponding muon mode is used.
}
\label{tab:penguin:LU_extrapolations}
\centering
\begin{tabular}{lrrrrr}
\hline
Yield  & Run~1 result & 9\invfb & 23\invfb & 50\invfb & 300\invfb \\
\hline
 \decay{\Bu}{\Kp\epem} 		& $254 \pm 29$\cite{LHCb-PAPER-2014-024} & 1\,120 & 3\,300 & 7\,500 & 46\,000 \\
 \decay{\Bd}{\Kstarz\epem}	& $111 \pm 14$\cite{LHCb-PAPER-2017-013} & 490 & 1\,400 & 3\,300 & 20\,000 \\
 \decay{\Bs}{\phi\epem}		& -- & 80 & 230 & 530 & 3\,300 \\
 \decay{\Lb}{\proton\kaon\epem} & -- & 120 & 360 & 820 & 5\,000 \\
\decay{\Bp}{\pip\epem}		& -- & 20 & 70 & 150 & 900 \\ 
\hline
$R_X$ precision &  Run~1 result & 9\invfb & 23\invfb & 50\invfb & 300\invfb \\
\hline
$R_{\kaon}$					& $0.745 \pm 0.090\pm 0.036$\cite{LHCb-PAPER-2014-024} & 0.043 & 0.025 & 0.017 & 0.007 \\
$R_{\Kstarz}$					& $0.69 \pm 0.11\pm 0.05$\cite{LHCb-PAPER-2017-013} & 0.052 & 0.031 & 0.020 & 0.008 \\
$R_{\phi}$					& -- & 0.130 & 0.076 & 0.050 & 0.020 \\
$R_{\proton\kaon}$  & -- & 0.105 & 0.061 & 0.041 & 0.016 \\
$R_{\pi}$						& -- & 0.302 & 0.176 & 0.117 & 0.047 \\ 
\hline
\end{tabular}
\end{table}

\begin{figure}[t!]
\centering
\includegraphics[width=0.49\textwidth]{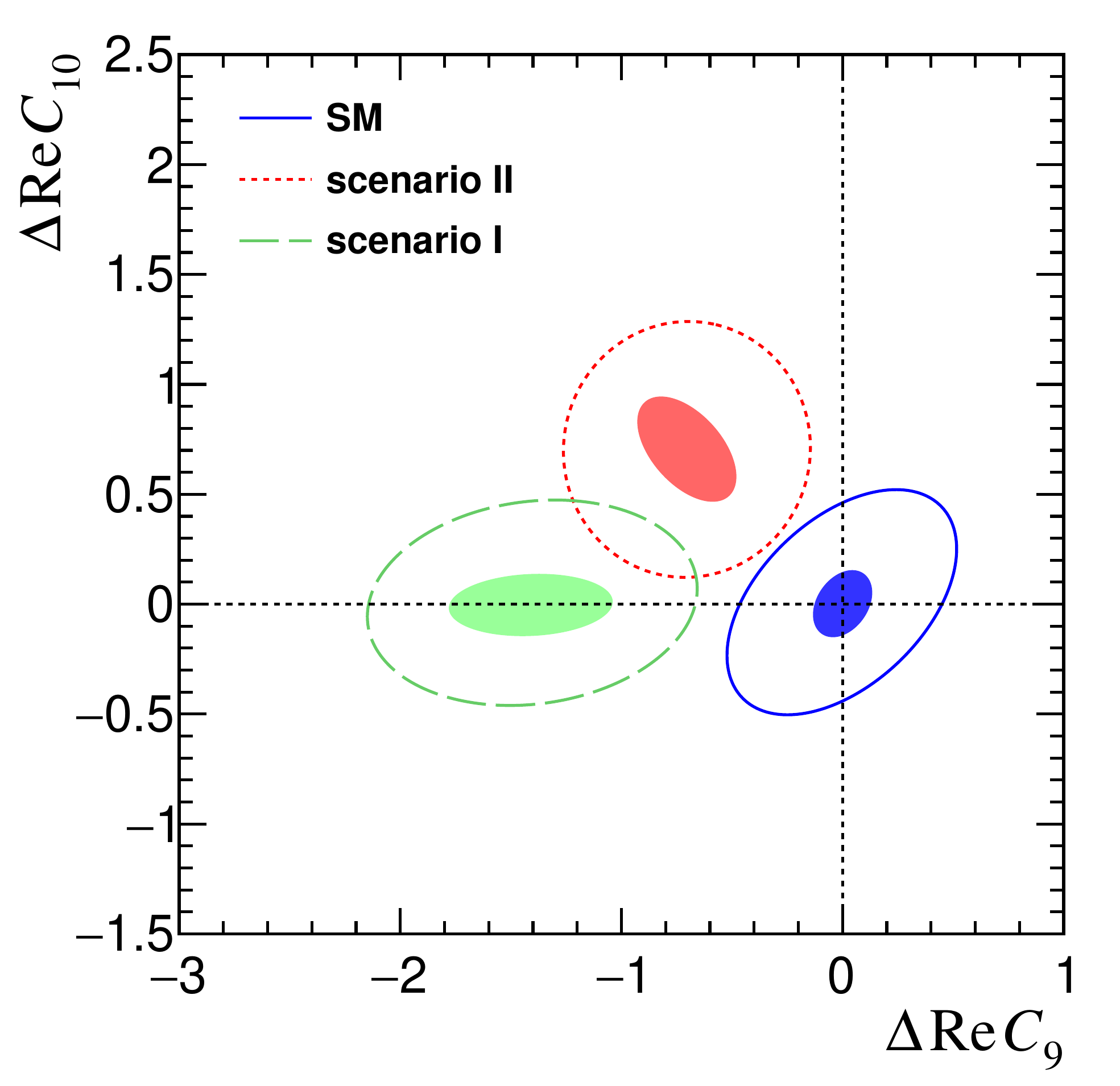}
\caption{
  Constraints on the difference in the $C_9$ and $C_{10}$ Wilson coefficients from electron and muon modes with the Run~3 and \upgradetwo data sets.
  The $3\sigma$ regions for the Run~3 data sample are shown for the SM (solid blue), a vector-axial-vector new physics contribution (red dotted) and for a purely vector new physics contribution (green dashed). 
  The shaded regions denote the corresponding constraints for the \upgradetwo data set. 
}
\label{fig:penguin:DeltaC}
\end{figure}

In addition to improvements in the $R_{X}$ measurements, the enlarged \upgradetwo data set will give access to new observables.
For example, the data will allow precise comparisons of the angular distribution of dielectron and dimuon final-states. 
Differences between angular observables in \decay{\B}{X\mumu} and \decay{\B}{X\ep\en} decays are theoretically pristine~\cite{Capdevila:2016ivx,Serra:2016ivr}
and are sensitive to different combinations of Wilson coefficients compared to the $R_X$ measurements. 
Figure~\ref{fig:penguin:DeltaC} shows that an upgraded \lhcb detector will enable such decays to be used to discriminate between different NP models, for example separating between Scenarios~I and~II~\cite{Mauri:2018vbg}. 
Excellent NP sensitivity can be achieved irrespective of the assumptions made about the hadronic contributions to the decays.

In the present \lhcb detector, electron modes have an approximately factor five lower efficiency than the corresponding muon modes, owing to the tendency for the electrons to lose a significant fraction of their energy through bremsstrahlung in the detector. 
This loss impacts on the ability to reconstruct, trigger and select the electron modes.
The precision with which observables can be extracted therefore depends primarily on the electron modes and not the muon modes.
In order for $R_{X}$ measurements to benefit from the large \upgradetwo data samples, it will be necessary to reduce systematic uncertainties to the percent level.
These uncertainties can be controlled by taking a double ratio between $R_{X}$ and the decays \decay{\B}{\jpsi X}, where the \jpsi decays to \mumu and $\ep\en$.
This approach is expected to work well, even with very large data sets.  

Other sources of systematic uncertainty can be mitigated through design choices for the upgraded detector.
The recovery of bremsstrahlung photons is inhibited by the ability to find the relevant photons in the ECAL (over significant backgrounds) and by the energy resolution.
A reduced amount of material before the magnet would reduce the amount of bremsstrahlung and hence would increase the electron reconstruction efficiency and improve the electron momentum resolution. 
Higher transverse granularity would aid signal selection and help reduce the backgrounds. 
With a large number of primary $pp$ collisions, the combinatorial background will increase and will need to be controlled with the use of timing information.
However, the Run~1 data set indicates that it may be possible to tolerate a significant (\ie\ larger than a factor two) increase in combinatorial backgrounds without destroying the signal selection ability.

%% file: CONTRIBUTIONS/7_Rare_and_radiative_decays/7.3.6.tex
\subsection{Time-dependent analysis of $\Bs \to \phi\mup\mun$ and $\Bz \to \rho^0 \mup\mun$}
\label{sec:penguin:timedependent} 

Time-dependent analyses of rare decays into \CP eigenstates can deliver orthogonal experimental information to time-integrated observables. 
So far, no time-dependent measurement of the $\decay{\Bs}{\phi\mumu}$ decay has been performed due to the limited signal yield of $432\pm 24$ in the Run~1 data sample~\cite{LHCb-PAPER-2015-023}.
However, the larger data samples available in \upgradetwo will enable time-dependent studies. 
The framework describing $\Bbar$ and $\B\to V\ellell$ transitions to a common final-state is discussed in  Ref.~\cite{Descotes-Genon:2015hea}, 
where several observables are discussed that can be accessed with and without flavour tagging.  
Two observables called $s_8$ and $s_9$, which are only accessible through a time-dependent flavour-tagged analysis, are of particular interest. 
These observables  are proportional to the mixing term $\sin\left(\Delta m_s t\right)$ and provide information that is not available through flavour specific decays.
Assuming a time resolution of around $45\fs$ and an effective tagging power of $5\%$ 
results in an effective signal yield of  $2000$ decays for the \upgradetwo data set. 

As a first step towards a full time-dependent analysis, the effective lifetime of the decay $\decay{\Bs}{\phi\mumu}$ can be studied. 
The untagged time-dependent decay rate is given by
\begin{align}
  \frac{\deriv\Gamma}{\deriv t} &\propto  e^{-\Gamma_s}\left[\cosh\left(\frac{\Delta\Gamma_st}{2}\right)+A^{\Delta\Gamma}\sinh\left(\frac{\Delta\Gamma_st}{2}\right)\right].
\end{align}
The observable $A^{\Delta\Gamma}$ can be related to the angular observables $F_{\rm L}$ and $S_3$ via $A^{\Delta\Gamma}=2S_3-F_{\rm L}$. 
Due to the significant lifetime difference $\Delta\Gamma_s$ in the \Bs\ system, even an untagged analysis can probe right-handed currents. 
For the combined low- and high-\qsq regions, preliminary studies suggest a statistical sensitivity to $A^{\Delta\Gamma}$ of $0.05$ can be achieved with a $300\invfb$ data set. 

With the \upgradetwo data set it will also be possible to perform a time-dependent angular analysis of the $\bquark \to \dquark$ process \decay{\Bd}{\rho^0\mumu}. 
This process differs from \decay{\Bs}{\phi\mumu} in two important regards: it is CKM suppressed and therefore has a smaller SM branching fraction; and $\Delta \Gamma_d \approx 0$, removing sensitivity to $A^{\Delta\Gamma}$. 
The uncertainties on the angular observables are expected to be on the order of $0.1$ for this case. 

The time-dependent angular analyses will still be statistically limited even with $300\invfb$. 
It will be important to maintain good decay-time resolution and the performance of the particle identification will be crucial to control backgrounds, as well as to improve flavour tagging performance.

%% file: CONTRIBUTIONS/7_Rare_and_radiative_decays/7.4.tex
\section{Radiative beauty and charm decays}
\input{CONTRIBUTIONS/7_Rare_and_radiative_decays/7.4.0.tex}

\input{CONTRIBUTIONS/7_Rare_and_radiative_decays/7.4.1.tex}
\input{CONTRIBUTIONS/7_Rare_and_radiative_decays/7.4.2.tex}
\input{CONTRIBUTIONS/7_Rare_and_radiative_decays/7.4.3.tex}
\input{CONTRIBUTIONS/7_Rare_and_radiative_decays/7.4.4.tex}

%% file: CONTRIBUTIONS/7_Rare_and_radiative_decays/7.4.0.tex
In the SM, $W^\pm$ bosons couple only left-handedly to quarks.
As a consequence, photons emitted in FCNC \decay{b}{s\gamma} transitions are predominantly left handed, but several extensions of the SM, see \eg Ref.~\cite{Pati:1974yy}, predict that the photon can acquire a significant right-handed component through a new heavy particle that couples right-handedly to quarks.
Indirect constraints on right-handed currents
leave a large fraction of the space for non-SM right-handed currents unexplored.
The photon polarisation---the asymmetry between the right- and left-handed components of the photon---as well as measurements of \CP violation or the ratio between \decay{b}{s\gamma} and \decay{b}{d\gamma} decays can be used to constrain these right-handed currents.

Experimentally, the study of these transitions at \lhcb is dominated by the limitations of the electromagnetic calorimeter, which result in (a)\,a limited mass resolution that prevents good separation of partially reconstructed backgrounds, (b)\,a limited ability to distinguish signal from charmless backgrounds in which the photon is replaced by a \piz meson decaying into two photons, which are reconstructed as a single cluster in the \ecal, and (c)\,a degraded decay-time resolution coming from the photon momentum resolution. The \upgradetwo detector is expected to improve the mass resolution for \piz mesons thanks to a better spatial segmentation, which would be crucial to match with the expected statistical precision of several of the measurements discussed in this section.

A precise measurement of the photon polarisation has not been performed yet, but several methods to achieve the result have been proposed:

\begin{itemize}
  \item The time-dependent \CP asymmetry of \decay{\BdorBs}{f_{\CP}\gamma}, where $f_{\CP}$ is a \CP eigenstate, discussed in Sec.~\ref{sec:7.4.1}.
  \item The angular correlations among the three-body decay products of a kaonic resonance in \decay{\Bp}{K_{\mathrm{res}}^+(\to \Kp\pim\pip)\gamma}, presented in Sec.~\ref{sec:7.4.1}.
  \item The transverse asymmetries of \decay{\Bz}{\Kstarz e^+ e^-}, highly sensitive to the \decay{b}{s\gamma} process at low dielectron invariant mass squared ($q^2$), and which will be discussed in Sec.~\ref{sec:7.4.2}.
  \item The angular distributions of radiative $b$-baryon decays, detailed in Sec.~\ref{sec:7.4.3}.
    
\end{itemize}

Each of these measurements constrains right-handed currents differently~\cite{Paul:2016urs} and they thus give complementary constraints in global fits of Wilson coefficients. 

%% file: CONTRIBUTIONS/7_Rare_and_radiative_decays/7.4.1.tex
\subsection{Radiative \B decays\label{sec:7.4.1}}


The time dependent \CP asymmetry of \decay{\BdorBs}{f_{\CP}\gamma} arises from the interference between decay amplitudes with and without $\Bds-\Bdsb$ mixing and is predicted to be small in the SM\ \cite{Atwood:1997zr,Ball:2006cva,Matsumori:2005ax}.
As a consequence, a large asymmetry due to interference between the \B mixing and decay diagrams can only be present if the two photon helicities contribute to both \B and \Bb decays.
From the time dependent decay rate
\begin{align}
    \Gamma(\decay{\Bds(\Bdsb)}{f_{\CP}\gamma})(t) \sim e^{-\Gamma_s t}
    &\big[
        \cosh\left(\frac{\Delta\Gamma_{(s)}}{2}\right)
        - \mathcal{A}^{\Delta}\sinh\left(\frac{\Delta\Gamma_{(s)}}{2}\right) \pm \nonumber\\
        & \pm \mathcal{C}_\CP\cos\left(\Delta m_{(s)} t\right)
        \mp \mathcal{S}_\CP\sin\left(\Delta m_{(s)} t\right)
    \big],
\end{align}
where $A^\Delta$, $\mathcal{C}_\CP$ and $\mathcal{S}_\CP$ depend on the photon polarisation~\cite{Muheim:2008vu}.
Two strategies can be devised:
one studying the decay rate independently of the flavour of the \B meson, which allows $A^\Delta$ to be accessed, and one tagging the flavour of the \B meson, which accesses $\mathcal{S}_\CP$ and $\mathcal{C}_\CP$.
The first strategy has been exploited at \lhcb to study the $4000$ \decay{\Bs}{\phi\gamma} candidates collected in Run 1 to obtain $A^\Delta=-0.98^{+0.46}_{-0.52}\stat ^{+0.23}_{-0.20}\syst$~\cite{LHCb-PAPER-2016-034}, compatible at two standard deviations with the prediction of $A^\Delta_{\text{SM}}=0.047^{+0.029}_{-0.025}$.

With $\sim60$k signal candidates expected with $50\,\invfb$, the full analysis, including flavour tagging information, will improve the statistical uncertainty on $A^\Delta$ to $\sim0.07$, and will need a careful control of the systematic 
uncertainties.
An analysis performed with $\sim800$k signal decays expected with $300\,\invfb$, with a statistical uncertainty to $\sim0.02$, requires some of the possible improvements in \piz reconstruction of the \upgradetwo detector 
to be able to use the full statistical power of the data.

In addition to studying the \Bs system, \lhcb can study the time-dependent decay rate of \decay{\Bz}{\KS\pip\pim\gamma} decays, which permits access of the photon polarisation through the $\mathcal{S}_\CP$ term.
With $\mathcal{O}(1000)$ signal events in Run 1, around $35$k and $200$k are expected at the end of \upgradeone and \upgradetwo, respectively
($1.75$k and $10$k when considering the flavour tagging efficiency), opening the doors to a very competitive measurement of $\mathcal{S}_\CP$ in the \Bz system.

Another way to study the photon polarisation is through the angular correlations among the three-body decay products of a kaonic resonance in \decay{\Bp}{K_{\mathrm{res}}^+(\to \Kp\pim\pip)\gamma},
which allows the direct measurement of the photon polarisation parameter in the effective radiative weak Hamiltonian~\cite{Gronau:2002rz}. 
As a first step towards the photon polarisation measurement, \lhcb observed nonzero photon polarisation for the first time by studying the photon angular distribution in bins of $\Kp\pim\pip$ invariant mass\ \cite{LHCb-PAPER-2014-001}, but the determination of the value of this polarisation could not be performed due to the lack of knowledge of the hadronic system.
To overcome this problem, a method to measure the photon polarisation using a full amplitude analysis of \decay{\Bp}{\Kp\pim\pip\gamma} decays is currently under development, 
with an expected statistical sensitivity on the photon polarisation parameter of $\sim5\%$ in the charged mode with the Run 1 dataset. 
The extrapolation of the precision to $300\,\invfb$  results in a statistical precision better than $1\%$, and hence control of the systematic uncertainties will be crucial.


%% file: CONTRIBUTIONS/7_Rare_and_radiative_decays/7.4.2.tex
\subsection{Angular analysis with $B^0 \to K^{\ast 0}e^+e^-$ decays in the low-$q^2$ region\label{sec:7.4.2}}

The polarisation of the photon emitted in $b\to s\gamma$ transitions can also be accessed via semileptonic $b\to s\ell\ell$ transitions, for example in the decay $B^0\to\Kstarz\ell^+\ell^-$. Indeed, as mentioned in 
previous sections,
at very low $q^2$ these decays are dominated by the electromagnetic dipole operator ${\cal O}_7^{(\prime)}$. 
Namely, the longitudinal polarisation fraction ($\FL$) is expected to be below $20\%$ for $q^2<0.2\gevgevcccc$.
In this $q^2$ region, the angle $\phi$ between the planes defined by the dilepton system and the $\Kstarz\to K^+\pi^-$ decay is sensitive to the $b\to s\gamma$ photon helicity.

While the $\Kstarz\mumu$ final state is experimentally easier to select and measure at LHCb,
the $\Kstarz\epem$ final state allows $q^2$ values below $4m_\mu^2$ to be probed, 
where the sensitivity to the photon helicity is maximal.
Compared to the radiative channels used for polarisation measurements, the $B^0\to\Kstarz\epem$ final state is fully charged and gives better mass resolution and therefore better separation from partially reconstructed backgrounds. 

The sensitivity of this decay channel at LHCb was demonstrated by an angular analysis performed with Run 1 data~\cite{LHCb-PAPER-2014-066}. The angular observables most sensitive to the photon polarisation at low $q^2$ are $A_{\rm T}^{(2)}$ and $A_{\rm T}^{\rm Im}$, as defined in Ref.~\cite{LHCb-PAPER-2014-066}. 
Indeed, in the limit $q^2\to 0$, these observables can be expressed by the following functions of $C_{7,7^{\prime}}$ (assuming NP contributions to be much smaller than $|C_7^{\rm SM}|$):
\begin{equation}
  \label{eq:qSqToZero} 
  A_{\rm T}^{(2)} (\qsq \to 0) \simeq 2 \frac{\Real (C_7^{' \ast}) } {| C_7 |}  \quad {\rm and} \quad  A_{\rm T}^{\rm Im} (\qsq \to 0) \simeq 2 \frac{\Imag (C_7^{' \ast}) } {| C_7 |}.
\end{equation}
In order to maximise the sensitivity to the photon polarisation, 
the angular analysis should be performed as close as possible 
to the low $q^2$ endpoint. However, the events at extremely low $q^2$ have worse $\phi$ resolution (because the two electrons are almost collinear) and are polluted by $B^0\to\Kstarz\gamma$ decays with the $\gamma$ converting in the VELO material. In the Run 1 analysis~\cite{LHCb-PAPER-2014-066} the minimum required $m(\epem)$ was set at 20\mevcc, but this should
be reduced as the \upgradetwo VELO detector will have a significantly lower material budget (multiple scattering is the main effect worsening the $\phi$ resolution). 
Similarly, the background from $\gamma$ conversions will be reduced with a lighter RF-foil or with the complete removal of it in \upgradetwo~\cite{LHCb-PII-EoI}.

Assuming the signal yields given in Table~\ref{tab:penguin:LU_extrapolations} leads to the following statistical sensitivities to $A_{\rm T}^{(2)}$ and $A_{\rm T}^{\rm Im}$: 11\% with 9\invfb, 7\% with 23\invfb and 2\% with 300\invfb.
The theoretical uncertainty induced when this observable is translated into a photon polarisation measurement is currently at the level of $2\%$ but should improve by the time of the \upgradetwo analyses.
The current measurements performed with Run 1 data have a systematic uncertainty of order $5\%$ coming mainly from the modelling of the angular acceptance and from the uncertainty on the angular shape of the combinatorial background. The acceptance is independent of $\phi$ at low $q^2$ and its
modelling can be improved with larger simulation samples and using the proxy channel $\Bd\to\Kstarz\jpsi(\to\epem)$. 



%% file: CONTRIBUTIONS/7_Rare_and_radiative_decays/7.4.3.tex
\subsection{Radiative $b$-baryon decays\label{sec:7.4.3}}

Weak radiative decays of \bquark baryons are largely unexplored, with the best limits coming from CDF: $\mathcal{B}(\decay{\Lb}{\Lz\gamma}) < 1.3\times10^{-3}$ at $90\%\,$CL~\cite{Acosta:2002fh}.
They offer a unique sensitivity to the photon polarisation through the study of their angular distributions, and will constitute one of the main topics in the radiative decays programme in the \lhcb \upgradetwo.

With predicted branching fractions of $O(10^{-5}-10^{-6})$, the first challenge for \lhcb will be their observation, as the production of long-lived particles in their decay, in addition to the photon, means in most cases that the $b$-baryon secondary vertex cannot be reconstructed. This makes their separation from background considerably more difficult than in the case of regular radiative \bquark decays.

The most abundant of these decays is \decay{\Lb}{\Lz(\to p\pim)\gamma}, which is sensitive to the photon polarisation mainly\footnote{In the following, we assume that the \Lb (and any other beauty baryon) polarisation is zero~\cite{LHCb-PAPER-2012-057}, removing part of the photon-polarisation dependence.} through the distribution of the angle between the proton and the \Lz momentum in the rest frame of the \Lz ($\theta_p$), 
\begin{equation}
    \frac{\operatorname{d}\Gamma}{\operatorname{d}\cos\theta_p} \propto 1 - \alpha_\gamma \alpha_{p,1/2}\cos\theta_p, 
\end{equation}
where $\alpha_\gamma$ is the asymmetry between left- and right-handed amplitudes and $\alpha_{p,1/2}=0.642\pm0.013$~\cite{PDG2018} is the \decay{\Lz}{p\pim} decay parameter.
Using specialised trigger lines for this mode
$15-150$ signal events are expected using the Run 2 dataset.
Preliminary studies show that a statistical sensitivity to $\alpha_\gamma$ of $(20-25)\%$ is expected with these data, which would be reduced to $\sim15\%$ with $23\,\invfb$ and below $4\%$ with $300\,\invfb$.
In the \lhcb \upgradetwo, the addition of timing information in the calorimeter will be important to be able to study this combinatorial-background dominated decay;
additionally, improved downstream reconstruction would allow the use of downstream \Lz decays, which make up more than $2/3$ of the total signal.

The \decay{\Xib}{\Xi^-(\to\Lz(\to p\pim)\pim)\gamma} decay presents a richer angular distribution, with dependence to the photon polarisation in both the \Lz angle ($\theta_\Lz$) and proton angle ($\theta_p$),
\begin{equation}
    \frac{\operatorname{d}\Gamma}{\operatorname{d}\cos\theta_\Lz\cos\theta_p} \propto 1 - \alpha_\gamma \alpha_\Xi \cos\theta_\Lz + \alpha_{p,1/2}\cos\theta_p\left(\alpha_\Xi-\alpha_\gamma\cos\theta_\Lz\right),
\end{equation}
but the lower production cross-section for these baryons, 
combined with a lower reconstruction efficiency due to the presence of one extra track, results in an order of magnitude fewer events than in the \Lb case, making the increase of 
statistics from the \upgradetwo even more relevant. With a similar sensitivity to the photon polarisation to that of \decay{\Lb}{\Lz(\to p\pim)\gamma}, \decay{\Xib}{\Xi^-\gamma} decays
will allow this parameter to be probed with a precision of $40\%$ and $10\%$ with $23$ and $300\,\invfb$, respectively.

%% file: CONTRIBUTIONS/7_Rare_and_radiative_decays/7.4.4.tex
\subsection{Radiative $D$ decays}
\label{sec:radCharm}

Charm decays with a photon in the final state are complementary to radiative beauty and strange decays, 
due to the different quark content of the underlying loop amplitudes. 
Radiative charm decays, although rare (\BF$\sim 10^{-5}$) and overwhelmed by long-distance QCD effects, still offer 
theoretically clean probes of the SM, like for instance the \CP asymmetry or photon polarisation.
The most promising channels 
are $D^0 \to V \gamma$, where $V$ stands 
for a light vector meson, $\rho^0$ or $\phi$.
The SM predictions for \CP violation in these decays
is small at the order of a few $10^{-3}$, however $A_{\CP}$ could be enhanced by NP contributions to up to $10\%$~\cite{deBoer:2017que}. 
The best sensitivities so far achieved have been obtained with the full Belle data set and are limited by the statistical uncertainty, 
reaching a precision of  $(7-15)\%$~\cite{Abdesselam:2016yvr}. No observation of photon polarisation has 
yet been performed in charm decays.

Charm decays involve photons of significantly lower energies than their beauty counterparts  
and thus suffer from higher combinatorial background 
and lower energy resolution. 
Moreover, $D^0 \to V \gamma$ decays are also subject to a significant, 
peaking and irreducible background from the $D^0 \to V \pi^0$ decays, whose branching fractions exceed the radiative ones by two orders of magnitude. In particular, merged $\pi^0$ mesons, where both photons form a single energy cluster, are difficult 
to suppress. Nevertheless, a feasibility study performed for the $D^0 \to \phi \gamma$ decay has shown that LHCb 
will improve Belle sensitivities by factor $2-3$ using Run 2 data.

Having an improved \piz reconstruction in the \upgradetwo 
would certainly allow for background reduction.  
This could allow studies of decays with more complicated topologies \eg $D^+ \to K \pi \pi \gamma$. 
Such four-body processes give a direct access to $P$-parity and the photon polarisation, which can be measured 
through an up-down asymmetry of the photon direction relative to the $K\pi\pi$ decay plane. 
In the $\Dz \to V \gamma$ decays the photon polarisation can be accessed indirectly, by probing a pattern 
of the $\Dz-\Dzb$ oscillations and, thus, requires a time-dependent analysis. Such measurements should
become feasible with the \upgradetwo data sample. 


%% file: CONTRIBUTIONS/7_Rare_and_radiative_decays/7.5.tex
\section{Rare (semi-)leptonic charm decays}

The study of very rare FCNC charm decays is a unique probe for NP in the up-quark sector and a relatively unexplored area of research, both theoretically and experimentally.
Moreover, the anomalies seen in the \bquark-sector make progress on charm rare decays involving $c\to u$ transitions even more pressing.

\input{CONTRIBUTIONS/7_Rare_and_radiative_decays/7.5.1.tex}
\input{CONTRIBUTIONS/7_Rare_and_radiative_decays/7.5.2.tex}

\input{CONTRIBUTIONS/7_Rare_and_radiative_decays/7.5.3.tex}

%% file: CONTRIBUTIONS/7_Rare_and_radiative_decays/7.5.1.tex
\def\brdmumu{\ensuremath{\mathcal{B}(D^0 \to \mu^{+} \mu^{-})}\xspace}
\def\dmumu{\ensuremath{D^0 \to \mu^{+} \mu^{-}}\xspace}
\def\dgammagamma{\ensuremath{D^0 \to \gamma \gamma}\xspace}


\subsection{Search for $\Dz \to \mumu$ decays}

%
The decay of a $D$ meson to a dimuon pair is a key very rare charm decay. 
Within the Standard Model two kinds of contributions to the \dmumu decay are present: the \emph{long-distance} (LD) contribution due to long-range propagation of intermediate states, of non-perturbative nature, and the \emph{short-distance} (SD) contribution due to perturbatively calculable amplitudes. 
In the down-type quark sector the GIM suppression is less effective than in the up-quark sector, due to the presence of $top$-quarks running in the loops which leads to an effective decoupling with the other diagrams, owing to the large top quark mass. In the \emph{charm} sector FCNCs are instead more effectively suppressed due to the absence of a large mass down-type quark and therefore, within the SM, branching fractions of rare $D$ decays have very small values. However, these processes  can be enhanced in New Physics scenarios by up to several orders of magnitude when compared to the SM.  Predictions for the SD to \brdmumu are of $\mathcal{O}(10^{-18})$ while the LD contribution from $\gamma\gamma$ recombination, based on the current limit for \dgammagamma\cite{Nisar:2015gvd}, brings the expected branching fraction up to $10^{-11}$.

The world's best limit was obtained by LHCb with $0.9\invfb$ of 2011 data~\cite{LHCb-PAPER-2013-013}, resulting in
\begin{equation}
\brdmumu< 6.2 \times 10^{-9}\text{ at 90\% CL. }
\end{equation}
%
Extrapolating the current detector performance, the expected limit is about $5.9 \times 10^{-10}$ with $23~\invfb$ and $1.8 \times 10^{-10}$ with $300~\invfb$ of integrated luminosity,
covering a large part of the unambiguous space to search for NP without being affected by LD uncertainties in the SM predictions.

%% file: CONTRIBUTIONS/7_Rare_and_radiative_decays/7.5.2.tex
\subsection{Search for $D \to h\mumu$ and $\Lc \to p\mumu$ decays}

As already stated \decay{\bquark}{\squark \ellell} electroweak penguin transitions have proven to be powerful tools to search for New Physics effects~\cite{Capdevila:2017bsm,Altmannshofer:2017yso}. One can study the same New Physics effects in the up-quark sector with the \decay{\cquark}{\uquark \ellell} transitions. 
Until recently, limits on the branching fractions of these types of decays were only set at the ${\cal O}(10^{-6})$ level~\cite{Lees:2011hb}. 
Significant progress has also been made in the SM predictions for $c\to u$ FCNCs providing more robust calculations~\cite{Fajfer:2015mia,deBoer:2015boa,Meinel:2017ggx}. 

\begin{figure}[tb]

\includegraphics[width=0.5\textwidth]{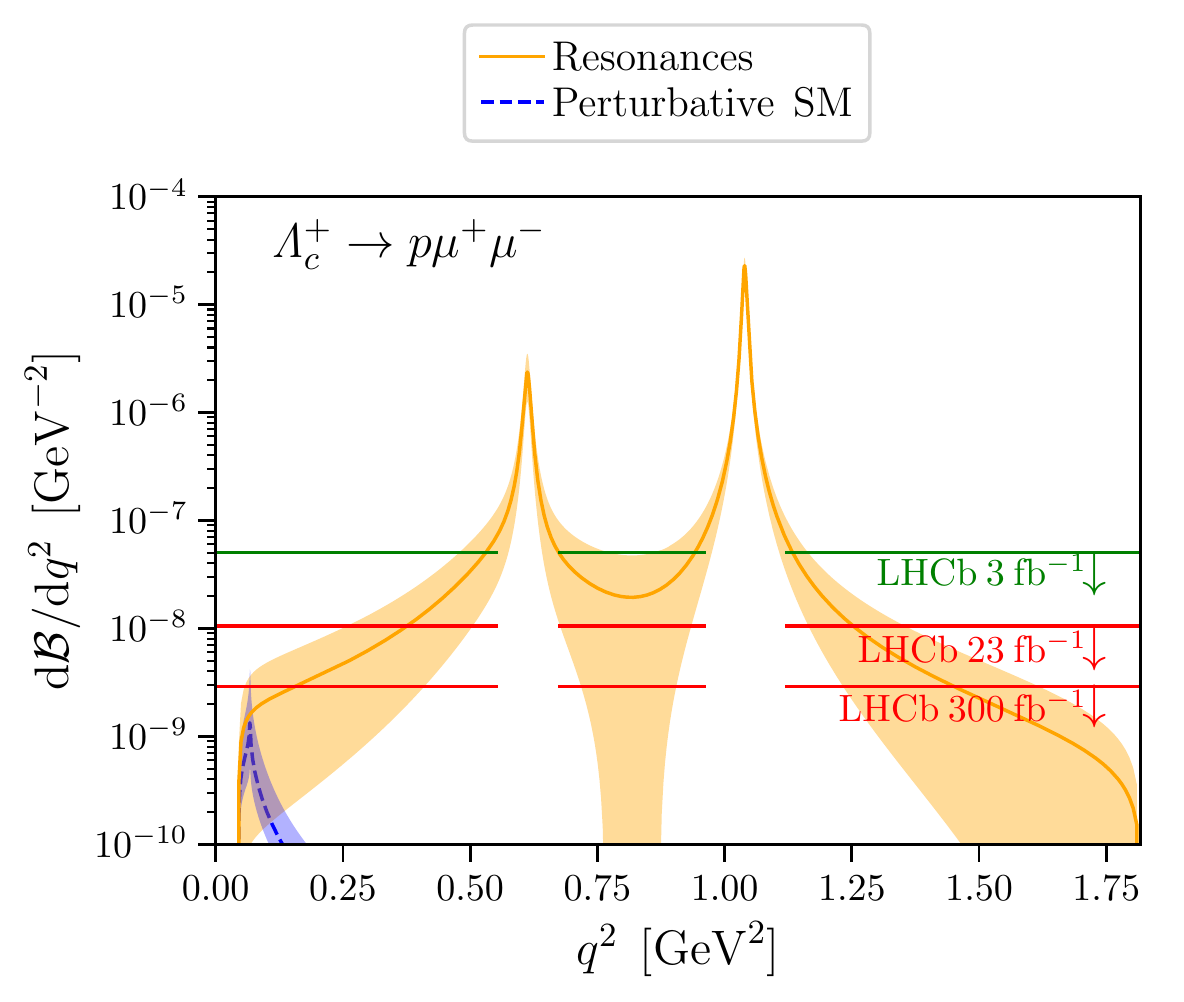}
\includegraphics[width=0.5\textwidth]{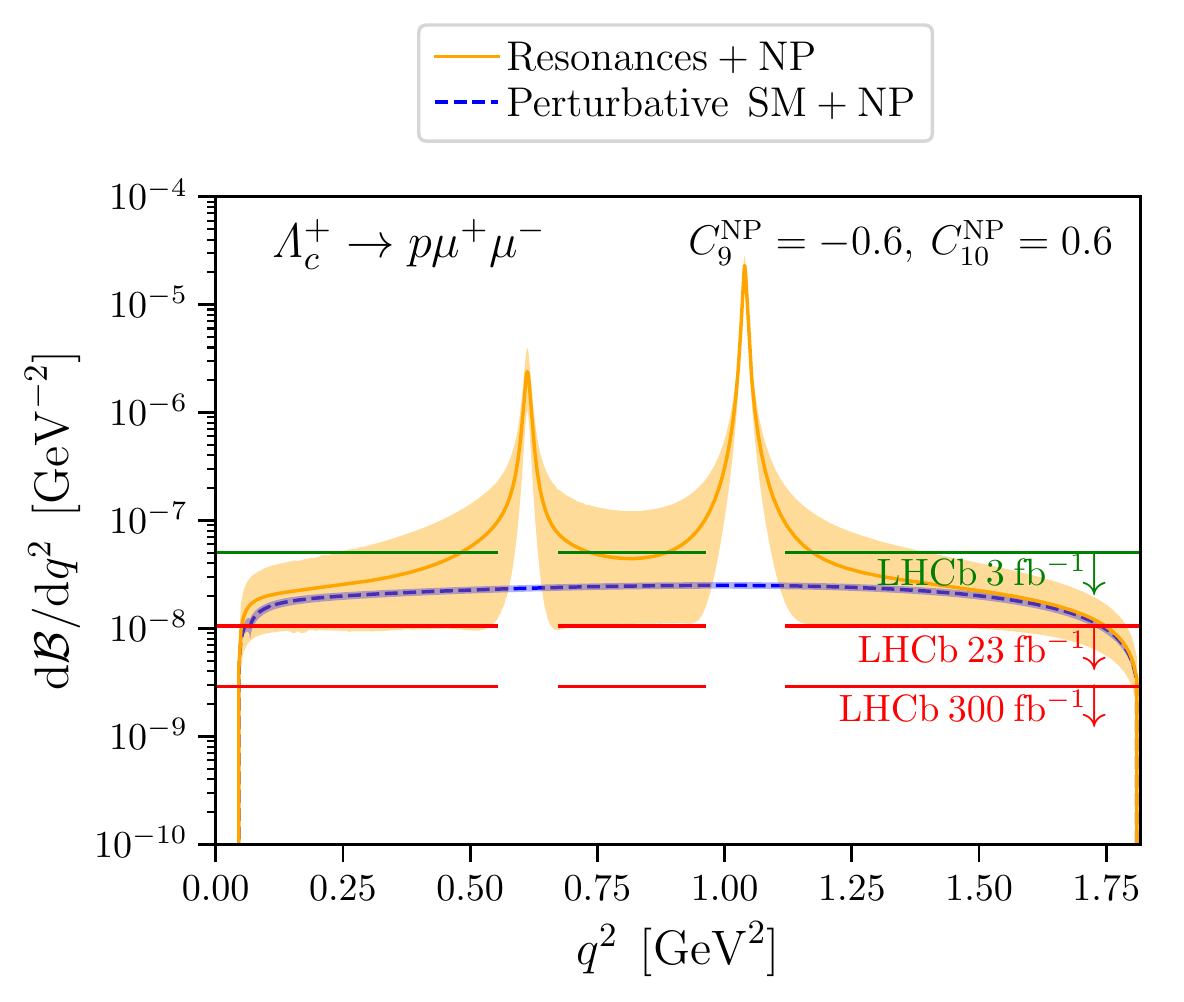}
\caption{Differential branching fraction for the decay $\Lc \to p \mumu$ as a function of $q^2$ (left) in the SM and (right) in a New Physics scenario with ${\cal C}_9^{\rm NP}=-0.6$ and ${\cal C}_{10}^{\rm NP}=-0.6$~\cite{Meinel:2017ggx}. The current LHCb limit is marked with green. The extrapolated upper limit with the $23$ and $300~\invfb$ data sets is marked in red.\label{fig:LC2PMUMU}}
\end{figure}

\begin{figure}[tb]
\includegraphics[width=0.33\textwidth]{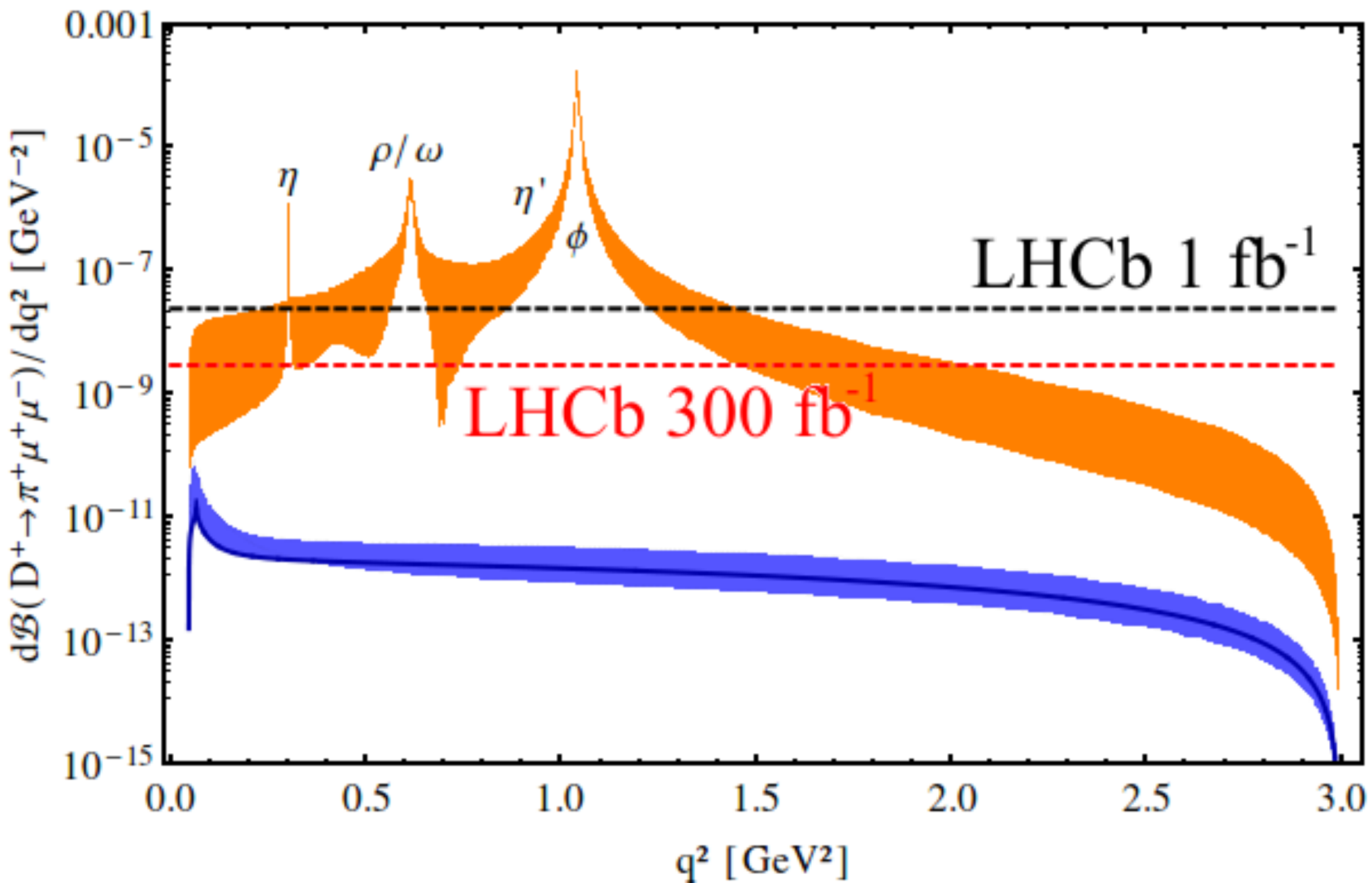}
\includegraphics[width=0.66\textwidth,clip=true,trim=0mm 60mm 0mm 0mm]{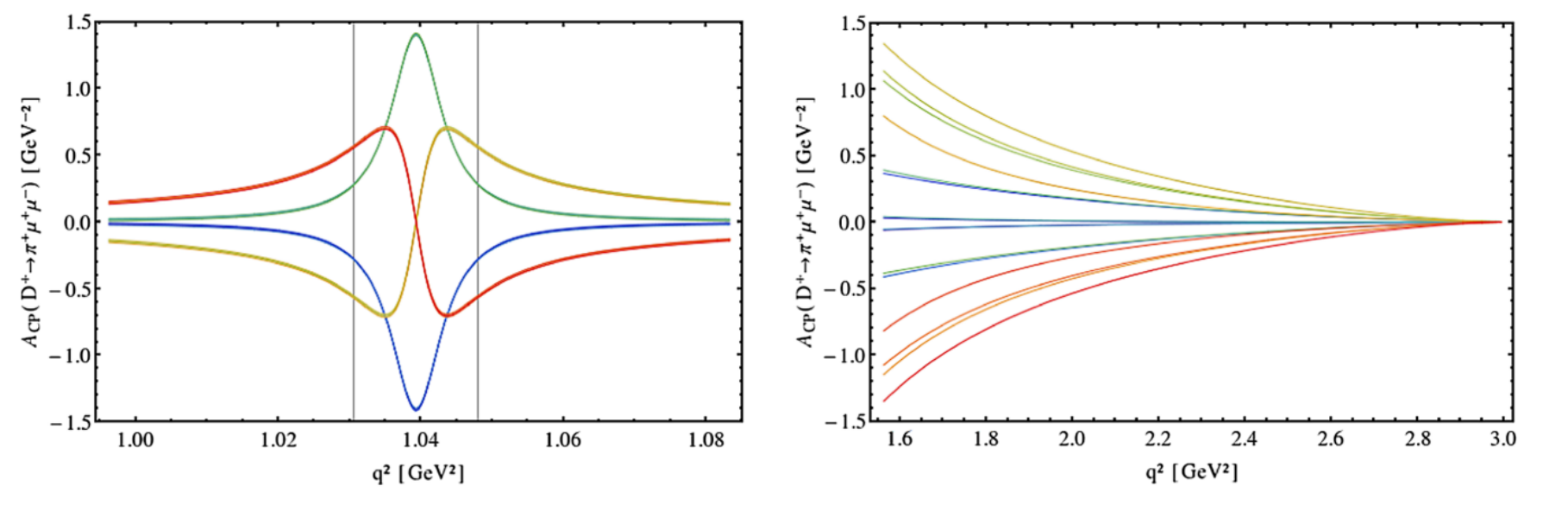}
\caption{
    (Left) Differential branching fraction of $\D^+ \to \pi^+ \mumu$ as a function of $q^2$ in the SM. The current LHCb limit is marked with a dashed black line. (Middle and right) Predictions for $A_{\CP}$ in a NP scenario with different strong phases $\delta_{\rho,\phi}$~\cite{deBoer:2015boa}.\label{fig:D2PIMUMU}
}
\end{figure}

Particularly suited for experiments at hadron colliders is the study of decays with a pair of leptons in the final state, such as $D^{+}_{(s)}/ \Lc \to h^+ \ellell$ and $D^0 \to h^+h^- \ellell$. The SD contributions to branching fractions of these decays involves FCNC processes which are heavily suppressed in the SM through the GIM mechanism and are at the level of $10^{-9}$ or below, depending on the decay~\cite{Meinel:2017ggx,deBoer:2015boa}. On the other hand, LD contributions dominate the branching fraction via vector resonances $\left( \rho/\omega/\phi\right)$ decaying into a dilepton pair. In some New Physics scenarios the SD contributions can be enhanced by several orders of magnitude allowing NP to manifest as an enhancement of the branching fraction. An example of such a model is shown in Fig.~\ref{fig:LC2PMUMU}. In the benchmark scenario, the modification of Wilson coefficients are: $C_9^{\rm NP} =-0.6$ and $C_{10}^{\rm NP} =0.6$.
Outside of the resonance regions the SD contributions are enhanced to  similar levels as the LD effects. This benchmark point is currently just below the current LHCb limit~\cite{LHCb-PAPER-2017-039}
\begin{equation}
\BF( \Lc \to p \mumu) < 5.9 \times 10^{-8}~{\rm at}~90\%~{\rm CL}
\end{equation}
In the \upgradetwo era LHCb is expected to improve the limit to:
\begin{equation}
\BF( \Lc \to p \mumu) < 4.4 \times 10^{-9}~{\rm at}~90\%~{\rm CL}.
\end{equation}

An order of magnitude improvement is expected for the limit on $\BF(D^{+}_{(s)}\to\pi^+\mu^+\mu^-)$ decays, which currently is set at $7.3\times10^{-8}$
at 90\%
CL \cite{LHCb-PAPER-2012-051}. The expected upper limits are about $1.3\times10^{-8}$ with $23 \invfb$ and $0.37\times10^{-8}$  with $300 \invfb$.
In addition, LHCb will have the ability to measure angular observables such as the forward-background asymmetry $A_{\rm FB}$ or time integrated $A_{\CP}$, 
which will provide additional handles to separate the LD from the SD and for which some theoretical predictions in NP scenarios already exist \cite{deBoer:2015boa},
as shown in Fig.~\ref{fig:D2PIMUMU}.

Last but not least with a sample corresponding to an integrated luminosity of $300~\invfb$ LHCb will be able to perform searches for the LFV decays $D^{+}_{(s)}/ \Lc\to h^+ \ell^+ \ell^{\prime -}$
and perform tests of lepton universality via the ratios ${\cal B} (\decay{D^{+}_{(s)}/\Lc}{h^+ \mu^+ \mu^{-}})/{\cal B}(\decay{D^{+}_{(s)}/\Lc}{h^+ e^+ e^{-}})$.

%% file: CONTRIBUTIONS/7_Rare_and_radiative_decays/7.5.3.tex
\subsection{Measurements with $D \to h^+h^-\ellell$ decays}

The decays $D^0\to h^+h^-\ellell$ have a richer dynamics compared to 
two- and three-body decays allowing for a variety of differential distributions to be investigated.
The SD contributions to the branching fraction proceeds through FCNC processes, which are very suppressed in the SM through the GIM mechanism at the level of $10^{-9}$ or below~\cite{PaulBigi:2011}. On the other hand, LD contributions dominate with branching fractions up to $\mathcal{O}(10^{-6})$ via intermediate vector resonances ($\rho/\omega/\phi$) decaying into a dilepton pair~\cite{Fajfer:2007,PaulBigi:2011,Cappiello}. 
In some BSM scenarios the SD contribution can be enhanced by several orders of magnitude allowing NP to manifest as an enhancement of the branching fraction away from resonances or in asymmetries such as \CP and angular asymmetries~\cite{Bigi:2012,Paul:2012ab,Fajfer:2015mia,Cappiello,Fajfer:2005ke,deBoer:2015boa,deBoer:2018}. 

Due to the huge charm production cross-section at the LHC,
and LHCb's ability to trigger on low \pt\ dimuons, 
LHCb has unique physics reach in studying these decays. 
In fact, significant progress has already been made with the observation of the Cabibbo-favoured decay $\decay{\Dz}{\Km\pip\mun\mup}$ (with the dimuon mass in the $\rho/\omega$ region) with a branching fraction $(4.17\pm0.42)\times10^{-6}$~\cite{LHCb-PAPER-2015-043}  and the singly Cabibbo-suppressed decays $\decay{\Dz}{\pip\pim\mup\mun}$ and $\decay{\Dz}{\Kp\Km\mup\mun}$ with branching fractions of $(9.64\pm1.20)\times10^{-7}$ and $(1.54\pm0.33)\times10^{-7}$, respectively~\cite{LHCb-PAPER-2017-019}. For the latter, also the differential branching fraction as a function of the dimuon mass squared, $q^2$, was measured.
Furthermore, LHCb has performed a first measurement of \CP- and angular asymmetries in $\decay{\Dz}{\pip\pim\mumu}$ and $\decay{\Dz}{\Kp\Km\mumu}$ decays using Run~2 data. 
This resulted in the first determination of the forward-backward asymmetry $A_{\rm FB}$, the triple-product asymmetry $A_{2\phi}$ and the \CP-asymmetry $A_{\rm CP}$ with uncertainties at the percent-level~\cite{LHCb-PAPER-2018-020}. 
Moreover, the implementation of triggers for dielectron modes 
opens the possibility of measuring branching-fraction ratios between dimuon and dielectron modes.
Since the main limit for these studies comes from the available statistics, excellent prospects are foreseen for the LHCb \upgradeone. 
Projected signal yields for the muonic modes of $\mathcal(10^4)$ will allow more sensitive studies of angular asymmetry and first amplitude analyses to attempt to disentangle SD and LD components.
However, it is with the $300$\invfb upgrade that the full potential for these decays will be exploited.
The enormous event yields and improved calorimetry of \upgradetwo 
will allow to perform studies of both the dimuon and dielectron modes with high precision.

%% file: CONTRIBUTIONS/7_Rare_and_radiative_decays/7.6.tex
\section{Rare strange decays}

Rare strange decays can probe BSM physics that could escape all other experimental tests. Indeed, if BSM physics is beyond the LHC direct searches reach and thus only accessible through indirect measurements, searches for non-Minimal Flavour Violation BSM physics become more relevant. In this context, $s\rightarrow d$ transitions are of paramount importance, since they have the strongest CKM suppression. Examples of these transitions are \Ksmm and \Kspizmm.

\input{CONTRIBUTIONS/7_Rare_and_radiative_decays/7.6.1.tex}
\input{CONTRIBUTIONS/7_Rare_and_radiative_decays/7.6.2.tex}

%% file: CONTRIBUTIONS/7_Rare_and_radiative_decays/7.6.1.tex
\subsection{Rare kaon decays}
\begin{figure}[t!]
\centering
\includegraphics[width=0.46\textwidth]{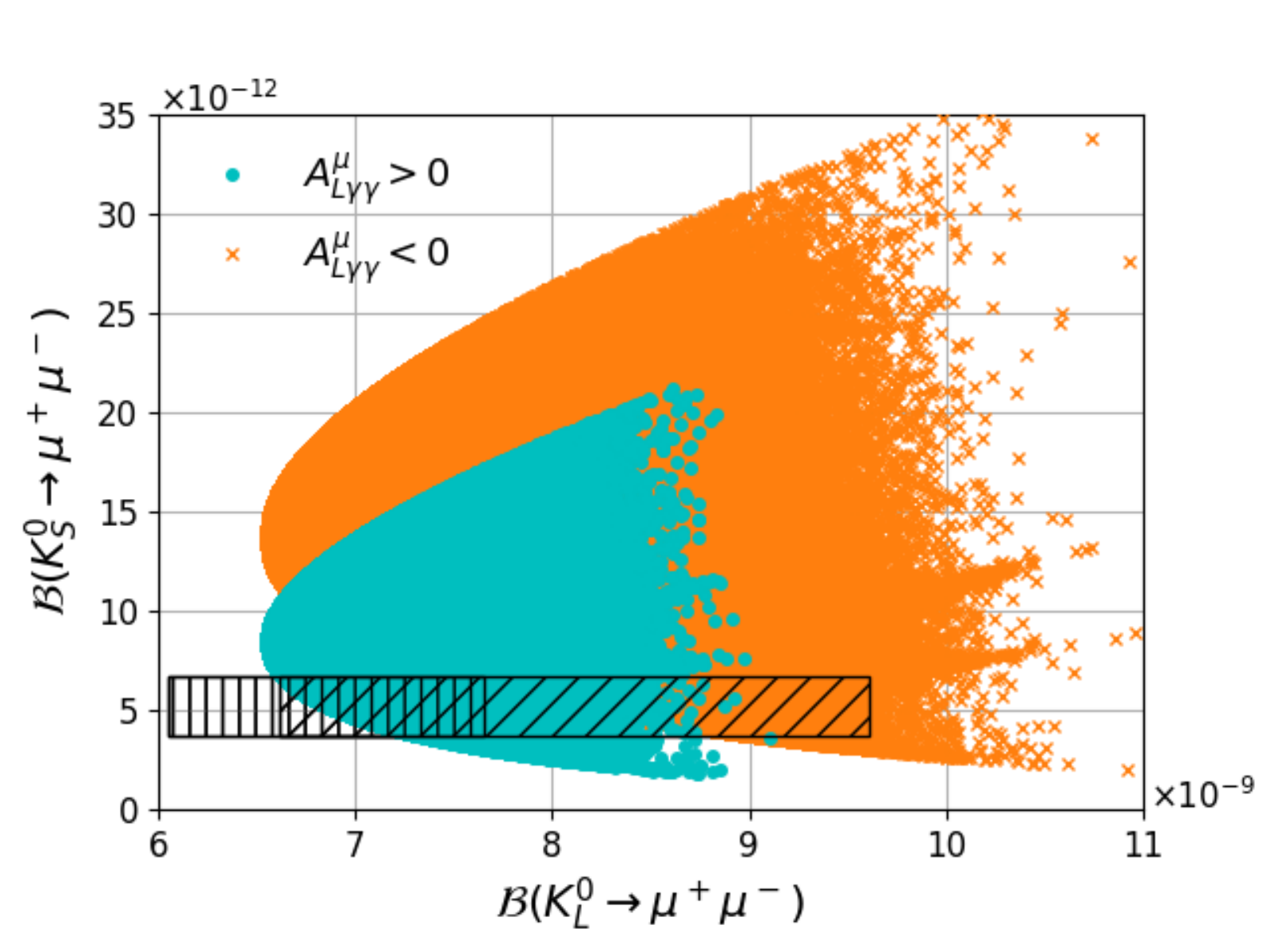}
\includegraphics[width=0.46\textwidth]{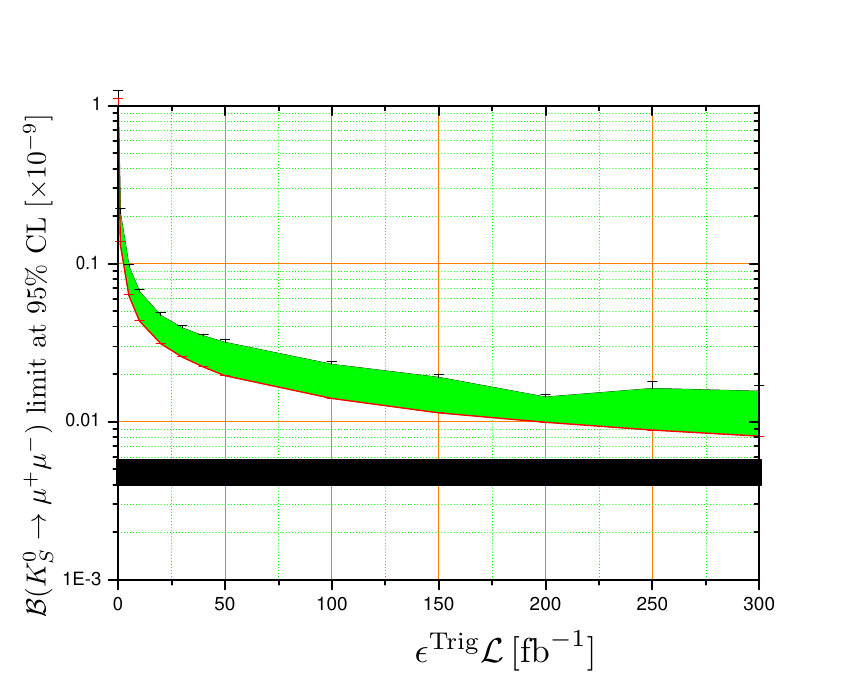}
\caption{\label{fig:Kmm} (Left) $\mathcal{B}(K_S^0\rightarrow\mu^+\mu^-)$ vs $\mathcal{B}(K_L^0\rightarrow\mu^+\mu^-)$ in an MSSM scenario, from Fig. 2 of Ref.~\cite{Chobanova:2017rkj}. The cyan dots correspond
  to predictions using positive sign for the long distance contribution to $\mathcal{B}(K_L^0\rightarrow\mu^+\mu^-)$ (i.e. $A^\mu_{L\gamma \gamma} > 0$) and the orange crosses to predictions using $A^\mu_{L\gamma \gamma} < 0$. The vertically hatched area corresponds to the SM prediction for $A^\mu_{L\gamma \gamma} > 0$ and the inclined hatched area corresponds to the SM prediction for $A^\mu_{L \gamma \gamma} < 0$.
(Right) Expected limit in \BRof\Ksmm from LHCb and upgrades, as a function of integrated luminosity times trigger efficiency.
  LHCb \upgradetwo\ will collect at least $300\invfb$. 
  Thus, it can be seen that if the trigger efficiency is high,
as expected from a full software trigger, LHCb can exclude branching fractions down to near the SM prediction}
\end{figure}

In the SM, the \Ksmm decay is long-distance dominated, with subdominant short-distance contributions. However, the long-distance contribution is still very small in absolute terms, and the decay
rate is very suppressed. For example, the SM prediction~\cite{Ecker:1991ru,Isidori:2003ts,DAmbrosio:2017klp}
$\BRof\Ksmm_{\rm SM} = (5.18\pm1.50_{\rm LD}\pm 0.02_{\rm SD})\times 10^{-12}$ can be compared with the current experimental upper limit~\cite{LHCb-PAPER-2017-009}
$\BRof\Ksmm_{\rm Exp} < 8 \times 10^{-10} ~{\rm{at}}~90\%~{\rm{CL}}$.
Therefore, even small BSM contributions and BSM--SM interferences can compete with the SM rate. This has been proven
to be the case in leptoquark models ~\cite{Dorsner:2011ai,Bobeth:2017ecx} as well as supersymmetric models ~\cite{Chobanova:2017rkj}.
In the latter, \BRof\Ksmm can have values anywhere in the range $[0.78-35]\times 10^{-12}$ (see \figref{fig:Kmm}, left) or even saturate the current
experimental bound in certain narrow regions of the parameter space ~\cite{Chobanova:2017rkj}. The \CP asymmetry of the $K^0\rightarrow\mu^+\mu^-$ decay
is also sensitive to BSM contributions and experimentally accessible by means of a tagged analysis. 

The LHCb prospects for the search for \Ksmm decays are excellent. With 2011 data the experiment overtook the previous world best upper limit
by a factor of thirty~\cite{LHCb-PAPER-2012-054}, and has recently gained another order of magnitude~\cite{LHCb-PAPER-2017-009}.
The right hand side of \figref{fig:Kmm} shows the expected upper limit
for \BRof\Ksmm as a function of the integrated luminosity multiplied by the trigger efficiency. 
It can be seen that if the trigger efficiency is high, 
as expected from a full software trigger, LHCb can explore branching fractions down to near the SM prediction.

Currently, the only existing measurement of  \BRof\Kspizmm comes from the NA48 experiment~\cite{Batley:2004wg}
\begin{equation}
\BRof\Kspizmm = (2.9_{-1.2}^{+1.5}\pm0.2)\times 10^{-9}. \nonumber
\end{equation}
In the \upgradetwo\ LHCb can achieve a statistical precision of $0.11\times 10^{-9}$ with $300\invfb$ of integrated luminosity, assuming a trigger efficiency of 100\%~\cite{Chobanova:2195218}. 
Assuming a trigger efficiency of $50\%$ LHCb will still be able to significantly improve on the NA48 measurement. 
Apart from the branching fraction, the differential decay rate in the dimuon mass contains interesting information about the form factor parameters $a_S$ and $b_S$ ~\cite{DAmbrosio:1998gur}.
The LHCb \upgradetwo can reach a $10\%$ statistical precision on the form factor term $|a_S|$ with free $b_S$~\cite{Junior:2018odx}.

Other kaon decays that can be studied at LHCb include $K^+\rightarrow\pi\mu\mu$ (both with opposite-sign and same-sign muon pairs), $\KS\rightarrow4\mu$, or decays involving electrons in order to test Lepton Universality~\cite{Junior:2018odx}. The LHCb acceptance is such that
it favours the sensitivity to \KS modes over \KL modes by about a factor 1000 due to the longer \KL lifetime.\footnote{However, it must be noted that the \KS--\KL interference component
has an effective lifetime of $\sim 2\tau_{\KS}$, with higher chances of decay inside LHCb tracking system. The interference component can be accessed in principle by means of a tagged analysis.}

%% file: CONTRIBUTIONS/7_Rare_and_radiative_decays/7.6.2.tex
\subsection{Rare hyperon decays}

\def\Sigmaplus{\ensuremath{\Sigmares^+}\xspace}
\def\sigmaplus{\Sigmaplus}
\def\Sigmaplusbar{\ensuremath{\bar \Sigmares^-}\xspace}
\def\sigmapmumu{\ensuremath{\Sigmares^+ \to \Pp \mu^+ \mu^-}\xspace}
\def\pmumu{\ensuremath{\Pp \mu^+ \mu^-}\xspace}
\def\mpmumu{\ensuremath{m_{\Pp \mu^+ \mu^-}}\xspace}
\def\sigmapgamma{\ensuremath{\Sigmares^+ \to \Pp \gamma}\xspace}
\def\sigmapxmumu{\ensuremath{\Sigmares^+ \to \Pp X^0 (\to \mu^+ \mu^-)}\xspace}
\def\sigmapmumulfv{\ensuremath{\Sigmares^+ \to \antiproton \mu^+ \mu^+}\xspace}
\def\pmumulfv{\ensuremath{\antiproton \mu^+ \mu^+}\xspace}
\def\sigmappiz{\ensuremath{\Sigmares^+ \to \Pp \pi^0}\xspace}
\def\sigmapgamma{\ensuremath{\Sigmares^+ \to \Pp \gamma}\xspace}
\def\sigmappizero{\sigmappiz}
\def\mppizero{\ensuremath{m_{\Pp \pi^0}}\xspace}
\def\pizero{\ensuremath{\pi^0}\xspace}
\def\sigmappizdalitz{\ensuremath{\Sigmares^+ \to \Pp \pi^0 (\to e^+ e^- \gamma)}\xspace}
\def\sigmapee{\ensuremath{\Sigmares^+ \to \Pp e^+ e^-}\xspace}
\def\sigmapemu{\ensuremath{\Sigmares^+ \to \Pp e^\pm mu^\mp}\xspace}
\def\kplus{\ensuremath{K^+}\xspace}
\def\kpipipi{\ensuremath{K^+ \to \pi^+ \pi^- \pi^+}\xspace}
\def\kpimumu{\ensuremath{K^+ \to \pi^+ \mu^- \mu^+}\xspace}
\def\qsquare{\ensuremath{q^2}\xspace}
\def\mmumu{\ensuremath{m_{\mu^+\mu^-}}\xspace}
\def\lambdappi{\ensuremath{\PLambda^0 \to \Pp \pi^-}\xspace}
\def\pdf{\emph{p.d.f.}\xspace}
\def\pdfs{\emph{p.d.f.s}\xspace}
\def\chisq{\ensuremath{\chi^2}\xspace}
\def\ipchisq{\ensuremath{\rm{IP}\chi^2}\xspace}

\def\bujpsikstar{\ensuremath{B^+ \to J/\psi K^{\ast +}(\to K^+ \pi^0) }\xspace}
\def\bujpsik{\ensuremath{B^+ \to J/\psi K^+}\xspace}

\def\ks{\ensuremath{K^0_S}\xspace}
\def\kspipi{\ensuremath{K^0_S \to \pi^+ \pi^-}\xspace}
\def\ksmumu{\ensuremath{K^0_S \to \mu^+ \mu^-}\xspace}
\def\ksemu{\ensuremath{K^0_S \to e^\pm \mu^\mp}\xspace}
\def\ksmumugamma{\ensuremath{K^0_S \to \mu^+ \mu^- \gamma}\xspace}
\def\ksee{\ensuremath{K^0_S \to e^+ e^-}\xspace}
\def\kspipigamma{\ensuremath{K^0_S \to \pi^+ \pi^- \gamma}\xspace}
\def\kspimunu{\ensuremath{K^0_S \to \pi^{\pm} \mu^{\mp} \nu}\xspace}
\def\kspienu{\ensuremath{K^0_S \to \pi^{\pm} e^{\mp} \nu}\xspace}
\def\kspipipiz{\ensuremath{K^0_S \to \pi^+ \pi^- \pi^0}\xspace}
\def\ksgammagamma{\ensuremath{K^0_S \to \gamma \gamma}\xspace}
\def\kspizmumu{\ensuremath{K^0_S \to \pi^0 \mu^+ \mu^-}\xspace}

\def\kspipiee{\ensuremath{K^0_S \to \pi^+ \pi^- e^+ e^-}\xspace}
\def\ksmumumumu{\ensuremath{K^0_S \to \mu^+ \mu^- \mu^+ \mu^-}\xspace}
\def\ksmumuee{\ensuremath{K^0_S \to \mu^+ \mu^- e^+ e^-}\xspace}
\def\kseeee{\ensuremath{K^0_S \to e^+ e^- e^+ e^-}\xspace}
\def\ksmumumue{\ensuremath{K^0_S \to \mu^+ \mu^- \mu^\pm e^\mp}\xspace}
\def\kseeemu{\ensuremath{K^0_S \to e^+ e^- e^\pm e^\mp}\xspace}
\def\ksmumueelfv{\ensuremath{K^0_S \to \mu^+ \mu^+ e^- e^-}\xspace}

\def\kpipipi{\ensuremath{K^+ \to \pi^+ \pi^- \pi^+ }\xspace}
\def\kpipimunu{\ensuremath{K^+ \to \pi^+ \pi^- \mu^+ \nu}\xspace}
\def\kpipienu{\ensuremath{K^+ \to \pi^+ \pi^- e^+ \nu}\xspace}
\def\kpimumu{\ensuremath{K^+ \to \pi^+ \mu^- \mu^+ }\xspace}
\def\kmumumu{\ensuremath{K^+ \to \mu^+ \mu^- \mu^+ \nu}\xspace}
\def\kpimumulfv{\ensuremath{K^+ \to \pi^- \mu^+ \mu^+ }\xspace}
\def\kpiee{\ensuremath{K^+ \to \pi^+ e^- e^+ }\xspace}

\def\sigmapmumu{\ensuremath{\Sigmares^+ \to p \mu^+ \mu^-}\xspace}
\def\sigmamumumu{\ensuremath{\Sigmares^+ \to \mu^+ \mu^+ \mu^-}\xspace}
\def\sigmapmumulfv{\ensuremath{\Sigmares^+ \to \bar p \mu^+ \mu^+}\xspace}
\def\sigmapee{\ensuremath{\Sigmares^+ \to p e^+ e^-}\xspace}
\def\sigmapemu{\ensuremath{\Sigmares^+ \to p e^\pm \mu^\mp}\xspace}
\def\sigmappiz{\ensuremath{\Sigmares^+ \to p \pi^0}\xspace}
\def\sigmappizero{\ensuremath{\Sigmares^+ \to p \pi^0}\xspace}
\def\sigmapgamma{\ensuremath{\Sigmares^+ \to p \gamma}\xspace}
\def\sigmalambdaenu{\ensuremath{\Sigmares^+ \to \Lz e^+  \nu}\xspace}
\def\sigmaminuslambdaenu{\ensuremath{\Sigmares^- \to \Lz e^-  \nu}\xspace}

\def\lambdappi{\ensuremath{\Lz \to p \pi^-}\xspace}
\def\lambdappiee{\ensuremath{\Lz \to p \pi^- e^+ e^-}\xspace}
\def\lambdappigamma{\ensuremath{\Lz \to p \pi^- \gamma}\xspace}
\def\lambdapmunu{\ensuremath{\Lz \to p \mu^- \nu}\xspace}
\def\lambdapenu{\ensuremath{\Lz \to p e^- \nu}\xspace}

The large production of strange hadrons at LHCb, with at least one strange meson or baryon per minimum bias event, 
gives rise to unprecedented datasets for the study of rare hyperon decays. 
The full spectrum of hyperons will be available~\cite{Junior:2018odx}. 
Studies of rare hyperon decays in the \upgradetwo would significantly profit from low momentum proton identification provided by the TORCH detector. 

\subsubsection{Rare decays of $\Sigmares$ hyperons}

LHCb has recently published a search for the \sigmapmumu decay~\cite{LHCb-PAPER-2017-049}, with a strong evidence 
for this decay with a significance of $4.0\,\sigma$; a measurement of the branching fraction was reported  and 
a dimuon invariant mass distribution found to be consistent with SM predictions, challenging the HyperCP anomaly~\cite{Park:2005eka}. 

This measurement will be improved in Run 2 with dedicated triggers~\cite{Aaij:2253050,Dettori:2297352}, 
leading most probably to a first observation of this channel and detailed differential branching fraction measurement. 
With the additional dataset of 
\upgradetwo a full 
angular analysis will be possible, probing 
complementary couplings to those probed by \kpimumu decays. 

Assuming a similar sensitivity, a search for the lepton and baryon number violating decay \sigmapmumulfv 
could be performed and it would be expected to reach a sensitivity of order $10^{-10}$ ($10^{-9}$)
on the upper limit of the correspondent branching fraction with 300 (50)~\invfb of integrated luminosity.
Similarly, a search for \sigmapee decays will be possible with dedicated triggers.
The latter mode is similar to the \sigmapmumu case, but owing to the lower mass of the electrons 
receives a larger contribution from long distance photon contributions, 
for a predicted branching fraction~\cite{He:2005yn} of $\mathcal{B}(\sigmapee) \in [9.1,10.1] \times 10^{-6}$. 
Experimentally only an upper limit of $7 \times 10^{-6}$ at 90\% CL is reported~\cite{Ang:1969hg}.
It is expected to reach the SM level already 
with the \upgradeone, but only with the statistics provided by the \upgradetwo it would be possible to perform a detailed study of the 
dielectron invariant mass distribution.
Analogously, the LFV decays \sigmapemu could be searched for with similar sensitivity, 
where the SM rate is negligible. 


\subsubsection{Rare decays of $\Lz$ hyperons}
%
An interesting study in the domain of $\Lz$ hyperons would be to measure 
the radiative $\Lz \to p \pi^- \gamma$ decay.  
The branching fraction of this decay is known only with an uncertainty of 16\% 
($\mathcal{B} = (8.4\pm 1.4)\times10^{-4}$~\cite{Baggett:1973qb})
and only for pion centre-of-mass momenta less than $95\mevc$.
LHCb can improve this measurement significantly.
In addition a first measurement of the FCNC decay $\Lz \to p \pi e^+ e^-$ will be possible.

Several baryon-number-violating $\Lz$ decays have been recently searched for by the CLAS collaboration, as reported in Ref.~\cite{McCracken:2015coa}. 
Most of these are in the form $\Lz \to m \ell$ where $m$ is a $K^+$ or  $\pi^+$ meson, and $\ell=e, \mu$ leptons. 
Typically, these decays have strong indirect constraints from the limits on nucleon decays,
but direct experimental limits in an interesting range of sensitivity were absent so far. 
The reported limits on the respective branching fractions are in the range $[10^{-7}, 10^{-6}]$. 
The LHCb experiment can certainly improve most of these limits, 
reaching a single event sensitivities of ${\cal O}(10^{-11})$ for an integrated luminosity of $300\invfb$.

%% file: CONTRIBUTIONS/8_Forward_and_high_pT_physics/8.tex
\label{chpt:qee}

\input{CONTRIBUTIONS/8_Forward_and_high_pT_physics/8.0.tex}

\input{CONTRIBUTIONS/8_Forward_and_high_pT_physics/8.1.tex}
\input{CONTRIBUTIONS/8_Forward_and_high_pT_physics/8.2.tex}

\input{CONTRIBUTIONS/8_Forward_and_high_pT_physics/8.3.tex}

\input{CONTRIBUTIONS/8_Forward_and_high_pT_physics/8.4.tex}

\input{CONTRIBUTIONS/8_Forward_and_high_pT_physics/8.5.tex}
\input{CONTRIBUTIONS/8_Forward_and_high_pT_physics/8.6.tex}
\input{CONTRIBUTIONS/8_Forward_and_high_pT_physics/8.7.tex}

%% file: CONTRIBUTIONS/8_Forward_and_high_pT_physics/8.0.tex
The unique forward spectrometer configuration of LHCb permits a rich programme of particle production studies
at high momentum, allowing precision measurements of Standard Model physics, and direct and indirect searches for new phenomena.
LHCb has excellent charged particle reconstruction and muon identification, which can be exploited in 
muonic $W$ and $Z/\gamma^{\ast}$ (denoted $Z$ for brevity) decays. 
However, the current ability of LHCb to reconstruct electrons with large transverse momenta is restricted, owing to the saturation of cells in the ECAL when the transverse energy deposited is above $\sim 10\gev$. 
The yields for high-\pt electron signatures, for example $Z \to e^+e^-$ decays, are up to a factor of 3 smaller than can be achieved for the equivalent muon channels.
With an upgraded ECAL, we expect similar precision for muon and electron reconstruction at high transverse momentum.
In addition, the integrated luminosity offered by \upgradetwo also offers important new opportunities.
A selection of the measurements made possible with \upgradetwo are discussed in this chapter, with the focus placed on the different distinct areas within the forward high-\pt programme of research at LHCb in turn.

Top quark production in the LHCb acceptance accesses larger Bjorken-$x$ values, and enhanced contributions from quark-initiated subprocesses, than at the central rapidities covered by ATLAS and CMS.
This results in a greater sensitivity to observables depending on the presence of sizeable $q\bar{q}$ amplitudes~\cite{Kagan:2011yx}. 
One such observable is the $t\bar{t}$ charge asymmetry which, in the Standard Model, rises from approximately 1\% in the central region to as high as 8\% within the LHCb acceptance~\cite{Gauld:2014pxa}.
A number of new physics models~\cite{AguilarSaavedra:2011hz} predicting extra contributions to quark-initiated top production result in enhancements in both the $t\bar{t}$ production cross-section and charge asymmetry in the forward region. While measurements of the asymmetry at ATLAS and CMS already provide limits on these models~\cite{Sirunyan:2017lvd}, high yield measurements in the forward region have the potential to provide even stronger constraints. In addition, as the dominant contribution to top pair production in the forward region still arises from gluon-gluon fusion, precise measurements of the $t\bar{t}$ cross-section at LHCb will place important constraints on the poorly known gluon parton distribution function (PDF) at high-$x$ values.

Measurements of gauge boson production at LHCb uniquely probe the PDFs at smaller and larger values of Bjorken-$x$ than are accessible by ATLAS and CMS~\cite{Thorne:2008am}. The LHCb vector-boson production measurements have been included within fits to data to determine PDFs~\cite{Harland-Lang:2014zoa,Ball:2017nwa,Dulat:2015mca,Alekhin:2013nda}. Detailed studies of the LHCb data have shown that the measurements of gauge-boson production have placed stringent constraints on the nonperturbative charm content of the proton and have halved the uncertainties associated with the distribution of valence quarks at large values of Bjorken-$x$ (above 0.1)~\cite{Rojo:2017xpe}. The physics of PDFs at large $x$ is an important study within QCD, but is also crucial to understand the production of any potential new high-mass states detected at ATLAS and CMS. 

Precision electroweak tests provide a powerful probe of new physics.
Two parameters stand out as demanding higher precision direct measurements, to better exploit the
high precision of their corresponding indirect determinations through global fits.
Those are the effective leptonic weak mixing angle, \ssqtwef, and the $W$ boson mass, $m_W$.
The forward rapidity design of LHCb presents several advantages with respect to the central detector experiments
in measurements of these fundamental parameters of the SM.

Studies performed thus far on the Higgs boson are consistent with the particle predicted by the Standard Model.
The vector couplings of the Higgs are already reasonably well measured.
However, the fermion couplings are more challenging to access.
New physics could easily be hiding in the, currently unmeasurable, second-generation couplings.
While no currently approved experiment is capable of reaching sensitivity to the Higgs-charm coupling at the Standard Model level, the unique charm-tagging capabilities of the LHCb detector mean that LHCb Upgrade II has the potential to make the most stringent constraint at the HL-LHC.

LHCb's excellent lepton and jet reconstruction, as well as the flexible trigger scheme, allow direct searches for new physics phenomena associated with long lived particles, considering, for example, dark photons or particles in hidden valley models. Existing searches at LHCb provide unique sensitivity to many exotic signatures, with LHCb having published some of the most sensitive searches in such models for new particles with masses below the electroweak scale. Sensitivities in such searches following Upgrade II are therefore also explored in this chapter.



%% file: CONTRIBUTIONS/8_Forward_and_high_pT_physics/8.1.tex

\section{Top physics in the forward region}
Three measurements of top production have been performed by LHCb during Runs 1 and 2 of the LHC with a precision of (20--40)\%, limited by the available data samples. 
The first observation in the forward region was made in the $\mu b$ final state, where the top quark is identified by the presence of a muon and a $b$-jet, using 3\invfb of data collected in Run 1~\cite{LHCb-PAPER-2015-022}. 
This final state has the highest signal yield, but suffers from the largest backgrounds, in particular from $W$ boson production in association with a \bquark-jet. It is also impossible to separate single top and top pair production, which both contribute to the final state. A measurement was also made in the $\ell b b$ final state\cite{LHCb-PAPER-2016-038} using 2\invfb of data, where a muon or electron (indicated by $\ell$), in addition to two \bquark-jets, is used to identify $t\bar{t}$ events. The first measurement in Run 2 was made in the $\mu e b$ final state, where a muon and electron, together with a \bquark-jet, were reconstructed~\cite{LHCb-PAPER-2017-050}. This final state has the highest purity, but the lowest yield  and was only made possible in Run 2 due to the increase of approximately a factor of ten, between $\sqrt{s}=13$~\tev and $\sqrt{s}=8$~\tev, in the top pair cross-section within the LHCb acceptance.

\begin{figure}
\begin{center}
\includegraphics[width=0.7\textwidth]{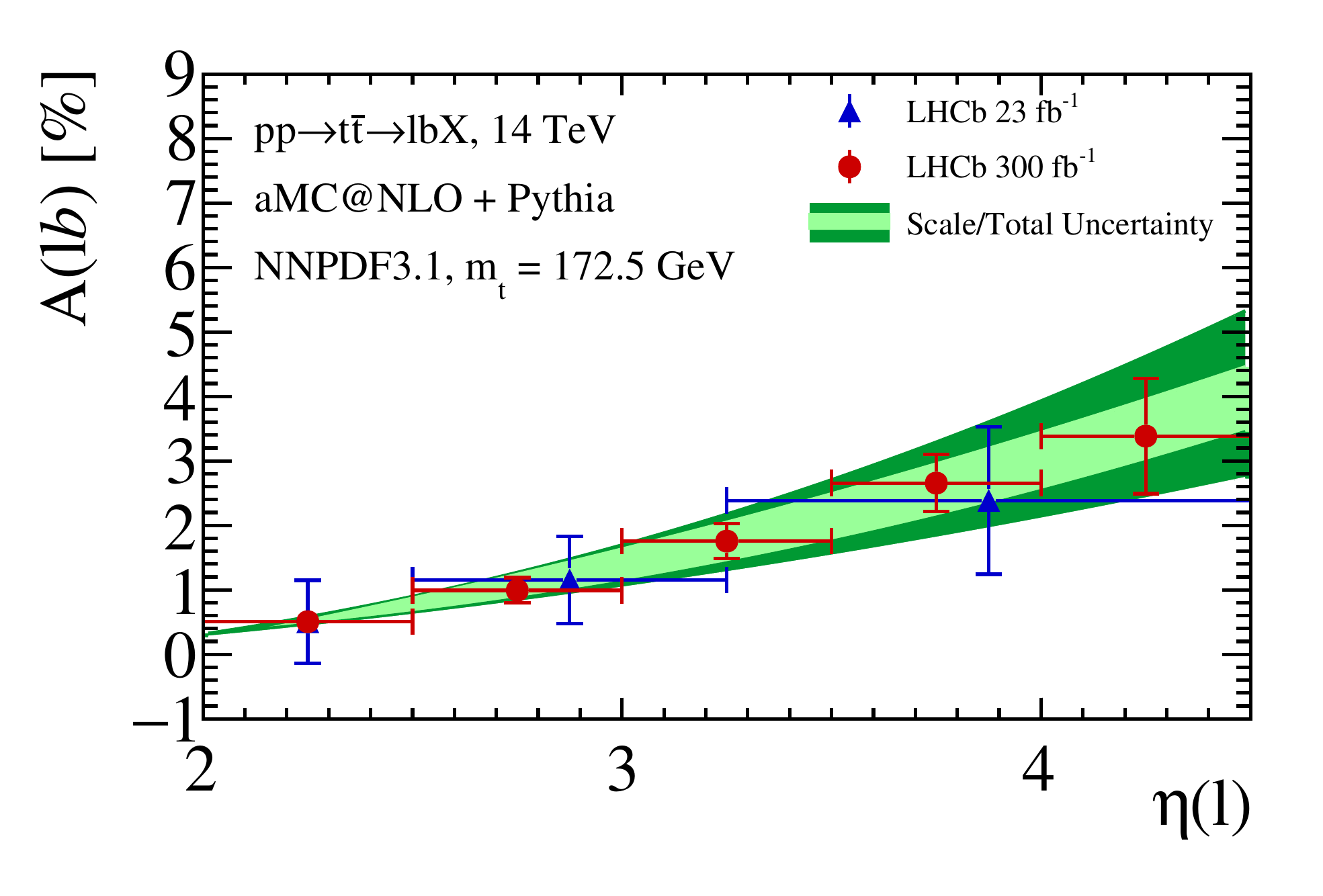}
\end{center}
\caption{
The predicted SM asymmetry at LHCb as a function of lepton pseudorapidity in the $\ell b$ final state at 14\:TeV. The bands show the uncertainty on the theoretical predictions due to scale variations (green) and due to combined scale, PDF and \as variations (yellow). The expected statistical precision on measurements performed by LHCb using 23 and 300\invfb of data is indicated by the uncertainty bars on the points.}
\label{fig:lhcb_top}
\end{figure}

\begin{table}
\begin{center}
\caption{
The  number of $t\bar{t}$ events reconstructed at LHCb per final state in current measurements and obtained using extrapolations to datasets corresponding to integrated luminosities of 23, 50 and 300\invfb. The mean value of  Bjorken-$x$ of the most energetic initiating parton is also shown for each final state.
}
\label{tab:topproj}
\begin{tabular}{c c c c c c }
\hline
final state & current & 23\invfb & 50\invfb & 300\invfb & $<x>$ \\
\hline
$\ell b$ & 220~\cite{LHCb-PAPER-2015-022}   & 54k & 117k & 830k & 0.295 \\
$\ell b \bar{b}$ & \phantom{0}24~\cite{LHCb-PAPER-2016-038} & 8k & 17k & 130k & 0.368 \\
$\mu e b$ & \phantom{0}38~\cite{LHCb-PAPER-2017-050} & 1k & 2k & 12k & 0.348 \\
$\mu e b \bar{b}$ & - & 120 & 260 & 1.5k & 0.415 \\
\hline
\end{tabular}
\end{center}
\end{table}

While uncertainties on current measurements in the top sector at LHCb are dominated by statistically uncertainties, the \upgradetwo data will permit precision measurements of both the top quark production cross-section and charge asymmetry. The expected number of top-pair events reconstructed at LHCb are given in Table~\ref{tab:topproj}, where the yields are obtained by extrapolating from current measurements, with increases of between 10 and 50\% in total efficiency assumed due to improvements in the \bquark-tagging algorithm and selection criteria. In addition, both muons and electrons are assumed to be employed for \upgradetwo due to anticipated improvements in electron performance. The cross-sections are calculated at next-to-leading order in perturbative QCD using the aMC@NLO generator with the parton shower provided by Pythia 8, and electroweak corrections are approximated as described in Ref.~\cite{Gauld:2014pxa}. They indicate that fiducial cross-section and asymmetry measurements will be made at subpercent statistical precision in the $\ell b$ final state, and at the percent level in the $\mu eb$ final state. 
The large yields will also allow differential measurements of the asymmetry to be made as a function of lepton pseudorapidity in the $\ell b$ and $\ell b\bar{b}$ final states to achieve the greatest sensitivity, with the former shown in Fig.~\ref{fig:lhcb_top}. A good knowledge of the background contributions and their asymmetry will also be required to reach the ultimate precision.

While the statistical precision of the $\mu eb$ final state will be lower, the higher purity of the sample, and the unambiguous identification of top pair events, presents the opportunity to make the most precise cross-section measurement at LHCb with an overall precision at the level of a few percent. An inclusive cross-section measurement at a precision of 4\% will provide reductions of over 20\% on the gluon PDF at large-$x$~\cite{Gauld:2013aja}, and more stringent constraints can be obtained by performing total and normalised differential $t\bar{t}$ production cross-section measurements, which will be possible with the \upgradetwo data.

%% file: CONTRIBUTIONS/8_Forward_and_high_pT_physics/8.2.tex
\section{Gauge-boson production and implications for PDFs}



LHCb has made an important series of measurements of \PW and \PZ boson production in proton-proton collisions at a range of energies~\cite{LHCb-PAPER-2012-008,LHCb-PAPER-2012-029,LHCb-PAPER-2012-036,LHCb-PAPER-2013-058,LHCb-PAPER-2013-062,LHCb-PAPER-2014-033,LHCb-PAPER-2014-055,LHCb-PAPER-2015-003,LHCb-PAPER-2015-021,LHCb-PAPER-2015-049,LHCb-PAPER-2016-011,LHCb-PAPER-2016-021,LHCb-PAPER-2017-024}. These include measurements using leptonic gauge boson decays (to final states including electrons, muons, or taus), and hadronic decays of the \PZ boson. Both inclusive and associated production measurements have been performed. These measurements rely on excellent luminosity determination (with the precision achieved below 1.5\%~\cite{LHCb-PAPER-2014-047}), powerful particle identification and tracking, jet reconstruction, and efficient jet-flavour identification~\cite{LHCb-PAPER-2015-016}. In the most precise LHCb cross-section measurements to date the knowledge of the luminosity determination is the dominant uncertainty, though this and other uncertainties cancel in measurements of cross-section ratios. In the following we explore the impressive reach of \upgradetwo in this important area of physics.

\begin{figure}[tp]
  \begin{center}
    \includegraphics[height=0.36\textheight]{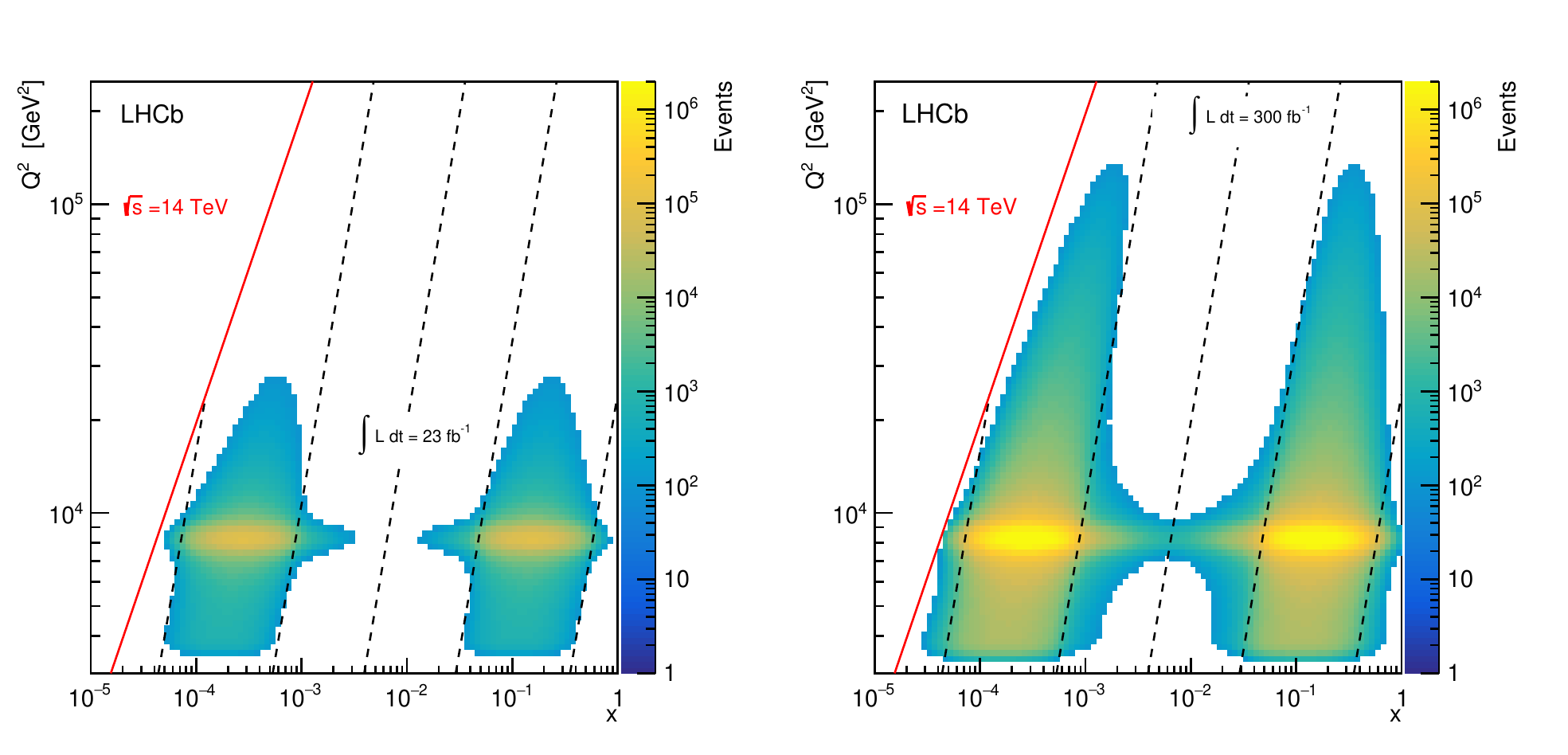} \\
  \end{center}
  \caption{The regions probed in the $x-Q^{2}$ plane with \PZ boson measurements at LHCb in 14~TeV proton-proton collisions. Integrated luminosities considered are (left) a dataset of $23\invfb$, considering only dimuon final states and (right) $300\invfb$, considering both dimuon and dielectron final states. Bins are only shown if the expected yield, determined using Pythia~8~\cite{Sjostrand:2014zea}, is greater than 100 events.
    The solid red line shows the kinematic limit set by the collision energy; the dashed lines show the values of $x$ and $Q^2$ associated with \PZ bosons with zero transverse momentum, and rapidities of 0, 2, and 4.5. 
     }
  \label{xQ2}
\end{figure}

The expected event yields and related information, are set out in Tables~\ref{tab:8.2.1_EW} and~\ref{tab:8.2.2_EW} for \PW and \PZ boson production, including high mass Drell-Yan, where one region ($m(\ell\ell)>200\gevcc$) is used to highlight typical quantities. Figure~\ref{xQ2} also shows the regions probed in the $x-Q^2$ plane for different integrated luminosities by measurements of (neutral) Drell-Yan production.
The \upgradetwo dataset is needed in order to make double-differential measurements as a function of the dilepton invariant mass and rapidity at high statistical precision if one wishes to maximise sensitivity in the regions where the PDFs have Bjorken-$x$ values close to 1, and to reach higher values of $Q^2$ than are currently accessible at high and low Bjorken-$x$. The same is also true for triple-differential measurements that probe the boson transverse momentum or measurements that probe high momentum jets. Differential measurements will also be made involving both $\PW^+$ and $\PW^-$ boson production and the ratio of their cross-sections, providing further constraints on PDFs. While such measurements of \PW boson production are already systematically limited, the new large dataset will allow the study and reduction of these uncertainties. In addition, following \upgradetwo a significantly improved ECAL (without low energy saturation) will allow better understanding of sample compositions in electron channels, further reducing systematic uncertainties on the ratio of the \PW boson cross-section in electron and muon channels, allowing a more precise test of lepton universality in \PW boson decays.

Measurements of gauge boson production in association with high momentum jets can also be used to place stringent constraints on the down-quark PDF at large-$x$, with a measurement at the 1\% level allowing a factor 3 reduction on the uncertainty~\cite{Farry:2015xha}. Differential measurements of such processes will give yet further sensitivity to higher values of $x$, possible with the high yields expected following Upgrade Ib and \upgradetwo.

\begin{table}[tb] \centering
\caption{\label{tab:8.2.1_EW}Expected yields in measurements at LHCb, and the higher of the two Bjorken-$x$ values probed. The yields for single boson processes are from NNLO calculations in pQCD, calculated for Ref.~\cite{LHCb-PAPER-2015-049}. The values of Bjorken-$x$ probed are determined using Pythia8. Ref.~\cite{LHCb-PAPER-2016-011} and Pythia8 are also used to extrapolate these cross-sections to those in the jet final states. The yields from electron and muon final states are summed for the 300\invfb yields (which assume the data collected is at 14~TeV). For the yields up to 50\invfb only muons are considered, as the current LHCb performance at high momentum is far superior for muons than it is for electrons. Other reconstruction effects are neglected, since the efficiencies are $\mathcal{O}(1)$.}
 \begin{tabular}{ccrrc}
\hline
  Process            &      & Additional                         &  Signal yields in       & $\langle x \rangle$  \\
  &      &  Criteria                          &   LHCb with 300\invfb     &  for $\sqrt{s} = 14\tev$  \\
                       &      &                 &   (50\invfb,23\invfb)     &  \\  \hline
  $Z/\gamma^{*}$     &      & $ 60<m(\ell\ell)<120\gev$           & $\sim 130\,(9,4)\times 10^6$     & 0.17 \\
      +jet               & & $p_\text{T}(\text{jet}) > 20\gev$   & $\sim 16\,(1,0.5)\times 10^6$      & 0.24\\
                     &      & $p_\text{T}(\text{jet}) > 100\gev$  & $\sim 500\,(30,14)\times 10^3$     & 0.38 \\\hline
  High-mass DY&      & $ m(\ell\ell)>200\gev$             & $\sim 120\,(9,4) \times 10^3$    & 0.32 \\\hline
  $W^{+}$             &     &                                     & $\sim 1400\,(109,50)\times 10^6$  & 0.16\\
  +jet                 &  & $p_\text{T}(\text{jet}) > 20\gev$   & $\sim 170\,(13,6)\times 10^6$    & 0.24 \\
                      &      & $p_\text{T}(\text{jet}) > 100\gev$   &  $\sim 4\,(0.4,0.2)\times 10^6$     & 0.35 \\\hline
  $W^{-}$           &      &                                     & $ \sim 1000\,(76,35)\times 10^6$   & 0.11\\
  +jet                &   & $p_\text{T}(\text{jet}) > 20\gev$    & $\sim 100\,(6.5,3)\times 10^6$     & 0.18 \\
                    &         & $p_\text{T}(\text{jet}) > 100\gev$ &$\sim 2\,(0.13,0.06)\times 10^6$         & 0.31\\
 \hline
\end{tabular}
\end{table}
\begin{table}[tb] \centering
\caption{\label{tab:8.2.2_EW}This table contains similar information to Table~\ref{tab:8.2.1_EW}, with the same dilepton invariant mass cuts, but considers inclusive measurements made differentially as a function of the $Z/\gamma^{*}$ boson rapidity. Bjorken-$x$ values probed in \PW boson events are similar to those shown here for Drell-Yan events at the Z peak.}
\begin{tabular}{ccrc}
\hline 
  Process      &   Rapidity Bin  &  Signal yields in  & $\langle x \rangle$ \\
               &                        &  LHCb with 300\invfb & \\ \hline
  $Z/\gamma^{*}$ & [2.0,2.5]       &         $\sim 20\times 10^6$                           &    0.08   \\
                 & [2.5,3.0]       &         $\sim 40\times 10^6$                           &    0.12   \\
                 & [3.0,3.5]       &         $\sim 50\times 10^6$                           &    0.18   \\
                 & [3.5,4.0]       &        $\sim 20\times 10^6$                            &    0.28   \\
                 & [4.0,4.5]       &        $\sim 4\times 10^6$                            &     0.42  \\\hline
  High-mass DY   & [2.0,2.5]  &           $\sim 30\times 10^3$                         &    0.21   \\
                 & [2.5,3.0]       &           $\sim 60\times 10^3$                         &    0.30   \\
                 & [3.0,3.5]       &           $\sim 30\times 10^3$                         &    0.43   \\
                 & [3.5,4.0]       &           $\sim 3\times 10^3$                         &     0.61  \\
                 & [4.0,4.5]       &               $10^1$                   &    0.84   \\\hline
\end{tabular}
\end{table}

Measurements of electroweak boson production in association with heavy-flavour jets (and mesons) can similarly be made. Existing LHCb measurements of these cross-sections have been performed, integrated over the full LHCb acceptance.  With the larger sample sizes expected following the LHCb upgrades, differential measurements can be made in these final states that probe the variation of strange PDFs with Bjorken-$x$. Measurements of \PZ+charm production at LHCb will also be made differentially. The ratio of the cross-sections of \PZ+charm to \PZ+jet events at LHCb varies between about 8\% and 3\% as a function of the \PZ boson rapidity; measurements made using the Run 3 dataset will typically be sensitive to intrinsic charm with a mean value of $x$ greater than 0.3\% (sea-like intrinsic charm) and 1\% (valence-like intrinsic charm)~\cite{Boettcher:2015sqn}. Improvements in sensitivity are expected following Upgrades Ib and II, since the larger sample yields will enable double-differential measurements that probe higher $p_\text{T}$ jets (placing constraints on higher $x$ values) while binning in the \PZ boson rapidity. Such measurements will allow the study of the $x$ and $Q^2$ dependence of the charm PDF and constrain the uncertainty from intrinsic charm on the Higgs boson production cross-section at the LHC to below 1\%. Knowledge of the proton's intrinsic charm content is also crucial for any measurement of the coupling of the Higgs boson to charm quarks using the Higgs + charm final state~\cite{Boettcher:2015sqn}.

With an improved ECAL following \upgradetwo, final states containing high-$E_\text{T}$ photons also become reconstructible at high precision. There are many such final states of interest, including $\gamma$+HF production. This final state is sensitive to the intrinsic charm and beauty content of the proton, with measurements at large-$x$ considered particularly useful and important~\cite{Rostami:2015iva,Boettcher:2015sqn}. This measurement at high precision and high photon momentum will only be possible with the upgraded detector. In addition, \upgradetwo will bring increased yields for diboson production within the LHCb acceptance. The cross-section for $WW$ production in the LHCb acceptance (with an electron-muon final state) is about 100\fb, with the cross-section for $WZ$ and $ZZ$ production roughly $10-100$ times smaller (depending on selection requirements and the final state considered). The full \upgradetwo dataset is therefore needed to measure these rare processes in the forward LHCb acceptance at high precision. Such measurements will rely on the improved calorimetry in order to identify the electrons accurately and suppress the backgrounds from misidentified hadrons.

%% file: CONTRIBUTIONS/8_Forward_and_high_pT_physics/8.3.tex
\section{Measurement of the effective weak mixing angle}

The effective weak mixing angle, \ssqtwef, is a fundamental parameter of the Standard Model. It defines the ratio of vector and axial-vector couplings of the \Z boson to fermions, and can be determined experimentally from a fit to the angular distribution (specifically the forward-backward asymmetry) of \Z boson final states. Determinations of the weak mixing angle have been made by experiments at LEP, SLC, the Tevatron and LHC. The two most precise determinations (from LEP and SLD) show an almost three standard deviation difference, which is important to resolve.
LHCb has two characteristics that benefit measurements of the weak mixing angle. At the higher \Z rapidities inside LHCb's forward acceptance the forward-backward asymmetry is larger and easier to measure than in the central region covered by the general purpose LHC detectors. In this forward kinematic region \Z boson production is also better constrained theoretically, owing to the predominance of up and down quark initial states. 

\subsection{Current precision}
\label{s2w_current}
LHCb has previously measured the forward-backward asymmetry of muon final states in 3 fb$^{-1}$ of $\sqrt{s}=7$ and 8 TeV data, and used this to determine a value for the effective weak mixing angle~\cite{LHCb-PAPER-2015-039}.
The forward-backward asymmetry measured in data is unfolded to particle level and fitted to a series of simulated templates generated at different values of effective weak mixing angle. A $\chi^2$ value is obtained for each fit. The minimum $\chi^2$ value defines the measured effective weak mixing angle. With the Run 1 dataset, the angle was determined to be $\ssqtwef = 0.23142 \pm0.00073 \pm 0.00052 \pm 0.00056$, where the first uncertainty is statistical, the second systematic (predominately momentum scale, with smaller contributions from background asymmetries, unfolding and simulated mass resolution), and the third theoretical (PDF, scale, $\alpha_s$ and FSR uncertainties). Both the systematic and theoretical uncertainties contain statistical components, which will reduce when larger datasets are analysed. The total uncertainty ($100 \times 10^{-5}$) is over six times larger than that of the combined LEP and SLD results.

\subsection{Future improvements}
\label{s2w_future}

The \Z boson production cross-section is approximately twice as large at Run 2 compared to Run 1.  If the current dimuon analysis is simply repeated, the statistical uncertainty on \ssqtwef should reduce to about $13 \times 10^{-5} \, (5 \times 10^{-5}$) by the end of Run 4 (Run 5). With the addition of dielectron final states (with which \Z production has been measured~\cite{LHCb-PAPER-2016-021} but no corresponding LHCb \ssqtwef measurement has yet been made) the statistical uncertainty should be further reduced, easily sufficient to probe the differences seen in the LEP and SLD determinations.

However, work must be done to reduce the systematic and theoretical uncertainties of the current analysis. 
The largest systematic uncertainty arises from the momentum scale and curvature biases, and its size reflects
the statistical precision of the dataset used to control it.
It is expected that this uncertainty will reduce as larger datasets are analysed (although the ultimate uncertainty is not yet known). The systematic uncertainty may reduce further if a combined dimuon and dielectron measurement is made, given that systematic uncertainties will be partially correlated between datasets. Studies are ongoing to quantify expected performance gains.

The largest theoretical uncertainty is due to the underlying parton density functions. ATLAS have studied {\it in situ} parton density constraining techniques to reduce the resulting uncertainty in their measurement.
Here the parton density function 68$\%$ uncertainty band is constrained to that portion which is consistent with the ATLAS data. Similar studies by CMS imply that parton density function uncertainty can be constrained to less than $10^{-4}$.
Given that parton uncertainties are smaller within LHCb's acceptance, the same statement should apply if these techniques are adopted. Studies are ongoing to quantify these effects for LHCb. Other sources of theoretical uncertainty are small, and will reduce when larger samples of simulated data are used to quantify them.

Finally, the analysis technique itself can be improved. Candidate events can be weighted, to improve sensitivity to the weak mixing angle, and it may be possible to improve sensitivity by binning and fitting events in rapidity as well as dilepton mass. The benefits of extending the analysis acceptance (from $\eta> 2$ to $>1.9$, and $\eta<4.5$ to $<5$) remain to be evaluated, 
and the impact of slightly increased pile-up in the upgrade needs to be quantified before final projections can be made.
However, it is clear that the \upgradetwo samples can enable an impressive sensitivity to \ssqtwef.

%% file: CONTRIBUTIONS/8_Forward_and_high_pT_physics/8.4.tex
\section{Prospects for $W$-mass measurement}

The current global EW fits yield an indirect determination of $m_W$ with an uncertainty of 8~\mev~\cite{Baak:2014ora}.
This is a factor of two more precise than the current world average of direct measurements,
$m_W = 80\,385 \pm 15$~\mev, which is dominated by measurements from the 
CDF~\cite{Aaltonen:2012bp} and D0~\cite{Abazov:2012bv,D0:2013jba} experiments.
This average does not yet include a recent first measurement by ATLAS of $m_W = 80\,370\pm 19$~\mev~\cite{Aaboud:2017svj}.
The CDF and D0 results were still statistically limited, being based on samples of $\mathcal{O}(10^{6})$ leptonic $W$ decays
and $\mathcal{O}(10^{5})$ leptonic $Z$ decays.
A natural target for the ultimate precision on the average of $m_W$ measurements from the LHC experiments is around 5~\mev~\cite{Baak:2014ora}.
The ATLAS and CMS experiments already have access to signal samples that are more than two orders of magnitude larger than those of the Tevatron experiments.
Their measurements of the $W$ mass are however limited by theoretical uncertainties in the modelling of $W$ and $Z$ boson production in $pp$ collisions.

In Ref.~\cite{Bozzi:2015zja} it was pointed out that a measurement of the $W$ mass with the LHCb experiment would be highly desirable due to the
complementary lepton acceptance $2 < \eta < 5$, which implies a partial anti-correlation between the PDF uncertainties as compared to the central-rapidity ATLAS and CMS experiments.
It was estimated that the $\mathcal{O}(10^{7})$ $W$ events of the Run 2 dataset could yield a $m_W$ measurement with a statistical precision of around 10 MeV.
A measurement at forward rapidities presents several additional advantages.
For example heavy-flavour annihilation contributions to $W$ and $Z$ production are substantially smaller than at central rapidities.
The boson \pt spectra tend to be softer which implies a more direct relation between $m_W$ and the charged lepton \pt spectrum.
While the large rapidity $W$ and $Z$ bosons carry high intrinsic physics value, they are produced with a smaller cross section compared to the central experiments. 

The \upgradetwo samples will permit a statistical precision of a few MeV and can impose tight {\em in situ} constraints
on the systematic uncertainties associated to the $W$ production model.
The upgraded ECAL will allow a similarly precise measurement with orthogonal experimental uncertainties using $W \to e\nu$ decays.
These capabilities will be crucial in the realisation of the ultimate precision on $m_W$ with the LHC.

%% file: CONTRIBUTIONS/8_Forward_and_high_pT_physics/8.5.tex
\section{Measurement of Higgs decays to $c\bar{c}$}

\begin{figure}[t]
  \begin{center}
    \includegraphics[width=0.49\textwidth]{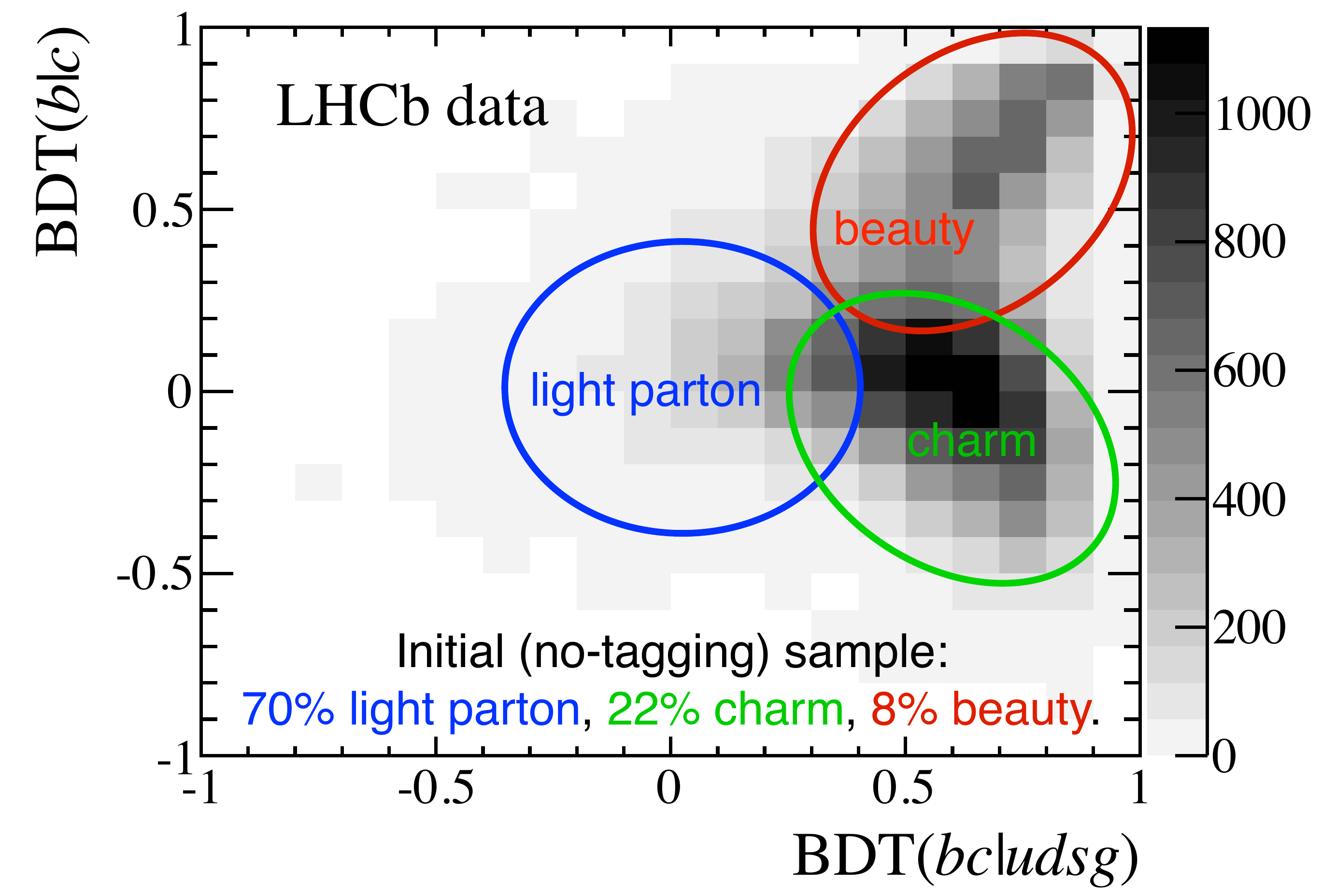}
    \includegraphics[width=0.50\textwidth]{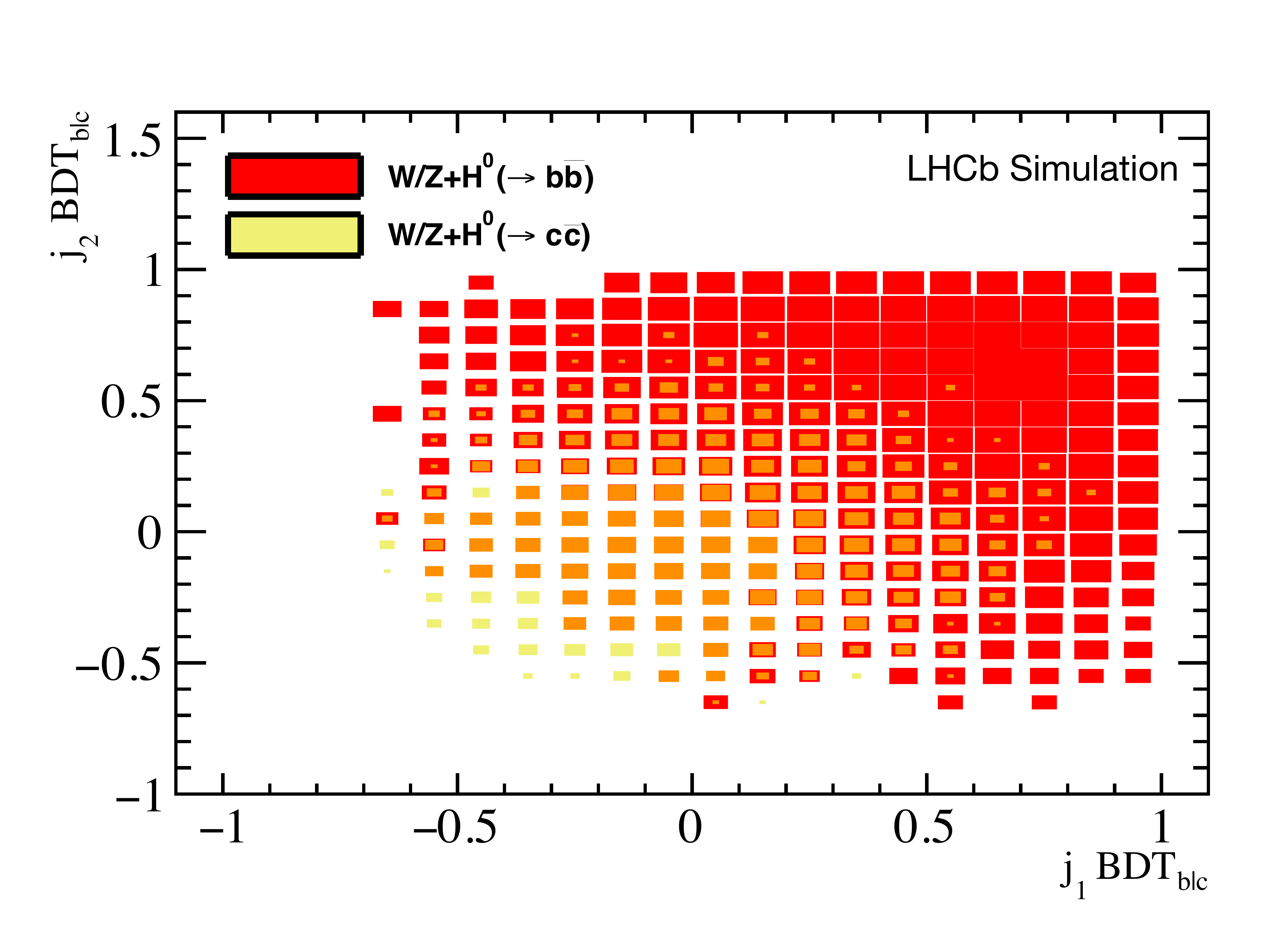}
  \end{center}
  \caption{\label{fig:Hcc}
  (left) From a $b$-jet and $c$-jet enriched data sample of Ref.\,\cite{LHCb-PAPER-2015-016}, the 2-D SV-tagger BDT response per-jet observed in data (annotation added here to show roughly where jets of each type are found).
  (right) From Ref.~\cite{LHCb-CONF-2016-006}, the SV-tagger BDT that separates $b$ and $c$ for the sub-leading versus leading jet for $VH(b\bar{b},c\bar{c})$ assuming SM-like Yukawa couplings.
  }
\end{figure}

In Run~1, LHCb developed a powerful heavy-flavour jet-tagging procedure~\cite{LHCb-PAPER-2015-016} that leveraged the world-leading performance of the VELO.
This method employed two Boosted Decision Trees (BDTs): one to separate heavy-flavour jets from those initiated by light partons, and one to separate beauty jets from charm jets.
The efficiency for identifying $b$-jets and $c$-jets is roughly 65\% and 25\%, respectively, with $\approx 0.3\%$ probability for misidentifying a light-parton jet as originating from heavy flavour.
More importantly for $H\to c\bar{c}$ sensitivity, there is clear separation between $b$-jets and $c$-jets in the 2-D BDT plane (see Fig.~\ref{fig:Hcc}).
This jet-tagging algorithm was used to study $Wb$ and $Wc$~\cite{LHCb-PAPER-2015-021},
$Wb\bar{b}$ and $Wc\bar{c}$~\cite{LHCb-PAPER-2016-038} production,
and to make the first observation of top-quark production in the forward region~\cite{LHCb-PAPER-2015-022}, all using Run~1 data,
and for a study of top-quark production using Run~2 data~\cite{LHCb-PAPER-2017-050}.
A similar measurement of $Zc$ production at LHCb, proposed in Ref.~\cite{Boettcher:2015sqn} to study the large-$x$ (intrinsic) charm content of the proton, is currently in progress.

This jet-tagging method was also used to set upper limits on $\sigma[pp\to VH(b\bar{b},c\bar{c})]$, where $V \equiv W$ or $Z$, using 2\invfb of 8\tev Run~1 data that translate into upper limits on the Yukawa couplings of $y^b < 7 y_{\rm SM}^b$ and $y^c < 80 y_{\rm SM}^c$, respectively~\cite{LHCb-CONF-2016-006}.
The $VH$ production cross section increases by a factor of 7 within the LHCb acceptance going from 8 to 14\tev; therefore, assuming no improvements in the detector performance or analysis, LHCb could set an upper limit on the charm Yukawa coupling of $\approx 7 y_{\rm SM}^c$ after collecting 300\invfb of data at 14\tev.
Ref.~\cite{LHCb-CONF-2016-006} required that both $c$-jets satisfied a stringent tagging criterion, resulting in a di-$c$-jet tagging efficiency of only about 2\%.
The improved VELO performance in LHCb Upgrade I is expected to increase the $c$-tagging efficiency from 25\% to 30\% for the same $b$-jet and light-parton mistagging probabilities.
If the di-$c$-jet tagging strategy is altered to require that (at least) one $c$-jet satisfies the standard tagging criteria, where the light-jet mistag probability is 0.3\%, while the other $c$-jet satisfies a looser tag, then the di-$c$-jet tagging efficiency can be dramatically increased to $\approx 30\%$ with minimal impact on the background composition.
Given that $\sigma(Wc\bar{c})$ and $\sigma(Wcj)$ are of the same order of magnitude when the di-jet mass is Higgs-like, the use of a much looser $c$-tag on one jet will not have a large impact on the total background yield.
Furthermore, since $\sigma(Wbc)$ is small compared to $\sigma(Wc\bar{c})$ and $\sigma(Wb\bar{b})$, it is only necessary to discriminate $b$ from $c$ for one jet.
Using this jet-tagging strategy improves the predicted sensitivity to $\approx 4 y_{\rm SM}^c$.
If the LHCb electron reconstruction performance can be improved in the upgrade such that it is close to that achieved for muons, the expected sensitivity is $\approx 3 y_{\rm SM}^c$.
Finally, if the separation between $b$ and $c$ jets can be further improved, such that only backgrounds consisting of $c$-jets are important, the ultimate sensitivity is about $\approx 2 y_{\rm SM}^c$.
This could be achieved, \eg by using deep-learning algorithms and including more low-level information than was done in the Run~1 tagging algorithm.
Large gains in performance have been achieved by ATLAS and CMS following this strategy, and we expect that LHCb could achieve similar gains in performance in the future.
In summary, LHCb UII has the potential to place the most stringent constraint on $y^c$ from the HL-LHC.

%% file: CONTRIBUTIONS/8_Forward_and_high_pT_physics/8.6.tex
\section{Searches for prompt and detached dark photons}

A new $U(1)$ dark force, analogous to the electromagnetic (EM) force, can be introduced into the SM, where the dark photon, $A'$, is the corresponding force mediator which couples to dark matter (or matter) carrying dark charge. However, the $A'$ can kinetically mix with the photon, allowing the $A'$ to be observed in the spectra of final states produced by the EM current. Results based on the $A'$ model can also be recast to any vector model~\cite{Ilten:2018crw}, including $B-L$, leptophobic, and protophobic currents.

\begin{figure}[t]
  \begin{center}
    \includegraphics[width=\textwidth]{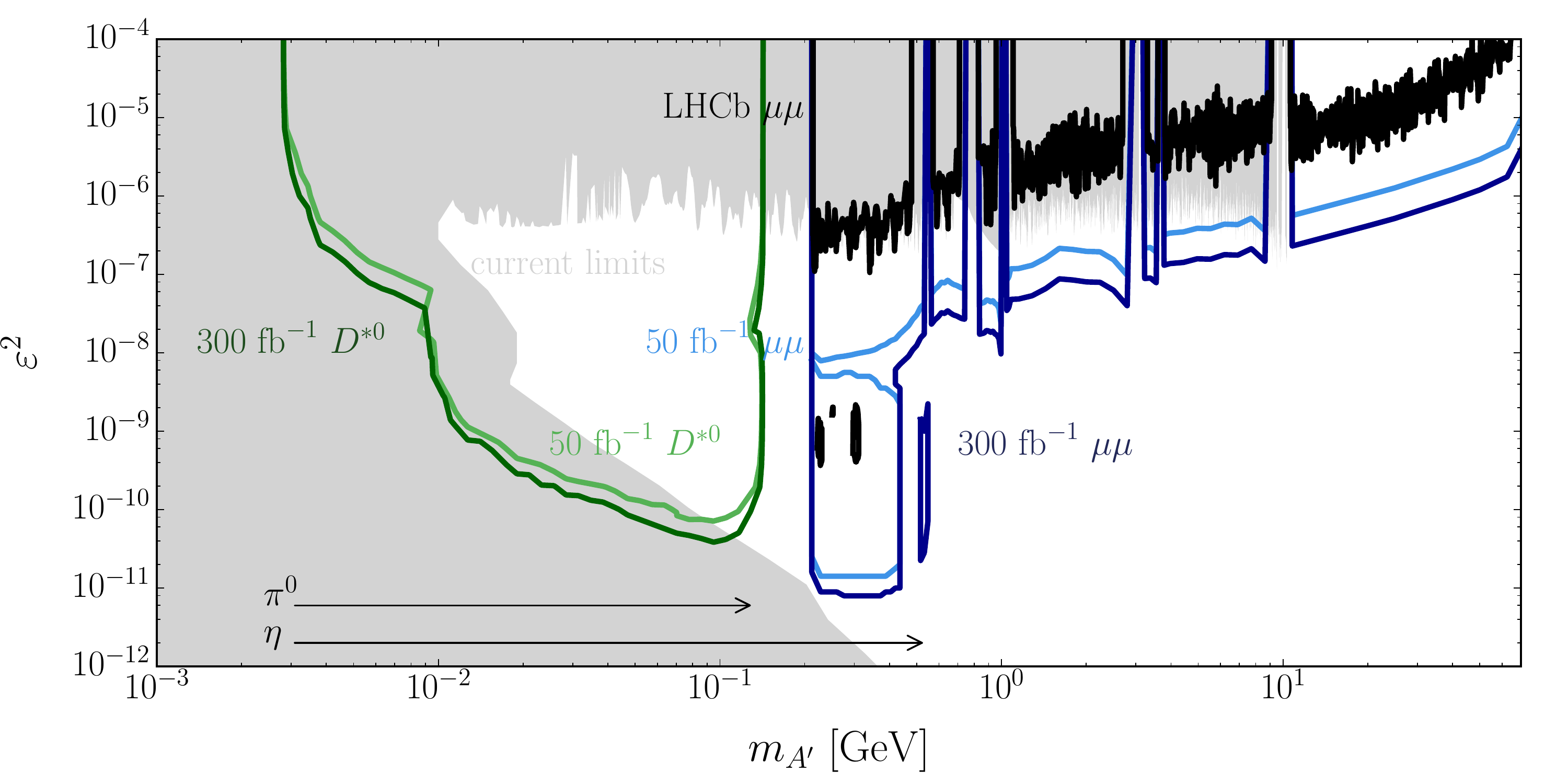}
  \end{center}
  \caption{Current limits (grey fills), current LHCb limits (black band), and proposed future
    experimental reach (coloured bands) on $A'$ parameter
    space. The arrows indicate the available mass range from light meson decays into $\ep \en \gamma$.\label{fig:DarkPhoton}}
\end{figure}

The $A'$ model has two free parameters: the mixing term $\varepsilon^2$ and the mass of the $A'$, $m_{A'}$, where $\epsilon^2$ is the ratio of the dark force strength to the EM force strength. In Fig.~\ref{fig:DarkPhoton} this parameter space is plotted with current limits (grey fills), current LHCb limits (black bands)~\cite{LHCb-PAPER-2017-038}, and possible LHCb future reach (coloured bands). The coloured light and dark bands correspond to discovery reach assuming $50$, and $300\invfb$ datasets, respectively.

LHCb can explore significant portions of unconstrained $A'$ parameter space. These searches are based on two strategies: prompt and displaced resonance searches using $\Dstarz \to \Dz \ep \en$ decays~\cite{Ilten:2015hya} (green bands); and inclusive dimuon production~\cite{Ilten:2016tkc} (blue bands). In both cases the lepton pair is produced from an EM current which kinetically mixes with the $A'$, producing a sharp resonance at the $A'$ mass. These search strategies depend on three core capabilities of LHCb: excellent secondary vertex resolution, particle identification, and real-time data-analysis. Production of $A'$ bosons from light meson decays (arrows), $\pi^0 \to \ep \en \gamma$ and $\eta \to \ep \en \gamma$, will cover the gap between the two primary search strategies.

A search for $A'$ bosons  decaying into muon pairs was performed by the LHCb collaboration \cite{LHCb-PAPER-2017-038} (black band), using a data sample corresponding to an integrated luminosity of 1.6~\invfb from $pp$ collisions taken at $\sqrt{s}=13$~\tev. This search already produced world-best upper limits in regions of $\varepsilon^2-m_{A'}$ space and is the first simultaneous prompt and displaced $A'$ search. 
By the end of \upgradetwo, LHCb will either confirm or reject the presence of a dark photon for nearly all relevant parameter space.



%% file: CONTRIBUTIONS/8_Forward_and_high_pT_physics/8.7.tex
\section{Searches for semileptonic and hadronic decays of long-lived particles}

\begin{figure}[t]
\begin{minipage}[b]{0.81\textwidth}
  \includegraphics[width=0.49\textwidth]{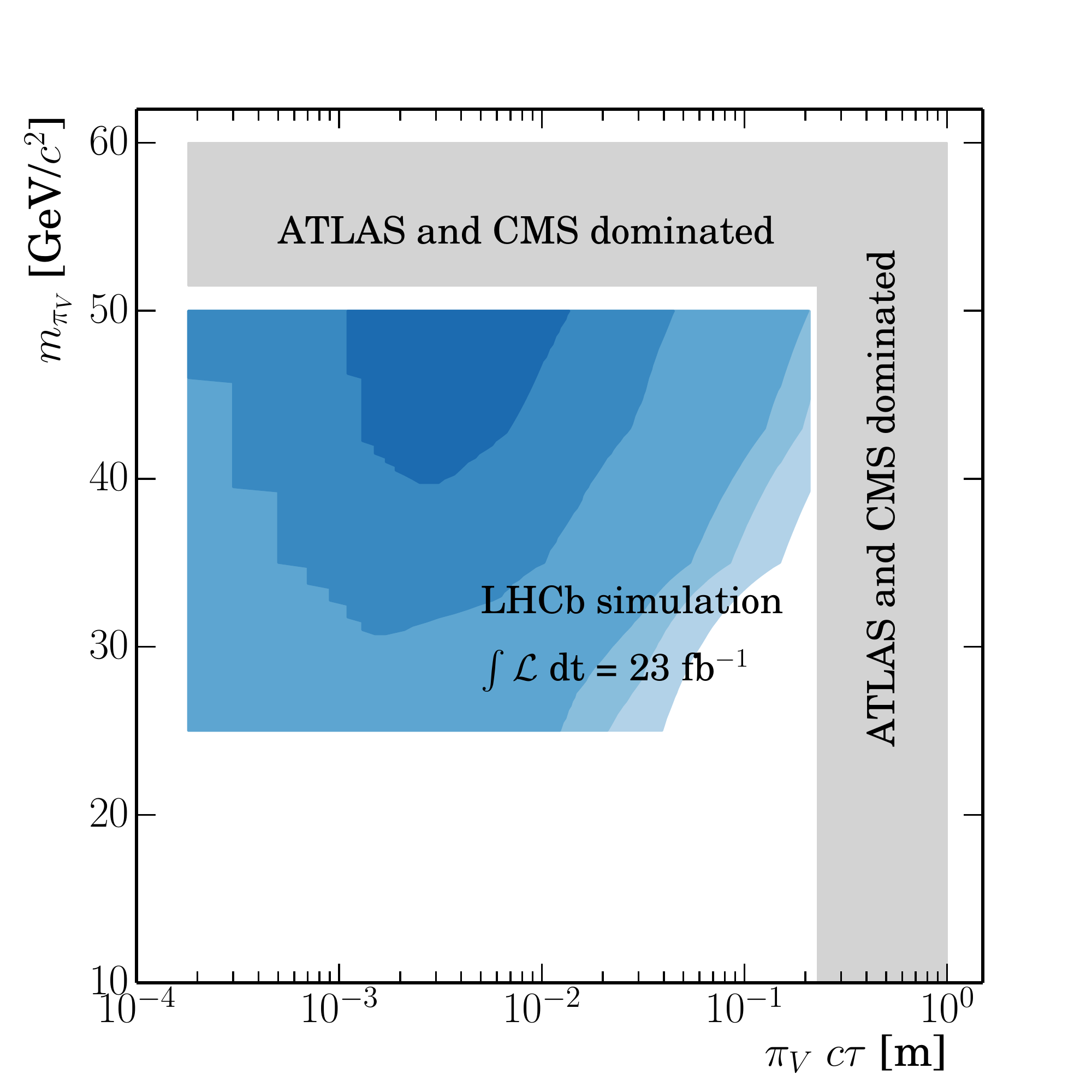}
  \includegraphics[width=0.49\textwidth]{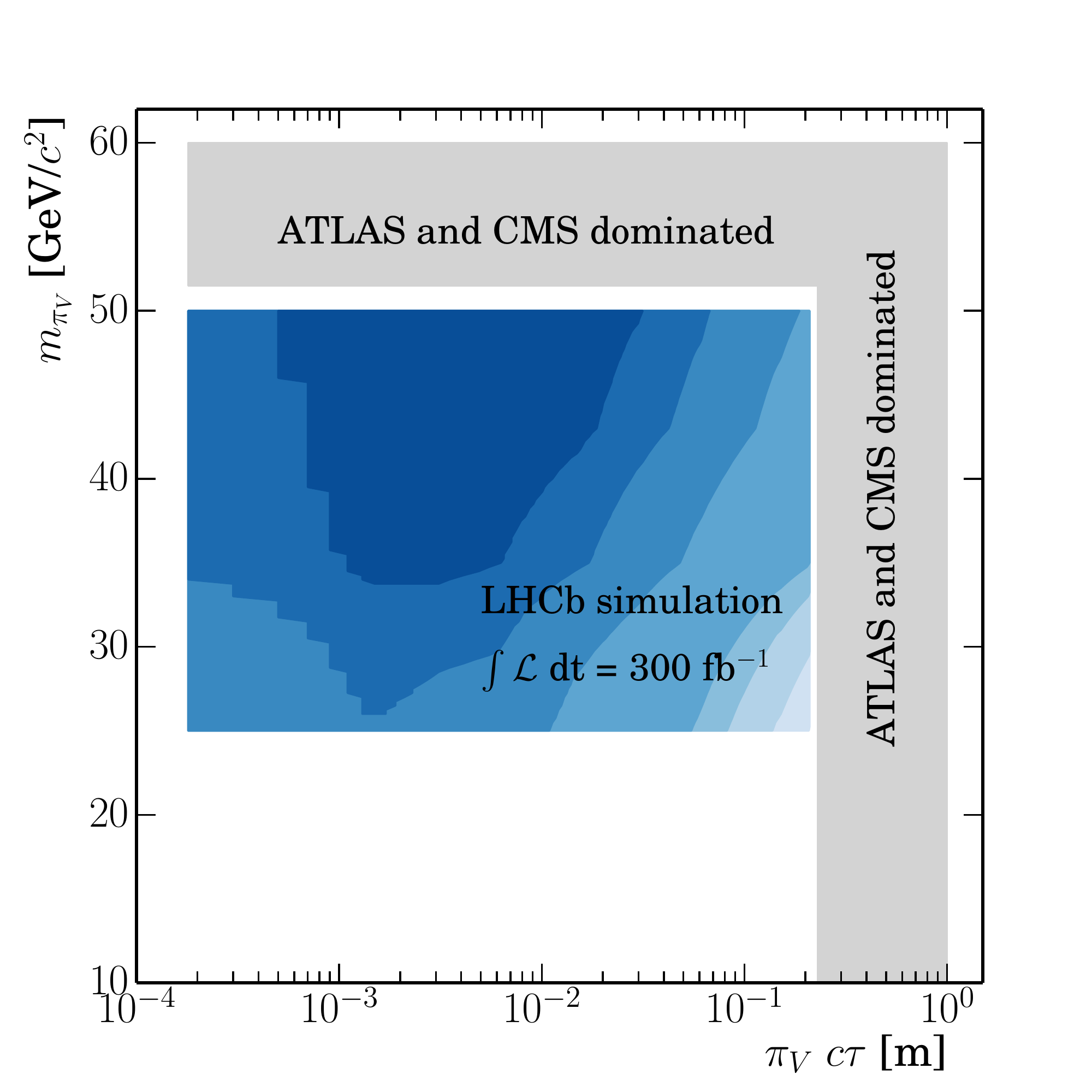}
\end{minipage}
\begin{minipage}[t]{0.18\textwidth}
  \vspace{-4.6cm}
  
  \hspace{-1.cm}
\includegraphics[width=1.2\textwidth]{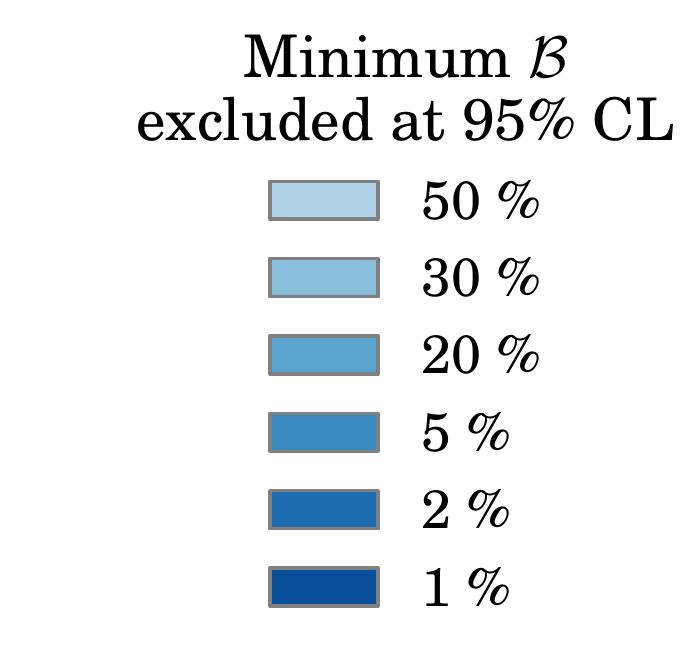}
\end{minipage}
  \caption{Projected sensitivities to a Hidden Valley model, extrapolated from Run 1 results with the displaced di-jet final state~\cite{LHCb-PAPER-2016-065}, for luminosities of (left) 23\invfb and (right) 300\invfb.}
  \label{fig:dvsearches_hllhc}
\end{figure}

Several NP scenarios predict the existence of long-lived particles (LLP) coupling to SM particles via different mechanisms. One interesting type of mechanism is that which involves exotic decays of the SM Higgs boson. In the context of this production mode, and depending on the decay of the LLP itself, different signatures may be considered: LHCb has searched, using Run 1 data, for supersymmetric neutralinos decaying semileptonically into a high-$\pt$ muon and a jet~\cite{LHCb-PAPER-2016-047}; and for Hidden Valley~\cite{Strassler:2006im, Pierce:2017taw} pions decaying hadronically into a pair of jets~\cite{LHCb-PAPER-2016-065}. These analyses have shown the potential of the LHCb experiment to search for these kind of signatures, especially in the low mass and low lifetime region, unexplored by other experiments at the present moment. Conservative extrapolations of these results to the integrated luminosity foreseen to be recorded by LHCb at the end of Run 5 are presented in Fig.~\ref{fig:dvsearches_hllhc}. 
%
%
Excellent reconstruction of displaced vertices and their associated tracks is crucial, as is the need to keep under control the dominant background contributions and pile-up effects.

Most of these searches use tracks which traverse the full LHCb spectrometer. These tracks have an excellent spatial and momentum resolution, and roughly correspond to LLP decaying within the VELO region. However, for long-lived candidates decaying outside or close to the boundaries of the VELO region, the tracks do not have hits in the VELO. 
Unfortunately, these so called ``downstream'' tracks have worse vertex and momentum resolution, limiting the capabilities of LHCb for this displacement range. Efforts to develop new trigger lines to select downstream tracks are summarised in Ref.~\cite{Aaij:2244312}. Among with these track types, upstream tracks are also considered useful for LLP searches: a proposal to add Magnet Stations (MS) inside the LHCb magnet to improve low momentum resolution can be found in Ref.
%

The  RF foil strongly affects the background composition of LLP searches in the LHCb experiment. Namely, for LLP candidates decaying below 5 mm from the beam line, the main source of background is due to heavy-flavour decays; while material interactions with the RF foil compose the main background contribution for those LLP decaying above 5 mm from the beam line. While the former is purely due to QCD processes and hence not reducible, the latter is kept under control by the use of a very detailed veto map~\cite{LHCb-PAPER-2017-038}. However, the removal of the RF foil would further enhance the sensitivity to this kind of analyses, since it would both significantly increase the impact parameter (IP) resolution and reduce the background due to material interactions. 
%
%

The impact of pile-up has to be studied in more detail (especially for those searches involving jets in the final state). 
There are some ideas under consideration, for example the removal of neutral particles, which are more pile-up sensitive than charged tracks, from the jet reconstruction.
Machine-learning pile-up mitigation methods are also being studied.
The fast-timing capabilities of the TORCH will be invaluable in suppressing combinatoric background.

The search for exotic massive long-lived particles decaying into a pair of jets suffers for low efficiency in low-mass region, \ie below 20 GeV/$c^2$. At low masses the final state cannot always be reconstructed as two resolved jets but as a single jet (merged jet). Reference~\cite{Bokan:1753426} shows that the substructure of such merged jets can be exploited to improve the selection efficiency. More studies are needed, in particular for the development of new tagging algorithms for the identification of merged jets over the multi-jet background. CMS already showed promising results in this field~\cite{CMS-DP-2017-027} by employing machine learning techniques, in particular Boosted Decision Trees (BDT) and Deep Neural Networks. We expect that for the \upgradetwo similar techniques will be used at LHCb.
Ultimately \upgradetwo should be able to probe much of the currently unexplored parameter space.

%% file: CONTRIBUTIONS/9_Exotic_hadrons_and_spectroscopy_with_heavy_flavours/9.tex
\label{chpt:spect}

\def\Qbar    {{\kern 0.2em\overline{\kern -0.2em \PQ}{}}\xspace}

The particle zoo of 1960s led to the quark hypothesis,
which brought 
the interpretation of hadrons as colourless states with a minimal quark content of \qqbar and \quark\quark\quark for mesons and baryons respectively. A decade later, the discovery of charmonium, and later of bottomonium, states
solidified the quark model and reinforced the \qqbar picture of mesons. The quark model~\cite{GellMann:1964nj,Zweig:352337} conceived also of the existence of states made of four and five quarks --- referred to as tetraquarks and pentaquarks hereafter --- collectively named exotic hadrons.
Although the observed number of light scalar mesons exceeds by far the expected \qqbar states, it is still not proved which (or if any) of them are indeed multiquark states. Pentaquarks states, such as $\PZ_0(1780)$ or $\Theta(1540)^+$ baryons, have been claimed in the light baryon sector but none of them has been firmly confirmed~\cite{Hicks:2012zz}.

In 2003 the discoveries of two puzzling states, $\PD^*_{\squark0}(2317)^+$~\cite{Aubert:2003fg} and $\PX(3872)$~\cite{Choi:2003ue}, renewed the interest for exotic spectroscopy in the heavy quark sector. The following years have been witness to a proliferation of new states
with masses at and above the open heavy flavour thresholds 
 and with odd properties or explicit sign of four-quark content, $\PQ\Qbar\quark\quarkbar$ ($\PQ=\cquark$ or \bquark, and $\quark=\uquark$, \dquark or \squark).  
The discovery of \jpsi\proton structures by \lhcb~\cite{LHCb-PAPER-2015-029}, with \cquark\cquarkbar \uquark\uquark\dquark quark content, broadened the crisis to baryon spectroscopy. 


While the existence of hadrons with multi-quark content 
had been anticipated from the early years of the quark model, 
an understanding of their dynamics is presently lacking.
Many states appear near the meson-antimeson (also meson-baryon) thresholds
giving raise to molecular interpretations~\cite{Guo:2017jvc,Karliner:2017qhf,Oset:2016lyh}, or models involving hadron
rescattering (cusps or triangle singularities)~\cite{Lebed:2016hpi,Bayar:2016ftu}.
However, some exotic hadron candidates cannot be explained in this manner. 
 Models binding many quarks in the same confining volume have also been proposed~\cite{Esposito:2016noz,Ali:2017jda}.
In fact, the structure of QCD suggests attractive forces in antisymmetric quark-quark colour-antitriplet combinations,
which can replace the simplest colour-antitriplet \ie antiquark,
as a building block of hadrons.
Similarly, an antiquark-antiquark pair with the right symmetry can take the place of a quark.
The diquarks can play important role in the internal structure of ordinary baryons made of three quarks.
They also naturally lead to emergence of tightly bound diquark-antidiquark tetraquarks,
diquark-diquark-antiquark pentaquarks or structures with even more quarks in one confining volume.
Due to the nonperturbative regime of QCD,
it is difficult to make first-principle predictions of excitation spectrum and widths of such multiquark states.
Therefore, it is essential for experimental results to inform QCD-motivated phenomenology.
Such studies may also impact upon the phenomenology of any strongly coupled extension of the Standard Model,
or our understanding of possible states of matter in the early universe or under extreme gravitational conditions.

The \lhcb experiment has already strongly contributed to the field by (\eg) determining the quantum numbers of the $\PX(3872)$ meson~\cite{LHCb-PAPER-2013-001,LHCb-PAPER-2015-015} and observing  two pentaquark candidates, $\PP_\cquark(4380)^+$ and $\PP_\cquark(4450)^+$~\cite{LHCb-PAPER-2015-029}. The large data set collected in the \upgradetwo era will boost sensitivity in searches for heavy states with small production cross sections and/or suppressed decay rates. For example the search for charmonium-like states decaying to the \etac meson will become attractive. 
If history is of any guide, we expect many more surprising observations, which will challenge phenomenologists. 

\input{CONTRIBUTIONS/9_Exotic_hadrons_and_spectroscopy_with_heavy_flavours/9.1.tex}

\input{CONTRIBUTIONS/9_Exotic_hadrons_and_spectroscopy_with_heavy_flavours/9.2.tex}
\input{CONTRIBUTIONS/9_Exotic_hadrons_and_spectroscopy_with_heavy_flavours/9.3.tex}

\input{CONTRIBUTIONS/9_Exotic_hadrons_and_spectroscopy_with_heavy_flavours/9.4.tex}
\input{CONTRIBUTIONS/9_Exotic_hadrons_and_spectroscopy_with_heavy_flavours/9.5.summary.tex}

%% file: CONTRIBUTIONS/9_Exotic_hadrons_and_spectroscopy_with_heavy_flavours/9.1.tex
\section{Measurements of tetraquark and pentaquark properties}

Because the physics of exotic quarkonia-like states have not yet been understood~\cite{Olsen:2017bmm}, 
it is difficult to make precise projections for what can be accomplished with the future large data set from the upgraded \lhcb detectors.
Nevertheless, we can point out a number of measurements on the known structures that will help clarify their dynamical nature.

\subsection{The $\PX(3872)$ meson}
The $\PX(3872)$ state\footnote{The $\PX(3872)$ state is referred to as the $\Pchi_{c1}(3872)$ meson in the 2018 edition of the Review of Particle Physics~\cite{PDG2018}. The former name is used in this document.} is the best studied exotic meson candidate with a \cquark\cquarkbar pair content.
Some properties, like the enhancement of isospin-violating \jpsi\rhoz decays,
the mass right at the \Dz\Dstarzb threshold and the large fall-apart rate to \Dz\Dstarzb
above this threshold
find a natural explanation in the loosely bound molecular model.
However, its large prompt production rates at \tevatron~\cite{Abazov:2004kp} and
\lhc~\cite{LHCb-PAPER-2011-034,Chatrchyan:2013cld,Aaboud:2016vzw}
suggests a compact size.
The double-differential production cross-section (as function of \pt and rapidity),
which follows closely that of the \psitwos meson,
as well as its preference for phase-space suppressed \g\psitwos
decays over \g\jpsi, match the expectation for the $\Pchi_{\cquark1}(2P)$ state.
Yet, the rate of isospin-violating  \jpsi\rhoz decays, as compared to the radiative decays, is much too big
for this state to be ordinary  $\Pchi_{\cquark1}(2P)$ meson.  Models suggesting mixing between  the $\Pchi_{\cquark 1}(2P)$
and  the \Dz\Dstarzb  molecule face the challenge of explaining how a compact \cquark\cquarkbar state can have
a large mixing with a loosely bound molecule.
Tetraquark or $\cquark\cquarkbar+$glue hybrid explanations have also been proposed~\cite{Li:2004sta}.

To advance our understanding of this puzzling state, it will be very important to learn even more about its decay pattern.
In particular, if it really has a strong $\Pchi_{\cquark 1}(2P)$ component, it should have \pip\pim transitions to the $\Pchi_{\cquark1}(1P)$ state.
Unfortunately at \lhcb, the reconstruction efficiency for the dominant $\Pchi_{\cquark 1}(1P)$ decay  to \g\jpsi decay is low, making this prediction hard to test.
The very large data set of \upgradetwo  will allow this problem to be overcome
by detecting or refuting the existence of such transitions.
Studies of the $\PX(3872)$ lineshape by a simultaneous fit to all detected channels are important for pinning down the location
of its resonant pole and determining its natural width.
Both are very important inputs in helping with the understanding of the state
but require excellent statistical precision and the reconstruction of $\PX(3872)$ decays to \Dz\Dstarzb.

\subsection{Measurements of exotic hadron production}
Searching for prompt production of any known exotic hadron candidates at \lhc remains an important task,
since its detection would signify a compact component, either conventional quarkonium,  or a tightly bound tetraquark or pentaquark.
To date, the $\PX(3872)$ state is the only exotic hadron candidate with $\PQ\Qbar$ content that has been confirmed to be produced promptly. This anecdotally speaks against compact interpretations for the other states.
However, it will be important to quantify the upper limits in negative searches to allow more rigorous phenomenological analysis. 

\subsection{Radiative decays and baryon magnetic moments}
The prediction of magnetic moments for the light baryon states was a key success of the quark model.
As such these observables provide an additional handle on the structure of baryons and can potentially be used to
distinguish between several of the models that attempt to explain the structure of the observed pentaquarks~\cite{Wang:2016dzu}.

Experimentally the measurement of magnetic moments is challenging.
Access can be gained
through radiative decays of pentaquark states. The observation of radiative decays
involving exotic baryons would thus provide a new window into the structure of the
pentaquark candidates. Interestingly, since the parity of the two observed states $\PP_\cquark(4380)^+$ and $\PP_\cquark(4450)^+$ have
opposite parities and their spins differ by one unit, the radiative transition
between the two states should be allowed and might be observable by performing an amplitude analysis of the
\jpsi\proton{}\g\kaon decay. The proposed improved ECAL is crucial to the
feasibility of such a measurement in a high-luminosity environment.

To get an idea of the typical suppression
of these decays with respect to hadronic decays we can look at the measured branching fractions for
ordinary baryons and for exotic mesons. The $\decay{\PX(3872)}{\jpsi  \g}$ radiative decay is suppressed
by about a factor 50 with respect to decays into open charm and a factor of 5 with respect to the \jpsi \pip\pim decay.
Branching fractions of the order of $1\%$ (\eg $\BF(\decay{\Lz(1520)}{\Lz \g})=0.85 \pm 0.15\% $)
are common for radiative decays of baryons, which again are around a factor $50$ lower than the largest hadronic branching fraction.

\subsection{Amplitude analyses of exotic hadrons}
\begin{figure}[th]
 \centering
\includegraphics[width=0.6\textwidth]{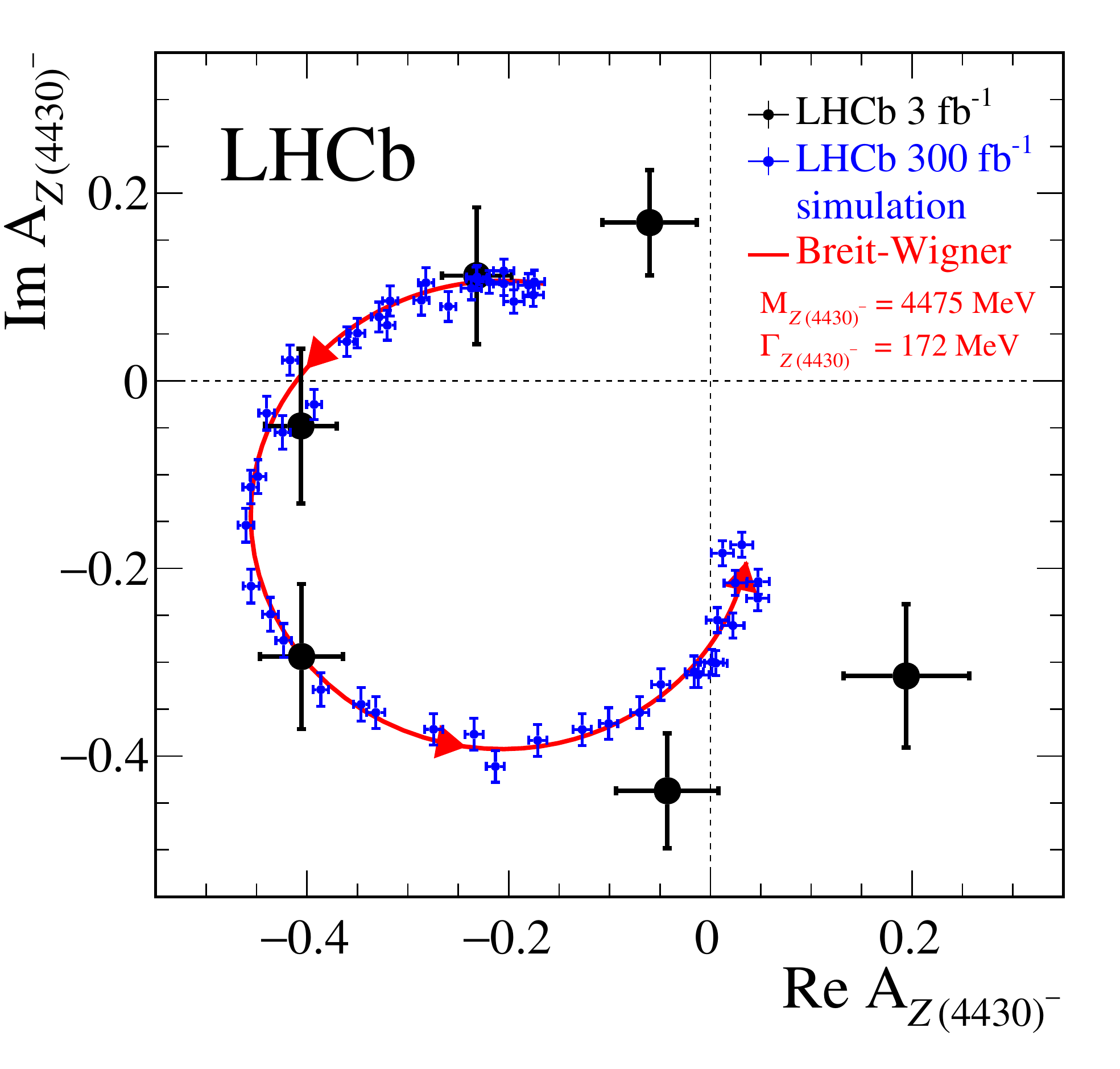}
\caption {
Argand diagram of the $\PZ(4430)^-$ amplitude (${\rm A}_{\PZ(4430)^-}$) in bins of $m^2_{\psitwos \pim}$ from a fit to the $\decay{\Bz}{\psitwos \Kp \pim}$ decays. The black points are the results based on Run 1 data~\cite{LHCb-PAPER-2014-014} while the blue points correspond to an extrapolation to an integrated luminosity of 300\invfb expected at the \lhcb \upgradetwo.
The red curve is the prediction from the Breit-Wigner formula with a resonance mass (width) of 4475 (172)\mev. Units are arbitrary.}
 \label{fig:argand}
\end{figure}

Many puzzling charged exotic meson candidates (\eg $\PZ(4430)^+$) decaying to \jpsi, \psitwos or $\Pchi_{\cquark 1}$ plus a charged pion have been observed in \B decays.
Some of them are broad, and none can be satisfactorily explained by any of the available phenomenological models.
The hidden-charmed mesons, observed in the $\jpsi\Pphi$ decay~\cite{LHCb-PAPER-2016-018,Chatrchyan:2013dma,Aaltonen:2009tz}, also belong to this category.
The determination of their properties, or even claim for their existence, relies on amplitude analyses,
which allow the exotic contributions to be separated from the (typically) dominant non-exotic components.
Further investigation of these $\PQ\Qbar\quark\quarkbar$ structures
will require much larger data samples and refinement of theoretical approaches to parametrisations of hadronic amplitudes.
 Similar comments apply to improvements in the determination of the properties of the pentaquark candidates
$\PP_\cquark(4380)^+$ and $\PP_\cquark(4450)^+$ and to the spectroscopy of excited \Lz baryons in $\decay{\Lb}{\jpsi \proton \kaon}$ decays. The large data set collected during the \lhcb \upgradetwo would allow the resonant character of the $\PP_\cquark(4380)^+$, $\PP_\cquark(4450)^+$ and $\PZ(4430)^+$ states to be tested further (Fig.~\ref{fig:argand}), while improvements in calorimetry would help in searching for new decay modes (\eg $\decay{\PP_\cquark^+}{\Pchi_{\cquark 1,2} (\to \jpsi \g) \proton}$) by amplitude analyses of $\decay{\Lb}{\Pchi_{\cquark 1,2}\proton\Km}$ decays~\cite{LHCb-PAPER-2017-011,Guo:2015umn}.

%% file: CONTRIBUTIONS/9_Exotic_hadrons_and_spectroscopy_with_heavy_flavours/9.2.tex
\section{Further tetraquark and pentaquark searches}

\subsection{Probing the $\rm{SU}_3$ multiplets of exotic hadrons}

Although the true nature of the $\PX(3872)$ meson is still unclear, both the molecular~\cite{Nieves:2012tt}
and tetraquark~\cite{Maiani:2004vq} models predict that a $C$-odd partner ($\PX(3872)^{C\rm{-odd}}$)
and charged partners ($\PX(3872)^\pm$) may exist and decay to $\jpsi \etaz / \g \Pchi_{\cquark J}$ and \jpsi \piz \pipm respectively.

\begin{figure}[tb]
 \centering
\includegraphics[width=0.32\textwidth]{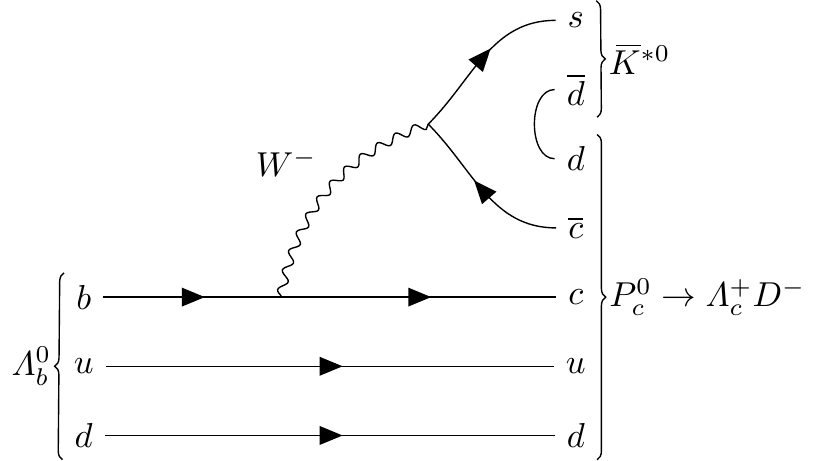}
\includegraphics[width=0.32\textwidth]{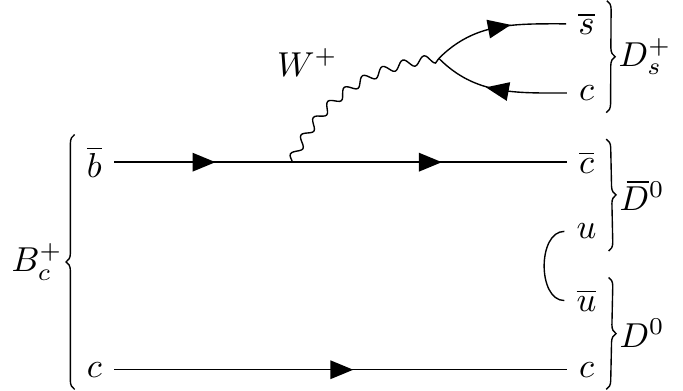}
\includegraphics[width=0.32\textwidth]{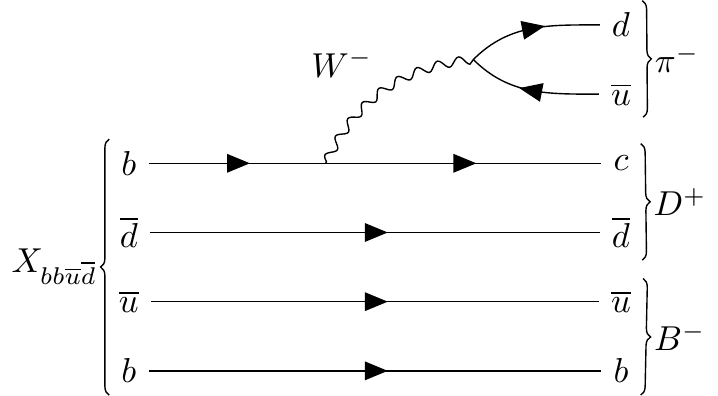}
\caption {Feynman diagrams of the (left) \decay{\Lb}{\Lc\Dm\Kstarzb},
  (middle) \decay{\Bc}{\Dz\Dzb\Ds} and
  (right) \decay{X_{bb\uquarkbar\dquarkbar}}{\Bm\Dp\pim} decays.}
 \label{fig:feynman}
\end{figure}

Similarly the existence of the $P_\cquark(4380)^+$ and $P_\cquark(4450)^+$ pentaquark states raises the question
of whether there is a large pentaquark multiplet. The observed states have an isospin-3 component of $I_3=+\frac{1}{2}$. They could be part of an isospin doublet with $I=\frac{1}{2}$ or a quadruplet with $I=\frac{3}{2}$.
In both cases there should be a neutral $I_3=-\frac{1}{2}$ state decaying into \jpsi \neutron. However this final state does
not lend itself well to observation. Instead the search for the neutral pentaquark candidate
 can be carried out using decays into pairs of open charm in particular in the process
\decay{\Lb}{\Lc\Dm\Kstarzb}, where the neutral pentaquark states would appear as resonances
in the \Lc\Dm subsystem (Fig.~\ref{fig:feynman}, left). Such decays can be very well reconstructed but the total reconstruction
efficiency suffers from the large number of tracks and the small branching fractions of $\Lc$ and $\Dm$ reconstructable final states; the total reconstruction
efficiency is about a factor 50
smaller than the efficiency for the \decay{\Lb}{ \jpsi \proton \Km} channel.
In the case of the existence of an isospin quadruplet, there is the interesting possibility to find doubly charged
pentaquarks decaying into $\Sigmares_\cquark^{++} \Dzb$. Channels such as these
require very large data sets to offset the low efficiency. The Magnet Side Stations will also improve the reconstruction efficiency of such decay modes with several tracks in the final states. 

The relative coupling of the pentaquark states to their decays into the double open-charm channels will
depend on their internal structure and the spin structure of the respective decay. For that reason
it is important to study decays involving \Dstarp resonances as well (\eg \decay{P_\cquark^+}{\Dstarm \Sigmares_\cquark^{++}}) 
 to investigate the internal structure of pentaquarks~\cite{Lin:2017mtz}. Since
these decays require the reconstruction of slow pions from the \Dstarp decays, the proposed tracking stations inside the magnet, enhancing the acceptance for low-momentum particles, will be highly beneficial for this study.

Invoking $\rm{SU}_3$ flavour symmetry,
 one would expect the existence of pentaquarks with strangeness, which would
decay into channels like \jpsi \Lz or \Lc\Dsm. To explore the potential of
the former case the decay \decay{\Xires_\bquark^-}{ \jpsi \Lz\Km} has been studied using Run~1 data.
About 300 signal decays have been observed~\cite{LHCB-PAPER-2016-053}.
Complementary information can be also achieved by a study of the \decay{\Lb}{ \jpsi \Lz \Pphi} decays.
An increase of the available integrated luminosity by a factor of 100 would allow detailed amplitude analyses to be performed for 
 these final states, with a similar sensitivity as was the case for the pentaquark discovery channel.



\subsection{Doubly charmed tetraquarks}
\newcommand{\slashed}[1]{\makebox[0pt][l]{/}#1}


The history of $\PX(3872)$ studies illustrates well the difficulty of distinguishing between exotic and conventional explanations for a hidden-charm state.
 Therefore it is appealing to search for  states with  an uncontroversial exotic signature.
A good candidate in this category would be a $\mathcal T_{\cquark\cquark}$ doubly charmed tetraquark~\cite{Ader:1981db,Moinester:1995fk,DelFabbro:2004ta,Carames:2011zz,Yuqi:2011gm,Hyodo:2012pm,Esposito:2013fma,Ikeda:2013vwa,Guerrieri:2014nxa,Maciula:2016wci,Richard:2016eis,Esposito:2016noz,Luo:2017eub,Eichten:2017ffp,Hyodo:2017hue,Cheung:2017tnt,Wang:2017dtg,Yan:2018gik}, being a meson with constituent quark content $\cquark\cquark\quarkbar\quarkbar^{\prime}$, where the light quarks \quark and $\quark^{\prime}$ could be \uquark, \dquark or \squark.


If the masses of the doubly charmed tetraquarks are below their corresponding open-charm thresholds, they would manifest as 
weakly decaying hadrons with properties including masses, lifetimes and decay modes not too different from the recently observed \Xiccpp baryons~\cite{LHCb-PAPER-2017-018},
and as for the $\Xires_{\cquark\cquark}$ baryons, the most promising searches are in prompt production. 

If the masses of $\mathcal T_{\cquark\cquark}$ states are instead above the open-charm threshold and their widths are broad, it will be very challenging to observe these states via prompt production. 
Instead, \Bcp decays to open-charm mesons can offer unique opportunity to test for their existence. 
In Run 5 the \Bc mesons will be copiously produced at the \lhc, because of the large production cross-sections of \bbbar and \ccbar pairs and of the enormous data sample.
Similarly to the amplitude analysis of the $\decay{\Lb}{\jpsi\proton\Km}$ decay, which led to the observation of the $P_\cquark^+$ pentaquark candidates~\cite{LHCb-PAPER-2015-029}, 
studying the angular distributions of the multi-body final states of the \Bc meson
has the potential of indicating new states, \eg $\mathcal T_{\cquark\cquark}$, inaccessible through decays of lighter hadrons. 
It also allows the spin-parity quantum numbers of any state to be determined.
 A good example is to study the $\mathcal T_{\cquark\cquark}^+$ state in the decay mode $\decay{\Bcp}{\Dsp \Dz \Dzb}$ (Fig.~\ref{fig:feynman}, middle)
through the decay chain $\decay{\Bc}{\mathcal T_{\cquark\cquark}^+ \Dzb}$ and 
$\decay{\mathcal T_{\cquark\cquark}^+}{\Dsp \Dz}$, as discussed in Ref.~\cite{Esposito:2013fma}.

The decay $\decay{\Bc}{\Dsp \Dz \Dzb}$ has not been observed with the Run 1 data,
and predictions on the branching fractions of \Bcp decays are subject to very large 
uncertainties. Estimates of the integrated luminosity needed to perform a full amplitude analysis
are therefore imprecise, and can only be formulated through considerations of other decay modes such as $\decay{\Bc}{\jpsi \Dsp}$.
The signal yield of $\decay{\Bc}{\jpsi \Dsp (\to \Pphi \pip)}$ decays observed in Run 1 data is $30 \pm 6$~\cite{LHCb-PAPER-2013-010}.
Considering the branching fraction of the decay of the additional charm hadron and the lower efficiency due to the higher track multiplicity,
the estimated number of signal of $\decay{\Bc}{\Dsp \Dz \Dzb}$ decays is $\mathcal O(10^2)$ 
in a future dataset corresponding to an integrated luminosity of $300$\invfb collected with $\mathcal O(100\%)$ trigger efficiency~\cite{LHCb-PAPER-2013-010}.
Since the \Dz and \Dsp mesons are pseudoscalars, the amplitude analysis simplifies,  
 and can provide conclusive results already with few hundreds decays.

Finally, strongly decaying doubly charmed tetraquarks with a narrow decay width as predicted by pure tetraquark models with 
spin-parity quantum numbers of $0^+$, $1^+$ and $2^+$, can also be searched for in prompt production.
The expected yields can be estimated by the associated production of open charm mesons measured with a fraction of the Run 1 data~\cite{LHCb-PAPER-2012-003}. 
With a data sample of $300$\invfb, the yield of \Dp\Dp (\Dp\Dsp) associated production is around $750$k ($150$k), which is a very promising sample in which to search for narrow $\mathcal T_{\cquark\cquark}$ states.


\subsection{Beautiful tetraquarks and pentaquarks}

If the coincidence of the $\Pchi_{\cquark 1}(2P)$ charmonium state with
the \Dz \Dstarzb threshold is responsible for the $\PX(3872)$ state,
 there is likely no bottomonium analogue of it, since  the $\Pchi_{\bquark 1}(3P)$
 state was detected well below the $\B\Bbar{}^{*}$ threshold,
and the $\Pchi_{\bquark 1}(4P)$ state is predicted to be too far above it.
However, if molecular forces dominate its dynamics, there could be an isosinglet state just below this threshold
decaying to \omegaz\OneS, where \omegaz could be reconstructed via the decay to \pip \pim \piz.
Unfortunately its prompt production would likely be very small unless
driven by tightly bound tetraquark dynamics.
The improved $\piz$ reconstruction in \upgradetwo will help for these searches.

Prompt production at \lhc remains the best hope for unambiguously establishing the existence of stable, weakly decaying $\bquark\bquark\uquarkbar\dquarkbar$ tetraquark predicted by both lattice QCD and phenomenological models, which accurately predicted the mass of the recently detected \Xiccpp baryon~\cite{LHCb-PAPER-2017-018}. 
 However, the inclusive reconstruction efficiencies for such states are tiny due to the small branching fractions of \B and \D mesons decays to low multiplicity final states (Fig.~\ref{fig:feynman}, right).

Recently there have  been several predictions for an exotic state with quark composition $\bquark\bquarkbar\bquark\bquarkbar$~\cite{Heller:1985cb,Berezhnoy:2011xn,Wu:2016vtq,Chen:2016jxd,Karliner:2016zzc,Bai:2016int,Wang:2017jtz,Richard:2017vry,Anwar:2017toa,Vega-Morales:2017pmm,Eichten:2017ual} with a mass below, the $2m_{\Peta_\bquark}$ threshold,
which implies that it can decay to  $\PUpsilon \mumu$. However lattice QCD calculations do not find evidence for such a state in the hadron spectrum~\cite{Hughes:2017xie}. Given the presence of four muons in the final state, \lhcb will have good sensitivity for observing the first exotic state composed of more than two heavy quarks~\cite{LHCb-PAPER-2018-027}.

Motivated by the discovery of the hidden-charm pentaquarks theorists have extended
the respective models for multiplet systems to include beauty quarks. In Ref.~\cite{Wu:2017weo} 
$\PQ\Qbar\quark\quark\quark$ ground states are investigated in an effective Hamiltonian
framework assuming a colour-magnetic interaction
between colour-octet \quark\quark\quark and $\PQ\Qbar$ subsystems. Several resonant states are predicted.
Such beautiful pentaquarks could be searched for in the
$\PUpsilon\proton$, $\PUpsilon\Lz$,  $\Bcpm \proton$ and $\Bcpm \Lz$ mass spectra.
In analogy with the popular $\Sigmares_\cquark \Lcbar$ molecular model,
Refs.~\cite{Yamaguchi:2017zmn} and~\cite{Shen:2017ayv} investigate similar dynamics in the hidden-bottom sector and predict
a large number of exotic resonances. Indeed in the hidden-beauty sector the theory calculations are
found to be even more stable than for the hidden charm, motivating searches for resonances close to the
$\B^*\Sigmares_\bquark, \B\Sigmares_\bquark^*, \B^*\Sigmares_\bquark^*$ and $\B \Lz^{*0}_\bquark, \B^*\Lb$ thresholds.

Another possibility is the existence of pentaquarks with open beauty and quark contents such as
\bquarkbar\dquark\uquark\uquark\dquark, \bquark\uquarkbar\uquark\dquark\dquark,
\bquark\dquarkbar\uquark\uquark\dquark and \bquarkbar\squark\uquark\uquark\dquark~\cite{Stewart:2004pd, Oh:1994np}.
If those states lie below the respective baryon-meson threshold containing beauty, then
they could be stable against strong decay and would decay through a weak
\decay{\bquark}{\cquark\cquarkbar\squark} transition. A search using a data set corresponding to $3\invfb$
in four decay channels $\jpsi\proton\hadron^+\hadron^-$ (\hadron = \kaon, \pion)  has been performed~\cite{LHCb-PAPER-2017-043}.
No signals have been found and $90\%$ confidence limits have been put on the production cross section times branching fraction
relative to the \Lb in the \jpsi\proton\Km mode. The obtained limits are of the order of $10^{-3}$, which
does not yet rule out the estimates for the production of such an object provided in Ref.~\cite{Stewart:2004pd}.
Similar searches in channels with open charm hadrons in the final state again lead
to large multiplicities and the respective small reconstruction efficiencies but could
profit from favoured branching fractions. Investigations of a large number of channels
will maximise sensitivity for weakly decaying exotic hadrons.

It has  also been proposed to search for  exotic \Omegab states~\cite{Liang:2017ejq}
in analogy to the recently discovered excited \Omegac states~\cite{LHCb-PAPER-2017-002}.
Such open-beauty exotic states could be searched for in decays to a \Xib\kaon final state.


%% file: CONTRIBUTIONS/9_Exotic_hadrons_and_spectroscopy_with_heavy_flavours/9.3.tex
\section{Study of conventional doubly heavy baryons in charm and beauty}
\label{sec:9.3}

The quark model of hadrons has been relatively successful in describing hadrons with one heavy quark, using, 
for example, Heavy Quark Effective Theory~\cite{Khoze:1983yp,Bigi:1991ir,Bigi:1992su,Blok:1992hw,Blok:1992he,Neubert:1997gu,Uraltsev:1998bk,Bigi:1995jr}, 
QCD sum rules~\cite{Shifman:1978bx,Shifman:1978bw,Shifman:1978by}, or potential models. Baryons with two heavy quarks probe the
QCD potential in a different way than baryons with a single heavy quark~\cite{Gershtein:1998un,Kiselev:1999zj,Likhoded:2009zz,Karliner:2014gca}.
With the recent discovery of the \Xiccpp baryon~\cite{LHCb-PAPER-2017-018}, quantitative comparisons can begin. However, there is a vast spectrum of additional
doubly heavy baryonic states that are yet to be observed, such as the \Omegacc, $\Xires_{\bquark\cquark}^{0,+}$, $\Xires_{\bquark\bquark}^{-,0}$ baryons and their excitations.
Observations of these states and measurement of their properties will provide valuable insights into the QCD potential. Moreover, searches for these conventional states would also provide a
fertile ground for the search of exotic hadrons~\cite{Karliner:2013dqa,Karliner:2015jpa,Karliner:2016via,Karliner:2016joc}.

\subsection{Doubly charmed baryons: $\Xires_{\cquark\cquark}$ and \Omegacc}
\label{sec:9.3.1}

  The discovery of the \Xiccpp baryon has opened an exciting new line of research that
  \lhcb is avidly pursuing.
  Measurements of the lifetime and relative production cross-section of \Xiccpp,
  searches for additional decay modes, and searches for its isospin partner
  \Xiccp and their strange counterpart \Omegacc are underway.

  A signal yield of  $313 \pm 33$ \decay{\Xiccpp}{\Lc\Km\pip\pip} decays
  was observed in $1.7\invfb$ of Run 2 data~\cite{LHCb-PAPER-2017-018}.
  Improvements in the trigger for the upgraded \lhcb detector are projected to
  increase selection efficiencies by a factor two for most charm decays, with
  decays to high-multiplicity final states, such as those from the cascade
  decays of doubly charmed baryons, potentially benefiting much
  more~\cite{LHCb-TDR-012,LHCb-PAPER-2012-031}.
  Thus the  Run 5 sample  will contain more than 90\,000 decays of this mode.
  The branching fractions for \decay{\Xiccpp}{\Lc\Km\pip\pip} has been
  theoretically estimated to be up to 10\% making it one of the most abundant 
  nonleptonic decay modes, but several other lower multiplicity modes with
  predicted branching fractions of $\mathcal{O}(1\%)$ will yield samples of 
  comparable size~\cite{LHCb-PAPER-2018-026, Wang:2017mqp,Gutsche:2017hux,Sharma:2017txj}.

  The efficiency with which \lhcb can disentangle weak decays of doubly charmed
  baryons from prompt backgrounds depends on the lifetime of the
  baryon~\cite{LHCb-PAPER-2013-049}.
  Although the predicted lifetimes for the \Xiccp, \Xiccpp, and \Omegacc baryons span
  almost an order of magnitude, the relative lifetimes of \Xiccp and \Omegacc
  are expected to be approximately $1/3$ that of the
  \Xiccpp baryon~\cite{Kiselev:2001fw,Karliner:2014gca,Fleck:1989mb,Guberina:1999mx,Kiselev:1998sy,Chang:2007xa,Berezhnoy:2016wix,LHCb-PAPER-2018-019}.
  Assuming a relative efficiency of 0.25 with respect to \Xiccpp due to the
  shorter lifetimes and an additional production suppression of
  $\sigma(\Omegacc)/\sigma(\Xiccpp) \sim 0.2$ for
  \Omegacc~\cite{Kiselev:2001fw}, \lhcb will have Run 5 yields of around 25\,000 for \Xiccp and 4\,500 for \Omegacc in each of several decay modes.

  \lhcb will be the primary experiment for studies of the physics of doubly
  charmed baryons for the foreseeable future, and its potential will not be
  exhausted by the end of Run 5.
  With the data collected in Run 2, \lhcb should observe all three weakly decaying
  doubly charmed baryons and characterise their physical properties.
  Run 5 will supply precision measurements of doubly differential
  cross-sections that will provide insight into production
  mechanisms of doubly heavy baryons. In addition 
  Run 5 will allow the spectroscopy of excited states and bring studies of the
  rich decay structure of doubly charmed hadrons into the domain of precision
  physics.

\subsection{Baryons with beauty and charm quarks: $\Xires_{\bquark\cquark}$ and $\Omegares_{\bquark\cquark}^0$}

The production cross section of the $\Xires_{\bquark\cquark}$ baryons within the \lhcb acceptance is expected to be about 77\nb~\cite{Zhang:2011hi}.
This value is about 1/6 of the expected production cross-section of a \Bcp meson~\cite{Chang:2003cr,Gao:2010zzc}. It should be noted that the relative \Lb production rate is \pt dependent. In the typical \pt range in the 
\lhcb acceptance, a ratio of production rates, 
$\sigma(\decay{\proton\proton}{\Lb \PX})/\sigma(\decay{\proton\proton}{\Bzb X})\sim0.5$~\cite{LHCb-PAPER-2011-018,LHCb-PAPER-2014-004} is measured. It is therefore
conceivable that the $\Xires_{bc}$ production rates are also larger than predicted by the above calculations.

To observe and study $\Xires_{\bquark\cquark}$ and $\Omegares_{\bquark\cquark}^0$ baryons is quite challenging due to the low production rate, the small product of branching fractions
and the selection efficiency for reconstructing all of the final-state particles. To collect a large sample of $\Xires_{\bquark\cquark}$
baryons will require the higher integrated luminosity and detector enhancements planned for in the \upgradetwo. 
  Using the notation that $\PX_\cquark$ is a charmed baryon
containing a single charm quark, some of the most promising decay modes to detect $\Xires_{\bquark\cquark}$ and $\Omegares_{\bquark\cquark}^0$ baryons are:
\begin{itemize}
  \item $\jpsi \PX_\cquark$ modes: \jpsi\Xicp, \jpsi\Xicz, \jpsi\Lc, \jpsi\Lc\Km;
  \item $\Xires_{\cquark\cquark}$ modes: $\Xires_{\cquark\cquark}\pim$; 
  \item Doubly charmed modes: \Dz \Lc, \Dz \Lc \pim, \Dz\Dz \proton; 
  \item Penguin topologies: \Lc\Km, \Xicp\pim;
  \item $\Xires_\bquark$ , \Bd or \Lb modes: \Xib\pip, \Lb\pip, \Bz \proton, using fully reconstructed or semileptonic \Bz, \Lb or \Xib decays~\cite{LHCb-PAPER-2018-013};
  \item $W$-exchange between \bquark-\cquark quarks, which is not helicity suppressed. This can give rise to final state with just one charmed particle, $\eg$ \Lc\Km.
\end{itemize}

To put this in context, the \lhcb collaboration observed $30\pm6$ $\decay{\Bcp}{\jpsi \Dsp (\to\Pphi\pip)}$ decays with a 3\invfb data at 7 and 8\tev~\cite{LHCb-PAPER-2013-010}. 
With looser selections about $100$ signal decays can be obtained with reasonably good signal-to-background ratio.
The $\decay{\Xires^+_{\bquark\cquark}}{\jpsi\Xicp}$ decay is kinematically very similar.
A signal yield of about 600 $\decay{\Xires_{\bquark\cquark}^+}{\jpsi\Xicp}$ signal decays in expected in Run 5,\footnote{Assuming
$\frac{f_{\Xires_{\bquark\cquark}^+}}{f_{\Bcp}}\sim0.2$,
$\frac{\BR(\decay{\Xires_{\bquark\cquark}^+}{\jpsi\Xicp})}{\BR(\decay{\Bcp}{\jpsi\Dsp})}\sim~1$, 
$\frac{\BR(\decay{\Xicp}{p\Km\pip})}{\BR(\decay{\Dsp}{\Kp\Km\pip})}\sim0.1$, and
$\frac{\epsilon_{\Xires_{\bquark\cquark}^+}}{\epsilon_{\Bcp}}\sim~0.5$.
} albeit with sizeable uncertainties.
Other modes could also provide sizeable signal samples.
It is likely \lhcb will observe the $\Xires_{\bquark\cquark}$ baryons in Run 3/4, and further probe the spectrum of other doubly heavy baryons with the large samples accessible in the proposed \upgradetwo.

%% file: CONTRIBUTIONS/9_Exotic_hadrons_and_spectroscopy_with_heavy_flavours/9.4.tex
\section{Precision measurements of conventional quarkonium states}


Studies of the properties and production of quarkonia states at hadron colliders provide an
important testing ground for Quantum Chromodynamics~\cite{Brambilla:2010cs}.
According to the~current theoretical framework,
nonrelativistic QCD\,(NRQCD)~\cite{Caswell:1985ui,Bodwin:1994jh},
the~production of heavy quarkonium factorises into two steps,
separated by different time and energy scales, the creation of a heavy quark-antiquark pair, $\PQ\Qbar$,
and its subsequent nonperturbative evolution
into a~quarkonium state.  Calculations assuming an initial $\PQ\Qbar$ colour-singlet (CS)
state ~\cite{Kartvelishvili:1978id,Baier:1981uk,Berger:1980ni} show good
agreement~\cite{Campbell:2007ws,Gong:2008sn,Artoisenet:2008fc,Lansberg:2008gk,Han:2014kxa,Gong:2013qka}
with the experimental data~\cite{Brambilla:2010cs,Acosta:2001gv,
  Khachatryan:2010zg,
  LHCb-PAPER-2011-036,
  Aad:2012dlq,
  LHCb-PAPER-2013-016,
  LHCb-PAPER-2013-066,
  LHCb-PAPER-2015-045}
for production cross-sections and
the shapes of the transverse momentum spectra. However, this approach fails to describe
the spin-alignment (usually labelled as polarisation)
of $S$-wave charmonium states, \jpsi and \psitwos~\cite{LHCb-PAPER-2013-008,LHCb-PAPER-2013-067}.

The correct interpretation of the experimental polarisation results
for $S$-wave quarkonia requires a rigorous analysis of the feed-down contributions from
higher excited states ~\cite{LHCb-PAPER-2011-030,LHCb-PAPER-2014-031}.
The direct measurement of the polarisation for $\Pchi_\cquark$ and $\Pchi_\bquark$ states is necessary to decrease this 
model dependence. Since $P$-wave states are practically free from the feed-down from higher excited states,
any $\Pchi_\cquark$ and $\Pchi_{\bquark}$ polarisation measurements could be interpreted in a robust manner
 without additional model assumptions.

%

The recent discover of $\decay{\Pchi_{\cquark1,2}}{\jpsi\mumu}$ decays~\cite{LHCb-PAPER-2017-036} opens the possibility
to perform a detailed study of $\Pchi_{\cquark}$ production, allowing almost background-free measurements even for
very low transverse momentum of $\Pchi_\cquark$~candidates. 
Due to the excellent mass resolution, the vector state $\Pchi_{\cquark1}$  and the tensor state $\Pchi_{\cquark2}$ are well separated,
eliminating the possible systematic uncertainty  caused by the large overlap of
these states in the $\decay{\Pchi_{\cquark}(\Pchi_\bquark)}{\jpsi(\PUpsilon)\g}$ decay\cite{
  LHCb-PAPER-2011-019,
  LHCb-PAPER-2011-030,
  LHCb-PAPER-2012-015,
  LHCb-PAPER-2013-028,
  LHCb-PAPER-2014-031,
  LHCb-PAPER-2014-040}.

An integrated luminosity of 300\invfb will allow the high-multipole contributions 
to the  $\decay{\Pchi_{\cquark}}{\jpsi\mumu}$ amplitude to be probed, namely
the magnet-dipole contribution for $\Pchi_{\cquark1}$~decays and
the magnet-dipole and electrical-octupole contributions for $\Pchi_{\cquark2}$ decays. 

Use of Run 5 dataset is also necessary  to measure the $(\pt,y)$ dependence of $\Pchi_\cquark$ polarisation parameters. 
 In addition the effect of the form factor in
the~decays $\decay{\Pchi_{\cquark}}{\jpsi\mumu}$~\cite{Faessler:1999de,Luchinsky:2017pby}
could be probed with the precision of several percent from the shape of the $m(\mumu)$ spectra.

\subsection{Precise measurement of double quarkonia production}

Studies of the double quarkonia production allows independent tests for the  quarkonia production mechanism,
and in particular for the role of the colour-octet. So far the \lhcb collaboration has  analysed double \jpsi production at 7\tev and 13\tev data with relatively small
datasets~\cite{LHCb-PAPER-2011-013,LHCb-PAPER-2016-057}.
Using $280$\invpb of data collected at $\sqs=13\tev$,
$(1.05\pm0.05)\times10^3$ signal \jpsi\jpsi events are observed. 
 However, even with the larger sample of \jpsi\jpsi events now available, 
it is not possible to distinguish between 
the different theory descriptions of the single-parton scattering (SPS) mechanism~\cite{Sun:2014gca,Likhoded:2016zmk, Lansberg:2013qka, Lansberg:2014swa,Lansberg:2015lva,Shao:2012iz,Shao:2015vga, Baranov:2011zz,Baranov:1993qv} nor to separate the
  contributions from the SPS and double parton  scattering (DPS) mechanisms~\cite{Bansal:2014paa,Belyaev:2017sws}. 
%
Although the larger samples collected during  \upgradeone will allow some progress on these questions, the measurement of the correlation of \jpsi polarisation parameters will only be possible with the \upgradetwo data set.


In addition, while it is likely that  $\PUpsilon\PUpsilon$ and $\jpsi\PUpsilon$ production will be observed in the near future (assuming the dominance of the DPS mechanism),
the determination of the relative SPS and DPS contributions, as well as the discrimination between different SPS theory models, will require precision measurements only possible with  Run 5 data.

%% file: CONTRIBUTIONS/9_Exotic_hadrons_and_spectroscopy_with_heavy_flavours/9.5.summary.tex
\section{Summary}

The \upgradetwo\ detector will have a large impact on sensitivity in searches for heavy states: the
potential removal of the VELO RF foils, together with the improved particle identification provide by the TORCH, will enhance the reconstruction efficiency for multibody \B decays, such as \decay{\Bcp}{\Dsp \Dz \Dzb}; the selection of short-lived particles ($\eg$ \Bc, \Xicc, \Omegacc, $\Xires_{\bquark\cquark}$, \etc) will also benefit from an improved vertex resolution; the Magnet Side Stations will help in studying dipion transitions such as \decay{\PX(3872)}{\chicone \pip \pim} or \decay{\B_\cquark^{**+}}{\Bc \pip \pim}; improved \piz and \etaz mass resolutions will increase the sensitivity in searching for the $C$-odd and charged partners of the $\PX(3872)$ meson by \decay{\PX(3872)^{C\rm{-odd}}}{\jpsi \etaz} and \decay{\PX(3872)^\pm}{\jpsi \piz \pipm}. These improvements, together with the enormous sample sizes of Run 5, will ensure that \lhcb maintains its position at the forefront of spectroscopy studies, and will give it unique access to studies involving the production of \Bc, \Lb, \Xib hadrons. A summary of the expected yields in certain important modes, and a comparison with \belletwo, is given in Table~\ref{tab:sec9_summary}.


\begin{table}[tb]
\centering
\caption{\label{tab:yields}Expected data samples at \lhcb \upgradetwo and \belletwo for key decay modes for the spectroscopy of heavy flavoured hadrons. The expected yields at \belletwo are estimated by assuming similar efficiencies as at \belle.}
\begin{tabular}{l|ccc|c}
\hline
                      & \multicolumn{3}{c|}{\lhcb} & \belletwo\\
Decay mode & 23\invfb & 50\invfb & 300\invfb & 50\invab\\
\hline
\decay{\Bp}{\PX(3872) (\to \jpsi \pip \pim)\Kp} &14k &30k&180k&11k\\
\decay{\Bp}{\PX(3872) (\to \psitwos \g)\Kp}     &500 &1k& 7k & 4k \\
\decay{\Bz}{\psitwos \Km \pip}                  &340k &700k&4M& 140k\\
\decay{\Bcp}{\Dsp \Dz \Dzb}                     & 10 &20& 100 & ---\\
\decay{\Lb}{\jpsi \proton \Km}                  &340k &700k&4M& ---\\
\decay{\Xibm}{\jpsi \Lz \Km}                    &4k &10k&55k& ---\\
\decay{\Xiccpp}{\Lc\Km\pip\pip}                 &7k &15k&90k & $<$6k\\
\decay{\Xires_{\bquark\cquark}^+}{\jpsi \Xires_\cquark^+} & 50 &100& 600 &---\\
\hline
\end{tabular}
\label{tab:sec9_summary}
\end{table}


%% file: CONTRIBUTIONS/11_Summary_and_conclusions/11.tex
\label{chpt:summary}
\input{CONTRIBUTIONS/11_Summary_and_conclusions/11.1.tex}

\input{CONTRIBUTIONS/11_Summary_and_conclusions/11.2.tex}

%% file: CONTRIBUTIONS/11_Summary_and_conclusions/11.1.tex
\section{Sensitivity to key observables and physics reach in flavour}

\begin{figure}[htb!]
\begin{center}
\includegraphics[height=5.0cm,keepaspectratio]{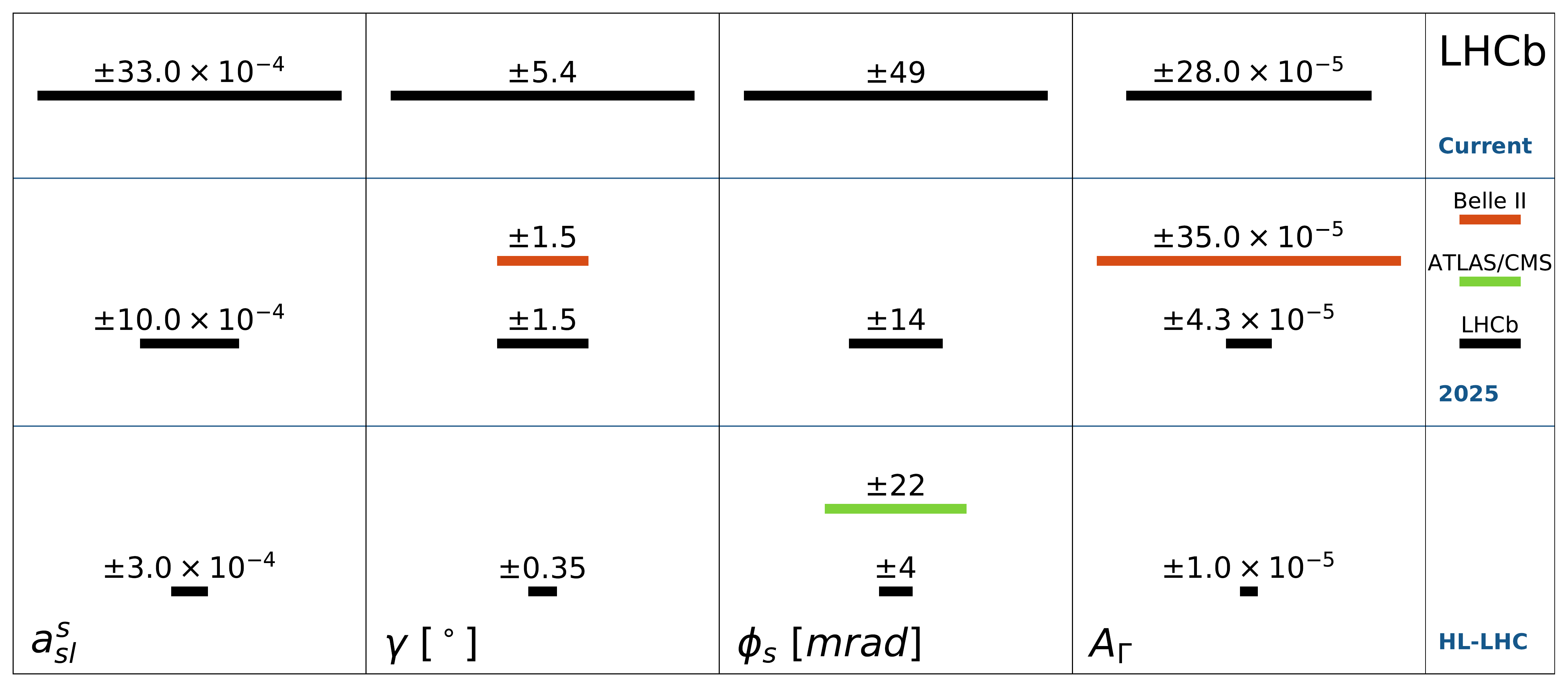}
\includegraphics[height=5.0cm,keepaspectratio]{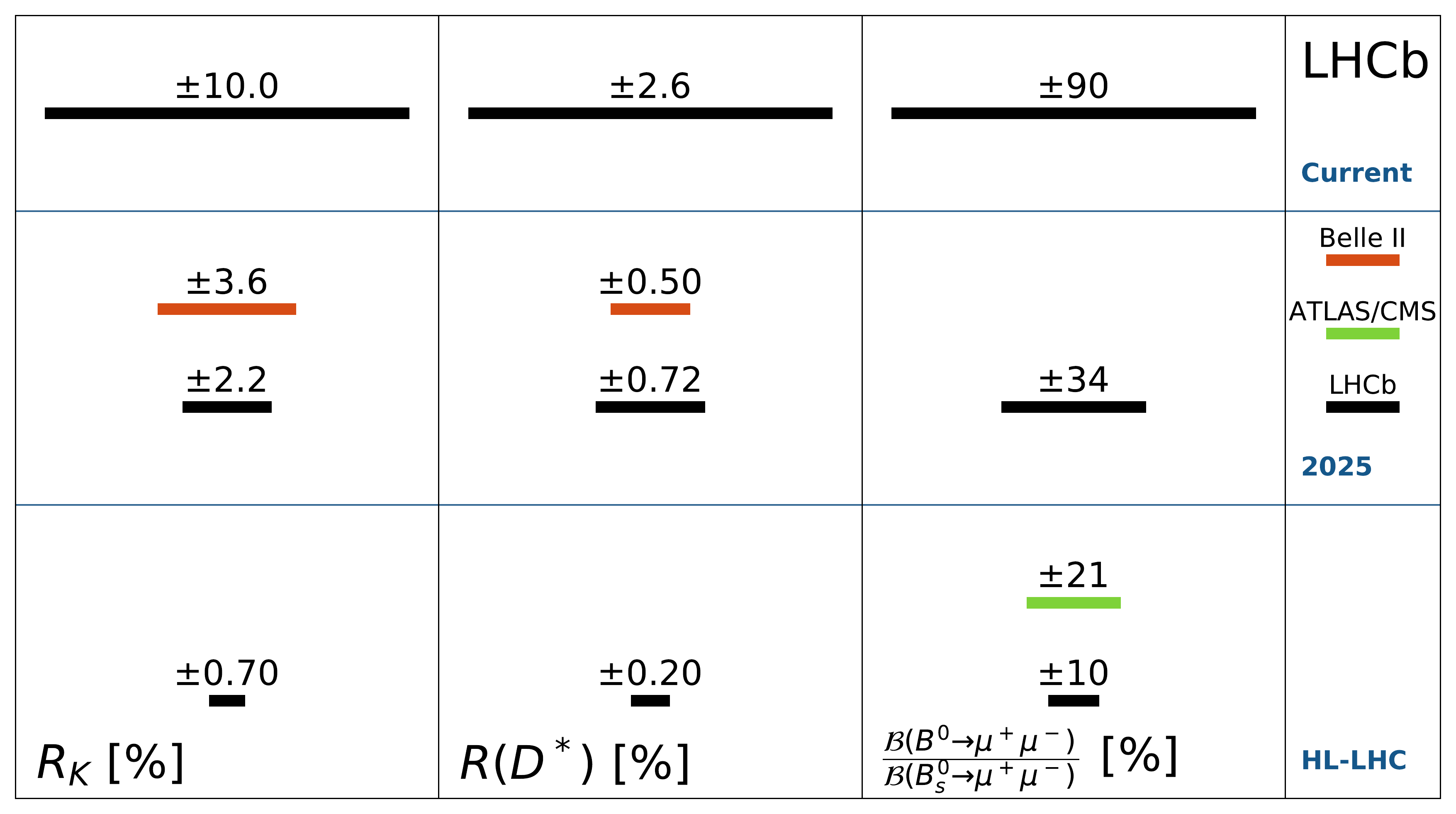}
\caption{\small  Projected improvement in sensitivities for selected \CP-violating observables (upper) and for rare decays and lepton-universality tests (lower).} 
\label{fig:SensitivitySummary}
\end{center}
\end{figure}

The Upgrade II of LHCb will enable a very wide range of flavour observables to be determined with unprecedented precision,
which will give the experiment sensitivity to New Physics scales several orders of magnitude above those accessible to direct searches. 
The expected uncertainties for a few key measurements  with 300\invfb are presented in Table~\ref{tab:physummary}. Also shown are the current uncertainties, those expected from LHCb in 2025, which is just before the start of the HL-LHC era, and for Belle II,  which is due to complete operation around this time.  In addition, and where available,  sensitivity estimates are given for ATLAS and CMS after their Phase-II Upgrades and with 3000\invfb of data.
A graphical representation of a subset of these entries is shown in Fig.~\ref{fig:SensitivitySummary}. The future LHCb estimates are all based on extrapolations from current measurements, and take no account of detector improvements apart from an approximate factor two increase in efficiency for hadronic modes, coming from the full software trigger that will be deployed from Run 3 onward.  

It is not possible to predict now, at the end of LHC Run 2, what will be the most important studies performed at Upgrade~II.  However, indicative examples can be given of measurements that are certain to be of the highest scientific interest.  These include:

\begin{itemize}
\item{A comprehensive measurement programme of observables in a wide range of $b \to s \ellell$ and $b \to d \ellell$ transitions, many not accessible in the current experiment or Upgrade I, employing both muon and electron modes;}
\item{Measurements of the \CP-violating phases $\gamma$ and $\phi_s$ with a precision of better than $0.35^\circ$ and $3\mrad$, respectively.  There will also be decisive contributions to the determination of other Unitarity Triangle parameters such as $|\Vub|/|\Vcb|$ that will greatly improve our knowledge of the $\bar{\rho}-\bar{\eta}$ plane;} 
\item{Measurement of $R\equiv {\cal{B}} (\Bd \rightarrow \mu^+ \mu^-)$/${\cal{B}} (\Bs \rightarrow \mu^+ \mu^-)$ with an uncertainty of  $10\%$, and the first precise measurements of associated $\Bs \to \mu^+\mu^-$ observables;}
\item{A wide-ranging set of lepton-universality tests in $b \to c \ell^-\bar{\nu_l}$ decays, exploiting the full range of $b$ hadrons;}
\item{\CP-violation studies  in charm with $10^{-5}$ precision.}
\end{itemize}

\begin{sidewaystable}[ptb!]
\caption{\small Summary of prospects for future measurements of selected flavour observables for LHCb, Belle II and Phase-II ATLAS and CMS.  The projected LHCb sensitivities take no account of potential detector improvements, apart from in the trigger.  The Belle-II sensitivities are taken from Ref.~\cite{BELLE2PHYSICS}.\vspace*{0.1cm}}\label{tab:physummary}
\centering
\begin{tabular}{lrrrrr} \hline
Observable  &   Current LHCb  &  LHCb 2025 &  Belle II  &  Upgrade II  &    ATLAS \& CMS \\ \hline
\underline{\bf EW Penguins}  &  & & &   & \\
$R_K$ $(1<q^2<6\,{{\rm GeV}^2c^4})$            & 0.1\cite{LHCb-PAPER-2014-024}  &  0.025  & 0.036  &   0.007 & -- \\
$R_{K^\ast}$ $(1<q^2<6\,{{\rm GeV}^2c^4})$  & 0.1\cite{LHCb-PAPER-2017-013}  &  0.031 & 0.032  &   0.008 & -- \\
$R_\phi$,  $R_{pK}$, $R_{\pi}$         & -- & 0.08, 0.06, 0.18  & -- & 0.02, 0.02, 0.05 & -- \\
\rule{0pt}{3ex}\underline{\bf CKM tests}  &   & & & &\\
$\gamma$, with $\Bs \to \Ds\Km$ & $(^{+17}_{-22})^\circ$~\cite{LHCb-PAPER-2017-047} & 4$^\circ$ & -- & 1$^\circ$ & -- \\
$\gamma$, all modes & $(^{+5.0}_{-5.8})^\circ$~\cite{LHCb-CONF-2018-002} & 1.5$^\circ$ & 1.5$^\circ$ & $0.35^\circ$ & -- \\
$\sin 2 \beta$, with $\Bd \to J/\psi\KS$ & $0.04$\cite{LHCb-PAPER-2017-029} & $0.011$ & $0.005$ & $0.003$ & -- \\
$\phi_s$, with $\Bs \to J/\psi \phi$ & 49 mrad\cite{LHCb-PAPER-2014-059}  & 14 mrad  & -- & 4 mrad & 22 mrad~\cite{ATL-PHYS-PUB-2013-010} \\
$\phi_s$, with $\Bs \to \Dsp\Dsm$   & 170 mrad\cite{LHCb-PAPER-2014-051}  & 35 mrad  & -- & 9 mrad & -- \\
$\phi_s^{s{\bar{s}s}}$, with $\Bs \to \phi\phi$  & 154 mrad\cite{LHCb-PAPER-2014-026} & 39 mrad & --  & 11 mrad  & Under study~\cite{CMS-PAS-FTR-16-006}\\
$a^s_{\rm sl}$ & $33 \times 10^{-4}$\cite{LHCb-PAPER-2016-013}  &  $10 \times 10^{-4}$   &  -- & $3 \times 10^{-4}$ & -- \\
$|\Vub|/|\Vcb|$  & 6\%\cite{LHCb-PAPER-2015-013}  & 3\%   &  1\%  &  1\%  &  -- \\
\rule{0pt}{3ex}\underline{\bf ${\bm{B^0_s, B^0} {\to} \bm{\mu^+}\bm{\mu^-}}$}  &  & & & \\
${\cal{B}} (\Bd \rightarrow \mu^+ \mu^-)$/${\cal{B}} (\Bs \rightarrow \mu^+ \mu^-)$ & 90\%\cite{LHCb-PAPER-2017-001}  & 34\%  &  -- &  10\%  & 21\%~\cite{CMS-PAS-FTR-14-005}\\
$\tau_{\Bs \to \mu^+\mu^-}$ &22\%\cite{LHCb-PAPER-2017-001} & 8\%  & --  &  2\% & -- \\
$S_{\mu\mu}$ & -- &  -- & -- &  0.2 & -- \\
\rule{0pt}{3ex}\underline{\bf\boldmath {${b \to c \ell^-\bar{\nu_l}}$} LUV studies}  &    &   & & \\
$R(D^\ast)$ & 0.026\cite{LHCb-PAPER-2015-025,LHCb-PAPER-2017-027}  & 0.0072 & 0.005 & 0.002 & -- \\
$R(J/\psi)$ & 0.24\cite{LHCb-PAPER-2017-035} & 0.071 &  --  & 0.02 &  -- \\
\rule{0pt}{3ex}\underline{\bf Charm}  &   &  & & &\\
$\Delta A_{\CP}(KK-\pi\pi)$ &  $8.5 \times 10^{-4}$\cite{LHCb-PAPER-2015-055}  & $1.7\times 10^{-4}$ & $5.4 \times 10^{-4}$ &  $3.0 \times 10^{-5}$ & -- \\
$A_\Gamma$ ($\approx x \sin\phi$) & $2.8 \times 10^{-4}$\cite{LHCb-PAPER-2016-063} & $4.3 \times 10^{-5}$ &  $3.5 \times 10^{-4}$ &  $1.0 \times 10^{-5}$ & -- \\
${x \sin \phi}$ from $\Dz \to K^+\pi^-$ & $13\times 10^{-4}$\cite{LHCb-PAPER-2017-046} & $3.2 \times 10^{-4}$ & $4.6 \times 10^{-4}$ & $8.0 \times 10^{-5}$ & -- \\
${x \sin \phi}$ from multibody decays & -- & ($K3\pi$) $4.0\times 10^{-5}$ & ($\KS\pi\pi$)  $1.2\times 10^{-4}$ & ($K3\pi$) $8.0 \times 10^{-6}$ & -- \\
\hline
\end{tabular}
\end{sidewaystable}

\begin{figure}[H]
\begin{center}
\includegraphics[width=0.82\textwidth]{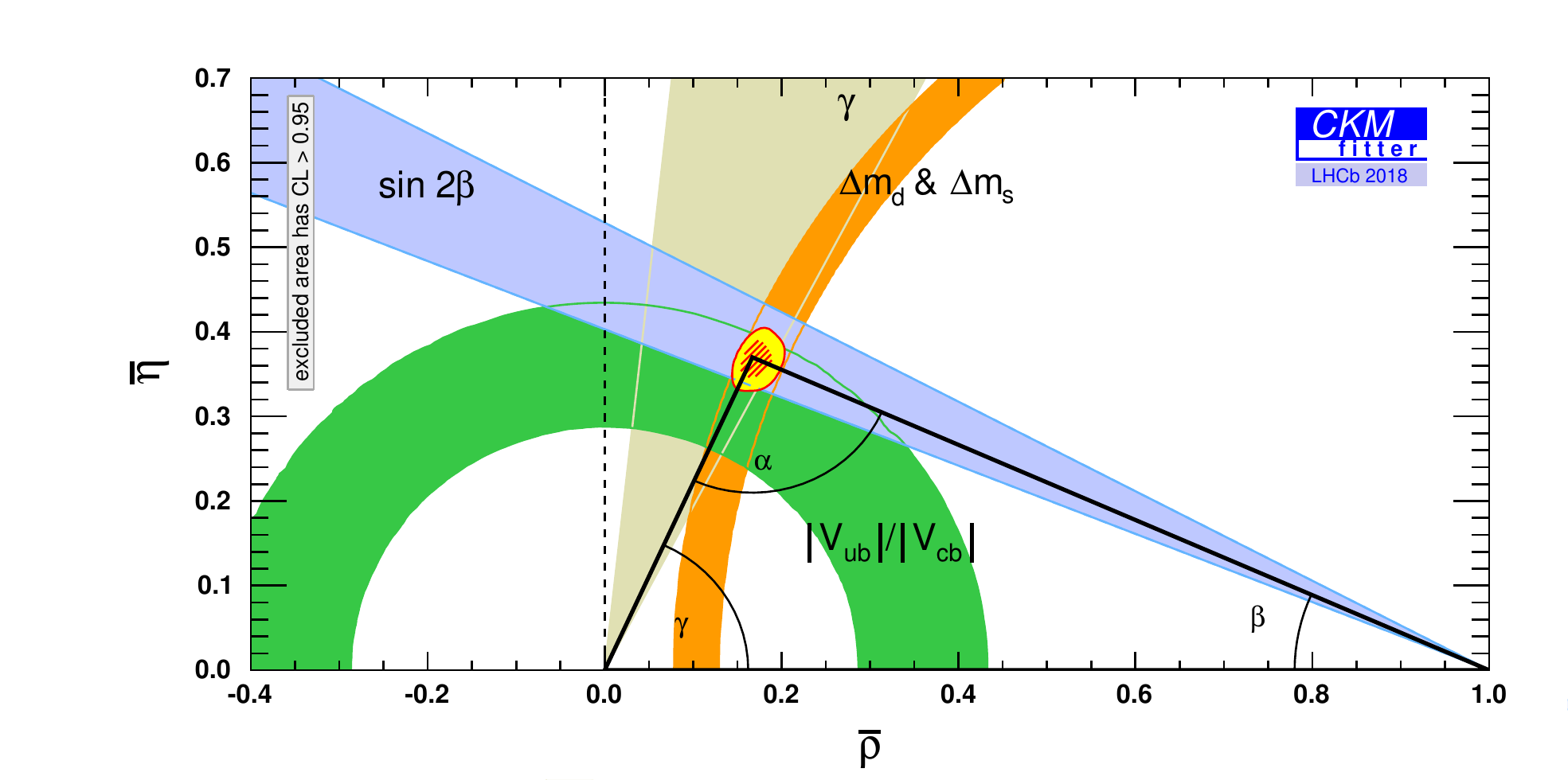}
\includegraphics[width=0.82\textwidth]{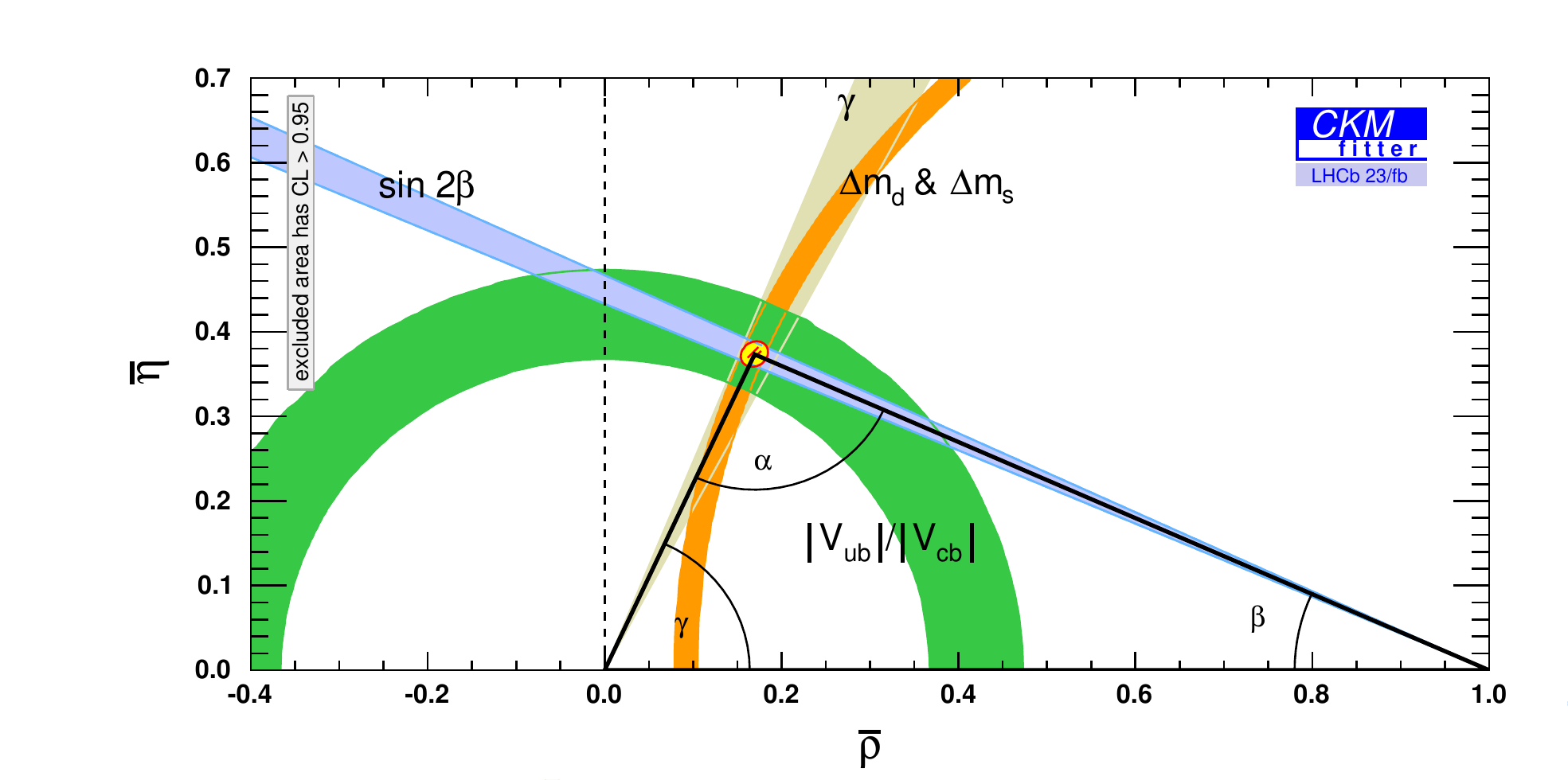}
\includegraphics[width=0.82\textwidth]{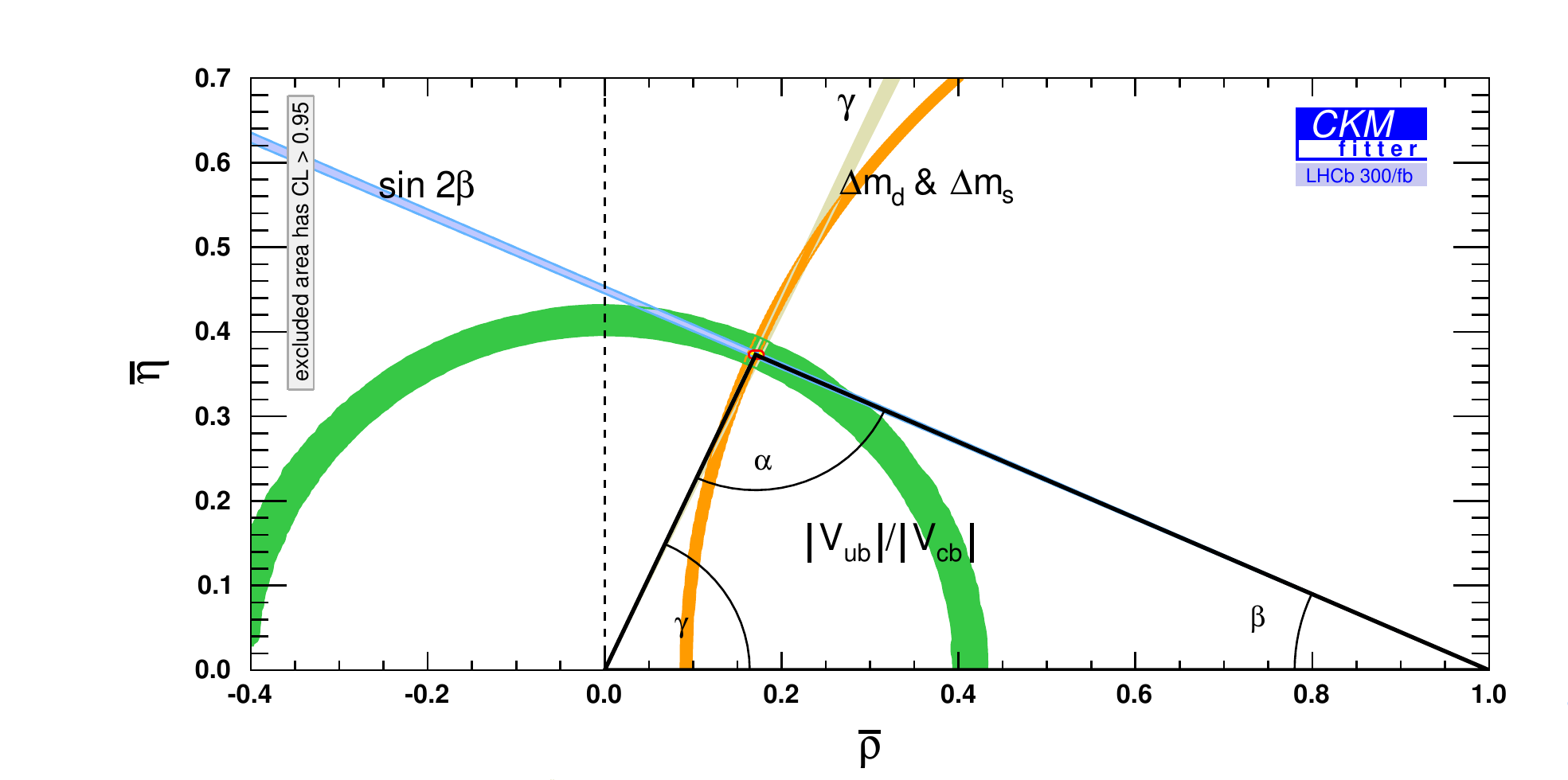}
\caption{\small  Evolving constraints in the $\bar{\rho}-\bar{\eta}$ plane from LHCb measurements and lattice QCD calculations, alone, with current inputs (2018), and the anticipated improvements from the data accumulated by 2025  (23\,fb$^{-1}$) and 2035 (300\,fb$^{-1}$), taking the values given in Table~\ref{tab:physummary}.
The hadronic parameter $\xi$ is a necessary input in the determination of the side opposite $\gamma$ and is assumed to be calculated with a precision of $0.6\%$ and $0.3\%$, in 2025 and 2035, respectively~\cite{XIGUESS}. In the future projections the central values of the inputs have been adjusted to provide internal consistency.
 }
\label{fig:UTprojection}
\end{center}
\end{figure}

\begin{figure}[H]
\begin{center}
\includegraphics[width=0.48\textwidth]{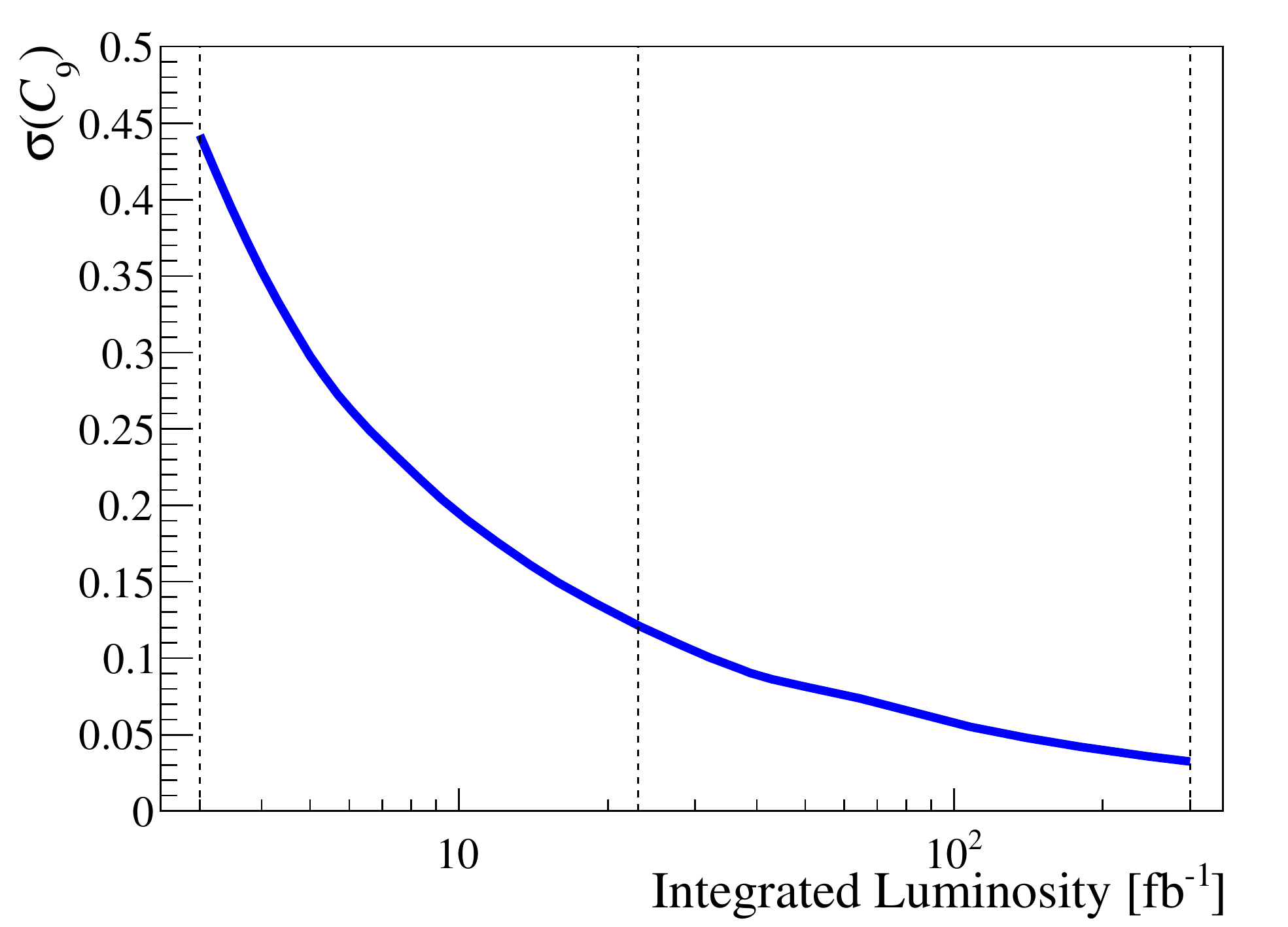}
\includegraphics[width=0.48\textwidth]{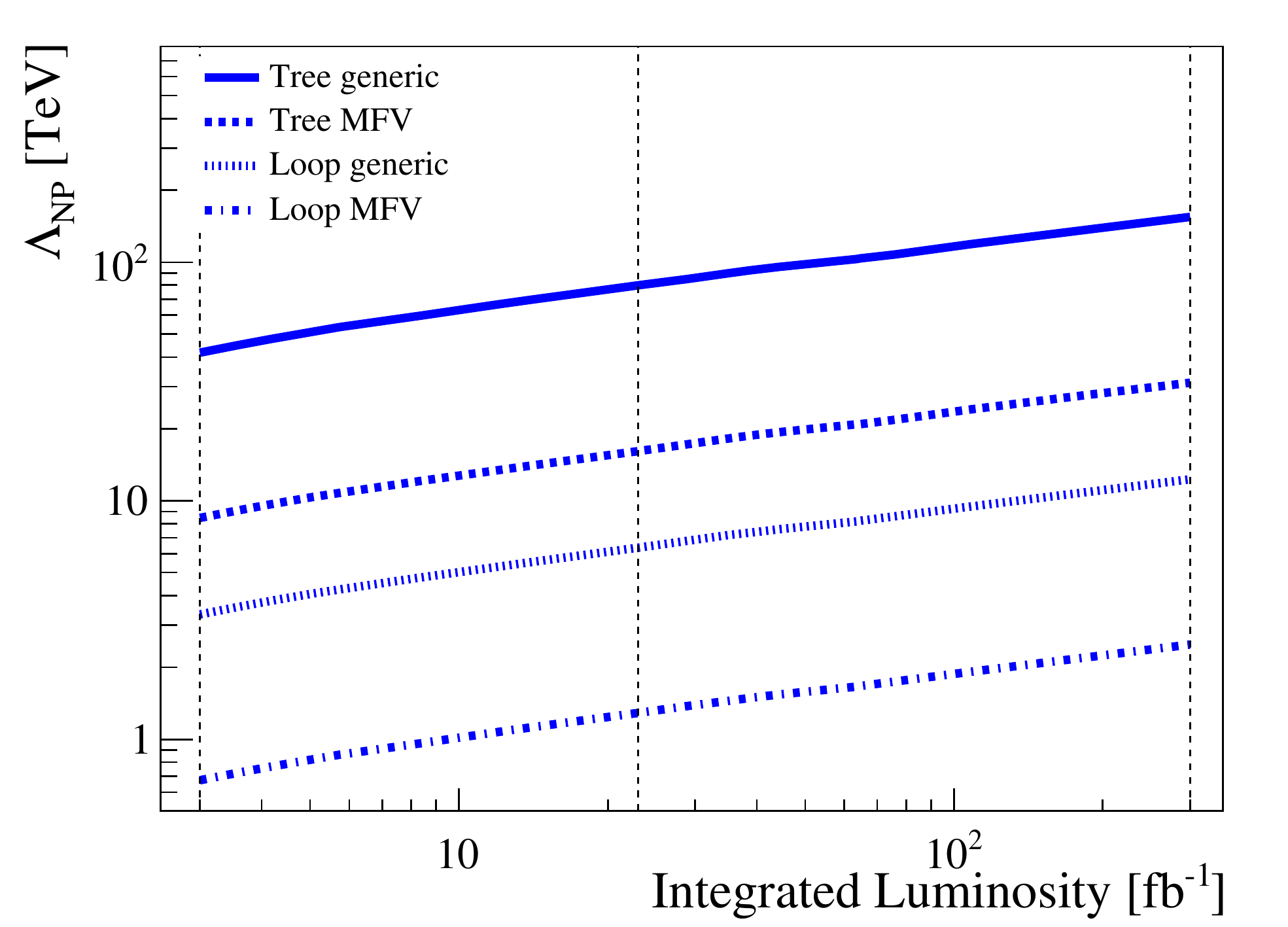}
\includegraphics[width=0.48\textwidth]{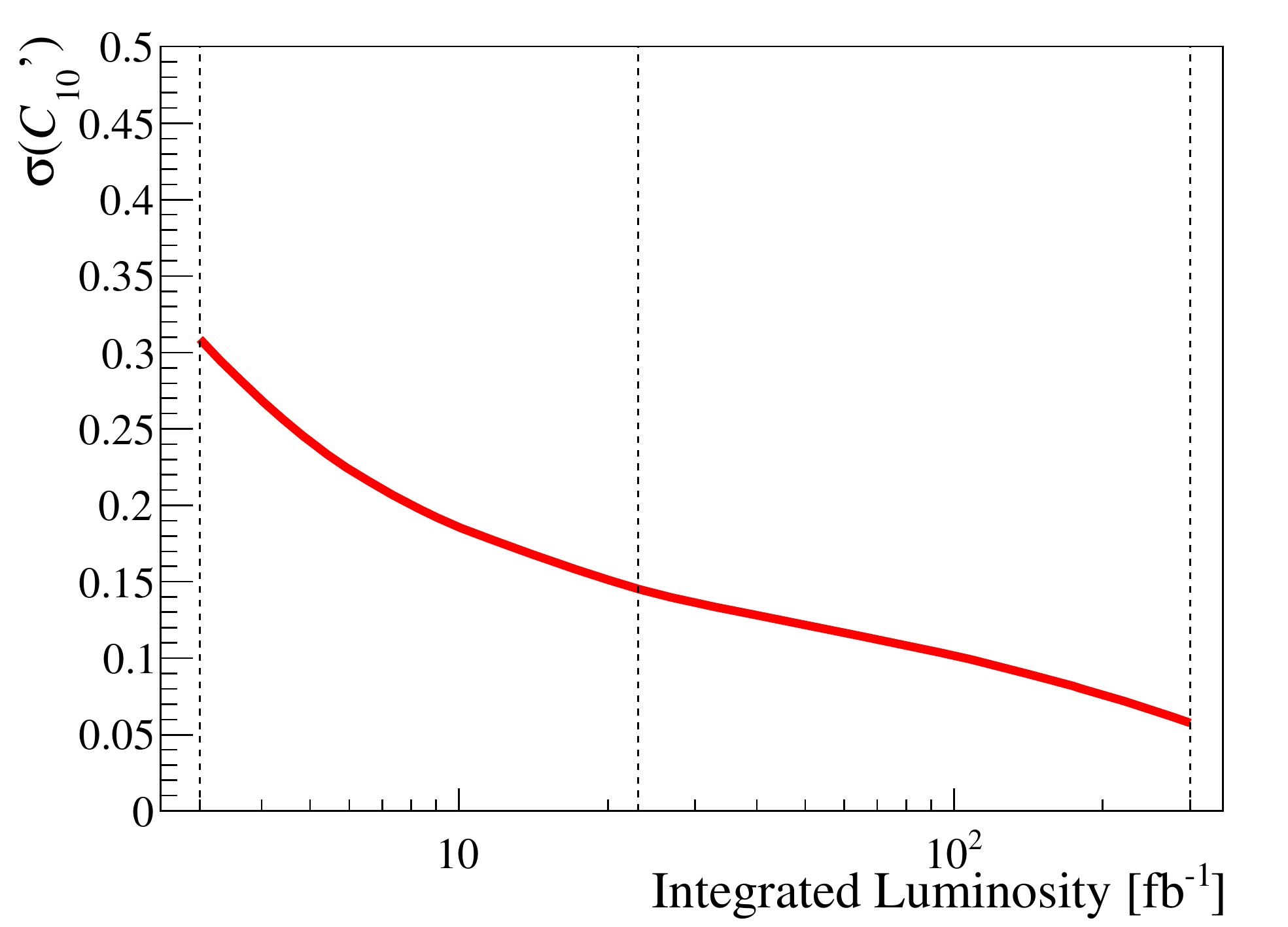}
\includegraphics[width=0.48\textwidth]{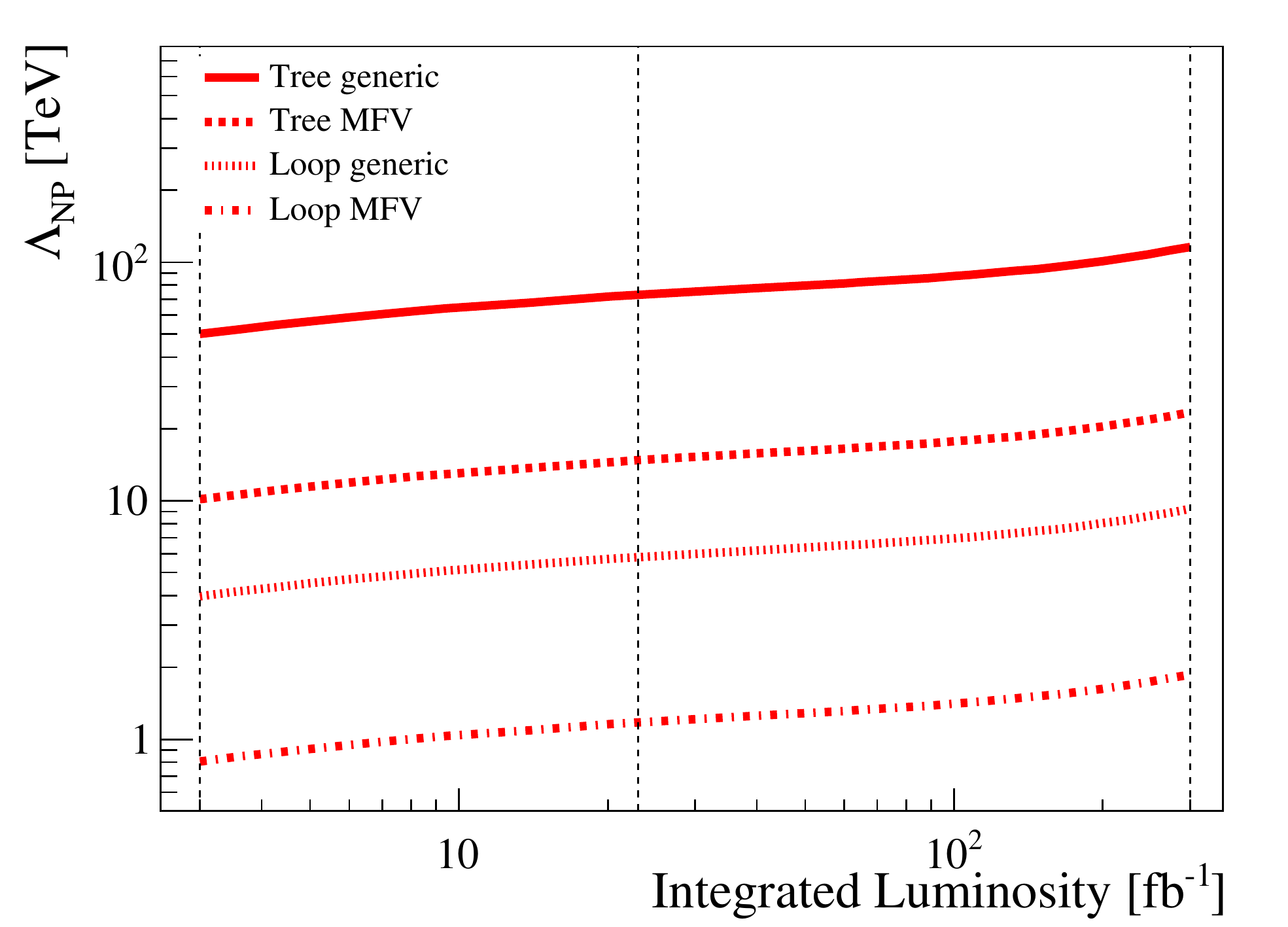}
\caption{\small Left, uncertainty on Wilson coefficients and, right, 90\% exclusion limits as a function of integrated luminosity.  The top plots present the $C_9$ analysis and use the ratio of branching fractions $R_K$ and $R_{K^\ast}$ in the range $1<q^2<6\,{\rm GeV}^2/c^4$.  The bottom plots show the $C^\prime_{10}$  study and use the angular observables $S_i$ from the decay $B^0 \to K^{\ast 0} \mu^+\mu^-$ in the ranges $1 < q^2 < 6\,{\rm GeV}^2/c^4$ and  $15 < q^2 < 19\,{\rm GeV}^2/c^4$.   The limits on the scale of New Physics, $\Lambda_{\rm NP}$, are given for the following scenarios: tree-level generic, tree-level minimum flavour violation, loop-level generic and loop-level minimum flavour violation.  Numerical results can be found in Table~\ref{tab:wilson_sum}. More information on the fits, particularly concerning the assumptions on the theory uncertainties in the $C^\prime_{10}$ study, may be found in Appendix~\ref{sec:wilson}.
}
\label{fig:wilson_sum}
\end{center}
\end{figure}

This list, though not exhaustive, illustrates well the two principal arguments that motivate the Upgrade II of LHCb, and the full exploitation of the HL-LHC for flavour physics.
\begin{enumerate}
\item
There is a host of measurements of {\it theoretically clean} observables, such as the \CP-violating phase $\gamma$, the lepton-universality ratios $R_K$, $R_{K^\ast}$ {\it etc.}, or the ratio of branching fractions $R\equiv {\cal{B}} (\Bd \rightarrow \mu^+ \mu^-)$/${\cal{B}} (\Bs \rightarrow \mu^+ \mu^-)$, where knowledge will still be statistically limited after Upgrade I.  The same conclusion applies for other observables such as $\phi_s$ and $\sin 2 \beta$, where strategies exist to monitor and control possible Penguin pollution. The HL-LHC and the capabilities of LHCb Upgrade II offer a unique opportunity to take another stride forward in precision for these quantities.  Advances in lattice-QCD calculations will also motivate better measurements of other critical observables, {\it e.g.} $|\Vub|/|\Vcb|$.  

The anticipated impact of the improved knowledge of Unitarity Triangle parameters can be seen in Fig.~\ref{fig:UTprojection}, which shows the evolving constraints in the $\bar{\rho}-\bar{\eta}$ plane from LHCb inputs and lattice-QCD calculations, alone.   The increased sensitivity will allow for extremely precise tests of the CKM paradigm. In particular, it will permit the tree-level observables, which provide Standard-Model benchmarks, to  be assessed against  those with loop contributions, which are more susceptible to New Physics.  In practice, this already very powerful ensemble of constraints will be augmented by complementary measurements from Belle II, particularly in the case of $|\Vub|/|\Vcb|$.

The increasing precision of observables from measurements of statistically-limited FCNC processes will provide significant improvements in sensitivity to the scale of New Physics.  As an example, Table~\ref{tab:wilson_sum} and  Fig.~\ref{fig:wilson_sum} show the expected improvement with integrated luminosity in the knowledge of the Wilson coefficients $C_9$ (vector current) and $C^\prime_{10}$ (right-handed axial-vector current), and the corresponding 90\% exclusion limits to the New Physics scale $\Lambda_{\rm NP}$ under various scenarios. These fit results take as input for $C_9$ the measurements in the low-$q^2$ region of $R_K$ and $R_{K^\ast}$,
and for $C^\prime_{10}$ angular observables  from the decay $B^0 \to K^{\ast 0} \mu^+\mu^-$.  The reach for generic New Physics at tree-level in Upgrade~II is found to exceed 100~TeV, and  for the $C_9$ study shows an approximate factor-of-two gain in progressing  from the 23\,fb$^{-1}$ to the 300\,fb$^{-1}$ data sets.  The gain is slightly less for $C^\prime_{10}$, a difference that can be attributed to conservative assumptions on potential hadronic charm-loop contributions.  Information on how the fits are performed, and the assumptions made, may be found  in Appendix~\ref{sec:wilson}.

\begin{table}[htb]
\caption{\small
Uncertainty on Wilson coefficients and 90\% exclusion limits on New Physics scales for different data samples.  The  $C_9$ analysis is based on the ratio of branching fractions $R_K$ and $R_{K^\ast}$ in the range $1<q^2<6\,{\rm GeV}^2/c^4$.  The $C^\prime_{10}$  analysis exploits the angular observables $S_i$ from the decay $B^0 \to K^{\ast 0} \mu^+\mu^-$ in the ranges $1 < q^2 < 6\,{\rm GeV}^2/c^4$ and  $15 < q^2 < 19\,{\rm GeV}^2/c^4$.   The limits on the scale of New Physics, $\Lambda_{\rm NP}$, are given for the following scenarios: tree-level generic, tree-level minimum flavour violation, loop-level generic and loop-level minimum flavour violation.  More information on the fits, particularly concerning the assumptions on the theory uncertainties in the $C^\prime_{10}$ study, may be found in Appendix~\ref{sec:wilson}} \label{tab:wilson_sum}
  \begin{center}
  \begin{tabular}{lrrr}\hline
    Integrated Luminosity & $3\invfb$ & $23\invfb$ & $300\invfb$\\ \hline\hline
    \multicolumn{4}{c}{$R_K$ and $R_{K^*}$ measurements}\\\hline
    $\sigma(C_9)$ & 0.44 & 0.12 & 0.03\\
    $\Lambda_\textrm{NP}^\textrm{tree\,generic}~[\tev]$ & 40 & 80 & 155\\
    $\Lambda_\textrm{NP}^\textrm{tree\,MFV}~[\tev]$ & 8 & 16 & 31\\
    $\Lambda_\textrm{NP}^\textrm{loop\,generic}~[\tev]$ & 3 & 6 & 12\\
    $\Lambda_\textrm{NP}^\textrm{loop\,MFV}~[\tev]$ & 0.7 & 1.3 & 2.5\\ \hline
    \multicolumn{4}{c}{$\decay{\Bd}{\Kstarz\mumu}$ angular analysis}\\\hline   
    $\sigma^\textrm{stat}(S_i)$ & 0.034--0.058 & 0.009--0.016 & 0.003--0.004\\
    $\sigma(C_{10}^\prime)$ & 0.31 & 0.15 & 0.06 \\
    $\Lambda_\textrm{NP}^\textrm{tree\,generic}~[\tev]$ & 50 & 75 & 115\\
    $\Lambda_\textrm{NP}^\textrm{tree\,MFV}~[\tev]$ & 10 & 15 & 23\\
    $\Lambda_\textrm{NP}^\textrm{loop\,generic}~[\tev]$ & 4 & 6 & 9\\
    $\Lambda_\textrm{NP}^\textrm{loop\,MFV}~[\tev]$ & 0.8 & 1.2 & 1.9\\
    \hline\end{tabular}
    \end{center}
\end{table}

\item
It will be essential to {\it widen the set of observables under study} beyond those accessible at the current experiment and at Upgrade I, {\it e.g.} by including additional important measurements involving  $b \to s/d \ellell$ and $b \to c \ell^-\bar{\nu_l}$ decays.  Improving our knowledge of the flavour sector both through better measurements and through new observables will be essential in searching for, and then characterising, New Physics in the HL-LHC era.
\end{enumerate}

\section{Comparison with Belle II and Phase-II ATLAS and CMS} 

Belle II will build on the success of the first generation $B$-factory experiments and allow $e^+e^-$ analyses to join LHCb at the frontier of precision flavour measurements.  The complementarity of the two experiments will be essential 
in exploring this territory, and making sense of any anomalies that emerge.
Belle II has certain capabilities that are unique, such as the ability to perform inclusive measurements and make precise studies of decay modes involving several neutral particles.  It will also attain similar sensitivity to Upgrade I for certain important measurements where LHCb is currently world-leading, for example that of the Unitarity Triangle angle $\gamma$.  However, as is clear from Table~\ref{tab:physummary}, the performance of Upgrade II will significantly exceed that of Belle II for the majority of observables of interest.  This is true not only for $B$ decays to charged final states involving hadrons and dimuons, and charm physics, but also for decays involving electrons, single neutrals, and semi-leptonic modes.  In addition, Upgrade II will allow LHCb to continue its exciting programme of studies in the \Bs sector, where opportunities are very limited at Belle II, and in the \Bc and $b$-baryon systems, which are uniquely accessible to the LHC.

The $B$-physics capabilities of ATLAS and CMS will be enhanced by their Phase II Upgrades.  These improvements, together with the very large data sets that are foreseen, will allow for precise measurements to be performed, in particular for final states including dimuons.  However, the sensitivities of these experiments will be lower than for those of LHCb Upgrade~II, which will also have a very much wider programme of flavour physics.

\section{Comparison with Upgrade I}

The improvement in physics reach between Upgrade I and Upgrade II will be greater than the simple factor of six in integrated luminosity would indicate (or factor 13, when integrating from Run 4 onward).  The intention to operate a flavour-physics experiment at luminosities of $10^{34}\,{\rm cm^{-2}s^{-1}}$ is already an ambitious one, but the planned improvements to the detector's capabilities will extend the physics gains still further.  These gains are not included in Table~\ref{tab:physummary} as full simulations have not yet been performed,  but the expected benefits are summarised below.

\begin{itemize}
\item{A high granularity, radiation-hard ECAL with fast-timing capabilities will be an essential component of the Upgrade-II experiment.  The conservative assumption is that this new subsystem will restore the capabilities of LHCb with electrons, photons and $\pi^0$ mesons to those of the existing detector in Run-1/-2 conditions, and improve upon the performance of this detector at Upgrade-I luminosities, which lie beyond its design specifications.  Depending on the exact technological solution adopted, further physics gains may arise from improved $\pi^0$ mass resolution and longitudinal segmentation.}
\item{Improvements to the tracking system will bring higher efficiencies for certain channels.  Most significant will be the installation of the Magnet Side Stations, which will be beneficial for multibody final states, decays involving `slow tracks', and for flavour tagging.}
\item{The hadron identification capabilities of the experiment will be enhanced.  Positive kaon and proton identification will be enabled below $10\,\gevc$ through the installation of the TORCH detector.  Improved Cherenkov angle resolution in the two RICH counters will increase $\pi-K$ separation at higher momenta.}
\item{Initial investigations yield several promising options for reducing, or even eliminating, the material before the first measured point in the VELO detector.  Such a development would bring great benefits for many flavour studies, in particular the reconstruction of semileptonic decays.}
\end{itemize}

\noindent As previously discussed, it is intended to take first steps towards some of these detector enhancements already in LS3, before the start of the HL-LHC , thereby improving the performance of the Upgrade-I experiment, and laying the foundations for Upgrade II.  

Finally, it must be emphasised that the raw gain in sample sizes during the HL-LHC era will have great consequences for the physics reach, irrespective of any detector improvements.   The energy scale probed by virtual loops in flavour observables will rise by a factor of up to 1.9 with respect to the pre-HL-LHC era, with a corresponding gain in discovery potential that is similar to what will apply for direct searches if the beam energy is doubled, as proposed for the HE-LHC.

\section{Opportunities beyond flavour}

LHCb has emphatically demonstrated its ability to make discoveries and perform unique measurements in topics beyond flavour physics.  Significant examples include high-impact results in spectroscopy, such as the discovery of pentaquarks, precise measurements in electroweak physics, and highly sensitive searches for dark photons. These achievements have been enabled by the forward acceptance of the spectrometer, and the specialised instrumentation and trigger of the experiment.  Upgrade II will provide the essential opportunity for this programme to continue with much larger data sets and ensure that the HL-LHC is exploited fully for the widest set of physics goals.  Table~\ref{tab:beyondflavour} recapitulates some of the main non-flavour topics that will be explored at Upgrade II.   Other exciting possibilities in physics beyond flavour, including AA and $p$A collisions, fixed-target physics and searches for long-lived particles, are discussed in Appendix~\ref{chpt:further}.

\begin{table}[htb!]
\caption{\small Opportunities beyond flavour at Upgrade II.}\label{tab:beyondflavour}
\begin{center}
\begin{tabular}{ll}\hline
Topic & Comment \\  \hline
Spectroscopy & Enormous yields in gold-plated final states \\
& {\it e.g.} 4M $\Lb \to J\psi p\Km$ decays (`pentaquark' mode) \\
Higgs &  Measure Higgs-charm Yukawa within factor 2 to 3 of SM value \\
$\sin^2\theta_W$ &  Uncertainty $<10^{-4}$, better than LEP/SLD\\
Proton structure & Precision probes at extremely low and high Bjorken-x values, \\
& with $Q^2 > 10^5\,  {\rm GeV}^2$ \\
Hidden sector & Sensitivity to most of relevant parameter space for dark-photon models \\
\hline
\end{tabular}
\end{center}
\end{table}

\section{Summary}

The principal goal of the HL-LHC is to search for, and then characterise, New Physics beyond the Standard Model.   The flavour programme proposed at Upgrade II of the LHCb experiment has exactly this focus,  and will allow the facility to be fully exploited for the precise measurement of observables that are sensitive to extremely high mass scales.  The detector will also have great capabilities in the areas of spectroscopy, electroweak measurements and dark-sector searches, building on the successes of the current experiment.   LHCb Upgrade II is a well motivated, ambitious, but achievable project, with outstanding discovery potential, and an assured programme of important and necessary measurements.

%% file: CONTRIBUTIONS/11_Summary_and_conclusions/11.2.tex

%% file: CONTRIBUTIONS/10_Further_opportunities/10.tex
\label{chpt:further}

 LHCb has very successfully 
expanded its physics program to topics beyond flavour physics
exploiting its unique forward geometry.
A very rich heavy ion physics programme has been developed,
with prominent results in proton-ion and ion-ion collision modes.
In addition, LHCb has successfully demonstrated  its ability to take data in configurations
like fixed target
or combined fixed target and collider mode
that are unique in the landscape of experiments at colliders.
These results have stimulated a number of proposals to exploit LHCb to develop
physics topics that, although not being part of the baseline HL-LHC operational
plan or requiring additional detector systems that are not part of the
core \upgradetwo detector, represent very exciting physics opportunities.

This appendix describes these proposals. The topics discussed are heavy-ion physics
(Sect.~\ref{sec:ion}), fixed target physics with gaseous targets
(Sect.~\ref{sec:fixedtarget}), measurement of magnetic and electric dipole moments of baryons
(Sect.~\ref{sec:dipolemoments}) and extending the programme for long-lived particle
searches (Sect.~\ref{codexb}).

\input{CONTRIBUTIONS/10_Further_opportunities/10.1.tex}

\input{CONTRIBUTIONS/10_Further_opportunities/10.2.tex}

%% file: CONTRIBUTIONS/10_Further_opportunities/10.1.tex
\section{Prospects with heavy-ion and fixed-target physics}
\input{CONTRIBUTIONS/10_Further_opportunities/10.1.1.tex}
\input{CONTRIBUTIONS/10_Further_opportunities/10.1.2.tex}
\input{CONTRIBUTIONS/10_Further_opportunities/10.1.3.tex}

%% file: CONTRIBUTIONS/10_Further_opportunities/10.1.1.tex
\subsection{Physics with $p$A and AA collisions}
\label{sec:ion}
Although there are no current plans for ion running beyond LS4, this possibility cannot be excluded if interesting results emerge in Runs 3 and 4.  
Therefore we present here  a selection of strong capabilities \upgradetwo would have in this domain.

The LHCb \upgradetwo with improved granularity and resolutions is inherently attractive for heavy-ion collision measurements. It offers the opportunity of a general purpose heavy-ion experiment suited from $p$A up to most central AA collisions at forward rapidity. 
The forward acceptance allows low-pT scales to be reached, with typical particle momenta of $3-100~\gevc$ in the laboratory frame. The LHCb \upgradetwo  therefore allows the study of the 
physics of nonperturbative scales.
The experimental set-up is unprecedented and unmatched by any current or planned detector for studies in heavy-ion collisions, thus representing a unique opportunity in this domain..

At \upgradetwo luminosity, the number of tracks in $pp$ collisions will be close to that observed in central nucleus-nucleus collisions. Therefore, the LHCb \upgradetwo, with its increased granularity, will provide sufficient tracking and calorimeter performance for precise measurements for a large range of observables in heavy ions collisions.
Occupancy estimates  for the most critical LHCb tracking detectors are given in Table~\ref{tab:occup}. 
In asymmetric $p$Pb and Pb$p$ collisions, the charged particle densities per primary vertex are about $3-6$ times larger than in $pp$ in the LHCb acceptance.  Hence, taking into account the lower pile-up at ion running luminosities, $p$Pb and Pb$p$ events will have drastically lower occupancies than is the case for standard pp collisions.

\begin{table}[tb]
\begin{center}
\caption{Estimated occupancies in different tracking detectors after \upgradeone and \upgradetwo.}
  \renewcommand{\arraystretch}{1.2}
 \begin{tabular}{lccc}
 \hline
 Detector                   & Maximum occupancy in most central  \\
                                 &  PbPb at $\sqrt{s_{NN}}=5$~TeV        \\ \hline
 VELO (\upgradeone)                   &         4~\%                                                               \\ 
 VELO upgrade (\upgradetwo)           &      1~\%                                                                          \\
  SciFi (\upgradetwo)                & 25\%                                                                     \\ 
 \hline
\end{tabular}
\renewcommand{\arraystretch}{1.0}
 \label{tab:occup}
\end{center}
\end{table}

Therefore, thanks to the  LHCb acceptance and instrumentation, these events will provide the best way to precisely measure the low-$x$ regime of QCD, beyond the reach of the proposed electron-ion collider~\cite{Accardi:2012qut}. 

In the following, a selection of LHCb \upgradetwo opportunities beyond the LHC Run 3 and 4 heavy-ion programme is given, highlighting areas that will benefit from the improved strength of the LHCb set-up, based on available experience. 



\subsubsection{Quarkonium and open heavy flavour}
\label{sec:hiquarkonium}

Heavy quarks are produced at early stages in ion collisions and experience the full system dynamics until hadronisation.   
Heavy quarkonia and open heavy-flavour hadrons are therefore ideally suited to study deconfinement at finite temperature~\cite{Mocsy:2013syh, Rapp:2018qla}.
 
Precise prompt production and flow coefficient measurements of the charmonium vector state  \jpsi are planned in Run 3 and 4 by ALICE. High-\pt prompt and nonprompt \jpsi nuclear modification measurements by ATLAS and CMS will probe energy loss with different quarks and \jpsi fragmentation in QCD matter~\cite{CMS:2013gga,CMS:2017dec}. Charm mesons and baryons will be reconstructed in the mid- and high-\pt range in nucleus-nucleus collisions. 

 However, in Run 3 and 4, \psitwos yield measurements at low \pt will remain statistically limited, with low signal over noise ratio.
 The statistical uncertainties of kinematic-integrated yields are projected to be at a 3$\% $(10$\%$) level at forward rapidity (mid rapidity) in 10\% most-central collisions at ALICE~\cite{Abelevetal:2014cna}. This precision will not allow precise differential measurements as a function of \pt and rapidity $y$ as in the \jpsi case~\cite{Abelevetal:2014cna}.  Fully reconstructed open heavy-flavour hadrons will be available only at mid rapidity in ALICE, CMS and ATLAS.  Finally, $P$-wave states like \eg the \chic states, which are particularly interesting to study colour charges in deconfined matter~\cite{Arleo:2012dxa}, will remain out of reach at ALICE~\cite{Weber} at low $\pt$.

 In the bottomonium sector, the CMS measurements of the $\PUpsilon$-states will become precise measurements.  However, the measurements by ALICE or CMS at forward rapidity for the excited states will be limited by the experimental resolutions. The natural normalisation to understand \bbbar~bound states in heavy-ion collisions is the integrated total \bbbar cross-section.  The measurements with fully reconstructed open beauty hadrons including \Lb will remain challenging  down to low-$\pt$~\cite{Abelevetal:2014cna,CMS:2013gga,CMS:2017dec}. More precise normalisations will rely on assumptions on the fragmentation patterns and the usage of nonprompt \jpsi~\cite{Winn:2016amg}. 

 In this panorama, LHCb \upgradetwo with its resolution and ion-ion capability offers 
 unique opportunities. Precise \psitwos and $\PUpsilon$ production measurements will
allow kinematic dependencies to be studied at forward rapidity down to $\pt=0$ with suitable open-charm and beauty normalisation channels.
The LHCb \upgradetwo is the only LHC detector with access to $P$-wave states at low \pt. 
In particular, the decay channel discovered by LHCb $\chic\to\jpsi+\mu^+\mu^-$~\cite{LHCb-PAPER-2017-036}
 may become accessible, although this will require higher integrated luminosities 
than currently foreseen  for the PbPb collision runs.


%


 The boost at forward rapidity and the vertex reconstruction performance will allow open charm and open beauty mesons to be accessed down to \pt = 0 in a low-background environment as proven in $p$Pb and Pb$p$ collisions by LHCb~\cite{LHCb-PAPER-2017-015,LHCb-PAPER-2017-014}.
 The study of baryonic heavy-quark states is of primary importance to understand hadronisation in ion-ion collisions. In addition, these states  can be used 
to determine the total charm and beauty cross sections as a necessary reference for quarkonium studies.  The baryon measurements profit similarly from the longitudinal boost in LHCb, enabling high signal-to-background measurements down to low-\pt as demonstrated in the $p$Pb and Pb$p$ measurement by LHCb~\cite{LHCb-PAPER-2018-021}. In particular, these heavy-ion measurements based on the reconstruction of small life-times and low-$q$ decays as those of the \Lc baryon, will naturally profit from the improved vertex performance. 




\subsubsection{Low-mass dileptons and photons}
The measurement of low-mass dilepton resonance production and their line-shape is sensitive to chiral symmetry restoration in the QGP phase at finite temperature~\cite{RappWambachHees_chiralrestorationreview_SPS} and to the thermal radiation of the QGP. Measurements at SPS and RHIC~\cite{CERES1_lowmassdileptons,NA60_LMDileptons,CERES_rhomelting,STAR_lowmass_ee_200,Phenix_200_lowmassfinal}, in particular the NA60 measurement of the $\rho$ spectral function, show a prominent modification compatible with partial chiral restoration~\cite{RappWambachHees_chiralrestorationreview_SPS} and with thermal radiation emission at higher masses. The ALICE mid rapidity upgrade in Runs 3--4 aims at a measurement of low-mass dielectrons. 
In addition, at the LHC the baryochemical potential $\mu_B$ is compatible with zero and contact can be be made to first principle lattice QCD calculations~\cite{Aarts:2014nba}.

However, the measurement of thermal radiation in the mass region between the $\phi$ and \jpsi mesons will remain systematically limited by the subtraction of the heavy-flavour background from semileptonic decays. 

\begin{figure}
  \includegraphics[width=1.0\textwidth]{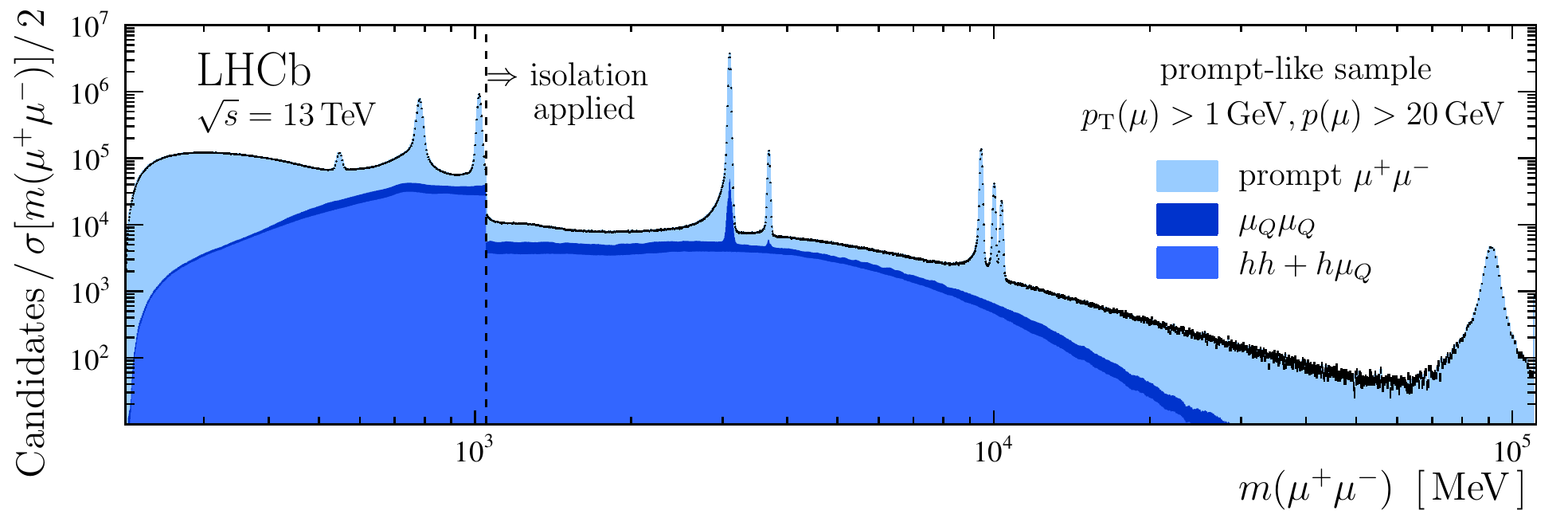}
  \caption{
Dimuon mass distrib. In the dark photon search at LHCb~\cite{LHCb-PAPER-2017-038}.  Note that the 
heavy-flavour background has been greatly suppressed.}
  \label{fig:lowmass}
  \end{figure}

LHCb \upgradetwo offers the unique potential to measure precisely dilepton production in the dimuon channel at the LHC. The measurement profits from the boost to remove the background muons from heavy-flavour semileptonic decays at forward rapidity, as successfully demonstrated 
in the context of dark-photon searches\cite{LHCb-PAPER-2017-038} in $pp$~collisions (see Fig.~\ref{fig:lowmass}).  
An extension of dilepton continuum measurements in ion-ion collisions with LHCb \upgradetwo down to the momentum cut-off of 3~\gevc (corresponding to \pt~thresholds between $200-400$~\mevc within the pseudorapidity range $2.5<\eta<4.0$) is doable.  Therefore, LHCb \upgradetwo has the potential for a textbook measurement in the $\rho$-mass region to probe chiral restoration.

In addition, thermal radiation can be measured in the real photon channel. The boosted configuration of LHCb allows low-\pt converted photons to be measured
at moderate track momenta. The momentum cut-off of the spectrometer amounts to around 2~GeV/$c$. This cut-off will be further reduced with the magnet stations on the side of the dipole magnet.  

\subsubsection{Drell-Yan, photons, \ccbar and \bbbar in $p$A collisions in view of nuclear PDFs and saturation}

The performance of LHCb is already near-optimal for $p$A collisions. Moreover, the capability of precise cross-section measurements of weak gauge bosons~\cite{LHCb-PAPER-2015-049} and unique low-mass Drell-Yan measurements down to the charmonium mass range~\cite{LHCb-CONF-2012-013} in $pp$ collisions show the potential for high-luminosity running in $p$A~collisions. Similar or better performance for the same signal yields  in $p$Pb collisions can be expected with LHCb \upgradetwo. Currently, the low-$x$ regime of nuclear parton distribution functions is largely unconstrained due to a lack of clean experimental data.
Because of the higher gluon density in the Lorentz-contracted nucleus compared to the proton, the low-$x$ regime of the nucleus is a good place to search for gluon saturation~\cite{Iancu:2003xm}. 

LHCb can perform measurements inaccessible even to the planned electron-ion 
collider. In the LHCb forward acceptance and at  hard momentum scales between 3 and 100~\gevc it is possible to probe the nucleus in the Bjorken-$x$ range $10^{-3}-10^{-6}$. This range is largely inaccessible at hard scales for the electron-ion collider that has a centre-of-mass energy of 140 GeV in its high-energy option. Hence, LHCb is a very good place to look for the break-down of linear parton evolution equations and, finally,  probe gluon saturation within a large kinematic range. 

In particular, precise low mass Drell-Yan (below masses of 10 GeV/$c^2$) or low-\pt below 10~GeV/$c^2$ photon measurements are world-leading unique opportunities. 
For the Drell-Yan measurement, the precision of the VELO and the forward acceptance allow the background from semileptonic heavy-flavour decays  to be controlled. 
For precise measurements, large luminosities will be required. An assessment of the possibilities for Run 3 and Run 4 will be given in the in the forthcoming HL-LHC report. Given the experience from $pp$ collisions, the measurement will strongly profit from yet larger accumulated luminosities. 
Photon measurements will benefit from improved calorimeter performances at high charged track multiplicity.
In addition, associated productions of $\gamma$+jet or correlation measurements will profit from the magnet stations.

Furthermore, the already pioneered  \ccbar and \bbbar production measurements in $p$Pb collisions, will be complemented with correlation measurements. There are  unique possibilities to exploit specific features of the theory in kinematic limits as the appearance of new transverse momentum dependent parton distribution functions in the back-to-back configuration that only appear for massive quarks~\cite{Marquet:2017xwy}. In particular, correlations in the \bbbar~sector will only become available with increased luminosities in the range of 10~pb$^{-1}$ of $p$Pb collisions.

%% file: CONTRIBUTIONS/10_Further_opportunities/10.1.2.tex
\subsection{Prospects with fixed-target physics}
\label{sec:fixedtarget}

The \lhcb experiment has pioneered the use of LHC beams in
fixed target mode since Run 2, using an internal gas
target. In this configuration, for which the forward
geometry of the detector is particularly well suited, collisions at an
energy scale $\sqsnn \sim 100$ GeV can be studied using targets of different
nuclear size  (currently He, Ne or Ar), with unique coverage of the
high-$x$ regime in the target nucleon. The samples collected during Run 2, corresponding
to integrated luminosities up to about 100~\invnb, allow for studies
of particle production in the soft QCD regime, of particular relevance
to cosmic ray physics~\cite{LHCb-PAPER-2018-031}, and to collect 
unprecedented samples of charmed mesons in fixed-target collisions at
this energy scale~\cite{LHCb-CONF-2017-001}. These data provide a unique test bench to
discriminate cold nuclear matter effects in heavy-flavour production
from the effect of deconfinement, and to study nuclear PDFs 
at large $x$. The physics reach of heavy-flavour studies is presently
limited by the size of these samples. 

The \lhcb collaboration is presently considering several proposals to
develop this programme with improved target systems. In the current
setup, the gas target is injected directly inside the beam pipe
surrounding the vertex detector, with a gas pressure limited  to a few
$10^{-7}$ mbar by beam safety considerations.
The density can be increased by at least one order of magnitude
by containing the injected gas inside a storage cell, which is being designed
and could be operational already during Run 3. The new setup would also allow 
other gases to be injected, notably hydrogen and deuterium, providing
$pp$ collisions in fixed-target mode as a reference for all $p$A
collision samples, and extending the physics case to the study
of the three-dimensional structure functions of the nucleon through
spin-independent observables~\cite{3dpdf}. 

More ambitious projects have also been proposed for a possible installation
in LHCb on the time scale of HL-LHC. These are beyond the baseline
\upgradetwo detector considered in this document.
A polarised gas target similar to that used in
HERMES~\cite{Airapetian:2004yf}, installed upstream of the \lhcb vertex
detector, would make \lhcb a key contributor to spin physics. This
would enable a rich programme in spin physics in a
unique kinematic range for a variety of final states, including unique
measurements with quarkonium and Drell-Yan lepton pairs~\cite{Kikola:2017hnp}. 
The study of collisions of Pb beams on heavy nuclei, that could also
be possible with the gas target (\eg using Xenon), are
presently limited by the detector tracking capabilities and would
greatly profit from the higher granularity and better coverage of the
low-momentum region offered by the \upgradetwo detector.

Assuming that about 10\% of the beam intensity can be exploited for
fixed-target physics, either in synergy with $pp$ data taking, using
the beam bunches not colliding in \lhcb, or through dedicated runs,
data corresponding to integrated luminosities of order 0.1~\invfb per year could be collected 
using the proposed targets, also profiting from the increased beam
intensity provided by the HL-LHC.
Samples of this size would allow copious production of Drell-Yan
and heavy flavour states, including \bbbar mesons, 
opening many novel possibilities, which have been studied in the
last years, notably by the AFTER initiative~\cite{Brodsky:2012vg,
after}. Substantial advancements in the understanding of parton
distributions for gluons, antiquark and heavy-quarks 
at $x>0.5$, where PDFs are now essentially unconstrained, can be
expected. These studies can clarify the extent of the intrinsic heavy
quark distributions in the nucleon, and of the 
modifications of the nuclear PDFs (anti-shadowing, EMC effect, Fermi motion),
which are relevant to the understanding of initial state effects 
in heavy-ion collisions.
Signatures of deconfinement, such as sequential suppression of \jpsi, 
\psitwos and \chic production, can be investigated at an energy scale
between the SPS and RHIC/LHC, complementing the studies in AA collisions
discussed in Sect.~\ref{sec:hiquarkonium}.
This programme would be unrivalled in terms of achievable energy and
signal yields in fixed target configuration.


%% file: CONTRIBUTIONS/10_Further_opportunities/10.1.3.tex
\subsection{Magnetic and electric dipole moments of heavy and strange baryons}
\label{sec:dipolemoments}

The magnetic (MDM) and electric (EDM) dipole moments are important static properties of elementary particles.
To date, these properties have not been accessible experimentally for heavy (charm and beauty) baryons, and $\tau$
leptons, due to their short lifetime, $\sim 10^{-13} - 10^{-12}\sec$, with decay lengths of few \cm for \tev energies. 
%
For strange baryons from the baryon octet $J^P=\frac12^+$, with lifetimes about three orders of magnitude higher,
the MDMs have been measured with accuracies at the percent level or
slightly better\cite{PDG2018}. 
Direct upper bounds for the EDM of strange baryons only exist for
the case of the \Lz hyperon, $1.5\times 10^{-16}~e\cm$\cite{Schachinger:1978qs,Pondrom:1981gu}.

Measurements of the MDM of heavy baryons,  would provide valuable
information for low-energy QCD calculations, and if a percent-level
accuracy is achieved, they could be used to  
 discriminate between different
 models\cite{Sharma:2010vv,Dhir:2013nka}, improving the current
 understanding of the internal structure of hadrons.
Searches for EDM, arising from flavour-diagonal \CP violation and strongly suppressed within the SM, 
provide a very clean tool to search for BSM physics. From general considerations, a precision
at $10^{-17}-10^{-18}$~$e$\cm level would be required to access the electroweak scale (1--10\tev)~\cite{Pospelov:2005pr,Pospelov:2018}.
%
Current indirect bounds are set from different experimental measurements, \eg neutron and electron EDM, and span
several orders of magnitude,
$\sim 10^{-15}-10^{-17}~e\cm$~\cite{Sala:2013osa,Zhao:2016jcx,Grozin:2009jq,Escribano:1993xr,Blinov:2008mu} for charm
($\sim 10^{-12}-10^{-17}~e\cm$~\cite{Grozin:2009jq,Escribano:1993xr,Blinov:2008mu,CorderoCid:2007uc} for beauty),
depending on the different models and assumptions. Indirect bounds on strange EDM based on neutron EDM are much more 
stringent, $\sim 10^{-23}~e\cm$\cite{Guo:2012vf,Atwood:1992fb}.

\begin{figure*}[tb]
\centering
{\includegraphics[width=0.7\linewidth]{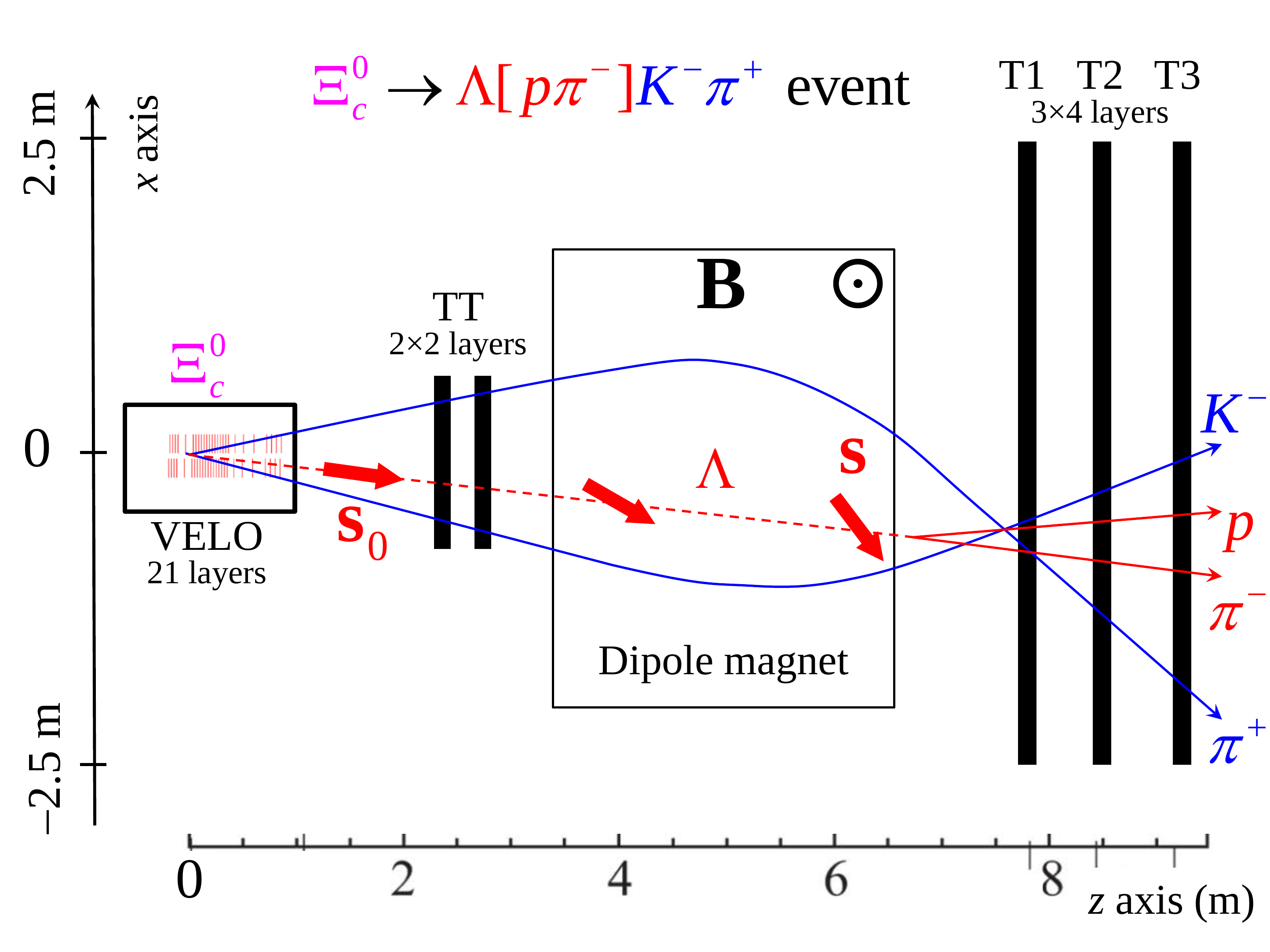} }\\
{\includegraphics[width=\linewidth]{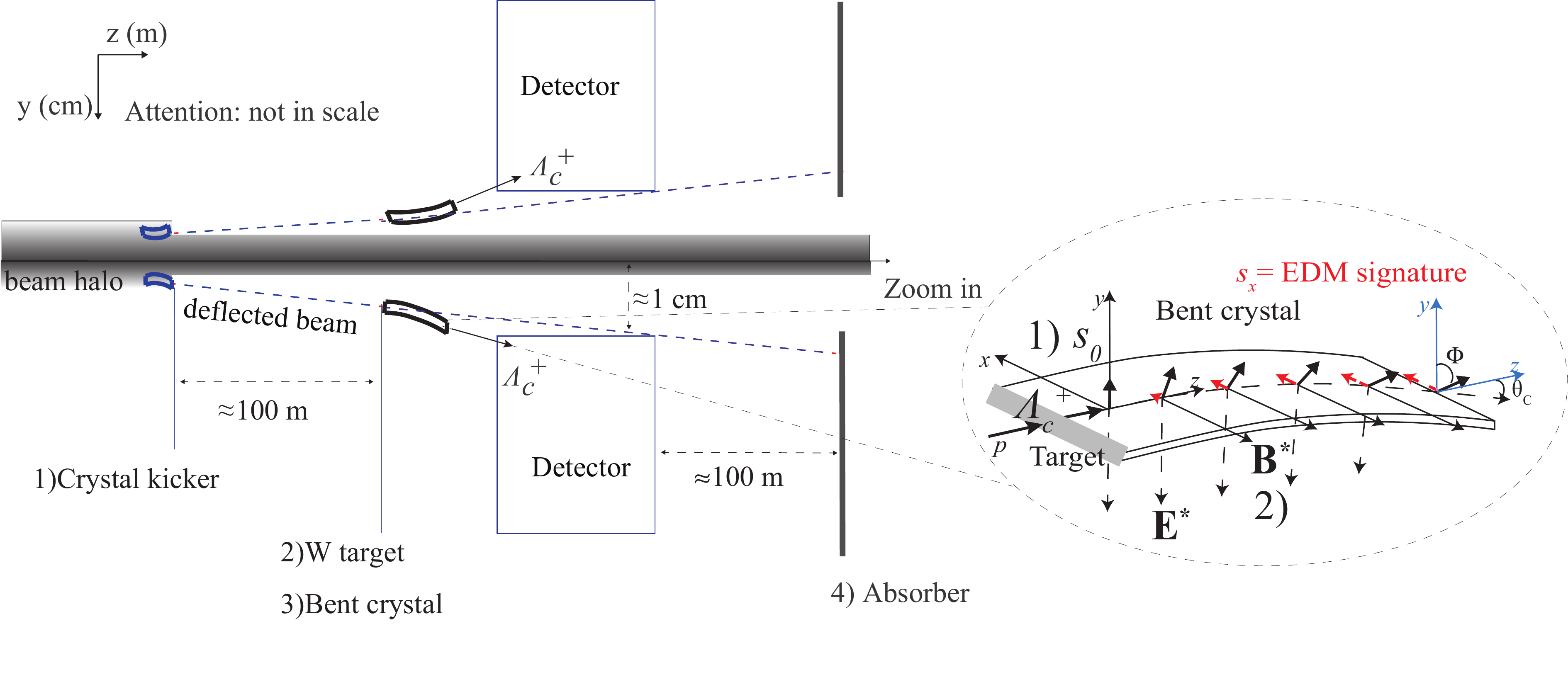} }
\caption{(Top) Sketch of a $\Xicz \to \Lz[\proton\pim]\Km\pip$ charm baryon decay. The spin precession induced by the \lhcb magnetic field is
  indicated by the rotation of the red arrow.
%
(Bottom) Layout of the fixed-target setup shown in side view with down- and up-bending crystals.
  The zoom in shows the baryon trajectory and spin precession in the down-bending crystal for channelled \Lc baryons induced by the MDM (black arrow) and EDM (red arrows).
}
\label{fig:MDM_EDM_Layouts}
\end{figure*}

The possibility to access the MDM and EDM of heavy and strange baryons at the \lhc has been explored in recent
years\cite{Baryshevsky:2016cul,Burmistrov:2194564,Bezshyyko:2017var,Baryshevsky:2017yhk,Botella:2016ksl,Bagli:2017foe,LHCb-INT-2017-011,Baryshevsky:2018dqq,Scandale:2016krl}.
The experimental approach to measure the electric and magnetic dipole moments of unstable
  baryons relies on the spin precession in external electromagnetic fields.
  For long-lived \Lz baryons produced longitudinally polarised from weak charm-baryon decays,
  the magnetic field of the LHCb dipole magnet can be exploited to induce
  spin precession, as illustrated in Fig.~\ref{fig:MDM_EDM_Layouts}~(top).
  For short-lived charm and beauty baryons,
  \ie~\Lc, \Xicp, \Xibbarp and \Omegabbar, the intense electromagnetic field
  between the atomic planes of bent crystals is necessary to induce spin precession.
  For this purpose, a fixed-target experiment has been proposed to produce and
  channel charged heavy baryons in bent crystals positioned in front of the \lhcb
  detector, as sketched in Fig.~\ref{fig:MDM_EDM_Layouts}~(bottom).
  %




Details on the feasibility of this proposal and on detailed
simulation studies are given in Refs.~\cite{Bagli:2017foe,Bezshyyko:2017var,LHCb-INT-2017-011}.
  More studies and R\&D are ongoing to assess the compatibility with the \lhc machine, the operation mode
  (synergetic \vs dedicated) and attainable proton flux, along with the feasibility of the 15\mrad bent crystal and double-crystal scheme\cite{FerroLuzzi:PBC2018,Redaelli:PBC2016}.
A potential installation of the proposed device is studied for the first
winter shut-down during Run~3.
Potential data taking schemes, compatible with the mainstream LHCb
running conditions in Run~3 and Run~4 have been studied, leading to MDM (EDM) sensitivities at $10^{-3}$, $10^{-1}$ and $10^{-3}$~$\mu_N$ ($10^{-17}$, $10^{-14}$ and $10^{-16}$~$e\cm$) level~\cite{Bagli:2017foe},
  for charm, beauty and strange baryons, respectively ($\mu_N$ is the nuclear magneton).
  %
An increase of the proton flux in Run 5, combined with detector
improvements would give better sensitivities by about one order of magnitude.
  %
  %
  For the long-lived \Lz baryon, a two orders of magnitude improved precision on the EDM compared to current measurements is expected by the end of Run 4,
  \ie $\sim 10^{-18}~e\cm$\cite{Botella:2016ksl}.
  With \Lz and \Lbar baryons produced in approximately equal amounts from charm baryon decays,
  a test of \CPT symmetry similar to that performed at the BASE experiment~\cite{Smorra:2016vxa,Nagahama:2017eqh}
  can be performed at $10^{-4}$ level by measuring separately the MDMs
  of both states. By profiting from detector improvements and luminosity increases, these accuracies
  could be improved further by about one order of magnitude by the end of Run 5.
  %


%% file: CONTRIBUTIONS/10_Further_opportunities/10.2.tex
\section{The CODEX-b way to long-lived particle searches at LHCb}
\label{codexb}
Analogously to long-lived SM particles such as the \Kz meson, BSM theories generically predict the existence of metastable states,
also known as long-lived particles (LLPs). LLPs with lifetimes up to the sub-second regime are broadly consistent with cosmological bounds,
and have been studied in a wide variety of BSM theories. Because the range of possible LLP lifetimes and masses is very large, many
different existing and future experiments have searched for or proposed to search for LLPs, including ATLAS and CMS, \belletwo, 
SHiP~\cite{Alekhin:2015byh}, FASER~\cite{Feng:2017uoz}, MATHUSLA~\cite{Curtin:2018mvb}, and DUNE~\cite{Acciarri:2016crz}. The CODEX-b detector, 
described more fully in Ref.~\cite{Gligorov:2017nwh}, 
has been proposed
to augment LHCb's ability to search for LLPs. This is an additional detector system which is not part of the baseline LHCb \upgradetwo programme.  Its key features are a relatively compact detector volume, simple shielding to ensure a zero-background
environment, and the ability to integrate into LHCb's triggerless DAQ and potentially tag or characterise observed LLP candidates based on activity within LHCb.

\begin{figure*}[t]
	\includegraphics[width = 0.49\linewidth]{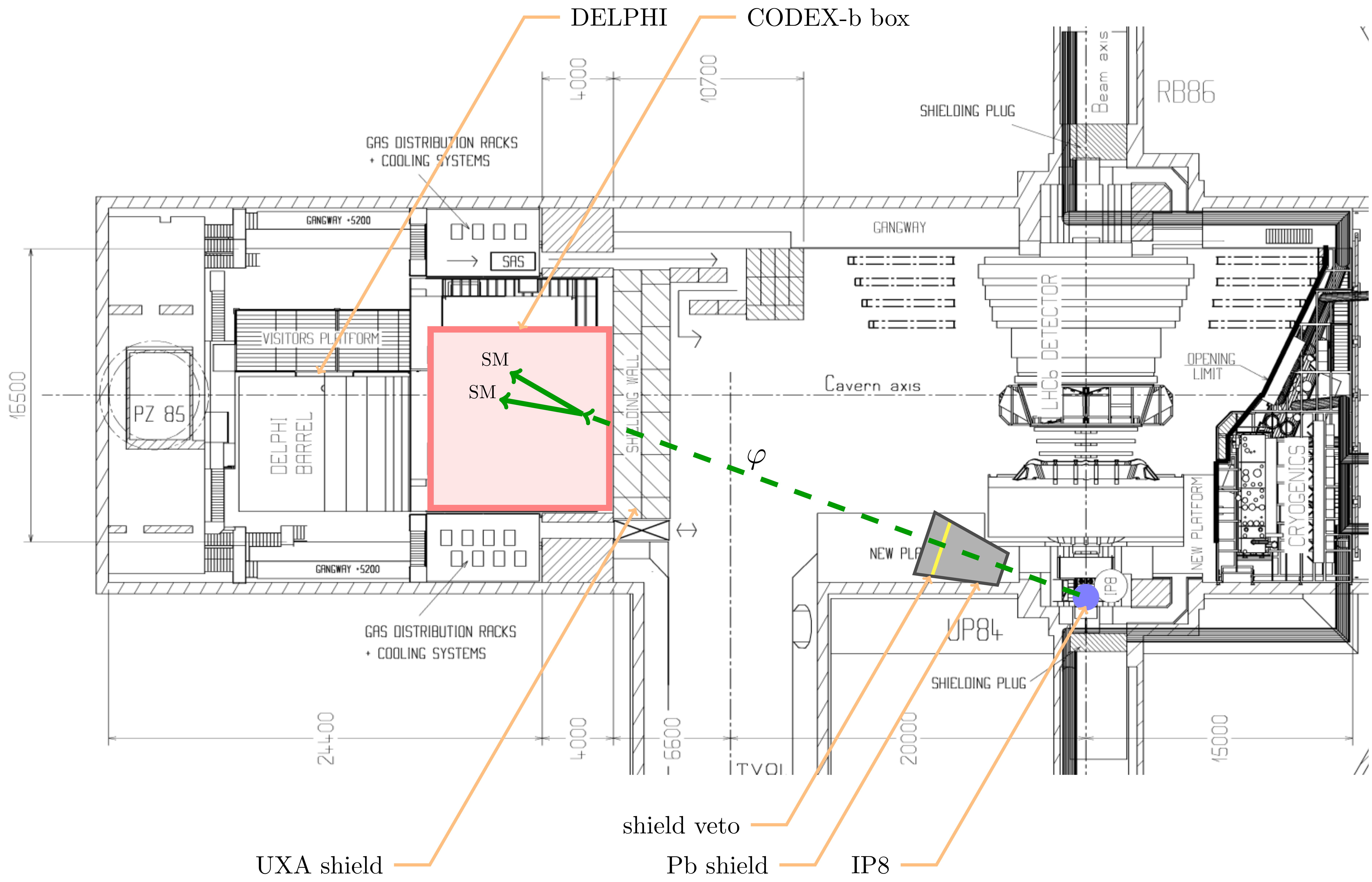}
	\includegraphics[width =  0.49\linewidth]{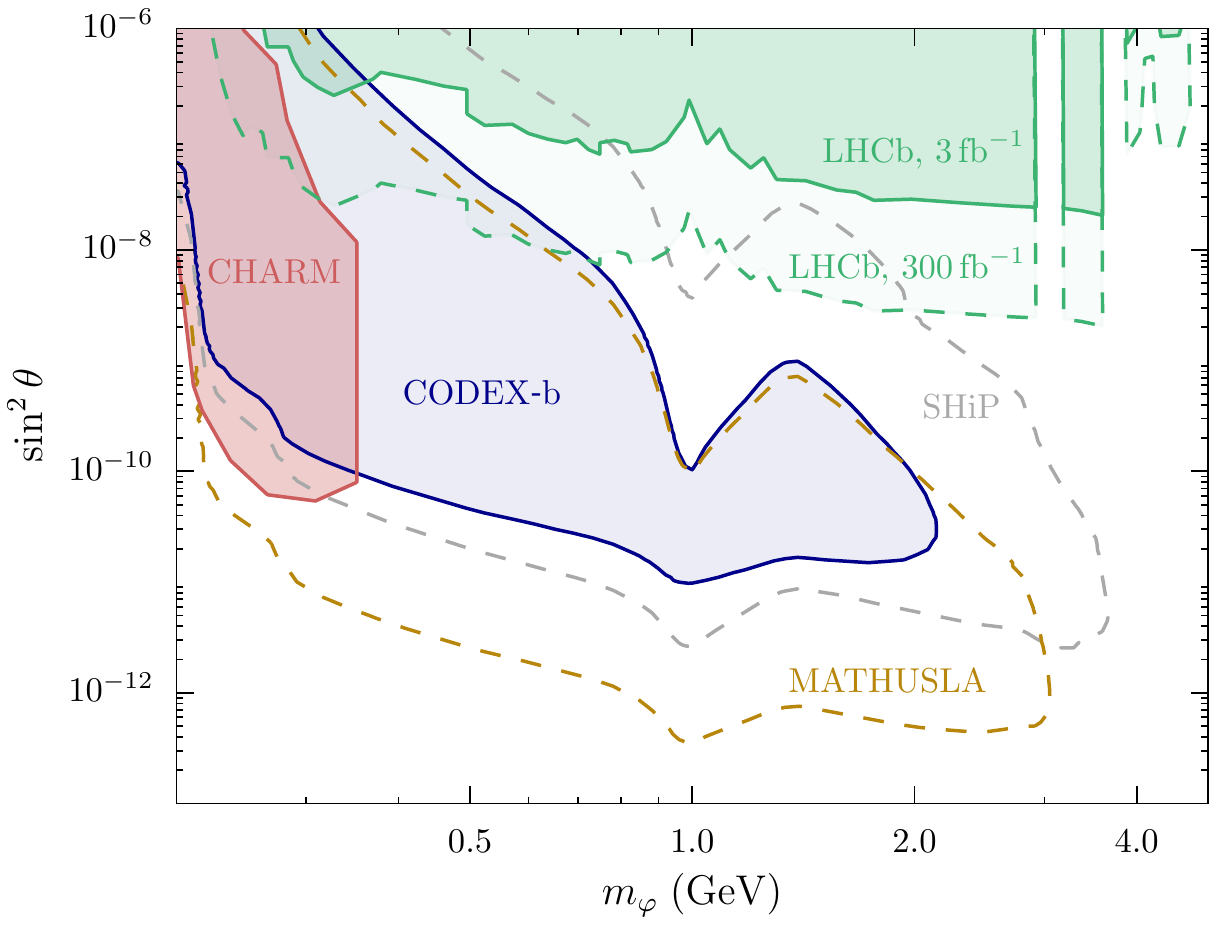}
	\caption{(Left) Layout of the LHCb experimental cavern UX85 at point 8 of the LHC, overlaid with the CODEX-b apparatus. 
Reproduced from Ref.~\cite{Gligorov:2017nwh}. 
(Right) CODEX-b reach for $B\to \phi X_s$ in the $s^2_\theta$--$m_\phi$ plane, assuming $100\%$ tracking efficiency and 300~fb$^{-1}$.} 
	\label{fig:LHCbCavReach}
\end{figure*}

The CODEX-b geometry and the reach of CODEX-b for one particular class of LLPs produced
in the $B\to \phi X_s$ decay (where $\phi$ is the LLP) is shown in Fig.~\ref{fig:LHCbCavReach}. 
The reuse of the DELPHI cavern, following the move of LHCb's DAQ to the surface, gives access to an
already shielded decay volume with readily available infrastructure for powering and reading out the CODEX-b detector. Additional shielding is placed
between LHCb and CODEX-b in order to reduce the flux of backgrounds, in particular neutrons and \Kz mesons; an instrumented shield layer is used
to veto secondary production of neutral particles inside the shield itself. GEANT-based simulations have been used to validate that this design should reduce
backgrounds to a negligible level, and data-driven measurements of background levels in the DELPHI cavern are underway to validate this simulation.

In order to maintain a high efficiency across the full range of LLP masses
and decay modes, all six sides of the decay volume are instrumented with tracking layers, with the sides closest to LHCb also serving as vetoes
against residual charged backgrounds such as muons penetrating into the decay volume. The reach shown in Fig.~\ref{fig:LHCbCavReach} is representative of the typical CODEX-b reach, 
compared to other experiments, over a broad range of other LLP models. As can be seen, CODEX-b
significantly extends LHCb's existing reach and gives access to longer LLP lifetimes, making LHCb's overall reach very competitive with any other experiment.
Due to its low-rapidity coverage, CODEX-b would also extend the reach of the LHC programme for LLPs produced through high mass portals, like exotic Higgs decays.

\begin{figure}[t]
	\includegraphics[width = 0.49\linewidth]{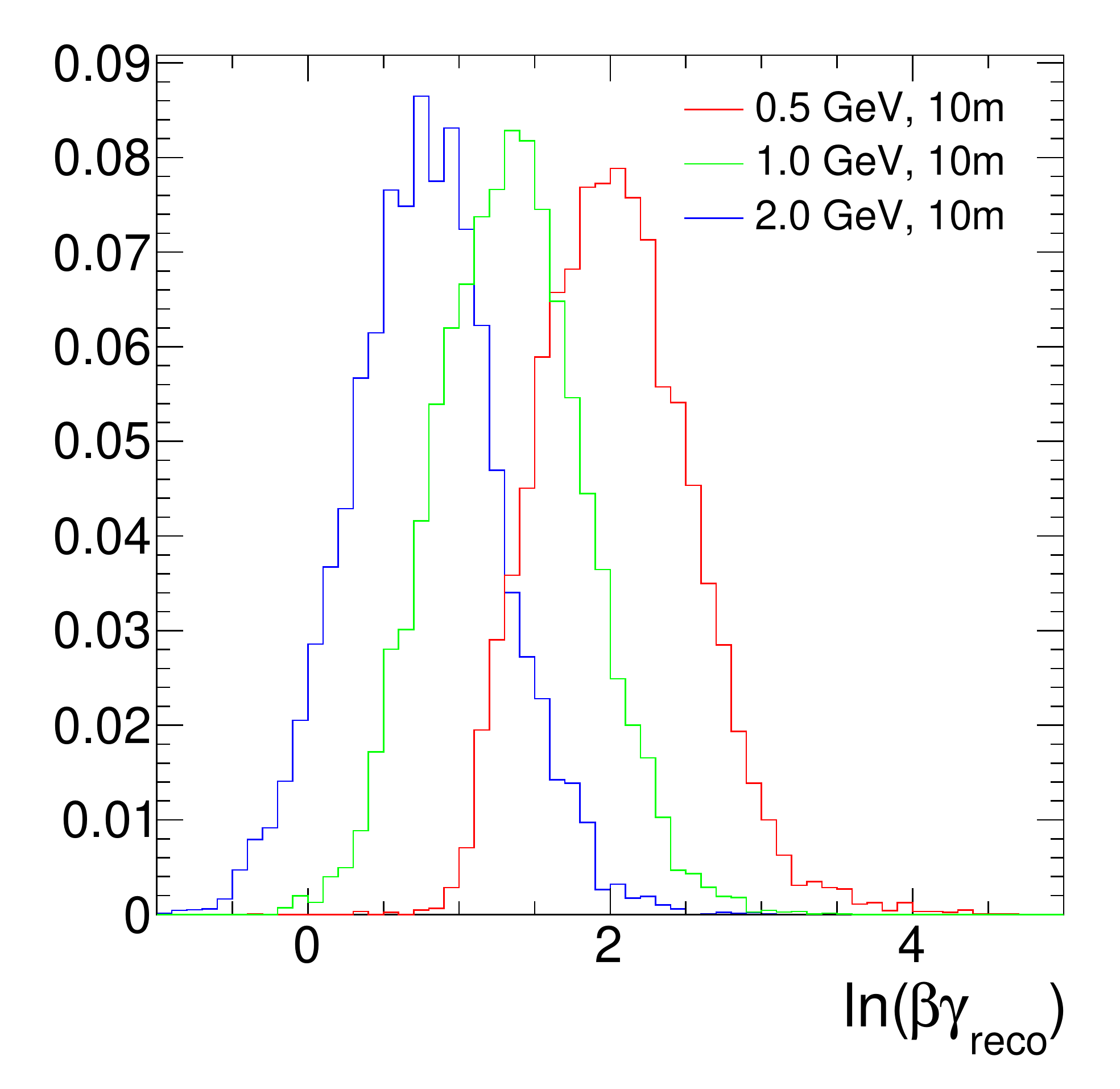}
	\includegraphics[width = 0.49\linewidth]{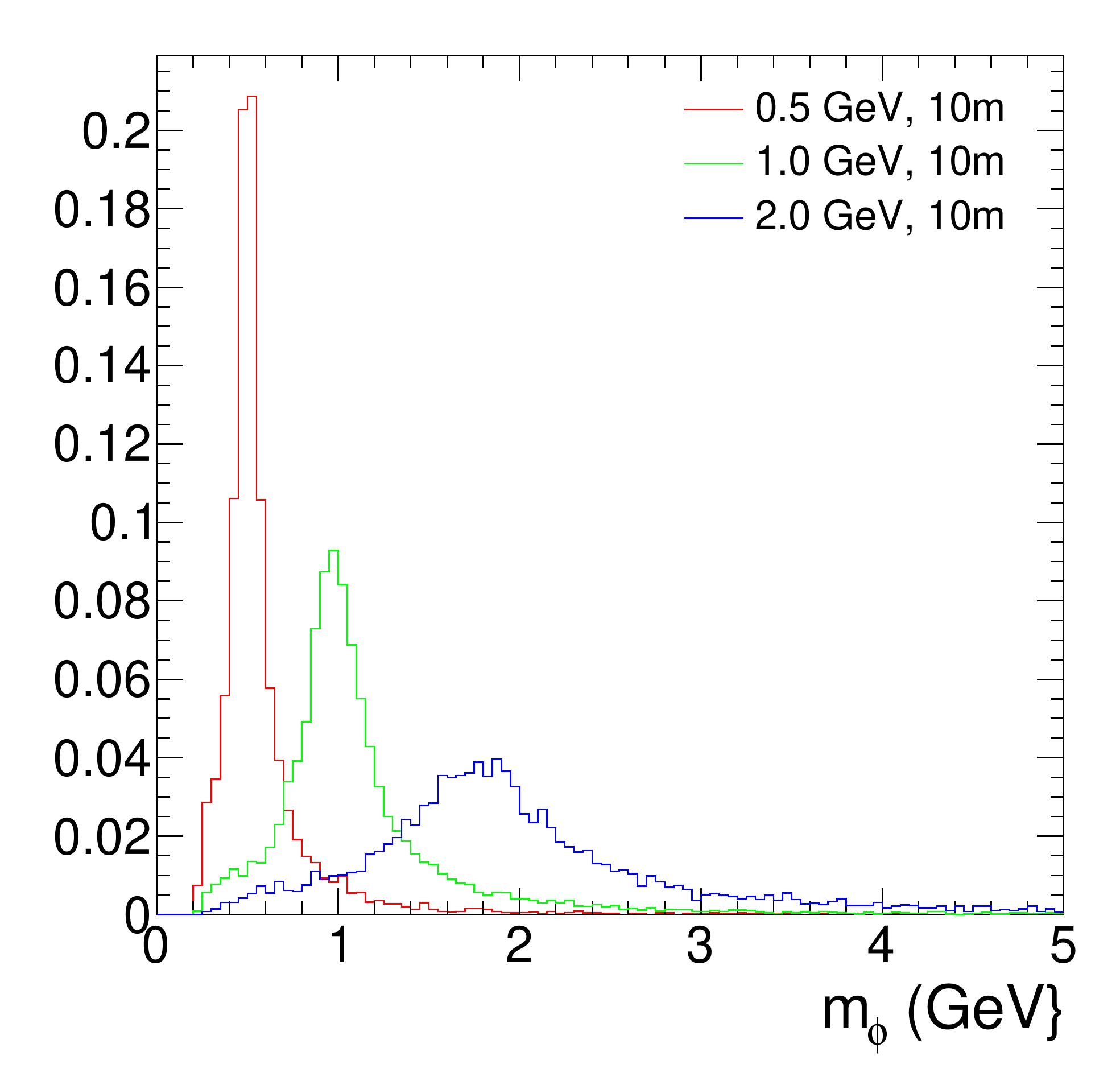}
	\caption{Reconstructed LLP (left) boost and (right) mass for different LLP masses, for $c\tau =10$\,m.
Reproduced from Ref.~\cite{Gligorov:2017nwh}.} 
	\label{fig:recoboosts}
\end{figure}

A particular benefit of CODEX-b's compact size and low-radiation operating environment is the possibility to cheaply instrument the decay volume with
more sophisticated detector technologies. The most basic CODEX-b design adds tracking stations inside the decay volume itself, improving the resolution
on the reconstructed LLP decay vertex and enabling the LLP boost to be measured. A more sophisticated detector design would include timing layers in the
tracker, enabling the LLP momentum and therefore mass to be reconstructed using time-of-flight information. Potentially achievable resolutions are shown
in Fig.~\ref{fig:recoboosts}, assuming 50~ps tracker timing resolution. It may even be desirable to fully instrument the CODEX-b detector volume as a tracking calorimeter,
for example using the NOVA experiment modules. This would further extend the reach of the experiment to light axion-like particles decaying two photons. 

%% file: CONTRIBUTIONS/11_Summary_and_conclusions/wilson.tex
\chapter{Fits to Wilson coefficients and determination of New Physics reach}
\label{sec:wilson}

To determine the exclusion limits for New Physics presented in Fig.~\ref{fig:wilson_sum}
the expected uncertainties for the angular observables in the decay $\decay{\Bd}{\Kstarz\mumu}$, and the uncertainties on $R_K$ and $R_{K^*}$
are translated into uncertainties on single Wilson coefficients using the {\it flavio} software~\cite{FLAVIO}. 
The uncertainties on the Wilson coefficients (here ${C}_9$ and ${C}_{10}^{\prime}$)
are then used to determine lower limits on the New Physics scale $\Lambda_\textrm{NP}$ at the $90\%$ confidence level. 

The study is performed by using the effective Hamiltonian 
\begin{align}
{\cal H}_\textrm{eff} &= -\frac{4 G_\textrm{F}}{\sqrt{2}} V_{tb}V_{ts}^* \frac{e^2}{16\pi^2} \sum_i {C}_i {O}_i + \Delta {\cal H}_\textrm{NP}\\
\Delta{\cal H}_\textrm{NP} &= \frac{\kappa}{\Lambda_\textrm{NP}^2}{O}_i,
\end{align}
where $C_i$ are the Wilson coefficients, ${O}_i$ are the local operators,
$\Lambda_\textrm{NP}$ the New Physics scale, and  $\kappa$ the flavour violating coupling. 
There are several common choices for $\kappa$ depending on the suppression of New Physics. 
For generic tree-level New Physics we have $\kappa=1$, a tree-level Minimum Flavour Violating (MFV) coupling gives $\kappa=V_{tb}V_{ts}^*$,
loop-level generic coupling results in $\kappa=1/(16\pi^2)$, and loop-level MFV yields $\kappa=V_{tb}V_{ts}^*/(16\pi^2)$. 
The corresponding New Physics scale (assuming New Physics contributes to the Wilson coefficients ${C}_{9,10}^{(\prime)}$) is then given by
\begin{align}
  \frac{4 G_\textrm{F}}{\sqrt{2}} V_{tb}V_{ts}^*\sigma({C}_{9,10}^{(\prime)}) \frac{\alpha}{4\pi} &= \frac{\kappa}{\Lambda_\textrm{NP}^2}\\
  \to \Lambda_\textrm{NP} &= \sqrt{\kappa\frac{\sqrt{2}}{4 G_\textrm{F}}\frac{4\pi}{\alpha V_{tb}V_{ts}^* \sigma({C}_{9,10}^{(\prime)})}},
\end{align}
depending on $\kappa$ and the uncertainty on the Wilson coefficients $\sigma({C}_{9,10}^{(\prime)})$.
The 90\% exclusion limits (double sided) are given by 
\begin{align}
  \Lambda_\textrm{NP}^{90\%} &= \sqrt{\kappa\frac{\sqrt{2}}{4 G_\textrm{F}}\frac{4\pi}{\alpha V_{tb}V_{ts}^* \times 1.64485\times \sigma({C}_{9,10}^{(\prime)})}}
\end{align}

When determining the uncertainty on a Wilson coefficient, the \textit{simple fit} method in flavio is used and only one Wilson coefficient is varied.
The $\decay{\Bd}{\Kstarz\mumu}$ form-factor uncertainties are assumed to reduce by a factor $0.25$ by the Upgrade-II era. 
The reduction in uncertainty of the theory nuisance parameters is assumed to scale linearly with integrated luminosity. 
Other theory uncertainties, for example those concerned with charm loops, are conservatively assumed to stay at the current level. 
All studies assume Standard Model central values for the experimental observables.   